\documentclass[%
reprint,
superscriptaddress,
bibnotes,longbibliography
amsmath,amssymb,
aps,
floatfix
]{revtex4-2}
\usepackage[dvipdfmx]{graphicx}
\usepackage{graphicx}
\usepackage{color}
\usepackage{natbib}
\usepackage{amsmath,amssymb,amsthm,mathrsfs,amsfonts,dsfont}
\usepackage{subfigure, epsfig}
\usepackage{braket}
\usepackage{bm}
\usepackage{bbm}
\usepackage{enumerate}
\usepackage{here}
\usepackage{comment}
\usepackage{appendix} 
\usepackage[colorlinks,linkcolor=blue,citecolor=blue]{hyperref}
\newcommand{\tr}{\mathrm{tr}}

\usepackage{makecell} 
\usepackage{booktabs} 

\newtheorem{theorem}{Theorem}

\newtheorem{lemma}{Lemma}

\theoremstyle{definition}
\newtheorem{dfn}{Definition}
\theoremstyle{remark}

\theoremstyle{example}

\newcommand{\Exnu}{\mathbb{E}_{U \sim\nu}}

\newcommand{\polylog}{\mathrm{polylog}}

\usepackage[utf8]{inputenc}
\usepackage{array}       
\usepackage{cellspace}   
\usepackage{graphicx}    
\usepackage{xcolor}
\usepackage{multirow}

\newcolumntype{C}[1]{>{\centering\arraybackslash}m{#1}}
\addparagraphcolumntypes{C}   

\allowdisplaybreaks[4]

\begin{document}
\title{Optimal Scaling of Unitary Design Formation in \(U(1)\)-Symmetric Random Circuits: \\ A Bottleneck Slower than Charge Transport}

\author{Toshihiro Yada}
\email{toshihiroyada.physics@gmail.com}
\affiliation{Analytical Quantum Complexity RIKEN Hakubi Research Team, RIKEN Center for Quantum Computing (RQC), Wako, Saitama 351-0198, Japan}


\begin{abstract}
Understanding randomness formation under conservation laws is important both for characterizing chaotic phenomena in isolated quantum systems and for symmetry-constrained quantum information processing.
$U(1)$-symmetric random circuits on an $N$-qubit system provide a minimal tractable setting for this problem, in which the formation rate of unitary designs was conjectured to be governed by charge transport in the underlying circuit geometry.
However, whether charge transport indeed sets the slowest relaxation mode has remained an open problem, due to the difficulty in deriving a tight lower bound on the symmetric design formation rate.
In this work, we resolve this problem by determining the optimal scaling of the unitary $k$-design formation rate for $k\leq O(\sqrt{\log N})$ across a wide range of circuit geometries, including lattices in general spatial dimension, expander graphs, and all-to-all interactions.
In particular, for $k\geq 2$, we find that the formation rate is governed by a two-particle encounter rate---the rate at which two random walkers encounter each other in a given circuit geometry.
This encounter rate is parametrically smaller than the single-particle transport rate in many geometries, thereby disproving the previous conjecture.
We further establish that this encounter mechanism indeed sets the design formation rate by deriving its matching lower bound with a novel proof technique based on an auxiliary subgroup ensemble.
Although this encounter mode imposes a robust bottleneck on design formation in $U(1)$-symmetric random circuits, we also show that this bottleneck can be circumvented by using symmetry-breaking local gates at intermediate stages and provide an explicit protocol that achieves faster design formation.
These results establish a new picture of randomness formation under conservation laws and provide guiding principles for characterizing and efficiently generating symmetry-constrained randomness.
\end{abstract}
\maketitle

\section{Introduction} \label{s:introduction}

\subsection{Background}

Random quantum circuits have played a significant role both as tractable models of chaotic quantum many-body dynamics~\cite{fisher2023random,nahum2017quantum,nahum2018operator,von2018operator,brandao2021models,haferkamp2022linear,oszmaniec2022saturation} and as useful tools in quantum information processing, including randomized benchmarking~\cite{helsen2022general}, quantum tomography~\cite{elben2023randomized,huang2020predicting}, and random circuit sampling~\cite{bouland2019complexity,arute2019quantum,zhu2022quantum}.
A central question in this model is how rapidly the statistical properties of such a circuit approach those of the global Haar-random unitary ensemble.
This convergence is naturally characterized through the moments of a unitary ensemble: an ensemble is called an approximate unitary $k$-design if its $k$-th moment is sufficiently close to the corresponding moment of the Haar measure.
For random circuits without constraints such as conservation laws or symmetries, recent studies have led to a detailed understanding of the unitary design formation rate~\cite{harrow2009random,brown2010convergence,brandao2016local,haferkamp2022random,chen2024incompressibility,haferkamp2021improved,hunter2019unitary,harrow2023approximate,mittal2023local,belkin2023approximate,schuster2024random,laracuente2024approximate,yada2025non,baer2026random}.

\begin{figure}[t!]
\begin{center}
\includegraphics[width=0.44\textwidth]{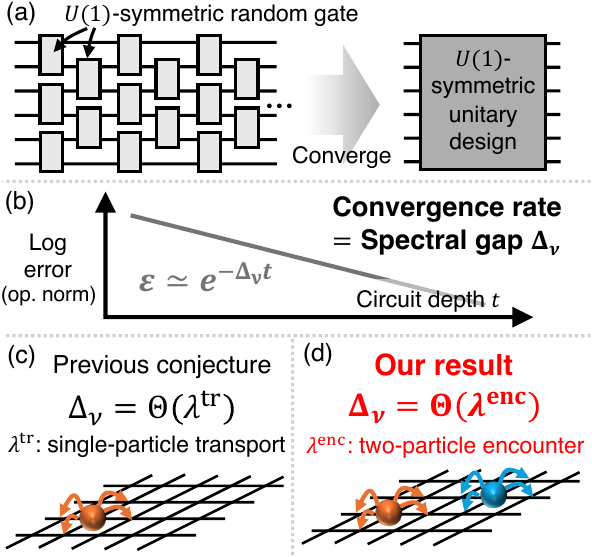}
\caption{
Background and main result.
(a) The low-order moments of a $U(1)$-symmetric random circuit converge to those of the $U(1)$-symmetric Haar-random unitary ensemble in the infinite-depth limit~\cite{mitsuhashi2024unitary,mitsuhashi2024characterization,liu2024unitary}.
(b) The convergence rate is characterized by the spectral gap $\Delta_{\nu}$, which controls the exponential decay with circuit depth $t$ of the operator-norm distance between the circuit moment and its infinite-depth limit (Lemma~\ref{lem:spectral_gap_to_design_formation}).
(c) Previous work~\cite{hearth2025unitary} conjectured that $\Delta_{\nu}$ is governed by the single-particle transport rate $\lambda^{\rm tr}$.
(d) This work shows that, for the second and higher moments, $\Delta_{\nu}$ is instead governed by the two-particle encounter rate $\lambda^{\rm enc}$, which is parametrically smaller than $\lambda^{\rm tr}$ in many circuit geometries.
}
\label{fig:spectral_gap}
\end{center}
\end{figure}

However, quantum systems of physical interest are typically subject to nontrivial constraints, most notably conservation laws, that are absent from such unconstrained random circuits.
These constraints are known to qualitatively modify dynamical phenomena such as operator spreading~\cite{chen2017otoc,bohrdt2017scrambling,luitz2017information} and R\'enyi entanglement entropy growth~\cite{rakovszky2019sub,zhou2020diffusive}, motivating the need for models that explicitly capture their effects.
For this purpose, symmetric random circuits, in which each local gate is randomly chosen to respect symmetry, offer a particularly useful framework.
These models are known to capture key qualitative effects of conservation laws~\cite{khemani2018operator,rakovszky2018diffusive,rakovszky2019sub,zhou2020diffusive,huang2019dynamics,huang2020dynamics}, while retaining the analytical tractability afforded by their explicit randomness.
Symmetry-preserving random unitaries also find diverse applications in quantum information processing, including classical-shadow tomography~\cite{hearth2024efficient,bringewatt2024randomized,bringewatt2026classical}, covariant quantum error correction~\cite{kong2022near}, and quantum machine learning~\cite{li2025d}.

For symmetric random circuits, the infinite-depth limit of the moments is by now well understood~\cite{marvian2022restrictions,marvian2024theory,marvian2024rotationally,hulse2024framework}.
For instance, the low-order moments of $U(1)$-symmetric random circuits are known to converge to those of the global $U(1)$-symmetric Haar-random unitary ensemble~\cite{mitsuhashi2024unitary,mitsuhashi2024characterization,liu2024unitary}, as illustrated in Fig.~\ref{fig:spectral_gap}(a).
On the other hand, how fast this convergence occurs, or equivalently, how rapidly $U(1)$-symmetric unitary designs are formed, remains much less understood, despite recent progress~\cite{hearth2025unitary,li2024efficient}.
This convergence rate is characterized by a spectral gap, which determines the slowest relaxation rate of the moments, as shown in Fig.~\ref{fig:spectral_gap}(b).
In a seminal work~\cite{hearth2025unitary}, an upper bound on this gap was obtained from a slow mode associated with the diffusive transport of the conserved $U(1)$ charge.
The authors further conjectured that this single-particle transport mode indeed determines the gap up to constant factors [Fig.~\ref{fig:spectral_gap}(c)].
However, establishing a matching lower bound for the spectral gap remained an open problem, since the proof techniques developed for unconstrained random circuits~\cite{brandao2016local,haferkamp2022random,chen2024incompressibility} encounter serious obstacles in the presence of symmetry constraints~\cite{hearth2025unitary,li2024designs}.

\subsection{Summary of our main results}

In this work, we determine the optimal scaling of the formation rate of $U(1)$-symmetric unitary $k$-designs for $k\leq O(\sqrt{\log N})$ across a broad range of circuit geometries, as summarized in Table~\ref{tab:gap_summary}, by establishing matching upper and lower bounds on the spectral gap up to constant factors.
Contrary to the previous conjecture, we show that the $k$-design formation rate is not generally controlled by single-particle transport for $k\geq2$, but instead by a slower mode with a simple physical interpretation.
This slow mode corresponds to a \emph{two-particle encounter process}, in which two random walkers move on the underlying circuit geometry until they encounter each other, as illustrated in Fig.~\ref{fig:spectral_gap}(d).
The slow relaxation due to this encounter mechanism can be observed in the dynamics of standard physical quantities such as subsystem purity and entanglement entropy.
We further show that this bottleneck persists for random circuits with any constant gate size $\ell=O(1)$, demonstrating the robustness of this encounter mechanism in $U(1)$-symmetric random circuits.

We further prove that this two-particle encounter mode indeed sets the design formation rate for $2\leq k\leq O(\sqrt{\log N})$ by establishing a matching lower bound on the spectral gap up to a constant factor.
Our proof develops a new framework for evaluating randomness formation rates by comparing the moment operators of local random circuits with suitably chosen auxiliary subgroup ensembles.
The resulting components can be analyzed through tractable structures, such as classical stochastic and substochastic dynamics, allowing us to rule out any mode that relaxes parametrically more slowly than the two-particle encounter process.
Together, these results establish a scaling-optimal characterization of randomness formation under $U(1)$ symmetry and revise the previous picture based solely on single-particle transport.

Having established this robust bottleneck for circuits composed of $U(1)$-symmetric gates, we further show that it can be circumvented by using symmetry-breaking local gates at intermediate stages.
Specifically, we propose an explicit protocol to construct a circuit unit that has a constant spectral gap, with a smaller depth than the symmetric random circuit in the same geometry. 
For moment order $2\leq k\leq O(\log N/\log\log N)$, our protocol requires a smaller circuit depth than the corresponding symmetric random circuit does, on lattices in general spatial dimension and in the all-to-all geometry.
In particular, on lattices, this depth is parametrically smaller even than the single-particle transport timescale.
This result demonstrates that locally breaking symmetry can enable substantially more efficient generation of global symmetry-preserving randomness.

\begin{table}[t]
\centering
\footnotesize
\setlength{\tabcolsep}{3.0pt}
\renewcommand{\arraystretch}{1.12}

\begin{tabular}{c|c|c|c}
\hline
\multicolumn{2}{c|}{Circuit structures}
& Previous conjecture~\cite{hearth2025unitary}
& Our result
\\
\hline\hline

\multirow{5}{*}{\shortstack{Single-\\edge}}
    & 1D
    & $\Theta(N^{-3})$
    & $\Theta(N^{-3})$
    \\
    & 2D
    & $\Theta(N^{-2})$
    & {\color{red}$\Theta\!\left(N^{-2}(\log N)^{-1}\right)$}
    \\
    & $\alpha$D $(\alpha\geq3)$
    & $\Theta(N^{-1-2/\alpha})$
    & {\color{red}$\Theta(N^{-2})$}
    \\
    & Expander
    & ---
    & {\color{red}$\Theta(N^{-2})$}
    \\
    & All-to-all
    & ---
    & {\color{red}$\Theta(N^{-2})$}
    \\
\hline\hline

\multirow{5}{*}{Parallel}
    & 1D
    & $\Theta(N^{-2})$
    & $\Theta(N^{-2})$
    \\
    & 2D
    & $\Theta(N^{-1})$
    & {\color{red}$\Theta\!\left(N^{-1}(\log N)^{-1}\right)$}
    \\
    & $\alpha$D $(\alpha\geq3)$
    & $\Theta(N^{-2/\alpha})$
    & {\color{red}$\Theta(N^{-1})$}
    \\
    & Expander
    & ---
    & {\color{red}$\Theta(N^{-1})$}
    \\
    & All-to-all
    & ---
    & {\color{red}$\Theta(N^{-1})$}
    \\
\hline\hline

\multirow{4}{*}{\shortstack{Fixed-\\arch.}}
    & 1D
    & $\Theta(N^{-2})$
    & $\Theta(N^{-2})$
    \\
    & 2D
    & $\Theta(N^{-1})$
    & {\color{red}$\Theta\!\left(N^{-1}(\log N)^{-1}\right)$}
    \\
    & $\alpha$D $(\alpha\geq3)$
    & $\Theta(N^{-2/\alpha})$
    & {\color{red}$\Theta(N^{-1})$}
    \\
    & Expander
    & ---
    & {\color{red}$\Theta(N^{-1})$}
    \\
\hline

\end{tabular}

\caption{
Spectral gap scaling of $N$-qubit $U(1)$-symmetric random circuits for moment order $2\leq k\leq O(\sqrt{\log N})$.
Previous work~\cite{hearth2025unitary} conjectured that the scaling of the spectral gap is governed by the single-particle transport rate, whereas our results show that it is governed by the two-particle encounter rate.
Our results are highlighted in red for geometries in which the two-particle encounter rate is parametrically smaller than the single-particle transport rate.
Here, $\alpha$D denotes an $\alpha$-dimensional lattice, the expander entries represent regular bipartite and vertex-transitive expanders, and all-to-all denotes the all-to-all interaction geometry.
We also note that for $k=1$, $\Delta_{\nu_G}^{(1)}=\lambda_G^{\rm tr}$ holds exactly for any connected graph $G$, so the spectral gap is governed by the single-particle transport mode in this case.
}
\label{tab:gap_summary}

\end{table}

These results have rich implications for both quantum information science and many-body physics.
The formation rate of low-order $U(1)$-symmetric unitary designs determined in this work quantifies the circuit depth required for applications such as symmetric classical-shadow tomography \cite{hearth2024efficient,hearth2025unitary} and covariant quantum error-correcting codes \cite{kong2022near}.
Our proof framework also opens a route to determining formation rates of constrained randomness under other symmetries, such as $SU(2)$ or more general $SU(d)$ symmetries.
From the perspective of many-body physics, our results reveal an unexpected and robust slow relaxation mechanism associated with the two-particle encounter process in symmetric random circuits.
Given the close connection between symmetric random circuits and generic isolated quantum systems under conservation laws~\cite{khemani2018operator,rakovszky2018diffusive,rakovszky2019sub,zhou2020diffusive,huang2019dynamics,huang2020dynamics}, our results suggest that analogous slow phenomena may also emerge more broadly in quantum many-body dynamics.

We conclude this introduction by outlining the structure of the paper.
In Sec.~\ref{s:preliminary}, we introduce the basic concepts needed in this work, including the moment operator and the spectral gap, and describe the random circuit architectures considered throughout the paper.
In Sec.~\ref{s:gap_upper_bound}, we derive our upper bound on the spectral gap and explain its connection to the two-particle encounter process.
In Sec.~\ref{s:gap_lower_bound}, we present a matching lower bound on the spectral gap and provide an overview of its proof.
In Sec.~\ref{s:efficient_construction}, we present an efficient protocol for generating $U(1)$-symmetric unitary designs by allowing symmetry-breaking gates at intermediate stages.
In Sec.~\ref{s:add_rel_error_design}, we derive bounds on the circuit depths required to form approximate unitary designs under $U(1)$ symmetry, as direct consequences of the spectral gap bounds obtained in Secs.~\ref{s:gap_upper_bound}, \ref{s:gap_lower_bound}, and \ref{s:efficient_construction}.
In Sec.~\ref{s:demonstration_slow_decay}, we demonstrate the slow relaxation of subsystem purity and entanglement entropy in all-to-all random circuits and provide an intuitive physical picture explaining why these quantities do not fully relax on the charge-transport timescale.
Finally, in Sec.~\ref{s:summary_outlook}, we summarize our results and discuss future research directions.

\section{Preliminary} \label{s:preliminary}

\subsection{Moment operators of $U(1)$-symmetric unitary ensembles} 

In this work, we consider $N$-qubit unitary operations with $U(1)$ symmetry, or equivalently, unitary operations that conserve the total particle number.
The Hilbert space is decomposed into particle-number sectors as $\mathcal H=\bigoplus_{n=0}^{N}\mathcal H_n$, where $\mathcal H_n$ denotes the subspace with particle number $n$.
Accordingly, any $U(1)$-symmetric unitary can be written in the block-diagonal form $U=\bigoplus_{n=0}^{N}U_n$, where $U_n$ is a unitary operator on $\mathcal H_n$.
We denote the group of such unitaries by $\mathcal U_{U(1)}$, and the projector onto $\mathcal H_n$ by $\Pi_n$.

For an ensemble $\nu$ of $U(1)$-symmetric unitaries, its $k$-th moment is characterized by the $k$-th moment operator
$M_{\nu}^{(k)} \equiv \Exnu [U^{\otimes k,k}]$, where $\Exnu$ denotes the average over $U\sim\nu$ and $U^{\otimes k,k} \equiv U^{\otimes k} \otimes U^{* \otimes k}$, with $U^{*}$ denoting the complex conjugate of $U$ in the computational basis.
This operator acts on the $2k$-copy moment space $\mathcal H^{\otimes k} \otimes \overline{\mathcal H}^{\otimes k}$, where $\overline{\mathcal H}$ denotes the complex-conjugate Hilbert space.
A computational-basis state in this space is specified by $2kN$ bits.
For each physical site $i\in[N]$, we collect the occupations in the $k$ ket and $k$ bra copies as $x_i=(x_i^1,\ldots,x_i^k)$ and $y_i=(y_i^1,\ldots,y_i^k)$, respectively, and define $u_i\equiv(x_i,y_i)\in\{0,1\}^{2k}$.
We call $u_i$ the \emph{local type} at site $i$, and denote a computational-basis state of the moment space by $\ket{u_1,\ldots,u_N}$.

Since every unitary in $\nu$ preserves the particle number separately in each ket and bra copy, the moment operator is block diagonal with respect to the corresponding charge sectors.
We label each charge sector by $(\bm n,\overline{\bm n})$, where $\bm n=(n_1,\ldots,n_k)^T$ and $\overline{\bm n}=(n_{\bar 1},\ldots,n_{\bar k})^T$, with $n_\alpha$ and $n_{\bar\alpha}$ denoting the particle numbers in the $\alpha$-th ket and bra copies, respectively.
Introducing the projector $\Pi_{\bm n,\overline{\bm n}} \equiv \left(\bigotimes_{\alpha=1}^{k} \Pi_{n_\alpha}\right) \otimes \left(\bigotimes_{\alpha=1}^{k}\Pi_{n_{\bar\alpha}} \right)$, we have $[M_{\nu}^{(k)},\Pi_{\bm n,\overline{\bm n}}]=0$, and hence
\begin{equation}
    M_{\nu}^{(k)} = \sum_{\bm n,\overline{\bm n}} M_{\nu;\bm n,\overline{\bm n}}^{(k)},\qquad
    M_{\nu;\bm n,\overline{\bm n}}^{(k)} \equiv \Pi_{\bm n,\overline{\bm n}}M_{\nu}^{(k)}\Pi_{\bm n,\overline{\bm n}}.
    \label{eq:moment_op_general_U(1)_symmetric_unitary}
\end{equation}
Here, $M_{\nu;\bm n,\overline{\bm n}}^{(k)}$ denotes the moment operator restricted to the subspace $\mathcal H_{\bm n,\overline{\bm n}} \equiv \left(\bigotimes_{\alpha=1}^{k} \mathcal H_{n_\alpha} \right)\otimes \left(\bigotimes_{\alpha=1}^{k}\overline{\mathcal H}_{n_{\bar\alpha}}\right)$.

We next consider the $k$-th moment operator of the Haar measure on $\mathcal U_{U(1)}$, denoted by $P^{(k)}_{U(1)\mathrm{Haar}}$.
The Haar average of a unitary representation of a compact group is the orthogonal projector onto its invariant subspace, and we therefore use $P$ rather than $M$ for this moment operator.
It is similarly decomposed into charge sectors as
\begin{equation}
\begin{split}
    &P^{(k)}_{U(1)\mathrm{Haar}} = \sum_{\bm n,\overline{\bm n}} P^{(k)}_{U(1)\mathrm{Haar};\bm n,\overline{\bm n}},\\
    &P^{(k)}_{U(1)\mathrm{Haar};\bm n,\overline{\bm n}}
    \equiv
    \Pi_{\bm n,\overline{\bm n}}
    P^{(k)}_{U(1)\mathrm{Haar}}
    \Pi_{\bm n,\overline{\bm n}}.
\end{split}
\label{eq:moment_op_U(1)_symmetric_Haar_random_unitary}
\end{equation}
The charge-sector block $P^{(k)}_{U(1)\mathrm{Haar};\bm n,\overline{\bm n}}$ is nonzero if and only if the ket and bra charge vectors coincide up to a permutation, namely, if there exists $\sigma\in S_k$ such that $\overline{\bm n}=\sigma\bm n$ \cite{hearth2025unitary}.
Here, $S_k$ denotes the symmetric group on $k$ elements, and the action of $\sigma$ on a charge vector is defined by $(\sigma\bm n)_i\equiv n_{\sigma^{-1}(i)}$.

\subsection{$U(1)$-symmetric unitary design} \label{ss:approx_unitary_design}

The main goal of this work is to reveal the rate at which the $k$-th moment of a $U(1)$-symmetric random circuit approaches that of the $U(1)$-symmetric Haar-random unitary ensemble.
Denoting the unitary ensemble corresponding to each circuit unit by $\nu$, the essential quantity characterizing this convergence rate is the $k$-th spectral gap, defined as
\begin{equation}
\label{eq:spectral_gap_def}
    \Delta_{\nu}^{(k)} \equiv 1- \left\| M_{\nu}^{(k)} - P^{(k)}_{U(1) \mathrm{Haar}} \right\|_{\infty}.
\end{equation}
This quantity satisfies $0\leq \Delta_{\nu}^{(k)}\leq 1$, with a larger spectral gap corresponding to faster convergence to the Haar measure under repeated circuit iterations.
Since both $M_{\nu}^{(k)}$ and $P^{(k)}_{U(1)\mathrm{Haar}}$ are block diagonal with respect to the charge sectors, we can equivalently describe it as
\begin{equation}
\begin{split}\label{eq:spectral_gap_block_diagonal_def}
    &\Delta_{\nu}^{(k)} = \min_{\bm n, \overline{\bm n}} \Delta_{\nu;\bm n, \overline{\bm n}}^{(k)} ,\\
    &\Delta_{\nu;\bm n, \overline{\bm n}}^{(k)} \equiv 1- \left\| M_{\nu;\bm n, \overline{\bm n}}^{(k)} - P^{(k)}_{U(1) \mathrm{Haar};\bm n, \overline{\bm n}} \right\|_{\infty},
\end{split}
\end{equation}
where $\Delta_{\nu;\bm n, \overline{\bm n}}^{(k)}$ is the spectral gap restricted to a fixed charge sector $(\bm n, \overline{\bm n})$.

Using the spectral gap, we can characterize the convergence rate to the Haar measure as follows:
\vspace{-1.5mm}
\begin{lemma} \label{lem:spectral_gap_to_design_formation}
For any ensemble $\nu$ of $U(1)$-symmetric unitaries, denote its $t$-fold convolution ($t$ iterations) by $\nu^{* t}$.
Then, we have
\begin{equation}\label{eq:spectral_gap_to_design_formation_general}
    \left\|M_{\nu^{*t}}^{(k)} -P_{U(1)\rm Haar}^{(k)} \right\|_\infty \leq \exp\left(-\Delta_{\nu}^{(k)} t\right).
\end{equation}
Furthermore, when the moment operator $M_{\nu}^{(k)}$ is Hermitian, we have
\begin{equation}
    \label{eq:spectral_gap_to_design_formation_Hermitian}
    \left\|M_{\nu^{*t}}^{(k)} -P_{U(1)\rm Haar}^{(k)} \right\|_\infty = \left(1- \Delta_{\nu}^{(k)} \right)^t.
\end{equation}
\end{lemma}
\vspace{-1.5mm}
\noindent
The Hermiticity condition on the moment operator is satisfied for various types of random circuits considered later.

This relation has been widely used in previous works~\cite{brandao2016local,haferkamp2022random,chen2024incompressibility,haferkamp2021improved,mittal2023local,belkin2023approximate,hearth2025unitary} to quantify unitary design formation rates.
The derivation of Eq.~\eqref{eq:spectral_gap_to_design_formation_general} proceeds as follows:
\begin{align}
    \left\|M_{\nu^{*t}}^{(k)} -P_{U(1)\rm Haar}^{(k)} \right\|_\infty &= \left\| \left(M_{\nu}^{(k)} - P_{U(1)\rm Haar}^{(k)}\right)^t \right\|_\infty\nonumber \\
    &\leq \left\| M_{\nu}^{(k)} -P_{U(1)\rm Haar}^{(k)} \right\|_\infty^t \nonumber \\
    &= \left(1- \Delta_{\nu}^{(k)} \right)^t \leq  \exp\left(-\Delta_{\nu}^{(k)} t\right), \nonumber
\end{align}
where in the first line, we used the fact that the moment operator for the convoluted ensemble is given by $M_{\nu_1*\nu_2}^{(k)} = M_{\nu_2}^{(k)}M_{\nu_1}^{(k)}$, and the left and right invariance of the Haar measure, which leads to $M_{\nu}^{(k)}P_{U(1)\rm Haar}^{(k)} = P_{U(1)\rm Haar}^{(k)}M_{\nu}^{(k)} = P_{U(1)\rm Haar}^{(k)}$.
In the second line, we used the submultiplicativity of the operator norm.
Equation~\eqref{eq:spectral_gap_to_design_formation_Hermitian} follows similarly because the submultiplicativity inequality above becomes an equality when $M_{\nu}^{(k)}$ is Hermitian.
Thus, Lemma~\ref{lem:spectral_gap_to_design_formation} indicates that the spectral gap indeed characterizes the convergence rate of the $k$-th moment to the Haar measure.

When the $k$-th moment of the unitary ensemble $\nu$ is sufficiently close to that of the $U(1)$-symmetric Haar-random unitary ensemble, we call such an ensemble a $U(1)$-symmetric approximate unitary $k$-design.
While there are several definitions of approximate unitary designs depending on how the difference between $\nu$ and the Haar measure is quantified, we primarily use the following definition, which is directly connected to the spectral gap.
\begin{dfn}[$U(1)$-symmetric approximate unitary design]\label{def:approx_design}
For $\varepsilon \geq 0$ and $k \in \mathbb{N}$, a unitary ensemble $\nu$ supported on $\mathcal U_{U(1)}$ is called a $U(1)$-symmetric $\varepsilon$-approximate unitary $k$-design in the operator-norm sense if and only if
\begin{equation}
\label{eq:def_approx_design}
\left\| M_{\nu}^{(k)} - P^{(k)}_{U(1) \mathrm{Haar}} \right\|_{\infty} \leq \varepsilon.
\end{equation}
\end{dfn}
\noindent
In this manuscript, for notational simplicity, we simply refer to such an ensemble as an $\varepsilon$-approximate unitary design.
From Lemma~\ref{lem:spectral_gap_to_design_formation}, we immediately see that the operator-norm approximation error after $t$ circuit iterations is bounded as $\varepsilon \leq e^{-\Delta_{\nu}^{(k)}t}$.
We note that Definition~\ref{def:approx_design} is known to be quantitatively related to other standard definitions of approximate unitary designs, such as additive-error and relative-error $\varepsilon$-approximate $k$-designs \cite{brandao2016local,schuster2024random,li2024designs}.
Therefore, these relationships allow us to translate our results for Definition~\ref{def:approx_design} into corresponding bounds for these other notions of approximate designs.
We discuss upper and lower bounds on the circuit depths required for additive- and relative-error approximate designs separately in Sec.~\ref{s:add_rel_error_design}.

We also note that several previous works~\cite{mitsuhashi2024unitary,mitsuhashi2024characterization} have shown that, for $k\geq 2(N-1)$, the $k$-th moment of a random circuit composed of two-qubit $U(1)$-symmetric gates does not coincide with that of the $U(1)$-symmetric Haar-random unitary ensemble even in the infinite-depth limit.
However, since this work addresses moments satisfying $k\leq O(\log N)$, this discrepancy does not arise in the regime considered here.

\subsection{Circuit architectures}

\begin{figure}[t]
\begin{center}
\includegraphics[width=0.5\textwidth]{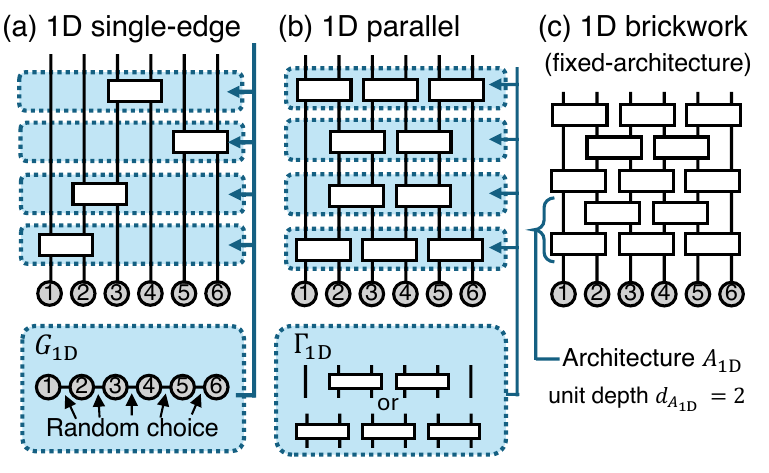}
\caption{Schematics for (a) 1D single-edge circuit, (b) 1D parallel circuit, and (c) 1D brickwork circuit. In (a) and (b), the gate locations are randomly chosen at each layer, while they are predetermined in (c). Each two-qubit gate is independently chosen to be $U(1)$-symmetric Haar random.}
\label{fig:setup_circuit_structure}
\end{center}
\end{figure}

In this work, we mainly consider random circuits composed of two-qubit gates, each independently drawn from the $U(1)$-symmetric Haar measure.
Each such gate takes the form
\begin{equation}
\label{eq:two_qubit_gate_matrix}
U_{2 \rm qubit} =
\begin{pmatrix}
        \! e^{i\theta_{00}} \!\! &0&0&0\\
        0&a&b &0\\
        0&c&d &0\\
        0&0&0 &\!e^{i\theta_{11}} \!\!
\end{pmatrix},
\end{equation}
in the computational basis ordered as $\ket{00},$$\ket{01},$$\ket{10},$$\ket{11}$.
Here, $\theta_{00}$ and $\theta_{11}$ are independent and uniformly distributed over $[0,2\pi)$, while the $2\times2$ block matrix is independently chosen as a Haar-random unitary.
Throughout this work, we classify the random circuit architectures into three types: single-edge, parallel, and fixed-architecture random circuits.
In the following, we introduce each architecture in turn.

A single-edge random circuit is specified by an $N$-vertex connected graph $G=(V,E)$, where $V$ is the set of vertices with $|V|=N$, and $E$ is the set of edges.
In each layer of the single-edge circuit, one of the edges $\{i,j\}\in E$ is chosen uniformly at random, and a two-qubit $U(1)$-symmetric Haar-random gate is applied to sites $i$ and $j$.
We denote the ensemble corresponding to one layer by $\nu_G$.
An example of such a circuit is illustrated in Fig.~\ref{fig:setup_circuit_structure}(a).

For later use in the spectral gap lower bound analysis, we introduce two graph-theoretic properties.
First, we call $G$ a bounded-degree graph if its maximum degree is bounded by a constant independent of the system size $N$.
Second, we introduce the notions of perfect and near-perfect matchings.
A matching $\gamma$ on the vertex set $V$ is a set of mutually vertex-disjoint edges, which can be written as $\gamma \equiv \{\{i_1,i_2\},\ldots,\{i_{2|\gamma|-1},i_{2|\gamma|}\}\}$, with $|\gamma|\leq \left\lfloor\frac{N}{2}\right\rfloor$.
We say that $G=(V,E)$ admits a perfect matching if $N$ is even and there exists such a matching $\gamma\subseteq E$ of size $N/2$, and that $G$ admits a near-perfect matching if $N$ is odd and there exists such a matching $\gamma\subseteq E$ of size $(N-1)/2$.

A parallel random circuit is specified by a matching family $\Gamma$, i.e., a set of matchings.
In each layer of the parallel circuit, one of the matchings $\gamma \in \Gamma$ is chosen uniformly at random, and independent two-qubit $U(1)$-symmetric Haar-random gates are applied to all edges $\{i_{2j-1},i_{2j}\}\in\gamma$.
We denote the corresponding ensemble by $\nu_{\Gamma}$.
For each parallel circuit, we define its associated graph as $G_\Gamma=(V,E_\Gamma)$, where $E_\Gamma \equiv \bigcup_{\gamma\in\Gamma}\gamma$, and throughout this work we assume that $G_\Gamma$ is connected.
For example, for the 1D parallel circuit shown in Fig.~\ref{fig:setup_circuit_structure}(b), the associated graph is the one-dimensional lattice with open boundary conditions.
When $|\Gamma|$ is a constant independent of $N$, we call $\Gamma$ a constant-size matching family.
Under the connectivity assumption on $G_\Gamma$, the average matching size is $\Theta(N)$ for such $\Gamma$.

A fixed-architecture random circuit consists of repetitions of a predetermined circuit unit $A$, in which the locations and ordering of the two-qubit gates are fixed, while each gate is independently chosen to be $U(1)$-symmetric Haar random.
We denote the unitary ensemble corresponding to one circuit unit by $\nu_A$, and the depth of the unit by $d_A$.
A representative example is the 1D brickwork circuit, for which $d_A=2$, as illustrated in Fig.~\ref{fig:setup_circuit_structure}(c).
For each circuit architecture $A$, we define its associated graph as $G_A=(V,E_A)$, where $E_A$ is the set of edges appearing in the circuit unit $A$.
Throughout this work, we assume that $G_A$ is connected.
When $d_A$ is a constant independent of $N$, we refer to $A$ as a constant-depth fixed architecture, and in this case, the total number of gates in the unit $A$ is $\Theta(N)$.

Finally, we briefly describe the representative interaction geometries appearing in Table~\ref{tab:gap_summary}.
An $\alpha$-dimensional lattice refers to the nearest-neighbor interaction graph in spatial dimension $\alpha$, where the number of sites along each spatial dimension is $\Theta (N^{1/\alpha})$.
The corresponding single-edge, parallel, and brickwork circuits are the natural higher-dimensional generalizations of those shown in Figs.~\ref{fig:setup_circuit_structure}(a)--(c).
For fixed spatial dimension $\alpha=O(1)$, these lattice graphs have bounded degree and admit a perfect or near-perfect matching.
Moreover, the corresponding matching family for a parallel circuit has constant size, and the brickwork architecture has constant depth.

An expander graph refers to a bounded-degree graph that nevertheless remains highly connected, so that a standard random walk on the graph mixes rapidly \cite{hoory2006expander}.
In particular, throughout this work, we consider regular bipartite expanders \cite{marcus2015interlacing} and vertex-transitive expanders \cite{godsil2001algebraic} that admit at least one perfect or near-perfect matching.
Since such a graph has bounded degree, its edge set can be partitioned into a constant number of matchings \cite{vizing1964estimate}.
These matchings define a constant-size matching family and, by applying them in a fixed order, a constant-depth architecture.
We refer to the corresponding parallel and fixed-architecture circuits as the expander parallel circuit and the expander fixed-architecture circuit, respectively.

We also consider the all-to-all interaction geometry, in which a two-qubit gate may act on any pair of sites.
The all-to-all single-edge circuit is the single-edge circuit on the complete graph denoted by $G_{\rm all}$, and the all-to-all parallel circuit is defined by the matching family consisting of all matchings of size $\lfloor N/2\rfloor$ on the vertex set $V$.

\section{Upper bound on the spectral gap} \label{s:gap_upper_bound}

In this section, we derive upper bounds on the spectral gap of $U(1)$-symmetric random circuits.
In previous work~\cite{hearth2025unitary}, a spectral gap upper bound associated with the diffusive transport of the conserved $U(1)$ charge (or equivalently, the Goldstone mode in the associated replicated-Hamiltonian description) was derived for circuits on $\alpha$-dimensional lattices.
Here, the charge transport process on a graph $G=(V,E)$ is the lazy random walk of a single particle in which, at each step, an edge is chosen uniformly at random and the occupations of its two endpoints are swapped with probability $1/2$ and left unchanged otherwise.
Denoting the spectral gap of this classical stochastic process by $\lambda_G^{\rm tr}$, the corresponding upper bound for the single-edge random circuit is given by \footnote{
For qubits on the $\alpha$-dimensional lattice graph $G_{\alpha\rm D}=(V,E_{\alpha\rm D})$, the auxiliary Hamiltonian $\hat H_k$ in Eq.~(54) of Ref.~\cite{hearth2025unitary} corresponds to $|E_{\alpha\rm D}|(\mathbb I-M_{\nu_{G_{\alpha\rm D}}}^{(k)})$ in our notation.
Equation~(61) of that reference shows that each local moment projector acts on their density excitations as the corresponding single-edge update of the transport process.
Thus, the lowest nonzero energy within their variational subspace is $|E_{\alpha\rm D}|\lambda_{G_{\alpha\rm D}}^{\rm tr}$, which upper bounds the spectral gap of the Hamiltonian and yields Eq.~\eqref{eq:gap_upper_bound_prev_work_transport} for the lattice graphs.
We note that the two-particle encounter mode constructed in this work is orthogonal to their Goldstone-mode ansatz.
}
\begin{equation}
\label{eq:gap_upper_bound_prev_work_transport}
    \Delta_{\nu_{G}}^{(k)} \leq \lambda_{G}^{\rm tr},
\end{equation}
for $k<N$.
They further conjectured that this upper bound is tight up to constant factors, i.e., $\Delta_{\nu_{G}}^{(k)}=\Theta(\lambda_{G}^{\rm tr})$, since the diffusive relaxation of the conserved charge is expected to set the slowest timescale.
Indeed, we can show that $\Delta_{\nu_{G}}^{(k=1)}=\lambda_{G}^{\rm tr}$ exactly holds for the first moment $k=1$ for any connected graph $G$ \cite{supplement}.
However, it has remained an open problem whether this single-particle transport mode is indeed the slowest mode for general $k$, or whether another mechanism can lead to even slower convergence.

In this work, we prove that for $k\geq 2$, the $k$-th moment of $U(1)$-symmetric random circuits in fact contains a mode that decays more slowly than the single-particle transport mode in many circuit geometries, contrary to this conjecture.
This slower mode can also be understood in terms of simple classical dynamics, which we call the two-particle encounter process.
In this process, two distinguishable particles occupy distinct vertices of $G$.
At each step, an edge is chosen uniformly at random from $E$; if the chosen edge connects the two particles, the process is killed, whereas otherwise the occupations at its two endpoints are swapped with probability $1/2$ and left unchanged with probability $1/2$.
Since the process can be killed, its transition matrix is substochastic and the survival probability decays with time.
We denote the transition matrix of this process by $K_G^{\rm enc}$ and define its principal decay rate as $\lambda_G^{\rm enc}\equiv 1-\rho(K_G^{\rm enc})$, where $\rho(\cdot)$ denotes the spectral radius.

Using this process, we obtain spectral gap upper bounds tighter than Eq.~\eqref{eq:gap_upper_bound_prev_work_transport}.
\vspace{-1mm}\noindent
\begin{theorem}[Spectral gap upper bound] \label{thm:gap_upper_bound}
Let $G$ be any connected $N$-vertex graph, let $\Gamma$ be any constant-size matching family with associated graph $G_\Gamma$, and let $A$ be any constant-depth fixed architecture with associated graph $G_A$.
Then, we have
\begin{equation}
\begin{split}\label{eq:gap_upper_bounds_this_work}
    &\Delta_{\nu_G}^{(2)} \leq \lambda_G^{\rm enc}, \\
    &\Delta_{\nu_\Gamma}^{(2)} \leq O(N \lambda_{G_\Gamma}^{\rm enc}), \\
    &\Delta_{\nu_A}^{(2)} \leq O(N \lambda_{G_A}^{\rm enc}).
\end{split}
\end{equation}
\end{theorem}
\noindent
The all-to-all parallel circuit defined by $\Gamma_{\rm all}$ is not covered by this theorem because $\Gamma_{\rm all}$ is not constant-size.
Nevertheless, the same upper bound $\Delta_{\nu_{\Gamma_{\rm all}}}^{(2)} \leq O(N\lambda_{G_{\rm all}}^{\rm enc})$ can be established by a separate argument. 
The additional factor of $N$ in Eq.~\eqref{eq:gap_upper_bounds_this_work} reflects the $\Theta(N)$ two-qubit gates applied on average per parallel layer and per fixed-architecture circuit unit.
Since the scaling of $\lambda_{G}^{\rm enc}$ can be determined for the representative geometries considered in this work \cite{aldous2002reversible,levin2026markov}, these bounds directly yield the spectral gap upper bound for these geometries.
Furthermore, since the spectral gap is nonincreasing with the moment order, i.e., $\Delta_{\nu}^{(k)}\leq\Delta_{\nu}^{(2)}$, the upper bounds obtained in this theorem also apply to all higher moments $k\geq2$.

\begin{figure}[t]
\begin{center}
\includegraphics[width=0.5\textwidth]{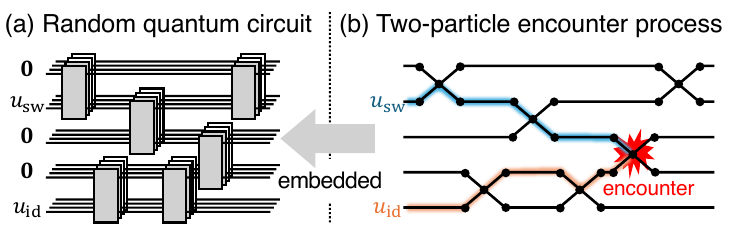}
\caption{
Schematic illustration of the correspondence between a $U(1)$-symmetric random circuit and the two-particle encounter process.
(a) Schematic of the action of the second-moment operator $M_{\nu_G^{*t}}^{(2)}$ on a particular initial configuration. 
(b) The corresponding two-particle encounter process $K^{\rm enc}_G$, as described by Eq.~\eqref{eq:mom_op_enc_correspondence}.
The orange and blue paths illustrate one possible pair of trajectories of the particles associated with $u_{\rm id}$ and $u_{\rm sw}$, respectively, until they encounter each other.
}
\label{fig:encounter_mapping}
\end{center}
\end{figure}

Although the full proof of Theorem~\ref{thm:gap_upper_bound} is provided in the Supplemental Material, we here explain the essential mechanism behind the two-particle encounter mode in the simplest subspace in which it appears.
For this purpose, we consider a subspace of the moment space in which the local types $u_{\rm id}\equiv1010$ and $u_{\rm sw}\equiv0110$ each appear at exactly one site, while all remaining sites carry $\bm 0\equiv0000$.
More specifically, we define
\begin{align*}
    \mathcal H_{\rm enc} &\equiv \operatorname{span}\left\{\ket{\psi_{ij}}:i,j\in[N],\ i\neq j\right\}, \\
    \ket{\psi_{ij}} &\equiv \ket{u_{\rm id}}_i\ket{u_{\rm sw}}_j \otimes \bigotimes_{x\in[N],\,x\neq i,j}\ket{\bm 0}_x,
\end{align*}
where $\ket{u}_i$ denotes the local-type basis state $u$ at site $i$.
The subspace $\mathcal H_{\rm enc}$ lies in the charge sector with $(n_1,n_2,n_{\overline{1}},n_{\overline{2}})=(1,1,2,0)$.

The moment operator corresponding to the graph $G=(V,E)$ is given by
\begin{equation*}
    M_{\nu_G}^{(2)} = \frac{1}{|E|}\sum_{\{i,j\}\in E} P^{(2)}_{i,j} \otimes \mathbb{I}_{\overline{i,j}},
\end{equation*}
where $P^{(2)}_{i,j}$ denotes the second moment operator of the two-qubit $U(1)$-symmetric Haar-random unitary ensemble on sites $i$ and $j$, and $\mathbb{I}_{\overline{i,j}}$ is the identity on all other sites. It therefore suffices to examine the action of $P^{(2)}_{i,j}$ on $\mathcal H_{\rm enc}$.
A direct calculation gives
\begin{align*}
    &P^{(2)}_{i,j}\ket{u_{\rm id}}_i\ket{u_{\rm sw}}_j = 0,\qquad
    P^{(2)}_{i,j}\ket{\bm 0}_i\ket{\bm 0}_j = \ket{\bm 0}_i\ket{\bm 0}_j, \\
    &P^{(2)}_{i,j}\ket{\bm 0}_i\ket{u_{a}}_j
    = \frac{1}{2}\left(\ket{\bm 0}_i\ket{u_{a}}_j+\ket{u_{a}}_i\ket{\bm 0}_j\right),
    \quad a\in\{\mathrm{id},\mathrm{sw}\},
\end{align*}
together with the same equations under the exchange $i\leftrightarrow j$.
These equations show that $\mathcal H_{\rm enc}$ is invariant under each two-qubit projector and that the restricted action exactly reproduces the two-particle encounter process.
Indeed, identifying $\ket{\psi_{ij}}$ with the classical configuration in which the two particles occupy sites $i$ and $j$, we obtain
\begin{equation}
\label{eq:mom_op_enc_correspondence}
    \left.M_{\nu_G}^{(2)}\right|_{\mathcal H_{\rm enc}} = K_G^{\rm enc}.
\end{equation}
Furthermore, since the ket and bra charge vectors $(1,1)$ and $(2,0)$ are not related by a permutation, the $U(1)$-symmetric Haar moment operator vanishes on this charge sector.
Therefore, this correspondence shows that the two-particle encounter process is embedded in the moment operator as illustrated in Fig.~\ref{fig:encounter_mapping}, which leads to $\Delta_{\nu_G}^{(2)} \leq \lambda_G^{\rm enc}$.

As a direct consequence of Theorem~\ref{thm:gap_upper_bound}, we obtain
\begin{equation}
\begin{split} \label{eq:gap_upper_bound_tr_N-2}
    \Delta_{\nu_G}^{(2)} &\leq \lambda_G^{\rm enc}
    \leq \min\left\{ \lambda_G^{\rm tr}, \frac{2}{N(N-1)} \right\}, \\
    \Delta_{\nu_\Gamma}^{(2)} &\leq O(N \lambda_{G_\Gamma}^{\rm enc})
    \leq \min\left\{ O(N\lambda_{G_\Gamma}^{\rm tr}), O\left(\frac{1}{N}\right)\right\}, \\
    \Delta_{\nu_A}^{(2)} &\leq O(N \lambda_{G_A}^{\rm enc})
    \leq \min\left\{ O(N\lambda_{G_A}^{\rm tr}), O\left(\frac{1}{N}\right)\right\}.
\end{split}
\end{equation}
Here, we used the fact that the two-particle encounter process always decays no faster than the single-particle transport process on the same graph, i.e., $\lambda_G^{\rm enc}\leq\lambda_G^{\rm tr}$.
We also used the geometry-independent bound $\lambda_G^{\rm enc}\leq 2/[N(N-1)]$.
Indeed, if the two particles are uniformly distributed over all $N(N-1)$ ordered pairs of distinct sites, only the two configurations occupying its two endpoints are killed for any chosen edge.
Since \(K_G^{\rm enc}\) is real symmetric, its largest eigenvalue is at least this uniform-state expectation value, giving \(\lambda_G^{\rm enc}\le 2/[N(N-1)]\).
Thus, Eq.~\eqref{eq:gap_upper_bound_tr_N-2} provides a universal spectral gap upper bound arising from a mechanism distinct from charge transport, which yields an asymptotically slower mode for many circuit geometries.

We further show that this slow mode mechanism is not confined to the particular charge sector considered above, but appears in a broad class of charge sectors.
Specifically, consider a charge sector satisfying $n_1+n_2=n_{\overline{1}}+n_{\overline{2}}\leq N$ and $n_i\geq2$ for all $i\in\{1,2,\overline{1},\overline{2}\}$.
Then, for any connected graph $G$, for any constant-size matching family $\Gamma$, and for any constant-depth fixed architecture $A$, we have
\begin{equation}
\begin{split}\label{eq:gap_upper_bound_robust_n}
    & \Delta_{\nu_G;(n_1,n_2)^T,(n_{\overline{1}},n_{\overline{2}})^T}^{(2)} \leq2\left(\frac{n_1+n_2}{N}\right)^2, \\
    & \Delta_{\nu_\Gamma;(n_1,n_2)^T,(n_{\overline{1}},n_{\overline{2}})^T}^{(2)} \leq O \left[ \frac{\left(n_1+n_2\right)^2}{N} \right], \\
    & \Delta_{\nu_A;(n_1,n_2)^T,(n_{\overline{1}},n_{\overline{2}})^T}^{(2)} \leq O\left[ \frac{\left(n_1+n_2\right)^2}{N} \right].
\end{split}
\end{equation}
These equations highlight that the slow modes are not restricted to the particular low-particle-number sector in which the exact correspondence to the two-particle encounter process occurs, but persist in a broad class of sufficiently dilute charge sectors.
In particular, when $n_1+n_2=o(\sqrt{N})$, the bounds in Eq.~\eqref{eq:gap_upper_bound_robust_n} are $o(N^{-1})$ for the single-edge circuit and $o(1)$ for the parallel and fixed-architecture circuits. 
These bounds imply an asymptotically slower mode than the single-particle transport mode, for example in expander graph geometry.
We also note that by applying the particle-hole transformation $0\leftrightarrow1$, an analogous statement holds for sufficiently dense charge sectors with $n_1+n_2=2N-o(\sqrt{N})$.

Moreover, we show that the slow-mode mechanism cannot be avoided merely by increasing the gate locality to $\ell>2$.
Specifically, for random circuits composed of $\ell$-local $U(1)$-symmetric Haar-random gates, we obtain
\begin{equation}
\begin{split}\label{eq:gap_upper_bound_robust_l}
    &\Delta_{\nu_{G,\ell\mathrm{-loc}}}^{(2)}
    \leq 2\left(\frac{\ell}{N}\right)^2, \\
    &\Delta_{\nu_{\Gamma,\ell\mathrm{-loc}}}^{(2)}
    \leq \frac{2\ell}{N}, \\
    &\Delta_{\nu_{A,\ell\mathrm{-loc}}}^{(2)}
    \leq O\left(\frac{\ell}{N}\right).
\end{split}
\end{equation}
Here, $G$ denotes any $\ell$-uniform hypergraph, and $\nu_{G,\ell\mathrm{-loc}}$ denotes the corresponding single-hyperedge circuit, in which a single $\ell$-local gate is applied in each layer.
The ensemble $\nu_{\Gamma,\ell\mathrm{-loc}}$ denotes the corresponding parallel circuit, in which mutually disjoint $\ell$-local gates are applied in parallel, with at most $\left\lfloor N/\ell\right\rfloor$ gates in a single layer, while $\nu_{A,\ell\mathrm{-loc}}$ denotes the corresponding fixed-architecture circuit with constant depth.
These inequalities show that the slow-mode mechanism cannot be removed by increasing the gate locality to any fixed $\ell$.

\section{Lower bound on the spectral gap} \label{s:gap_lower_bound}

In the previous section, we showed that the spectral gap is upper bounded by the decay rate of the two-particle encounter process, which can be parametrically smaller than the rate associated with single-particle transport in several circuit geometries.
It is then natural to ask whether the two-particle encounter mode actually sets the slowest relaxation timescale, or an even slower mode further delays unitary design formation. To answer this question, we need a matching lower bound on the spectral gap.

However, obtaining such tight lower bounds for symmetric random circuits has remained an open problem.
The major obstacle is that bootstrap methods such as Knabe's local gap threshold~\cite{knabe1988energy} and the martingale method~\cite{nachtergaele1996spectral}, which allow gap lower bounds for local subsystems to be lifted to the entire system, are not directly applicable in the presence of continuous symmetries~\cite{li2024designs}.
These proof techniques have been essential for deriving tight spectral gap lower bounds for random circuits without symmetry~\cite{brandao2016local,haferkamp2022random,chen2024incompressibility}.
As a consequence of this technical difficulty, the problem of establishing a matching lower bound for symmetric random circuits has remained largely untouched, despite its importance.

In this work, we develop a new proof strategy, which does not rely on the bootstrap methods, and establish the matching lower bounds for a broad range of circuit geometries:
\begin{theorem}[Spectral gap lower bound] \label{thm:gap_lower_bound}
Let $G$ be any bounded-degree graph that admits a perfect or near-perfect matching, let $\Gamma$ be any constant-size matching family whose associated graph $G_\Gamma$ admits a perfect or near-perfect matching, and let $A$ be any constant-depth fixed architecture whose associated graph $G_A$ admits a perfect or near-perfect matching. 
Then, for $2\leq k \leq O(\sqrt{\log N})$, we have
\begin{equation}
\label{eq:gap_lower_bound}
\begin{split}
    &\Delta_{\nu_G}^{(k)} \geq \Omega\left(\lambda_G^{\rm enc}\right), \\
    &\Delta_{\nu_\Gamma}^{(k)} \geq \Omega\left(N\lambda_{G_\Gamma}^{\rm enc}\right), \\
    &\Delta_{\nu_A}^{(k)} \geq \Omega\left(N\lambda_{G_A}^{\rm enc}\right).
\end{split}
\end{equation}
\end{theorem}
\noindent
The all-to-all single-edge and parallel circuits defined by $G_{\rm all}$ and $\Gamma_{\rm all}$, respectively, are not covered by this theorem.
Nevertheless, for the same range of $k$, the lower bounds $\Delta_{\nu_{G_{\rm all}}}^{(k)} \geq \Omega\left(\lambda_{G_{\rm all}}^{\rm enc}\right)$ and $\Delta_{\nu_{\Gamma_{\rm all}}}^{(k)} \geq \Omega\left(N\lambda_{G_{\rm all}}^{\rm enc}\right)$ can be established by separate arguments.

Combining Theorems~\ref{thm:gap_upper_bound} and~\ref{thm:gap_lower_bound}, we obtain, for $2\leq k \leq O(\sqrt{\log N})$,
\begin{equation}
\label{eq:gap_optimal_scaling}
\begin{split}
    &\Delta_{\nu_G}^{(k)} = \Theta\left(\lambda_G^{\rm enc}\right), \\
    &\Delta_{\nu_\Gamma}^{(k)} = \Theta\left(N\lambda_{G_\Gamma}^{\rm enc}\right), \\
    &\Delta_{\nu_A}^{(k)} = \Theta\left(N\lambda_{G_A}^{\rm enc}\right),
\end{split}
\end{equation}
which determines the asymptotic spectral gap scaling throughout this range of moment orders.
In particular, these results reveal that the two-particle encounter mode indeed sets the slowest relaxation timescale up to constant factors, for any fixed moment order $k\geq2$ in the large-system-size limit.
These equations directly yield the spectral gap scalings summarized in Table~\ref{tab:gap_summary}, since the scaling of $\lambda_G^{\rm enc}$ can be explicitly determined for these geometries \cite{aldous2002reversible,levin2026markov}.

\begin{figure*}[]
\begin{center}
\includegraphics[width=0.95\textwidth]{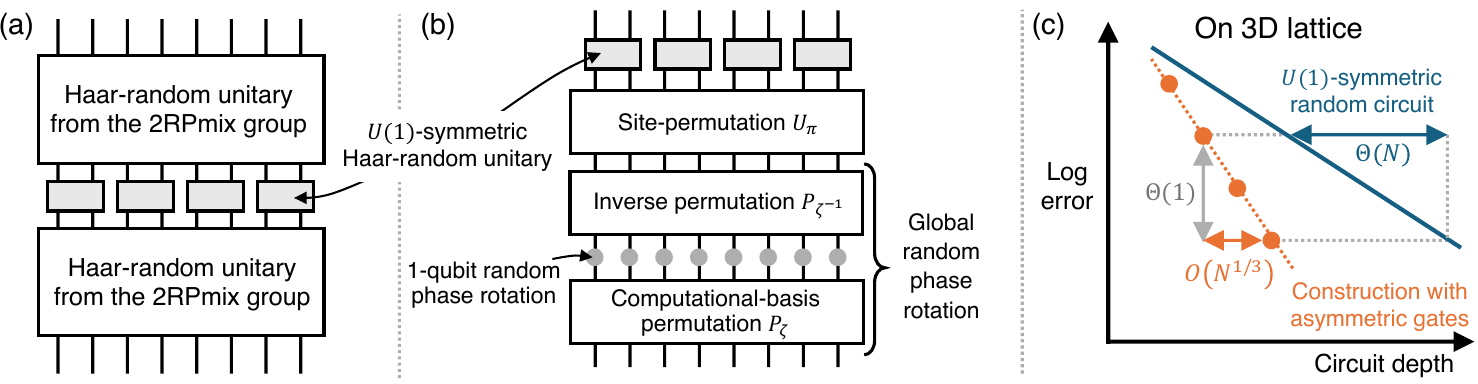}
\caption{
(a) Schematic of the auxiliary doped 2RPmix circuit used in the proof of the spectral gap lower bound (Theorem~\ref{thm:gap_lower_bound}).
(b) Efficient implementation of a constant-gap circuit unit using asymmetric local gates. The site-permutation and random-phase components approximately implement the RPmix ensemble, while the top layer consists of independent two-qubit $U(1)$-symmetric Haar-random gates. The computational-basis permutation $\zeta$ is randomly drawn from the ensemble $\eta$.
(c) Comparison of the circuit depths required to realize a constant-gap circuit unit with a $U(1)$-symmetric random circuit and with the asymmetric-gate construction shown in (b). On a three-dimensional lattice, the parallel random circuit with $U(1)$-symmetric gates requires a depth of $\Theta(N)$, whereas the construction with asymmetric gates given in Sec.~\ref{s:efficient_construction} requires only a depth of $O(N^{1/3})$ for a fixed moment order $k\geq 2$.
}
\label{fig:efficient_construction}
\end{center}
\end{figure*}

To prove this theorem, we develop a new proof technique that introduces an auxiliary subgroup of $\mathcal{U}_{U(1)}$ and separates the problem into two tractable ingredients.
Since the complete proof requires substantial technical preparation, including the introduction of several auxiliary concepts and intermediate steps, we defer the details to the Supplemental Material and describe only the essential ideas here.

As a first step of the proof, we establish a constant spectral gap lower bound for an auxiliary circuit unit that we call the \emph{doped 2RPmix circuit}.
As illustrated in Fig.~\ref{fig:efficient_construction}(a), one unit consists of a layer of $\lfloor N/2\rfloor$ disjoint $U(1)$-symmetric Haar-random two-qubit gates sandwiched between two independent Haar-random unitaries drawn from the 2RPmix group on the $N$-qubit system.
Here, the 2RPmix group is the subgroup of $\mathcal U_{U(1)}$ generated by permutations of the $N$ sites and diagonal two-qubit phase rotations acting on arbitrary pairs of sites.
We prove that this auxiliary circuit has a spectral gap bounded below by a positive constant independent of $N$ and $k$, for $k\leq O(\sqrt{\log N})$.
To establish this bound, we represent the moment operator in a basis adapted to the 2RPmix moment space and decompose it into matrix blocks.
The diagonal blocks are controlled by relating them to a classical random involution process on the $N$ sites~\cite{bernstein2018random}, while the off-diagonal blocks are bounded separately.

Second, we quantify how rapidly the global 2RPmix group can be generated using local two-qubit gates.
In each circuit geometry, the 2RPmix group can be generated locally using only two-site SWAP operations and two-site phase rotations.
Remarkably, we show that the rate of this local generation is governed by the same two-particle encounter process that appeared in the upper bound.
Combining these two ingredients yields Theorem~\ref{thm:gap_lower_bound}.

Thus, our proof is based on the comparison with an auxiliary subgroup-based ensemble.
Using tractable subgroup ensembles as building blocks for quantum randomness is itself a well-established strategy, as exemplified by permutation-phase-Clifford (PFC) constructions~\cite{metger2024simple} and doped Clifford circuits~\cite{haferkamp2023efficient,leone2026nonclifford}.
In previous spectral-gap analyses of random circuits~\cite{haferkamp2022random,chen2024incompressibility}, these subgroup-based comparisons did not directly yield the bound with optimal system-size scaling, and therefore were combined with the finite-size bootstrap arguments, which subsequently sharpened the bound.
In our proof, by contrast, the 2RPmix group is chosen so that its local generation rate can be determined sharply in terms of the two-particle encounter process, while the auxiliary doped 2RPmix circuit has a constant spectral gap; consequently, the subgroup comparison itself already yields the optimal system-size scaling without any finite-size bootstrap.
Therefore, our proof provides a novel framework for deriving tight spectral gap lower bounds by appropriately choosing auxiliary subgroups, which may also be useful for other constrained random circuits where standard bootstrap methods face limitations.

\section{Efficient construction of $U(1)$-symmetric unitary designs with asymmetric local gates} \label{s:efficient_construction}

In Sec.~\ref{s:gap_upper_bound}, we established a robust spectral gap upper bound arising from the two-particle encounter mechanism.
This slow mode persists across a broad class of circuit geometries and cannot be removed merely by increasing the gate locality to any fixed $\ell$.
For practical applications of $U(1)$-symmetric random unitaries, such as classical shadow tomography, it is therefore natural to ask whether this bottleneck can be avoided by allowing local gates that do not individually respect the $U(1)$ symmetry.
In this section, we answer this question affirmatively by providing an explicit protocol for efficiently constructing $U(1)$-symmetric unitary designs using asymmetric local gates.
Although the individual gates and intermediate operations in this protocol need not preserve the $U(1)$ symmetry, the overall unitary obtained from the protocol is always $U(1)$-symmetric.

More specifically, our goal is to implement a constant-gap circuit unit with as small a circuit depth as possible, using only local gates allowed in each circuit geometry.
Here, a constant-gap unit means a circuit unit whose spectral gap satisfies
\begin{equation}
\label{eq:constant_gap_lower_bound_unit}
    \Delta_{\nu_{\rm unit}}^{(k)} \geq c,
\end{equation}
where $\nu_{\rm unit}$ denotes the ensemble associated with the circuit unit, and $c$ is some positive constant.
Throughout this section, we count circuit depth by allowing mutually disjoint local gates to be applied simultaneously within a single layer.
Then, we obtain the following upper bounds on the circuit depth required to construct such a constant-gap unit:
\begin{theorem}[Efficient construction with asymmetric local gates] \label{thm:efficient_construction_with_asymmetric_gates}
For $k \leq O(\log N/\log\log N)$, the following constructions are possible.
On an $\alpha$-dimensional lattice with fixed $\alpha$, a constant-gap circuit unit can be implemented with circuit depth
\begin{equation}
\label{eq:alphaD_depth_efficient}
    d_{\alpha \rm D}^{\rm unit} = O(N^{1/\alpha} k^3 (\log k)^6).
\end{equation}
In the all-to-all interaction model, such a unit can be implemented with circuit depth
\begin{equation}
\label{eq:alltoall_depth_efficient}
    d_{\rm all}^{\rm unit}
    =
    O\left(
        (\log N)^3(\log\log N)^2
        k\,(\log k)^2
    \right).
\end{equation}
\end{theorem}
\noindent
These depth scalings for a fixed $k$ are summarized in Table~\ref{tab:constant_gap_depth}, together with the corresponding depths required for symmetric random circuits in the same geometries.
As illustrated in Fig.~\ref{fig:efficient_construction}(c), our construction circumvents the two-particle encounter bottleneck and yields parametrically shallower constant-gap circuit units than the symmetric random circuits.

While the full proof of Theorem~\ref{thm:efficient_construction_with_asymmetric_gates} is given in the Supplemental Material, we present its essential idea here.
The basic strategy is to approximately implement an auxiliary circuit unit that we call the \emph{doped RPmix circuit}.
This circuit is obtained from the doped 2RPmix circuit introduced in the previous section by replacing the two Haar-random unitaries drawn from the 2RPmix group with Haar-random unitaries drawn from the RPmix group.
Here, the RPmix group is generated by arbitrary permutations of the $N$ physical sites and arbitrary phase rotations diagonal in the computational basis, and contains the 2RPmix group as a subgroup.
We can show that the doped RPmix circuit also has a constant spectral gap for moment orders $k \leq O(\log N/\log\log N)$, using essentially the same proof techniques as in the analysis of the doped 2RPmix circuit.
Therefore, it suffices to approximate the $k$-th moment of the Haar-random RPmix ensemble using shallow circuits in each geometry.
For this purpose, we independently implement random site permutations and random diagonal phase rotations.

For the site-permutation part, we use routing via matchings~\cite{alon1994routing}, in which each layer consists of SWAP gates on mutually disjoint allowed edges.
With this strategy, any site permutation can be implemented exactly within depth $O(N^{1/\alpha})$ on an $\alpha$-dimensional lattice, and within two layers in the all-to-all interaction model.
Equivalently, in these geometries, the depth upper bound for an arbitrary site permutation has the same asymptotic scaling as the graph diameter, namely, the maximum shortest-path distance between any pair of sites.
Unlike a random walk generated by local SWAPs, this construction first samples a target permutation $\pi\sim S_N$ and then chooses the locations and ordering of SWAP gates to implement it.
This construction enables the implementation of an exact uniform site permutation without waiting for the single-particle transport timescale.

In the phase-rotation part, we use a suitably random permutation of the computational basis.
Let $\eta$ be an ensemble of permutations $\zeta$ of $\{0,1\}^N$, and denote the corresponding permutation unitary by $P_{\zeta}$, defined as $P_{\zeta}\ket{\bm x}=\ket{\zeta(\bm x)}$. Note that these are permutations of the $2^N$ computational-basis states and are different from permutations of $N$ physical sites generated by two-site SWAPs.
We then implement the phase rotation as
\begin{equation}
\label{eq:RPens_asym_gates}
    U(\zeta,\bm\theta)
    \equiv
    P_{\zeta^{-1}}
    \left[
        \bigotimes_{j=1}^{N}
        \left(
            \ket{0}\bra{0}_j
            +
            e^{i\theta_j}\ket{1}\bra{1}_j
        \right)
    \right]
    P_{\zeta},
\end{equation}
where $\bm\theta=(\theta_1,\ldots,\theta_N)$ denotes independently and uniformly chosen random phases.
As illustrated in Fig.~\ref{fig:efficient_construction}(b), this construction conjugates a parallel layer of single-qubit phase rotations by a permutation of the computational basis.
Although $P_{\zeta}$ and $P_{\zeta^{-1}}$ generally do not respect the $U(1)$ symmetry individually, the entire operation $U(\zeta,\bm\theta)$ is diagonal in the computational basis and is therefore $U(1)$-symmetric.

Furthermore, permutation ensembles $\eta$ that are sufficiently random for Eq.~\eqref{eq:RPens_asym_gates} to approximate the $k$-th moment of the random phase rotation ensemble can be efficiently generated by local reversible circuits in each geometry~\cite{brodsky2008simple,gretta2025more,gay2025pseudorandomness}.
Using these results, the random-phase operation can be implemented within the depths shown in Eqs.~\eqref{eq:alphaD_depth_efficient} and \eqref{eq:alltoall_depth_efficient} for the lattice and all-to-all geometries, respectively. 
Adding the site-permutation step and the Haar-random gate layer does not change these depth scalings, which yields Theorem~\ref{thm:efficient_construction_with_asymmetric_gates}.

This depth reduction compared with symmetric random circuits is enabled by two key ideas in our construction.
First, allowing symmetry-breaking local gates bypasses the two-particle encounter bottleneck.
In particular, in the implementation of the global random-phase rotation in Eq.~\eqref{eq:RPens_asym_gates}, a computational-basis permutation that can temporarily break $U(1)$ symmetry spreads single-qubit random phases over all $2^N$ computational-basis states.
Such an implementation of the global random-phase operation is not available under the $U(1)$-symmetry constraint on local gates.

Second, the single-particle transport bottleneck can also be circumvented by allowing target-dependent circuit construction, in which the target operation is sampled first and then compiled into local gates.
Specifically, by using routing via matchings, any site permutation can be implemented in this way within circuit depth proportional to the graph diameter, in the lattice and all-to-all geometries considered here.
This is in contrast to a random walk generated by local SWAPs, in which the single-particle transport bottleneck cannot be avoided.
On lattices, the random-phase rotation can also be implemented in depth parametrically smaller than the single-particle transport timescale; hence, the entire construction circumvents the single-particle transport bottleneck.

\begin{table}[t]
    \centering
    \footnotesize
    \setlength{\tabcolsep}{3.5pt}
    \renewcommand{\arraystretch}{1.15}
    \begin{tabular}{c|c|c}
        \toprule
        Geometry
        & \makecell{$U(1)$-symmetric\\ random circuit}
        & \makecell{Efficient construction\\with asymmetric gates}
        \\
        \midrule
        1D lattice
        & $\Theta(N^{2})$
        & $O(N)$
        \\
        2D lattice
        & $\Theta(N\log N)$
        & $O(N^{1/2})$
        \\
        $\alpha$D lattice ($\alpha\geq3$)
        & $\Theta(N)$
        & $O(N^{1/\alpha})$
        \\
        \midrule
        All-to-all
        & $\Theta(N)$
        & $O\left((\log N)^3(\log\log N)^2\right)$
        \\
        \bottomrule
    \end{tabular}
    \caption{
        Circuit depths required for a constant-gap unit with $U(1)$-symmetric parallel random circuits and the efficient construction shown in Fig.~\ref{fig:efficient_construction}(b), for a fixed moment order $k\geq 2$.
    }
    \label{tab:constant_gap_depth}
\end{table}

\section{Required circuit depth for additive-error and relative-error approximate unitary designs} \label{s:add_rel_error_design}

In the previous sections, we have analyzed the spectral gap for $U(1)$-symmetric random circuits and an efficient construction protocol.
In this section, we translate these results into upper and lower bounds on the depths required to form approximate unitary designs.
In particular, while the spectral gap is directly related to the formation depth of an operator-norm $\varepsilon$-approximate design (Definition~\ref{def:approx_design}) as shown in Lemma~\ref{lem:spectral_gap_to_design_formation}, its relationship to additive-error and relative-error approximate designs, which are other standard notions of approximate unitary designs, has not yet been discussed.
Here, we summarize the essential relationships between these notions and defer some of their derivations to the Supplemental Material.

First, we introduce the definitions of additive-error and relative-error approximate unitary designs under $U(1)$ symmetry.
For this purpose, we use the $k$-fold twirling channel
\begin{equation}
    \Phi_{\nu}^{(k)}(\cdot)
    \equiv
    \Exnu\left[
        U^{\otimes k}(\cdot)(U^\dagger)^{\otimes k}
    \right],
\end{equation}
where $U^\dagger$ denotes the Hermitian conjugate of $U$. The twirling channel contains the same information as the $k$-th moment operator, and therefore equivalently characterizes the $k$-th moment of the ensemble $\nu$.
Using this channel, additive-error and relative-error approximate designs are defined as follows:

\begin{dfn}[Additive-error and relative-error approximate unitary designs under $U(1)$ symmetry] \label{def:twirl_approx_unitary_design}
For $k\in\mathbb{N}$ and $\varepsilon\geq0$, a $U(1)$-symmetric unitary ensemble $\nu$ is an additive-error $\varepsilon$-approximate unitary $k$-design if and only if
\begin{equation}
\label{seq:def_add_t_des}
    \left\|
        \Phi_{\nu}^{(k)}
        -
        \Phi_{U(1)\mathrm{Haar}}^{(k)}
    \right\|_{\diamond}
    \leq
    \varepsilon,
\end{equation}
where $\Phi_{U(1)\mathrm{Haar}}^{(k)}$ denotes the twirling channel of the $U(1)$-symmetric Haar-random unitary ensemble, and
$\|\mathcal{E}\|_{\diamond}\equiv\sup_R\sup_{X\neq0}\|(\mathcal{E}\otimes\mathrm{id}_R)(X)\|_1/\|X\|_1$ is the diamond norm.
The ensemble $\nu$ is a relative-error $\varepsilon$-approximate unitary $k$-design if and only if
\begin{equation}
\label{seq:def_rel_t_des}
    (1-\varepsilon)\Phi_{U(1)\mathrm{Haar}}^{(k)}
    \preccurlyeq
    \Phi_{\nu}^{(k)}
    \preccurlyeq
    (1+\varepsilon)\Phi_{U(1)\mathrm{Haar}}^{(k)},
\end{equation}
where $\mathcal{E}\preccurlyeq\mathcal{F}$ means that the linear map $\mathcal{F}-\mathcal{E}$ is completely positive.
\end{dfn}
\noindent
These notions have operational interpretations in terms of indistinguishability~\cite{schuster2024random}, which we do not discuss further here.

We now evaluate the number of circuit unit repetitions required to form approximate $k$-designs under repeated convolution of a unitary ensemble $\nu$.
Specifically, we define $T_{\nu,\varepsilon,k}^{\rm op}$, $T_{\nu,\varepsilon,k}^{\rm add}$, and $T_{\nu,\varepsilon,k}^{\rm rel}$ as the minimum numbers of repetitions required to form a $U(1)$-symmetric $\varepsilon$-approximate unitary $k$-design in the operator-norm, additive-error, and relative-error senses, respectively.
When one application of $\nu$ corresponds to a single circuit layer, these quantities directly give the circuit depth, while when $\nu$ corresponds to multiple layers, the physical circuit depths are obtained by multiplying them by the depth of the unit.

Then, for $0<\varepsilon<1$ and $0<\Delta_{\nu}^{(k)}<1$, the spectral gap gives the following upper bounds:
\begin{equation}
\begin{split} \label{eq:depth_upper_bound}
    T_{\nu,\varepsilon,k}^{\mathrm{op}}
    &\leq
    \left\lceil
        \frac{1}{\Delta_{\nu}^{(k)}}
        \log\frac{1}{\varepsilon}
    \right\rceil, \\
    T_{\nu,\varepsilon,k}^{\mathrm{add}}
    &\leq
    \left\lceil
        \frac{1}{\Delta_{\nu}^{(k)}}
        \left(
            kN\log 2+\log\frac{1}{\varepsilon}
        \right)
    \right\rceil, \\
    T_{\nu,\varepsilon,k}^{\mathrm{rel}}
    &\leq
    \left\lceil
        \frac{1}{\Delta_{\nu}^{(k)}}
        \left(
            2kN\log 2+\log\frac{1}{\varepsilon}
        \right)
    \right\rceil .
\end{split}
\end{equation}
Here, the bound for $T_{\nu,\varepsilon,k}^{\mathrm{op}}$ follows directly from Eq.~\eqref{eq:spectral_gap_to_design_formation_general} of Lemma~\ref{lem:spectral_gap_to_design_formation}.
The bounds for $T_{\nu,\varepsilon,k}^{\mathrm{add}}$ and $T_{\nu,\varepsilon,k}^{\mathrm{rel}}$ follow from the standard quantitative relationships between operator-norm, additive-error, and relative-error approximate designs~\cite{brandao2016local,li2024designs}.
Combining these inequalities with the spectral gap lower bound of Theorem~\ref{thm:gap_lower_bound} gives upper bounds on the required depths in $U(1)$-symmetric random circuits.
Together with the construction of a constant-gap unit and its implementation depth in Theorem~\ref{thm:efficient_construction_with_asymmetric_gates}, they also give the upper bounds on the physical circuit depths required in the efficient construction protocol.

Furthermore, when the moment operator of the ensemble $\nu$ is Hermitian, we can also derive lower bounds for the required circuit depths, based on the spectral gap.
In particular, for $0<\varepsilon<\frac{1}{2}$ and $0<\Delta_{\nu}^{(k)}<1$, we have
\begin{equation}
\begin{split} \label{eq:depth_lower_bound}
    T_{\nu,\varepsilon,k}^{\mathrm{op}}
    &=
    \left\lceil
        \frac{\log(1/\varepsilon)}
        {-\log\left(1-\Delta_{\nu}^{(k)}\right)}
    \right\rceil
    \geq
    \left\lceil
        \frac{1-\Delta_{\nu}^{(k)}}{\Delta_{\nu}^{(k)}}
        \log\frac{1}{\varepsilon}
    \right\rceil, \\
    T_{\nu,\varepsilon,k}^{\mathrm{add}}
    &\geq
    \left\lceil
        \frac{\log(1/\varepsilon)}
        {-\log\left(1-\Delta_{\nu}^{(k)}\right)}
    \right\rceil
    \geq
    \left\lceil
        \frac{1-\Delta_{\nu}^{(k)}}{\Delta_{\nu}^{(k)}}
        \log\frac{1}{\varepsilon}
    \right\rceil, \\
    T_{\nu,\varepsilon,k}^{\mathrm{rel}}
    &\geq
    \left\lceil
        \frac{\log(1/(2\varepsilon))}
        {-\log\left(1-\Delta_{\nu}^{(k)}\right)}
    \right\rceil
    \geq
    \left\lceil
        \frac{1-\Delta_{\nu}^{(k)}}{\Delta_{\nu}^{(k)}}
        \log\frac{1}{2\varepsilon}
    \right\rceil ,
\end{split}
\end{equation}
where the rightmost inequalities follow from $-\log(1-x)\leq x/(1-x)$ for $0<x<1$.
Here, the bound for $T_{\nu,\varepsilon,k}^{\mathrm{op}}$ follows directly from Eq.~\eqref{eq:spectral_gap_to_design_formation_Hermitian} of Lemma~\ref{lem:spectral_gap_to_design_formation}.
Furthermore, the bound for $T_{\nu,\varepsilon,k}^{\mathrm{add}}$ follows from the spectral-radius lower bound on the additive error~\cite{ErratumHearthPRX2025}, while the bound for $T_{\nu,\varepsilon,k}^{\mathrm{rel}}$ follows from the fact that a relative-error $\varepsilon$-approximate $k$-design is always an additive-error $2\varepsilon$-approximate $k$-design~\cite{brandao2016local,li2024designs}.

We note that the derivation of these lower bounds assumes that the moment operator is Hermitian.
For a general fixed-architecture circuit, the moment operator can be non-Hermitian, so Eq.~\eqref{eq:depth_lower_bound} need not hold.
On the other hand, the Hermiticity condition holds for any single-edge and parallel circuits. 
Therefore, for these architectures, the spectral gap upper bounds of Theorem~\ref{thm:gap_upper_bound} directly give lower bounds on the required design formation depth.
Combining Eqs.~\eqref{eq:depth_upper_bound} and \eqref{eq:depth_lower_bound}, we see that the spectral gap bounds immediately translate into upper and lower bounds on the formation depths of approximate unitary designs.

\section{Slow relaxation of subsystem purity and entanglement entropy} \label{s:demonstration_slow_decay}

In Secs.~\ref{s:gap_upper_bound} and \ref{s:gap_lower_bound}, we established that the two-particle encounter rate governs the relaxation of low-order moments in $U(1)$-symmetric random circuits and can be parametrically smaller than the single-particle transport rate.
In this section, we demonstrate that this slow mode significantly contributes to the relaxation dynamics of familiar physical quantities.
Specifically, we consider the dynamics of the subsystem purity, second R\'enyi entanglement entropy, and von Neumann entanglement entropy for an $\lfloor N/2 \rfloor$-qubit subsystem in the single-edge all-to-all random circuit.
For simplicity, we take the initial state to be a uniformly random computational-basis state with particle number $n=2$, which is a minimal setting for demonstrating the role of the two-particle encounter mode.
We first analyze the dynamics of subsystem purity in Sec.~\ref{ss:purity} using a rigorous mode decomposition, and then consider the entanglement entropies in Sec.~\ref{ss:entanglement_entropy}, where we develop an intuitive physical picture that qualitatively explains their slow relaxation.

\begin{figure*}[]
\begin{center}
\includegraphics[width=0.95\textwidth]{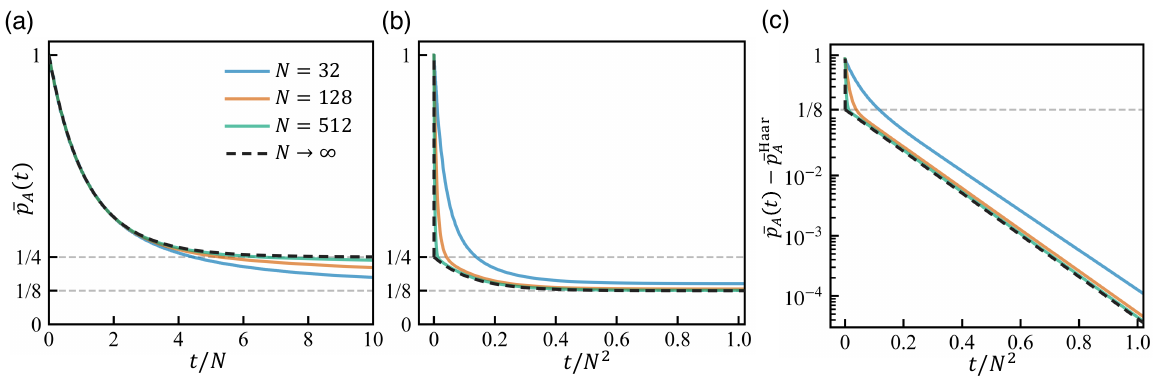}
\caption{
Time evolution of the averaged subsystem purity in the single-edge all-to-all random circuit on (a) the charge-transport timescale and (b) the two-particle encounter timescale. Figures (a) and (b) show the same dynamics, with the horizontal axis rescaled from $t/N$ to $t/N^2$.
(c) Difference between the averaged subsystem purity and its $U(1)$-symmetric Haar-random value, plotted on a logarithmic scale against $t/N^2$.
The curves are obtained from the exact expression for the averaged subsystem purity in Eq.~\eqref{eq:purity_decomposition_into_modes}, with the coefficients and decay rates determined for each system size $N$. The results are shown for $N=32,128,512$, illustrating their approach to the large-$N$ limit.
}
\label{fig:purity_dynamics}
\end{center}
\end{figure*}

\subsection{Subsystem purity} \label{ss:purity}

We first consider the dynamics of the subsystem purity.
Let $A$ denote the $\lfloor N/2 \rfloor$-qubit subsystem consisting of sites $1\leq i \leq \lfloor N/2 \rfloor$ and let $B$ denote its complement.
The subsystem purity of an $N$-qubit pure state $\ket{\psi}$ is defined as $p_{\!A}(\ket{\psi}) \equiv \tr_{A}\!\left[ \tr_{B}[\ket{\psi} \bra{\psi}]^2\right]$, where $\tr_{A}$ and $\tr_{B}$ denote the partial traces over $A$ and $B$, respectively.
The averaged subsystem purity is known to be expressed in terms of the second moment operator of the ensemble $\nu$.
Specifically, it is given by
\begin{align}
    &\underset{1\leq i<j\leq N}{\mathbb{E}}\, \underset{U\sim \nu^{*t}}{\mathbb{E}} [p_{\!A}(U\ket{i,j})]
    = \bra{F_{\!A}} M^{(2)}_{\nu^{*t}} \ket{\operatorname{unif}_2}, \label{eq:purity_rep_2nd_moment}\\
    &\ket{\operatorname{unif}_2}\equiv \underset{1\leq i<j\leq N}{\mathbb{E}} \left[\ket{i,j}^{\otimes 4} \right], \quad
    \ket{F_{\!A}} \equiv \bigotimes_{i\in A} \ket{F}_i \otimes \bigotimes_{i\in B} \ket{I}_i, \nonumber
\end{align}
where $\ket{i,j}$ for $1\leq i<j\leq N$ is the computational-basis state in which sites $i$ and $j$ are occupied and all other sites are unoccupied.
Here, $\ket{I}_i\equiv \sum_{x,y \in \{0,1\}} \! \ket{xyxy}_i$ and $\ket{F}_i\equiv \sum_{x,y \in \{0,1\}}\! \ket{xyyx}_i$ are the vectorizations of the identity and swap operators on two copies of the $i$-th qubit, respectively.

Furthermore, since the moment operator of the single-edge circuit is Hermitian and positive semidefinite, its $t$-fold iteration admits the spectral decomposition
\begin{equation}
\label{eq:2nd_moment_op_diagonalization}
    M^{(2)}_{\nu^{*t}} = P^{(2)}_{U(1)\mathrm{Haar}} + \sum_{i} e^{-\lambda_i t} \ket{v_{i}}\! \bra{v_i},
\end{equation}
where the sum runs over the normalized eigenvectors $\ket{v_i}$ orthogonal to the range of $P^{(2)}_{U(1)\mathrm{Haar}}$.
The decay rates satisfy $\lambda_i \geq \Delta_{\nu}^{(2)} = \Omega(\lambda^{\rm enc}_{G_{\rm all}})$, as follows from the spectral gap lower bound established in Sec.~\ref{s:gap_lower_bound}.
By substituting this spectral decomposition into Eq.~\eqref{eq:purity_rep_2nd_moment}, we see that the averaged purity can be expressed as a sum of a stationary contribution given by the Haar-random average and exponentially decaying modes with relaxation rates $\{\lambda_i\}_i$.

In particular, in the current setup, we can obtain the exact time evolution of the averaged purity defined as $\bar{p}_A(t) \equiv \underset{1\leq i<j\leq N}{\mathbb{E}}\,\underset{U\sim \nu_{G_{\rm all}}^{*t}}{\mathbb{E}} [p_{\!A}(U\ket{i,j})]$ as follows:
\begin{equation}
\label{eq:purity_decomposition_into_modes}
    \bar{p}_{A}(t) = \bar{p}_{A}^{\rm Haar} + C_{\rm slow} e^{-\lambda_{\rm slow} t} + \sum_{i\in\{1,2\}}C_{i} e^{-\lambda_{i} t},
\end{equation}
where $\bar{p}_{A}^{\rm Haar} \equiv \bra{F_{\!A}} P^{(2)}_{U(1)\mathrm{Haar}} \ket{\operatorname{unif}_2}$ is the $U(1)$-symmetric Haar-random average of the subsystem purity.
Here, the averaged purity is decomposed into only three decaying modes, independent of the system size $N$, due to the site-permutation symmetry of the all-to-all circuit and the choice of the initial state.
The detailed derivation of this representation is given in the Supplemental Material.
For any fixed system size $N$, we can exactly calculate all the coefficients and decay rates in Eq.~\eqref{eq:purity_decomposition_into_modes}.
For example, their asymptotic forms are given as 
\begin{align*}
    &\bar{p}_{A}^{\rm Haar}= 1/8 + O(N^{-1}), \\
    &C_{\rm slow} = 1/8+O(N^{-1}), \quad \lambda_{\rm slow} = 8N^{-2} + O(N^{-3}),\\
    &C_1 = 1/2+O(N^{-1}),\quad \lambda_{1} = (2/3)N^{-1}+O(N^{-2}), \\
    &C_2 = 1/4+O(N^{-1}), \quad \lambda_{2} = (4/3)N^{-1}+O(N^{-2}).
\end{align*}
Thus, $\bar{p}_{A}(t)$ consists of a single slow mode satisfying $\lambda_{\rm slow} = \Theta(\lambda^{\rm enc}_{G_{\rm all}}) =\Theta(N^{-2})$ and two fast modes satisfying $\lambda_{i} = \Theta(\lambda^{\rm tr}_{G_{\rm all}}) =\Theta(N^{-1})$.
In particular, since $C_{\rm slow}=1/8+O(N^{-1})$ remains nonzero in the large-system-size limit, this slow mode indeed dominates the late-time relaxation of the subsystem purity.

The resulting purity dynamics is shown in Fig.~\ref{fig:purity_dynamics} for system sizes $N=32,128,512$.
Figure~\ref{fig:purity_dynamics}(a) shows the dynamics on the charge-transport timescale $t=\Theta(1/\lambda^{\rm tr}_{G_{\rm all}}) =\Theta(N)$.
On this timescale, the slow mode barely decays, while the two fast modes relax, leading to an intermediate plateau at $\bar{p}_{A}^{\rm Haar} + C_{\rm slow}= 1/4 + O(N^{-1})$ for $N\ll t\ll N^2$.
As the system size $N$ increases, the relaxation dynamics approaches the dashed curve, which represents the asymptotic limit $N \to \infty$.
In Fig.~\ref{fig:purity_dynamics}(b), we show the same purity dynamics on the two-particle encounter timescale $t=\Theta(1/\lambda^{\rm enc}_{G_{\rm all}}) =\Theta(N^2)$.
On this timescale, the relaxation of the fast modes occurs in a vanishingly narrow region near $t/N^2=0$ as $N\to\infty$.
This figure shows that the relaxation of purity to the steady-state value $\bar{p}_{A}^{\rm Haar}$ indeed occurs on the two-particle encounter timescale.
We can also observe the dominance of the slow mode in Fig.~\ref{fig:purity_dynamics}(c), which plots $\bar{p}_{A}(t)-\bar{p}_{A}^{\rm Haar}$ on a logarithmic scale to highlight the slowest decay rate.

\begin{figure*}[]
\begin{center}
\includegraphics[width=0.99\textwidth]{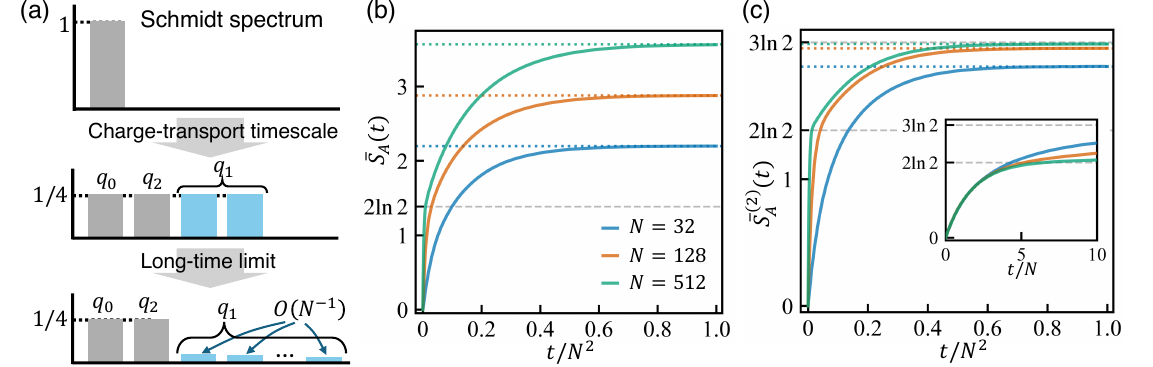}
\caption{
(a) Schematic of the two-step relaxation of the Schmidt spectrum in the single-edge all-to-all random circuit.
On the charge-transport timescale $t=\Theta(1/\lambda^{\rm tr}_{G_{\rm all}})=\Theta(N)$, the probabilities of finding $n_A=0,1,$ and $2$ particles in subsystem $A$ approach $q_0\simeq 1/4$, $q_1\simeq 1/2$, and $q_2\simeq 1/4$, respectively.
However, on this timescale, the entanglement structure of $\ket{\Psi_{11}}$ in Eq.~\eqref{eq:Schmidt_decomp_particle_number} remains constrained, with its squared Schmidt coefficients typically satisfying $\mu_1\simeq1/2$, $\mu_2\simeq1/2$, and $\mu_i=0$ for $i\geq 3$.
At sufficiently long times, the entanglement structure approaches that of a Haar-random state. Specifically, $\ket{\Psi_{11}}$ becomes highly entangled, with squared Schmidt coefficients typically satisfying $\mu_i=O(N^{-1})$ for $1\leq i\leq\lfloor N/2\rfloor$.
(b) Dynamics of the von Neumann entanglement entropy and (c) the second R\'enyi entanglement entropy. 
Each curve is obtained by averaging over $20{,}000$ independently sampled trajectories, with error bars omitted because they are smaller than the line widths.
The dotted horizontal lines indicate the corresponding Haar-random average values $\bar{S}_{\!A}^{\rm Haar}$ and $\bar{S}_{\!A}^{(2),\rm Haar}$ for each system size. The inset of (c) shows the dynamics on the charge-transport timescale.
Figures (b) and (c) demonstrate that, for large system sizes, the entanglement entropies exhibit two-step relaxation: they first grow toward the intermediate value $2\ln 2$ on the charge-transport timescale and then slowly approach their respective steady values on the two-particle encounter timescale.
These results, together with the purity dynamics shown in Fig.~\ref{fig:purity_dynamics}, are consistent with the physical picture of the evolution of the Schmidt spectrum illustrated in (a).
}
\label{fig:entanglement_dynamics}
\end{center}
\end{figure*}

\subsection{Entanglement entropy} \label{ss:entanglement_entropy}

We next consider the dynamics of entanglement entropy in this model.
Unlike purity, entanglement entropy is not a polynomial function of the state, and its ensemble average cannot be directly expressed in terms of a fixed-order moment operator.
It is therefore difficult to obtain an exact decomposition into decaying modes analogous to Eq.~\eqref{eq:purity_decomposition_into_modes}.
Instead, we provide an intuitive physical picture of entanglement relaxation by considering the Schmidt decomposition of the evolving pure state.
This picture qualitatively explains why the subsystem purity and entanglement entropy do not fully relax on the charge-transport timescale, but exhibit slow relaxation on the two-particle encounter timescale.
We further support this picture with numerical results at the end of this subsection.

A general pure state in the two-particle subspace can be written as $\ket{\psi}=\sum_{1\leq i<j\leq N} x_{ij} \ket{i,j}$, where $\{x_{ij}\}_{i<j}$ are the superposition amplitudes. 
To analyze the Schmidt coefficients of this state, it is useful to introduce the following decomposition according to the particle number in subsystem $A$:
\begin{equation}
\label{eq:Schmidt_decomp_particle_number}
\ket{\psi}= \sqrt{q_{0}} \ket{\bm 0}_{\!A} \ket{\psi_2}_{\!B} + \sqrt{q_{2}} \ket{\psi_2^\prime }_{\!A} \ket{\bm 0}_{\!B} + \sqrt{q_{1}}\ket{\Psi_{11}},
\end{equation}
where $q_{0}$, $q_{1}$, and $q_{2}$ denote the probabilities of finding $n_A=0,1,$ and $2$ particles in subsystem $A$, respectively, and $\ket{\bm 0}_{\!A}$ and $\ket{\bm 0}_{\!B}$ denote the vacuum states of the respective subsystems.
These probabilities and the normalized states are given by
\begin{align*}
&q_{0} = \sum_{i,j\in B} |x_{ij}|^2, \quad \ket{\psi_2}_{\!B}\equiv \sum_{i,j\in B} \frac{x_{ij}}{\sqrt{q_0}} \ket{i,j}_{\!B}, \\
&q_{1} = \sum_{i\in A,\,j\in B} |x_{ij}|^2,\quad \ket{\Psi_{11}} \equiv \sum_{i \in A,\,j \in B} \frac{x_{ij} }{\sqrt{q_1}} \ket{i,j}, \\
&q_{2} = \sum_{i,j\in A} |x_{ij}|^2,\quad \ket{\psi_2^\prime }_{\!A} \equiv\sum_{i,j\in A}\frac{x_{ij}}{\sqrt{q_2}} \ket{i,j}_{\!A},
\end{align*}
and $\{\mu_1,\mu_2,\dots,\mu_{\lfloor N/2 \rfloor}\}$ denotes the Schmidt spectrum of $\ket{\Psi_{11}}$, which is the set of eigenvalues of its reduced density operator.
Then, the squared Schmidt coefficients of the entire pure state $\ket{\psi}$ are given by $q_{0}$, $q_{2}$, and $q_{1}\mu_i$ for $1\leq i \leq \lfloor N/2 \rfloor$.
In particular, the subsystem purity is given by $p_{\!A}(\ket{\psi})=q_0^2+q_2^2+q_1^2\sum_i\mu_i^2$.

For example, when $\ket{\psi}$ is Haar-randomly chosen from the two-particle subspace, the probabilities typically satisfy $q_{0} \simeq 1/4$, $q_{1}\simeq 1/2$, and $q_{2}\simeq 1/4$.
Furthermore, $\ket{\Psi_{11}}$ typically has full Schmidt rank, with $\mu_i=O(N^{-1})$ for all $i$.
This is consistent with the Haar-random average of the subsystem purity, $\bar{p}_A^{\rm Haar} = 1/8 + O(N^{-1})$.

We now consider the time evolution of the Schmidt spectrum in the $U(1)$-symmetric random circuit.
On the charge-transport timescale, the two particles spread across the entire system, and the probabilities $q_0,q_1,$ and $q_2$ approach their respective steady-state values, $q_{0}\simeq 1/4$, $q_{1}\simeq 1/2$, and $q_{2}\simeq 1/4$.
However, the Schmidt coefficients of $\ket{\Psi_{11}}$ do not reach their Haar-random values on this timescale, since the amplitudes $\{x_{ij}\}_{i<j}$ satisfy nontrivial constraints.
Specifically, the squared Schmidt coefficients typically satisfy $\mu_1 \simeq 1/2$, $\mu_2 \simeq 1/2$, and $\mu_{i}=0$ for $i \geq 3$ on this timescale, as explained below.
At sufficiently long times, these constraints are relaxed, and the Schmidt spectrum of $\ket{\Psi_{11}}$ approaches that of the Haar-random state.
This two-step relaxation of the Schmidt spectrum is illustrated in Fig.~\ref{fig:entanglement_dynamics}(a).

The key to understanding the constraint on $\ket{\Psi_{11}}$ on the charge-transport timescale is that the two particles initially located at sites $i$ and $j$ evolve independently as long as their light cones do not overlap.
For a given circuit realization, let $\operatorname{LC}_t^i$ and $\operatorname{LC}_t^j$ denote the sets of sites reachable from the initial sites $i$ and $j$, respectively, through the gates applied up to depth $t$.
More precisely, starting from $\operatorname{LC}_0^i=\{i\}$ and $\operatorname{LC}_0^j=\{j\}$, each gate on an edge $e$ adds both endpoints of $e$ to a light cone if $e$ intersects that light cone.
As long as the two light cones remain disjoint, i.e., $\operatorname{LC}_t^i \cap \operatorname{LC}_t^j = \varnothing$, the amplitudes can be represented as
\begin{equation}
\label{eq:no_interaction_two_part_coeff}
x_{rs} = u_rv_s + u_s v_r \qquad (r<s),
\end{equation}
where $\bm u$ and $\bm v$ are normalized $N$-dimensional vectors supported on $\operatorname{LC}_t^i$ and $\operatorname{LC}_t^j$, respectively.
These vectors describe the single-particle amplitudes associated with the two initial particles.

To see why the factorized form in Eq.~\eqref{eq:no_interaction_two_part_coeff} holds, we note the following two facts.
First, the initial state satisfies this form with $u_l = \delta_{li}$ and $v_l = \delta_{lj}$.
Second, if this representation holds before the application of a two-qubit gate on an edge $e\equiv \{l_1,l_2\}$, it continues to hold even after the gate application, as long as the two light cones remain disjoint.
Indeed, under this condition, the edge $e$ intersects at most one of the two light cones.
Therefore, assuming $e\cap\operatorname{LC}_t^j=\varnothing$ without loss of generality and writing the two-qubit gate as Eq.~\eqref{eq:two_qubit_gate_matrix}, the vectors can be updated as $\bm v'=\bm v$, $u_l'=e^{i\theta_{00}}u_l$ for $l\not= l_1,l_2$, $u_{l_1}'=d u_{l_1}+c u_{l_2}$, and $u_{l_2}'=b u_{l_1}+a u_{l_2}$.
Thus, the factorized form in Eq.~\eqref{eq:no_interaction_two_part_coeff} is preserved.

When Eq.~\eqref{eq:no_interaction_two_part_coeff} holds, the state vector $\ket{\Psi_{11}}$ satisfies
\begin{equation}
\sqrt{q_1}\ket{\Psi_{11}} = \sum_{i \in A,\,j \in B} x_{ij} \ket{i}_{\!A} \ket{j}_{\!B}
=\ket{\bm u}_{\!A}\ket{\bm v}_{\!B}+\ket{\bm v}_{\!A}\ket{\bm u}_{\!B},
\end{equation}
where $\ket{i}$ denotes the computational-basis state with only site $i$ occupied.
Here, $\ket{\bm u}_{\!A} \equiv \sum_{i\in A} u_i \ket{i}_{\!A}$, $\ket{\bm u}_{\!B} \equiv \sum_{j \in B} u_j \ket{j}_{\!B}$, $\ket{\bm v}_{\!A} \equiv \sum_{i \in A} v_i \ket{i}_{\!A}$, and $\ket{\bm v}_{\!B} \equiv \sum_{j \in B} v_j \ket{j}_{\!B}$.
This representation shows that the Schmidt rank of $\ket{\Psi_{11}}$ is at most two.
In particular, when each of $\bm u$ and $\bm v$ has approximately equal weight in $A$ and $B$, the two nonzero squared Schmidt coefficients are approximately $1/2$.

In the single-edge all-to-all random circuit, the typical timescale for the two light cones to first overlap is $t_{\rm LC}=\Theta(N\log N)$, which is parametrically longer than the charge-transport timescale $1/\lambda^{\rm tr}_{G_{\rm all}}=\Theta(N)$.
Thus, the rank-two constraint of $\ket{\Psi_{11}}$ persists beyond the charge-transport timescale.
Even after the first overlap, the Schmidt coefficients do not immediately approach their Haar-random values, since most amplitudes $\{x_{ij}\}_{i<j}$ retain the approximately factorized structure in Eq.~\eqref{eq:no_interaction_two_part_coeff}.
For complete relaxation, the residual correlations among the amplitudes must be sufficiently randomized through repeated applications of two-qubit gates on randomly selected edges.
Since removing these correlations requires interactions that couple the two particles, we expect this relaxation to occur on the much longer two-particle encounter timescale $1/\lambda^{\rm enc}_{G_{\rm all}}=\Theta(N^2)$.

Finally, we examine the dynamics of the averaged von Neumann and second R\'enyi entanglement entropies, which can be understood based on this physical picture.
Specifically, we consider
\begin{align*}
&\bar{S}_{\! A}(t) = \underset{1\leq i<j\leq N}{\mathbb{E}}\, \underset{U\sim \nu_{G_{\rm all}}^{*t}}{\mathbb{E}}
\left[-\tr \left[\rho_{\!A}^{U;i,j}\log \rho_{\!A}^{U;i,j}\right]\right],\\
&\bar{S}_{\! A}^{(2)}(t) = \underset{1\leq i<j\leq N}{\mathbb{E}}\, \underset{U\sim \nu_{G_{\rm all}}^{*t}}{\mathbb{E}}
\left[-\log \tr\left[\left(\rho_{\!A}^{U;i,j}\right)^2\right]\right],
\end{align*}
where $\rho_{\!A}^{U;i,j} \equiv \tr_B[U\ket{i,j}\bra{i,j}U^\dagger]$ denotes the reduced density operator on subsystem $A$.
The corresponding $U(1)$-symmetric Haar-random averages have the following asymptotic forms \cite{murciano2022symmetry}:
\begin{align*}
&\bar{S}_{\! A}^{\rm Haar} = \frac{1}{2} \log N + \log 2-\frac{1}{4} + O\left(N^{-1} \log N \right), \\
&\bar{S}_{\! A}^{(2),\rm Haar} = 3 \log 2 + O\left(N^{-1} \right).
\end{align*}
As shown in Fig.~\ref{fig:entanglement_dynamics}(b), the von Neumann entanglement entropy first grows toward the intermediate value $2\log 2$ on the charge-transport timescale, and then slowly approaches the steady value $\bar{S}_{\! A}^{\rm Haar}$ on the two-particle encounter timescale.
This two-step relaxation is consistent with the physical picture illustrated in Fig.~\ref{fig:entanglement_dynamics}(a).
A similar two-step relaxation is also observed for the second R\'enyi entanglement entropy, as shown in Fig.~\ref{fig:entanglement_dynamics}(c).
These numerical results demonstrate that the slow relaxation associated with the two-particle encounter mechanism is clearly visible in standard measures of entanglement.

We conclude this subsection by commenting on the generality of the physical picture presented here.
Although we have focused on the two-particle sector $n=2$ for simplicity, the existence of slow modes in broader dilute and dense charge sectors suggests that similar relaxation behavior would occur more generally in sectors $n=o(\sqrt{N})$ and $n=N-o(\sqrt{N})$.
Furthermore, while we have only addressed the single-edge all-to-all circuit here, we expect the same mechanism to be relevant to other circuit architectures.
The essential physical intuition is that, on the charge-transport timescale, particles can spread throughout the system while retaining an approximately independent-particle structure, which imposes nontrivial constraints on the Schmidt spectrum. 
On the two-particle encounter timescale, repeated encounters allow local gates to randomize amplitudes across different particle configurations, gradually removing these constraints and driving the Schmidt spectrum toward that of a Haar-random state.

\section{Summary and outlook} \label{s:summary_outlook}

In this work, we determine the optimal scaling of the spectral gap for $U(1)$-symmetric random circuits across a broad range of circuit geometries by establishing matching upper and lower bounds.
In particular, for $2\leq k\leq O(\sqrt{\log N})$, we show that the slowest relaxation is governed by the two-particle encounter process, whose decay rate is parametrically smaller than the single-particle transport rate in many circuit geometries, as summarized in Table~\ref{tab:gap_summary}.
We first derive the spectral gap upper bound for the second moment $k=2$ in Theorem~\ref{thm:gap_upper_bound}, which also applies to all $k\geq2$.
We further show that this slow-mode mechanism is not restricted to a particular charge sector, but generically persists in sufficiently dilute and dense charge sectors with total particle or hole number $o(\sqrt{N})$, respectively, as shown in Eq.~\eqref{eq:gap_upper_bound_robust_n}.
We then establish a matching lower bound for $2\leq k\leq O(\sqrt{\log N})$ in Theorem~\ref{thm:gap_lower_bound}.
Together, these results determine the optimal scaling of the randomness formation rate in $U(1)$-symmetric random circuits.

Beyond characterizing the bottleneck associated with the two-particle encounter mode, we also show that it can be avoided by allowing symmetry-breaking local gates while preserving the $U(1)$ symmetry of the overall unitary.
In Sec.~\ref{s:efficient_construction}, we present an explicit protocol realizing this idea.
Theorem~\ref{thm:efficient_construction_with_asymmetric_gates} shows that, on lattices and in the all-to-all geometry, this protocol constructs a constant-gap circuit unit with a parametrically smaller circuit depth than that required by the corresponding symmetric random circuits.
In particular, on lattices, our construction even circumvents the single-particle transport bottleneck by using routing via matchings.
This result demonstrates that allowing asymmetric local gates can substantially speed up the generation of global symmetry-preserving randomness.

Thus, our results establish a scaling-optimal characterization of $U(1)$-symmetric unitary $k$-design formation for $k\leq O(\sqrt{\log N})$, a regime that includes any fixed moment order in the large-system-size limit.
Remarkably, across the broad range of circuit architectures, the two-particle encounter mechanism sets the slowest relaxation timescale up to constant factors, despite substantial differences in the underlying geometries.
Furthermore, our efficient construction protocol with symmetry-breaking local gates suggests that this bottleneck originates from the difficulty of efficiently generating global random phases using only symmetry-preserving local gates, whereas temporary symmetry breaking enables a much faster implementation.
Together, these results reshape the picture of randomness formation under symmetry constraints, and provide guiding principles both for its characterization and efficient generation.

This work opens numerous directions for future research.
First, it is important to understand the spectral gap scaling and the slowest relaxation mode at higher moment orders.
While our work established the optimal scaling of the spectral gap for $k\leq O(\sqrt{\log N})$, the behavior at higher moment orders remains open.
In particular, for $k\geq 2(N-1)$, the $k$-th moment of a $U(1)$-symmetric random circuit composed of two-qubit gates is known to fail to converge to that of the $U(1)$-symmetric Haar-random ensemble even at infinite depth~\cite{mitsuhashi2024characterization,mitsuhashi2024unitary,liu2024unitary}.
Understanding how this breakdown of Haar convergence is connected to the unitary design formation rate at lower moment orders would be an interesting direction for future work.

Another important direction is to develop shallower protocols for generating symmetric unitary designs.
Our efficient construction given in Sec.~\ref{s:efficient_construction} still yields only an $\widetilde{O}(N)$-depth construction for additive-error and relative-error $\varepsilon$-approximate unitary $k$-designs for fixed $k$ and $\varepsilon$, even in the all-to-all interaction geometry.
This remains substantially larger than the $O(\log N)$ depth achievable without symmetry constraints~\cite{schuster2024random,laracuente2024approximate}, motivating further improvements for practical applications.
Given the robust slow-mode mechanism found in this work, it may be promising to use additional resources, such as symmetry-breaking local gates, ancilla qubits, measurements, and feedback.

It would also be intriguing to consider randomness formation rates under other continuous symmetries.
For non-Abelian symmetries, most notably $SU(2)$ and more generally $SU(d)$, recent works have characterized the infinite-depth design properties of local symmetric circuits, while tight analytic bounds on their convergence rates and the corresponding slowest relaxation modes remain largely open~\cite{li2024designs,mitsuhashi2024characterization,liu2024unitary}.
It would therefore be interesting to determine whether few-body slow modes analogous to the two-particle encounter mode arise in these settings.
The proof framework for deriving lower bounds developed in this work, which does not rely on standard bootstrap methods, would also be useful for establishing matching spectral gap lower bounds for such non-Abelian symmetries.

Furthermore, our results may have implications for the study of low-energy properties of frustration-free Hamiltonians~\cite{gosset2016local,masaoka2024quadratic,masaoka2025rigorous,lemm2025critical}.
Recent works have revealed strong constraints on the low-energy behavior of gapless frustration-free systems, including a dynamical exponent $z\geq2$ under broad conditions~\cite{masaoka2024quadratic,masaoka2025rigorous,lemm2025critical}.
For local random circuits, the moment operator can be equivalently described by an associated frustration-free local Hamiltonian.
In this Hamiltonian description, our two-particle encounter mode provides an explicit mechanism for an excitation parametrically softer than the quadratic Goldstone mode: for the associated $k=2$ Hamiltonian on an $\alpha$-dimensional lattice with $\alpha\geq3$, the dynamical exponent is $z=\alpha$.
This example suggests that embeddings of substochastic dynamics into low-energy sectors may provide a useful perspective for understanding mechanisms that give rise to softer excitations beyond $z=2$ in frustration-free systems.

From the perspective of quantum many-body dynamics, it is also natural to ask whether the two-particle encounter mode has an analogue in isolated quantum systems.
Previous works have shown that symmetric random circuits can reproduce qualitative features of generic isolated quantum many-body dynamics arising from conservation laws, with representative examples including diffusive hydrodynamic tails in operator spreading~\cite{khemani2018operator,rakovszky2018diffusive} and sub-ballistic growth of R\'enyi entanglement entropies~\cite{rakovszky2019sub,zhou2020diffusive,huang2019dynamics,huang2020dynamics}.
These close connections suggest the possibility that slow modes first identified in random circuits may point to previously unnoticed relaxation mechanisms in isolated quantum systems.

\begin{acknowledgments}
The author thanks Yosuke Mitsuhashi and Ryotaro Suzuki for bringing concurrent related work to the author's attention and for valuable discussions.
The author also thanks Donghoon Kim and Tomotaka Kuwahara for fruitful discussions.
This work is supported by JSPS KAKENHI Grant No. JP26KJ0403.
\end{acknowledgments}

\vspace*{1em}
\noindent
\textbf{Data availability:}
The numerical data and code used to generate Figs.~\ref{fig:purity_dynamics} and \ref{fig:entanglement_dynamics} are available from the author upon reasonable request.

\vspace*{1em}
\noindent
\textbf{Use of generative AI:}
The author used OpenAI's ChatGPT (GPT-5.6) to explore certain properties of the 2RPmix group relevant to Theorem~\ref{thm:gap_lower_bound} and the applicability of routing via matchings in Theorem~\ref{thm:efficient_construction_with_asymmetric_gates}.
These AI-assisted ideas were subsequently examined and independently verified by the author.
ChatGPT was also used to check the mathematical arguments throughout the manuscript and to proofread and polish its English.

\vspace*{1em}
\noindent
\textbf{Note added:}
During the completion of this work, the author became aware of concurrent work by Schuster, Suzuki, McGinley, Mitsuhashi, Vardhan, and Preskill \cite{schustersuzuki2026}.
They establish the formation of approximate partial unitary $k$-designs in one-dimensional $U(1)$-symmetric random circuits with
$O(\log k+\log^{(t)}N)$-local gates for any fixed positive integer $t$, where $\log^{(t)}$ denotes the $t$-fold iterated logarithm.
Their work addresses the charge-sector blocks $(\bm n,\overline{\bm n})$ in which both the particle and hole numbers are extensive in every copy, i.e., $n_i,n_{\bar i},N-n_i,N-n_{\bar i}=\Omega(N)$ for all $i\in[k]$, and establishes $\varepsilon$-approximate partial $k$-design formation in circuit depth $O\left(N^2\polylog(Nk)\log(1/\varepsilon)\right)$.
Their result is complementary to ours: they establish design formation in the bulk-charge subspace on a nearly diffusive timescale, whereas our work determines the formation rate on the full moment space, including dilute and dense charge sectors outside their design guarantee.
Note that the two-particle encounter bottleneck identified in this work arises in these dilute and dense sectors.

\bibliography{biblio}

\clearpage

\appendix

\end{document}


\title{Supplemental Material for ``Optimal Scaling of Unitary Design Formation in \(U(1)\)-Symmetric Random Circuits: A Bottleneck Slower than Charge Transport''}

\author{Toshihiro Yada}
\email{toshihiroyada.physics@gmail.com}
\affiliation{Analytical Quantum Complexity RIKEN Hakubi Research Team, RIKEN Center for Quantum Computing (RQC), Wako, Saitama 351-0198, Japan}



\maketitle

\onecolumngrid

\tableofcontents

\vspace{4mm}\noindent
We first outline the structure of this Supplemental Material.
In Sec.~\ref{s:preliminary}, we introduce the basic notions of $U(1)$-symmetric unitary designs used throughout the manuscript.
In Sec.~\ref{s:main_result}, we describe the setup considered in this work, review the open problems left by previous work, and summarize our main results, while clarifying the correspondence between the theorems in the main text and those in this Supplemental Material.
In Sec.~\ref{s:auxiliary-permutation-phase}, we introduce several auxiliary groups that play an essential role in our proof.
In Sec.~\ref{s:gap_upper_bound}, we introduce the two-particle encounter process and show that it yields upper bounds on the spectral gap.
In Sec.~\ref{s:two_particle_encounter_rate}, we summarize the decay rates of the two-particle encounter process in representative graph geometries and briefly explain their derivations.
In Sec.~\ref{s:architecture_comparison_argument}, we present standard circuit architecture comparison arguments that relate the results for different circuit structures to one another.
In Sec.~\ref{s:lower_bound_spectral_gap}, we present the overall proof strategy for our spectral gap lower bounds.
Since the proof requires substantial technical preparation, we defer the two key ingredients to the following two sections.
In Sec.~\ref{s:generate_2RPmix_group_with_circuit}, we show that the local generation rate of the 2RPmix group is governed by the two-particle encounter process.
In Sec.~\ref{s:gap_doped_2RPmix_circuit}, we prove that the doped 2RPmix circuit has a constant spectral gap for $k\leq O(\sqrt{\log N})$.
In Sec.~\ref{s:gap_scaling_k=1}, we determine the spectral gap scaling for the first moment $k=1$.
In Sec.~\ref{s:efficient_const_asym}, we present an efficient construction of $U(1)$-symmetric unitary designs using symmetry-breaking local gates.
Finally, in Sec.~\ref{s:purity_all}, we derive an exact expression for the dynamics of subsystem purity in the single-edge all-to-all random circuit.

\section{Preliminary} \label{s:preliminary}

\subsection{Moment operators for $U(1)$-symmetric unitary ensembles}\label{subsec:u1_haar_moment}

In this work, we consider probability distributions over $U(1)$-symmetric unitaries acting on an $N$-qubit system.
The Hilbert space decomposes into particle-number sectors as
\begin{equation}
    \mathcal H=\bigoplus_{n=0}^{N}\mathcal H_n,
    \qquad
    \mathcal H_n\equiv\operatorname{span}\left\{\ket{x}:x\in\{0,1\}^{N},\ |x|=n\right\}.
\end{equation}
We denote the group of $U(1)$-symmetric unitaries by
\begin{equation}
    \mathcal U_{U(1)}
    \equiv
    \left\{
        \bigoplus_{n=0}^{N}U_n:
        U_n\in\mathrm U(\mathcal H_n)
    \right\}.
\end{equation}
Accordingly, any $U \in  \mathcal U_{U(1)}$ can be written in the block-diagonal form
\begin{equation}
    U=\bigoplus_{n=0}^{N}U_n,
    \qquad
    U_n\in\mathrm{U}(\mathcal H_n).
\end{equation}
We denote the dimension of $\mathcal{H}$ by $D\equiv 2^N$, and the dimension of $\mathcal H_n$ by
\begin{equation}
    D_n\equiv\dim\mathcal H_n=\binom{N}{n}.
\end{equation}

For an ensemble $\nu$ of $U(1)$-symmetric unitaries, we define its $k$-th moment operator as
\begin{equation}
\label{seq:def_moment_operator}
    M_{\nu}^{(k)}\equiv\mathbb E_{U\sim\nu}\left[U^{\otimes k,k}\right],
\end{equation}
where $U^{\otimes k,k}\equiv U^{\otimes k}\otimes U^{*\otimes k},$ and $U^*$ denotes the complex conjugate of $U$ with respect to the computational basis.
Since every $U\sim\nu$ preserves the particle number separately in each ket and bra copy, the moment operator is block diagonal with respect to the corresponding charge sectors.
We label these sectors by $(\bm n,\overline{\bm n})= \left((n_1,\ldots,n_k)^T,(n_{\overline{1}},\ldots,n_{\overline{k}})^T\right)$, and denote the corresponding Hilbert subspace as $\mathcal{H}_{\bm n,\overline{\bm n}}$.
We further define the projector onto that subspace as
\begin{equation}
    \Pi_{\bm n,\overline{\bm n}}
    \equiv
    \sum_{\substack{
    x^{1},\ldots,x^{k},\,y^{1},\ldots,y^{k}\in\{0,1\}^{N}\\
    |x^\alpha|=n_\alpha,\ |y^\alpha|=n_{\overline{\alpha}}\ \forall\alpha\in[k]
    }}
    \ket{\bm x,\bm y}\bra{\bm x,\bm y},
\end{equation}
where $\bm x=(x^{1},\ldots,x^{k})$ and $\bm y=(y^{1},\ldots,y^{k})$, and $[k]\equiv\{1,\ldots,k\}.$
Then, $\left[M_{\nu}^{(k)},\Pi_{\bm n,\overline{\bm n}}\right]=0$ for every charge sector $(\bm n,\overline{\bm n})$.
Therefore,
\begin{equation}
    M_{\nu}^{(k)}
    =
    \sum_{\bm n,\overline{\bm n}}
    \Pi_{\bm n,\overline{\bm n}}M_{\nu}^{(k)}\Pi_{\bm n,\overline{\bm n}}
    \equiv
    \sum_{\bm n,\overline{\bm n}}M_{\nu;\bm n,\overline{\bm n}}^{(k)},
\end{equation}
where
\begin{equation}
    M_{\nu;\bm n,\overline{\bm n}}^{(k)}
    \equiv
    \Pi_{\bm n,\overline{\bm n}}M_{\nu}^{(k)}\Pi_{\bm n,\overline{\bm n}}
\end{equation}
denotes the restriction of the moment operator to the charge sector $(\bm n,\overline{\bm n})$.

In particular, we denote the $k$-th moment operator of the Haar measure on the group of $U(1)$-symmetric unitaries by $P_{U(1)\mathrm{Haar}}^{(k)}$.
The moment operator of the Haar measure on any compact unitary group is an orthogonal projector, and throughout this work we use $P$ rather than $M$ for such projectors.
Moreover, for any ensemble $\nu$ supported on the group of $U(1)$-symmetric unitaries, the left and right invariance of the Haar measure gives
\begin{equation}
    P_{U(1)\mathrm{Haar}}^{(k)}M_{\nu}^{(k)}
    =
    M_{\nu}^{(k)}P_{U(1)\mathrm{Haar}}^{(k)}
    =
    P_{U(1)\mathrm{Haar}}^{(k)}.
\end{equation}
We similarly denote its restriction to a charge sector by
\begin{equation}
    P_{U(1)\mathrm{Haar};\bm n,\overline{\bm n}}^{(k)}
    \equiv
    \Pi_{\bm n,\overline{\bm n}}
    P_{U(1)\mathrm{Haar}}^{(k)}
    \Pi_{\bm n,\overline{\bm n}},
\end{equation}
which is itself an orthogonal projector.

We next characterize the range of $P_{U(1)\mathrm{Haar};\bm n,\overline{\bm n}}^{(k)}$.
For $n\in\{0,\ldots,N\}$, define
\begin{equation}
    \mu_n(\bm n)\equiv\#\left\{\alpha\in[k]:n_\alpha=n\right\},
    \qquad
    \mu_n(\overline{\bm n})\equiv\#\left\{\alpha\in[k]:n_{\overline{\alpha}}=n\right\}.
\end{equation}
Since the Haar measure is invariant under independent phase rotations of the particle-number blocks, $U_n\mapsto e^{i\theta_n}U_n$, it follows that
$P_{U(1)\mathrm{Haar};\bm n,\overline{\bm n}}^{(k)}=0$ unless
\begin{equation}
\label{eq:number_balance_condition}
    \mu_n(\bm n)=\mu_n(\overline{\bm n})
    \qquad
    \text{for every }n\in\{0,\ldots,N\}.
\end{equation}
In other words, the entries of $\bm n$ and $\overline{\bm n}$ must coincide as multisets, in order for $P_{U(1)\mathrm{Haar};\bm n,\overline{\bm n}}^{(k)}$ to be nonzero.

For any charge sector $(\bm n,\overline{\bm n})$, we define the following subset of the symmetric group $S_k$:
\begin{equation}
\label{eq:sigma_charge_sector}
    \Sigma(\bm n,\overline{\bm n})
    \equiv
    \left\{
    \sigma\in S_k:
    \overline{\bm n}=\sigma\bm n
    \right\},
\end{equation}
where the permutation acts as $(\sigma\bm n)_\alpha\equiv n_{\sigma^{-1}(\alpha)}.$
Then, condition~\eqref{eq:number_balance_condition} is equivalent to $\Sigma(\bm n,\overline{\bm n})\neq\varnothing.$
With this notation, the range of $P_{U(1)\mathrm{Haar};\bm n,\overline{\bm n}}^{(k)}$ is given by \cite{hearth2025unitary}
\begin{equation}
    \operatorname{Ran}P_{U(1)\mathrm{Haar};\bm n,\overline{\bm n}}^{(k)}
    =
    \operatorname{span}
    \left\{
    \ket{\bm n;\sigma}:
    \sigma\in\Sigma(\bm n,\overline{\bm n})
    \right\},
\end{equation}
where $\operatorname{Ran}$ denotes the range of an operator, and $\ket{\bm n;\sigma}$ is called a number-permutation state.
This number-permutation state is defined as
\begin{equation}
\label{eq:number_permutation_state}
    \ket{\bm n;\sigma}
    \equiv
    \frac{1}{\sqrt{Z_{\bm n}}}
    \sum_{x^1\in\mathcal X_{n_1},\ldots,x^k\in\mathcal X_{n_k}}
    \ket{\bm x,\sigma\bm x},
\end{equation}
where the permutation of $k$-copy states is defined as $(\sigma\bm x)^\alpha\equiv x^{\sigma^{-1}(\alpha)}$, 
\begin{equation}
\label{eq:comp_basis_set_weight_n}
    \mathcal X_n \equiv  \{x\in\{0,1\}^N:|x|=n\}
\end{equation}
is the set of computational-basis labels with particle number $n$, and 
\begin{equation}
\label{eq:normalizing_factor_number_permutation_basis}
    Z_{\bm n} \equiv \prod_{\alpha=1}^{k}D_{n_\alpha}
\end{equation}
is the normalization constant so that $\langle \bm n;\sigma |\bm n;\sigma \rangle =1$.

\subsection{Approximate unitary designs under $U(1)$ symmetry}\label{subsec:approximate_unitary_design}

We next introduce several notions of approximate unitary designs under $U(1)$ symmetry.
The notion used primarily in this work quantifies the difference between moment operators in the operator norm.
\vspace*{-0.5em}
\begin{definition}[$U(1)$-symmetric approximate unitary design in the operator-norm sense]\label{def:mom_approx_unitary_design}
Let $k\in\mathbb N$ and $\varepsilon\geq0$.
A probability distribution $\nu$ on $\mathcal U_{U(1)}$ is an $\varepsilon$-approximate unitary $k$-design in the operator-norm sense if
\begin{equation}
\label{eq:def_mom_approx_unitary_design}
\left\|M_{\nu}^{(k)}-P_{U(1)\mathrm{Haar}}^{(k)}\right\|_{\infty}\leq\varepsilon.
\end{equation}
\end{definition}
\noindent 
This notion is also referred to as an approximate unitary design in the $2$-to-$2$ norm sense.

Two other standard notions are additive-error and relative-error approximate unitary designs.
To define them, we introduce the $k$-fold twirling channel
\begin{equation}
\label{eq:twirling_channel}
\Phi_{\nu}^{(k)}(X)\equiv\mathbb E_{U\sim\nu}\left[U^{\otimes k}XU^{\dagger\otimes k}\right],
\end{equation}
acting on operators on $\mathcal H^{\otimes k}$.
We denote the corresponding Haar twirling channel by $\Phi_{U(1)\mathrm{Haar}}^{(k)}$.

\begin{definition}[Additive-error and relative-error approximate unitary designs under $U(1)$ symmetry]\label{def:twirl_approx_unitary_design}
For $k\in\mathbb N$ and $\varepsilon\geq0$, a probability distribution $\nu$ on $\mathcal U_{U(1)}$ is an additive-error $\varepsilon$-approximate unitary $k$-design if
\begin{equation}
\label{eq:def_add_k_des}
\left\|\Phi_{\nu}^{(k)}-\Phi_{U(1)\mathrm{Haar}}^{(k)}\right\|_{\diamond}\leq\varepsilon,
\end{equation}
where the diamond norm is defined by $\|\mathcal E\|_{\diamond} \equiv \sup_{R,X\neq 0} \frac{\left\|(\mathcal E\otimes\mathrm{id}_{R})(X)\right\|_1}{\|X\|_1}$.
For $\varepsilon\geq 0$, the ensemble $\nu$  is a relative-error $\varepsilon$-approximate unitary $k$-design if
\begin{equation}
\label{eq:def_rel_k_des}
(1-\varepsilon)\Phi_{U(1)\mathrm{Haar}}^{(k)}
\preccurlyeq
\Phi_{\nu}^{(k)}
\preccurlyeq
(1+\varepsilon)\Phi_{U(1)\mathrm{Haar}}^{(k)},
\end{equation}
where $\mathcal E\preccurlyeq\mathcal F$ means that $\mathcal F-\mathcal E$ is completely positive.
\end{definition}

It is known that these three definitions of approximate unitary designs are quantitatively related \cite{brandao2016local,li2024designs,hearth2025unitary}:
\begin{lemma}\label{lem:mom_to_design}
Let $\nu$ be a probability distribution on $\mathcal U_{U(1)}$.
If $\nu$ is a $D^{-2k}\varepsilon$-approximate $k$-design in the operator-norm sense, then it is also a relative-error $\varepsilon$-approximate $k$-design and an additive-error $D^{-k}\varepsilon$-approximate $k$-design.
Moreover, if $\nu$ is a relative-error $\delta$-approximate $k$-design, it is also a $2\delta D^{k/2}$-approximate $k$-design in the operator-norm sense, and an additive-error $2\delta$-approximate $k$-design.
\end{lemma}
\noindent
This lemma shows that an operator-norm error at most $D^{-k}\varepsilon$ is sufficient for an additive-error $\varepsilon$-approximate design, while an operator-norm error at most $D^{-2k}\varepsilon$ is sufficient for a relative-error $\varepsilon$-approximate design.

\subsection{Spectral gap}\label{subsec:spectral_gap}

We now introduce the quantity that controls the convergence of the moment operator under repeated circuit layers.
Let $\nu$ denote the ensemble corresponding to one layer of a random circuit, and let $\nu^{*t}$ denote its $t$-fold convolution.
The corresponding moment operator satisfies $M_{\nu^{*t}}^{(k)}=\left(M_{\nu}^{(k)}\right)^t$. 

The key quantity that determines the convergence rate of the moment operator is the $k$-th moment spectral gap, defined as
\begin{equation}
\label{eq:def_spectral_gap}
\Delta_{\nu}^{(k)}
\equiv
1-\left\|M_{\nu}^{(k)}-P_{U(1)\mathrm{Haar}}^{(k)}\right\|_{\infty}.
\end{equation}
Since $M_{\nu}^{(k)}P_{U(1)\mathrm{Haar}}^{(k)}=P_{U(1)\mathrm{Haar}}^{(k)}M_{\nu}^{(k)}=P_{U(1)\mathrm{Haar}}^{(k)}$, we have $0\leq\Delta_{\nu}^{(k)}\leq1$.
Furthermore, since both $M_{\nu}^{(k)}$ and $P_{U(1)\mathrm{Haar}}^{(k)}$ are block diagonalized into the charge sectors, we can also define the spectral gap for each charge sector $(\bm n,\overline{\bm n})$ as
\begin{equation}
\label{eq:def_spectral_gap_charge_sector}
\Delta_{\nu;\bm n,\overline{\bm n}}^{(k)}
\equiv
1-\left\|M_{\nu;\bm n,\overline{\bm n}}^{(k)}-P_{U(1)\mathrm{Haar};\bm n,\overline{\bm n}}^{(k)}\right\|_{\infty}.
\end{equation}
By definition, we have
\begin{equation}
\label{eq:spectral_gap_min_form_charge_sectors}
    \Delta_{\nu}^{(k)} = \min_{(\bm n,\overline{\bm n})} \Delta_{\nu;\bm n,\overline{\bm n}}^{(k)}.
\end{equation}

The following elementary lemma relates this spectral gap to the convergence of the moment operator:

\begin{lemma}[Moment convergence from the spectral gap]\label{lem:gap_to_moment_error}
For any probability distribution $\nu$ on $\mathcal U_{U(1)}$ and any $t\in\mathbb N$,
\begin{equation}
\label{eq:gap_to_error_ineq}
    \left\|M_{\nu^{*t}}^{(k)}-P_{U(1)\mathrm{Haar}}^{(k)}\right\|_{\infty}
    \leq
    \exp\left(-\Delta_{\nu}^{(k)}t\right).
\end{equation}
If $M_{\nu}^{(k)}$ is Hermitian, then more precisely
\begin{equation}
\label{eq:gap_to_error_eq}
    \left\|M_{\nu^{*t}}^{(k)}-P_{U(1)\mathrm{Haar}}^{(k)}\right\|_{\infty}
    =
    \left(1-\Delta_{\nu}^{(k)}\right)^t.
\end{equation}
\end{lemma}

\begin{proof}
Using $M_{\nu}^{(k)}P_{U(1)\mathrm{Haar}}^{(k)}=P_{U(1)\mathrm{Haar}}^{(k)}M_{\nu}^{(k)}= P_{U(1)\mathrm{Haar}}^{(k)},$
we obtain $M_{\nu^{*t}}^{(k)}-P_{U(1)\mathrm{Haar}}^{(k)}=\left(M_{\nu}^{(k)}-P_{U(1)\mathrm{Haar}}^{(k)}\right)^t.$
Therefore, by submultiplicativity of the operator norm,
\begin{align}
    \left\|M_{\nu^{*t}}^{(k)}-P_{U(1)\mathrm{Haar}}^{(k)}\right\|_{\infty}
    &\leq \left\|M_{\nu}^{(k)}-P_{U(1)\mathrm{Haar}}^{(k)}\right\|_{\infty}^{t} \nonumber \\
    &= \left(1-\Delta_{\nu}^{(k)}\right)^t \nonumber \\
    &\leq \exp\left(-\Delta_{\nu}^{(k)}t\right),
\end{align}
where the last inequality follows from $1-x\leq e^{-x}$.

If $M_{\nu}^{(k)}$ is Hermitian, then $M_{\nu}^{(k)}-P_{U(1)\mathrm{Haar}}^{(k)}$ is also Hermitian.
The spectral theorem therefore gives
\begin{equation}
    \left\|\left(M_{\nu}^{(k)}-P_{U(1)\mathrm{Haar}}^{(k)}\right)^t\right\|_{\infty}=\left\|M_{\nu}^{(k)}-P_{U(1)\mathrm{Haar}}^{(k)}\right\|_{\infty}^{t},
\end{equation}
which proves Eq.~\eqref{eq:gap_to_error_eq}.
\end{proof}

Thus, the spectral gap determines the exponential convergence rate of the $k$-th moment operator toward the $U(1)$-symmetric Haar moment operator.
Together with Lemma~\ref{lem:mom_to_design}, this relation can be converted into upper bounds on the depths required for the different notions of approximate unitary designs.

\subsection{Circuit depth required for approximate unitary designs}\label{subsec:design_depth_consequences}

In this subsection, we show the quantitative relationship between the spectral gap and the circuit depths required for approximate unitary designs.
We first explicitly define the required circuit depth for an $\varepsilon$-approximate unitary design under each definition as follows:
\begin{align}
    T_{\nu,\varepsilon,k}^{\mathrm{op}}
    &\equiv
    \min\left\{
        t\in\mathbb N:
        \left\|M_{\nu^{*t}}^{(k)}-P_{U(1)\mathrm{Haar}}^{(k)}\right\|_{\infty}\leq\varepsilon
    \right\}, \\
    T_{\nu,\varepsilon,k}^{\mathrm{add}}
    &\equiv
    \min\left\{
        t\in\mathbb N:
        \nu^{*t}\text{ is an additive-error }\varepsilon\text{-approximate unitary }k\text{-design}
    \right\}, \\
    T_{\nu,\varepsilon,k}^{\mathrm{rel}}
    &\equiv
    \min\left\{
        t\in\mathbb N:
        \nu^{*t}\text{ is a relative-error }\varepsilon\text{-approximate unitary }k\text{-design}
    \right\}.
\end{align}

Then, by combining Lemmas~\ref{lem:mom_to_design} and~\ref{lem:gap_to_moment_error}, we obtain the following upper bounds on these depths in terms of the spectral gap, for $0<\varepsilon<1$ and $0<\Delta_{\nu}^{(k)}<1$:
\begin{align}
    T_{\nu,\varepsilon,k}^{\mathrm{op}}
    &\leq
    \left\lceil
        \frac{1}{\Delta_{\nu}^{(k)}}
        \log\frac{1}{\varepsilon}
    \right\rceil, \\
    T_{\nu,\varepsilon,k}^{\mathrm{add}}
    &\leq
    \left\lceil
        \frac{1}{\Delta_{\nu}^{(k)}}
        \left(
            kN\log 2+\log\frac{1}{\varepsilon}
        \right)
    \right\rceil, \\
    T_{\nu,\varepsilon,k}^{\mathrm{rel}}
    &\leq
    \left\lceil
        \frac{1}{\Delta_{\nu}^{(k)}}
        \left(
            2kN\log 2+\log\frac{1}{\varepsilon}
        \right)
    \right\rceil.
\end{align}
Here, we consider an $N$-qubit system whose total dimension is $D=2^N$, which yields the dimension-dependent factors $kN\log 2$ and $2kN\log 2$ for additive- and relative-error designs, respectively.
Therefore, a lower bound on the spectral gap $\Delta_{\nu}^{(k)}$ immediately gives upper bounds on the required circuit depths.

Furthermore, in some cases, we can also derive lower bounds on the required circuit depths from the spectral gap.
In particular, to derive a lower bound on $T_{\nu,\varepsilon,k}^{\mathrm{add}}$, we use the following lemma:
\begin{lemma}\label{lem:depth_LB_with_spectral_gap}
Let $k \in \mathbb{N}$ and $\varepsilon>0$, and suppose that a unitary ensemble $\mu$ is a $U(1)$-symmetric additive-error $\varepsilon$-approximate unitary $k$-design.
Then,
\begin{equation}
    \varepsilon \geq \rho\left(M_{\mu}^{(k)} - P_{U(1)\mathrm{Haar}}^{(k)}\right).
\end{equation}
\end{lemma}
\noindent
A proof of this lemma is given, e.g., in Ref.~\cite{ErratumHearthPRX2025}.

Using this lemma, together with Lemma~\ref{lem:mom_to_design} and Eq.~\eqref{eq:gap_to_error_eq} of Lemma~\ref{lem:gap_to_moment_error}, we obtain the following lower bounds for $0<\varepsilon<1/2$ and $0<\Delta_{\nu}^{(k)}<1$, when the moment operator of the ensemble $\nu$ is Hermitian, i.e., $M_{\nu}^{(k)}=M_{\nu}^{(k),\dagger}$:
\begin{align}
    T_{\nu,\varepsilon,k}^{\mathrm{op}}
    &=
    \left\lceil
        \frac{\log(1/\varepsilon)}
        {-\log\left(1-\Delta_{\nu}^{(k)}\right)}
    \right\rceil
    \geq
    \left\lceil
        \frac{1-\Delta_{\nu}^{(k)}}{\Delta_{\nu}^{(k)}}
        \log\frac{1}{\varepsilon}
    \right\rceil, \label{eq:depth_LB_op_norm}\\
    T_{\nu,\varepsilon,k}^{\mathrm{add}}
    &\geq
    \left\lceil
        \frac{\log(1/\varepsilon)}
        {-\log\left(1-\Delta_{\nu}^{(k)}\right)}
    \right\rceil
    \geq
    \left\lceil
        \frac{1-\Delta_{\nu}^{(k)}}{\Delta_{\nu}^{(k)}}
        \log\frac{1}{\varepsilon}
    \right\rceil, \label{eq:depth_LB_additive}\\
    T_{\nu,\varepsilon,k}^{\mathrm{rel}}
    &\geq
    \left\lceil
        \frac{\log(1/(2\varepsilon))}
        {-\log\left(1-\Delta_{\nu}^{(k)}\right)}
    \right\rceil
    \geq
    \left\lceil
        \frac{1-\Delta_{\nu}^{(k)}}{\Delta_{\nu}^{(k)}}
        \log\frac{1}{2\varepsilon}
    \right\rceil, \label{eq:depth_LB_relative}
\end{align}
where we used $-\log (1-x) \leq x/(1-x)$ for $0<x<1$ in the second inequalities of each line.
Here, Eq.~\eqref{eq:depth_LB_op_norm} follows directly from Eq.~\eqref{eq:gap_to_error_eq} of Lemma~\ref{lem:gap_to_moment_error}.
Equation~\eqref{eq:depth_LB_additive} follows from Lemma~\ref{lem:depth_LB_with_spectral_gap} and the following equation:
\begin{align*}
    \rho\left(M_{\nu^{*t}}^{(k)} - P_{U(1)\mathrm{Haar}}^{(k)}\right)
    &= \left[\rho\left(M_{\nu}^{(k)} - P_{U(1)\mathrm{Haar}}^{(k)}\right)\right]^t \\
    &= \left\|M_{\nu}^{(k)} - P_{U(1)\mathrm{Haar}}^{(k)}\right\|_{\infty}^t \\
    &= \left[1-\Delta_{\nu}^{(k)}\right]^t,
\end{align*}
where the second equality follows from the Hermiticity of $M_{\nu}^{(k)}-P_{U(1)\mathrm{Haar}}^{(k)}$.
We further obtain Eq.~\eqref{eq:depth_LB_relative} from Eq.~\eqref{eq:depth_LB_additive} using the fact that a relative-error $\varepsilon$-approximate unitary $k$-design is also an additive-error $2\varepsilon$-approximate unitary $k$-design, as stated in Lemma~\ref{lem:mom_to_design}.

These inequalities show that, when the Hermiticity condition $M_{\nu}^{(k)}=M_{\nu}^{(k),\dagger}$ is satisfied, upper bounds on the spectral gap $\Delta_{\nu}^{(k)}$ yield lower bounds on the required circuit depths.

\section{Summary of main results} \label{s:main_result}

In this section, we introduce the setup addressed in this work and summarize our main results.
In Sec.~\ref{ss:circuit_structure}, we define the circuit architectures and notation used throughout this Supplemental Material.
In Sec.~\ref{ss:prev_work_open}, we review previous results on the formation rate of $U(1)$-symmetric unitary designs and summarize the current open problems. In Sec.~\ref{ss:main_result}, we state our main results.

\subsection{Setup}\label{ss:circuit_structure}

We distinguish three basic circuit architectures: single-edge random circuits, parallel random circuits, and fixed-architecture random circuits.
In the first two architectures, the locations of the gates are chosen randomly at each step, whereas in a fixed-architecture circuit the gate locations and their order are prescribed in advance.

We first introduce the single-edge random circuit on a connected graph $G=(V,E)$ with $|V|=N$.
At each step, an edge $\{i,j\}\in E$ is chosen uniformly at random, and an independent two-qubit $U(1)$-symmetric Haar-random gate is applied to the corresponding pair of qubits.
The moment operator of one step is
\begin{equation}
\label{eq:single_edge_Haar_G_def}
M_{\nu_G}^{(k)}\equiv\frac{1}{|E|}\sum_{\{i,j\}\in E} P_{i,j}^{(k)}\otimes\mathbb I_{\overline{i,j}},
\end{equation}
where $P_{i,j}^{(k)}$ is the $k$-th moment operator of the $U(1)$-symmetric Haar-random unitary ensemble on sites $i$ and $j$, and $\mathbb I_{\overline{i,j}}$ denotes the identity on all sites other than $i$ and $j$.

We briefly introduce several restrictions on the graph $G=(V,E)$ that will be required for some of our main results.
We call $G$ a \emph{bounded-degree graph} if the degree of every vertex is bounded above by a constant independent of the system size $N$, where the degree of a vertex is the number of edges incident to it.
We also sometimes require that $G$ admit at least one perfect or near-perfect matching.
A matching $\gamma$ of $G$ is a subset of mutually disjoint edges, which we write as
\begin{equation}
\gamma\equiv\left\{\{i_1^\gamma,i_2^\gamma\},\{i_3^\gamma,i_4^\gamma\},\ldots,\{i_{2|\gamma|-1}^\gamma,i_{2|\gamma|}^\gamma\}\right\},
\end{equation}
with $|\gamma|\leq\lfloor N/2\rfloor$.
For even $N$, a perfect matching is a matching satisfying $|\gamma|=N/2$, while for odd $N$, a near-perfect matching is a matching satisfying $|\gamma|=(N-1)/2$.
In particular, if $G$ has a Hamiltonian path, namely, a path that visits every vertex exactly once, then $G$ necessarily admits a perfect or near-perfect matching.

We next consider parallel random circuits, in which several disjoint two-qubit gates are applied simultaneously at each step.
Such an architecture is specified by a family of matchings $\Gamma =\{\gamma\}$. We call $\Gamma$ a constant-size matching family when the number of matchings is constant independent of system size, i.e., $|\Gamma|=O(1)$.
We can associate with $\Gamma$ the graph
\begin{equation}
G_\Gamma\equiv(V,E_\Gamma),\qquad E_\Gamma\equiv\bigcup_{\gamma\in\Gamma}\gamma.
\end{equation}
Throughout this work, we assume that $G_\Gamma$ is connected, so that information can propagate throughout the entire system.

At each step of the parallel random circuit, a matching $\gamma\in\Gamma$ is chosen uniformly at random, and independent two-qubit $U(1)$-symmetric Haar-random gates are applied simultaneously to all edges in $\gamma$.
For later use, we define the moment operator associated with a fixed matching $\gamma$ by
\begin{equation}
\label{eq:Pgamma_Haar_def}
P_{\gamma}^{(k)}\equiv\left[\bigotimes_{j=1}^{|\gamma|}P_{i_{2j-1}^\gamma,i_{2j}^\gamma}^{(k)}\right]\otimes\mathbb I_{\overline{\gamma}},
\end{equation}
where $\mathbb I_{\overline{\gamma}}$ acts on the sites not covered by $\gamma$.
The moment operator of one parallel layer is then given by
\begin{equation}
\label{eq:parallel_Haar_Gamma_def}
M_{\nu_\Gamma}^{(k)}\equiv\frac{1}{|\Gamma|}\sum_{\gamma\in\Gamma}P_{\gamma}^{(k)}.
\end{equation}

We next introduce a random circuit unit with a fixed architecture, denoted by $A$, in which the gate locations and their order are fixed in advance, while the local gates applied at these locations are independently Haar random.
The architecture is specified by an ordered sequence of matchings $A=(\gamma_1,\ldots,\gamma_{d_A})$, where $\gamma_i$ specifies the gate locations in the $i$-th layer and $d_A$ denotes the depth of the circuit unit $A$.
We define the graph associated with $A$ by
\begin{equation}
    G_A \equiv (V,E_A),\qquad E_A\equiv\bigcup_{i=1}^{d_A}\gamma_i.
\end{equation}
Throughout this work, we assume that $G_A$ is connected.
The moment operator corresponding to one circuit unit of the fixed architecture is
\begin{equation}
\label{eq:fixed_architecture_Haar_G_def}
M_{\nu_A}^{(k)}\equiv P_{\gamma_{d_A}}^{(k)}\cdots P_{\gamma_2}^{(k)}P_{\gamma_1}^{(k)}.
\end{equation}
Thus, unlike the single-edge and parallel circuits in Eqs.~\eqref{eq:single_edge_Haar_G_def} and \eqref{eq:parallel_Haar_Gamma_def}, respectively, the moment operator of a fixed-architecture circuit is an ordered product rather than an average over gate locations.
The brickwork random circuit is a representative example of this architecture.

\vspace{1em}
\begin{figure*}[]
    \centering
    \includegraphics[width=0.95\textwidth]{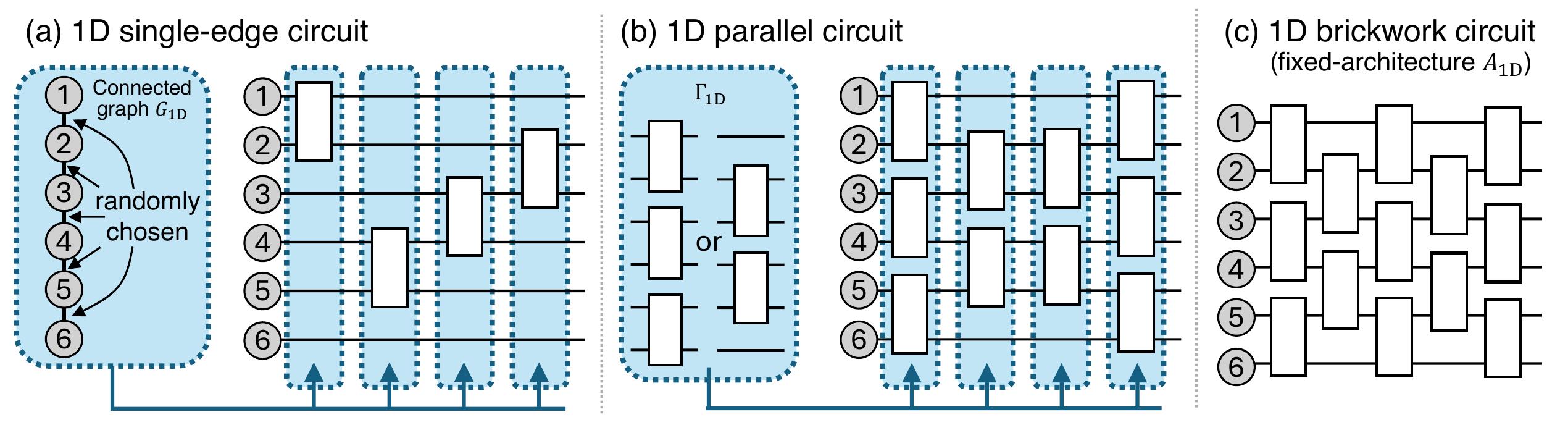}
    \caption{The schematics for (a) 1D single-edge circuit, (b) 1D parallel circuit, and (c) 1D brickwork circuit.}
    \label{fig:1D_circuit_structures}
\end{figure*}

Here, we provide representative examples of the circuit structures considered in this work.
In particular, we consider $\alpha$-dimensional lattice geometries, expander graphs, and the all-to-all interaction geometry discussed in the main text.

For the $\alpha$-dimensional lattice geometry, the associated graph is the $\alpha$-dimensional nearest-neighbor lattice, where the number of sites along each spatial dimension is $\Theta(N^{1/\alpha})$.
Although Fig.~\ref{fig:1D_circuit_structures}(a) illustrates the open-boundary case, periodic boundary conditions can be treated in the same manner, with the same asymptotic spectral-gap scaling.
For fixed spatial dimension $\alpha=O(1)$, the lattice is a bounded-degree graph and admits a perfect or near-perfect matching.
For the parallel circuit, we decompose the nearest-neighbor edges along each of the $\alpha$ spatial directions into two alternating matchings, as illustrated for the one-dimensional case in Fig.~\ref{fig:1D_circuit_structures}(b).
When the number of sites along each spatial direction is even, the resulting matching family therefore has size $|\Gamma|=2\alpha$, and is a constant-size matching family when $\alpha=O(1)$.
For the $\alpha$-dimensional brickwork circuit, these $2\alpha$ matchings are applied sequentially in a fixed order.
Thus, the depth of one circuit unit is $d_A=2\alpha$, so the architecture has constant depth when $\alpha=O(1)$.

We next consider expander graphs.
An expander graph is a bounded-degree graph that remains highly connected as the system size increases \cite{hoory2006expander}.
With the single-edge normalization used in this work, the single-particle transport rate on an expander graph $G_{\rm exp}$ scales as $\lambda_{G_{\rm exp}}^{\rm tr}=\Theta(N^{-1})$, which is the same scaling as for the complete graph.
Among the various classes of expanders, we mainly consider regular bipartite expanders \cite{marcus2015interlacing} and vertex-transitive expanders \cite{godsil2001algebraic} that admit at least one perfect or near-perfect matching.
For the parallel circuit associated with an expander graph $G_{\rm exp}=(V,E_{\rm exp})$, we decompose $E_{\rm exp}$ into a family of matchings.
By Vizing's theorem~\cite{vizing1964estimate}, the edge set of a graph of maximum degree $d_{G_{\rm exp}}$ can be decomposed into at most $d_{G_{\rm exp}}+1$ matchings.
Since $G_{\rm exp}$ is bounded degree, this gives a constant-size matching family $\Gamma_{\rm exp}$ satisfying $|\Gamma_{\rm exp}|=O(1)$.
The corresponding fixed-architecture circuit is obtained by applying the matchings $\gamma\in\Gamma_{\rm exp}$ sequentially in any chosen fixed order.
Consequently, the resulting architecture $A_{\rm exp}$ also has constant depth, $d_{A_{\rm exp}}=O(1)$.

For the all-to-all interaction geometry, the associated graph $G_{\rm all}$ is the complete graph, while the associated matching family $\Gamma_{\rm all}$ consists of all perfect matchings for even $N$ and all near-perfect matchings for odd $N$.
Accordingly, $G_{\rm all}$ is not bounded degree, and $\Gamma_{\rm all}$ is not a constant-size matching family.

\vspace{1em}

Although most of this Supplemental Material concerns random circuits composed of two-qubit gates, we also consider circuits composed of general $\ell$-qubit $U(1)$-symmetric gates when establishing the robustness of our spectral-gap upper bound.
For this purpose, let $G=(V,E)$ be an $\ell$-uniform hypergraph, where $E\subseteq\binom{V}{\ell}.$
Thus, every hyperedge $e \in E$ consists of exactly $\ell$ vertices. For a hyperedge $e=\{i_1,\ldots,i_\ell\}\in E$, we denote by $P_e^{(k)}\equiv P_{i_1,\ldots,i_\ell}^{(k)}$ the $k$-th moment projector of the $\ell$-qubit $U(1)$-symmetric Haar ensemble acting on $e$.
The moment operator of the corresponding single-hyperedge random circuit is
\begin{equation}
\label{eq:l_local_single_edge_def}
M_{\nu_{G,\ell\mathrm{-loc}}}^{(k)}\equiv\frac{1}{|E|}\sum_{e\in E}P_e^{(k)}\otimes\mathbb I_{\overline e}.
\end{equation}

The parallel and fixed-architecture versions are defined analogously.
In the $\ell$-local setting, each matching $\gamma$ is a collection of mutually disjoint hyperedges.
For a fixed matching $\gamma$, we define
\begin{equation}
P_{\gamma,\ell\mathrm{-loc}}^{(k)}\equiv\left[\bigotimes_{e\in\gamma}P_e^{(k)}\right]\otimes\mathbb I_{\overline{\gamma}}.
\end{equation}
The moment operator of one parallel layer is then
\begin{equation}
\label{eq:l_local_parallel_def}
M_{\nu_{\Gamma,\ell\mathrm{-loc}}}^{(k)}\equiv\frac{1}{|\Gamma|}\sum_{\gamma\in\Gamma}P_{\gamma,\ell\mathrm{-loc}}^{(k)}.
\end{equation}
Similarly, for a fixed architecture $A_G=(\gamma_1,\ldots,\gamma_d)$ associated with the hypergraph $G$, we define
\begin{equation}
\label{eq:l_local_fixed_def}
M_{\nu_{A_G,\ell\mathrm{-loc}}}^{(k)}\equiv P_{\gamma_d,\ell\mathrm{-loc}}^{(k)}\cdots P_{\gamma_2,\ell\mathrm{-loc}}^{(k)}P_{\gamma_1,\ell\mathrm{-loc}}^{(k)}.
\end{equation}

\subsection{Previous works and open problems}\label{ss:prev_work_open}

Although $U(1)$-symmetric unitary designs have recently been actively studied, most previous works have focused on characterizing the $k$-th moments obtained in the infinite-depth limit.
By contrast, the rate at which a finite-depth random circuit approaches these moments, namely, the formation rate of $U(1)$-symmetric unitary designs, remains much less understood.
In this subsection, we summarize the previous works most closely related to this question~\cite{hearth2025unitary,li2024efficient} and clarify the open problems addressed in this work.

First, Ref.~\cite{hearth2025unitary} analytically derived spectral-gap upper bounds for $U(1)$-symmetric random circuits on $\alpha$-dimensional lattices.
These upper bounds originate from the diffusion of the conserved charge and correspond to a single-particle transport mode.
The authors further conjectured that this diffusive mode gives the slowest relaxation mode of the circuit, or equivalently, that the spectral gap obeys a matching lower bound with the same asymptotic scaling.
Although Ref.~\cite{hearth2025unitary} mainly analyzes fixed-architecture circuits, in particular brickwork circuits on finite-dimensional lattices, 
the corresponding bounds for the single-edge and parallel random circuits can be immediately obtained from their result/conjecture through standard architecture-comparison arguments based on the detectability lemma~\cite{aharonov2009detectability,anshu2016simple} and the quantum union bound~\cite{gao2015quantum}.
Therefore, we also include these scalings in Table~\ref{tab:gap_summary}.

Second, Ref.~\cite{li2024efficient} derived a spectral-gap lower bound for the second moment, $k=2$, for Convolutional Quantum Alternating (CQA) circuits in one-dimensional and all-to-all geometries.
A layer of the CQA construction contains a two-qubit gate together with a Wick-projection step implemented by random diagonal phase rotations.
In particular, the phase rotation contains independent random two-body phases associated with all pairs of sites.
Therefore, a single such layer involves many mutually overlapping long-range interactions. On the other hand, the circuits considered in our work are built from local gates whose allowed supports are determined by the underlying circuit geometry.
Accordingly, their layer depths are not directly comparable.
Nevertheless, in the all-to-all setting, a comparison can still be made in terms of the total number of two-qubit gates: a naive implementation of a constant-gap circuit unit in the CQA construction requires $O(N^3)$ gates, whereas our $U(1)$-symmetric random-circuit construction requires $\Theta(N^2)$ gates.

Thus, while spectral-gap upper bounds associated with charge transport were already known, it remained unclear whether this mode is indeed the slowest relaxation mode.
In particular, establishing a matching spectral-gap lower bound for standard $U(1)$-symmetric random circuits was a fundamental open problem.
However, this lower-bound problem is known to be technically challenging.
As discussed in Ref.~\cite{li2024designs}, bootstrap methods such as the martingale method and Knabe-type bounds, which have been powerful tools for random circuits without symmetry, are difficult to apply directly in the presence of continuous symmetry.

\begin{table}[t]
    \centering
    \renewcommand{\arraystretch}{1.2}
    \setlength{\tabcolsep}{5pt}
    \resizebox{\textwidth}{!}{
    \begin{tabular}{cc|cc|cc}
    \hline
    \multicolumn{2}{c|}{} &
    \multicolumn{2}{c|}{Previous work~\cite{hearth2025unitary} ($k <N$)} &
    \multicolumn{2}{c}{Our results} \\
    \cline{3-6}
    \multicolumn{2}{c|}{Circuit structure} &
    Upper bound &
    Conjectured lower bound &
    Upper bound ($k\geq 2$) &
    \makecell{Lower bound \\ ($k\leq O( \sqrt{\log N})$)} \\
    \hline\hline

    \multirow{5}{*}{\makecell{Single-\\edge}}
    & 1D
    & $O(N^{-3})^{*}$
    & $\Omega(N^{-3})^{*}$
    & $O(N^{-3})$
    & $\Omega(N^{-3})$ \\

    & 2D
    & $O(N^{-2})^{*}$
    & $\Omega(N^{-2})^{*}$
    & \red{$O(N^{-2}(\log N)^{-1})$}
    & \red{$\Omega(N^{-2}(\log N)^{-1})$} \\

    & $\alpha$D ($\alpha\geq 3$)
    & $O(N^{-1-2/\alpha})^{*}$
    & $\Omega(N^{-1-2/\alpha})^{*}$
    & \red{$O(N^{-2})$}
    & \red{$\Omega(N^{-2})$} \\

    & Expander
    & $-$
    & $-$
    & \red{$O(N^{-2})$}
    & \red{$\Omega(N^{-2})$} \\

    & All-to-all
    & $-$
    & $-$
    & \red{$O(N^{-2})$}
    & \red{$\Omega(N^{-2})$} \\
    \hline

    \multirow{5}{*}{Parallel}
    & 1D
    & $O(N^{-2})^{*}$
    & $\Omega(N^{-2})^{*}$
    & $O(N^{-2})$
    & $\Omega(N^{-2})$ \\

    & 2D
    & $O(N^{-1})^{*}$
    & $\Omega(N^{-1})^{*}$
    & \red{$O(N^{-1}(\log N)^{-1})$}
    & \red{$\Omega(N^{-1}(\log N)^{-1})$} \\

    & $\alpha$D ($\alpha\geq 3$)
    & $O(N^{-2/\alpha})^{*}$
    & $\Omega(N^{-2/\alpha})^{*}$
    & \red{$O(N^{-1})$}
    & \red{$\Omega(N^{-1})$} \\

    & Expander
    & $-$
    & $-$
    & \red{$O(N^{-1})$}
    & \red{$\Omega(N^{-1})$} \\

    & All-to-all
    & $-$
    & $-$
    & \red{$O(N^{-1})$}
    & \red{$\Omega(N^{-1})$} \\
    \hline

    \multirow{4}{*}{\makecell{Fixed-\\architecture}}
    & 1D
    & $O(N^{-2})$
    & $\Omega(N^{-2})$
    & $O(N^{-2})$
    & $\Omega(N^{-2})$ \\

    & 2D
    & $O(N^{-1})$
    & $\Omega(N^{-1})$
    & \red{$O(N^{-1}(\log N)^{-1})$}
    & \red{$\Omega(N^{-1}(\log N)^{-1})$} \\

    & $\alpha$D ($\alpha\geq 3$)
    & $O(N^{-2/\alpha})$
    & $\Omega(N^{-2/\alpha})$
    & \red{$O(N^{-1})$}
    & \red{$\Omega(N^{-1})$} \\

    & Expander
    & $-$
    & $-$
    & \red{$O(N^{-1})$}
    & \red{$\Omega(N^{-1})$} \\
    \hline
    \end{tabular}
    }
    \caption{
    Summary of the spectral gap upper and lower bounds for $U(1)$-symmetric Haar-random circuits.
    The previous upper bounds and conjectured lower bounds are based on the single-particle transport mode identified in Ref.~\cite{hearth2025unitary}.
    The entries marked by an asterisk are not explicitly shown in that work, but can be immediately obtained from the corresponding fixed-architecture results through architecture-comparison arguments.
    Our results identify a slower two-particle encounter mode and establish matching lower bounds.
    The lower bounds hold for $k\leq O(\sqrt{\log N})$, while the upper bounds already hold for $k=2$ and hence for every $k\geq2$. 
    }
    \label{tab:gap_summary}
\end{table}

\subsection{Our main results}\label{ss:main_result}

In this work, we resolve the fundamental open problems described in Sec.~\ref{ss:prev_work_open}.
We obtain matching upper and lower bounds for the spectral gap for a wide range of circuit structures, including those listed in Table~\ref{tab:gap_summary}.
In this subsection, we summarize our main results in turn and clarify the correspondence between the theorem and equation numbers in the main text and those in this Supplemental Material.

First, we identify a two-particle encounter mode that relaxes more slowly than the single-particle transport mode in a broad class of geometries.
For any connected graph $G$, any constant-size matching family $\Gamma$, and any constant-depth fixed architecture $A$, we obtain
\begin{align}
    &\Delta_{\nu_G}^{(2)}\leq\lambda_G^{\rm enc}, \label{summaryeq:UB_sing_two_particle_encounter}\\
    &\Delta_{\nu_\Gamma}^{(2)}\leq O\left(N\lambda_{G_\Gamma}^{\rm enc}\right),\label{summaryeq:UB_par_two_particle_encounter} \\
    &\Delta_{\nu_{A}}^{(2)}\leq O\left(N\lambda_{G_A}^{\rm enc}\right),\label{summaryeq:UB_fix_two_particle_encounter}
\end{align}
which constitute Theorem~1 in the main text.
In the Supplemental Material, we prove these upper bounds in Sec.~\ref{s:gap_upper_bound}.
In particular, Eq.~\eqref{summaryeq:UB_sing_two_particle_encounter} is derived in Theorem~\ref{restatethm:gap_upper_bound_two_particle_encounter}, while Eqs.~\eqref{summaryeq:UB_par_two_particle_encounter} and \eqref{summaryeq:UB_fix_two_particle_encounter} are given in Corollary~\ref{restatecor:gap_upper_bound_two_particle_encounter_parallel_fixed}.
Furthermore, we derive charge-sector-wise upper bounds and upper bounds for random circuits composed of $\ell$-qubit gates, which are given as Eqs.~(13) and (14), respectively, in the main text.
Both of these upper bounds are consequences of Theorem~\ref{restatethm:universal_gap_upper_bound}, which is also proved in Sec.~\ref{s:gap_upper_bound}.

Second, we derive matching lower bounds for the spectral gap for a wide range of circuit structures.
Specifically, let $G$ be any bounded-degree graph that admits a perfect or near-perfect matching, let $\Gamma$ be any constant-size matching family (i.e., $|\Gamma|=O(1)$) whose associated graph $G_\Gamma$ admits a perfect or near-perfect matching, and let $A$ be any fixed architecture of constant depth $d_A=O(1)$ whose associated graph $G_A$ admits a perfect or near-perfect matching.
Then, for $2\leq k\leq O(\sqrt{\log N})$, we obtain
\begin{align}
    &\Delta_{\nu_{G}}^{(k)} \geq \Omega\left(\lambda_{G}^{\rm enc}\right), \label{summaryeq:lower_bound_single_edge_bounded_degree} \\
    &\Delta_{\nu_{\Gamma}}^{(k)} \geq \Omega\left(N \lambda_{G_\Gamma}^{\rm enc}\right), \label{summaryeq:lower_bound_parallel_bounded_degree} \\
    &\Delta_{\nu_{A}}^{(k)} \geq \Omega\left(N \lambda_{G_A}^{\rm enc}\right), \label{summaryeq:lower_bound_fixed_bounded_degree} 
\end{align}
which constitute Theorem~2 in the main text.
In the Supplemental Material, this result is given as Corollary~\ref{restatecor:gap_lower_bound_sing_fix} in Sec.~\ref{ss:lower_bound_spectral_gap_single_fixed}.
The proof of these lower bounds is developed in Secs.~\ref{s:lower_bound_spectral_gap}, \ref{s:generate_2RPmix_group_with_circuit}, and \ref{s:gap_doped_2RPmix_circuit}, where Sec.~\ref{s:lower_bound_spectral_gap} presents the overall proof strategy, while Secs.~\ref{s:generate_2RPmix_group_with_circuit} and \ref{s:gap_doped_2RPmix_circuit} provide the key ingredients for the proof.

Having established the optimal spectral gap scaling for $2\leq k\leq O(\sqrt{\log N})$, we further determine the spectral gap scaling for the first moment $k=1$ in Sec.~\ref{s:gap_scaling_k=1}.
The analysis in this case is considerably simpler than that for $2\leq k\leq O(\sqrt{\log N})$, and we prove in Lemma~\ref{lem:spectral_gap_1_design} that
$\Delta_{\nu_G}^{(1)}=\lambda_G^{\rm tr}$ exactly for any connected graph $G$.
The corresponding spectral gap scalings for parallel and fixed-architecture circuits then follow directly from the architecture comparison arguments.
Thus, in contrast to the higher-moment regime considered above, $U(1)$-symmetric unitary $1$-design formation is governed by the single-particle transport process.

Finally, we propose an efficient construction of $U(1)$-symmetric unitary designs using symmetry-breaking local gates, which avoids the slowdown caused by the two-particle encounter mode.
Using this construction, we show that a constant-gap circuit unit for $k \leq O(\log N/\log \log N)$ can be generated with circuit depth
\begin{equation}
    d_{\alpha\mathrm{D}}^{\rm unit}=O\left(N^{1/\alpha} k^3 (\log k)^6\right),
\end{equation}
on an $\alpha$-dimensional lattice, and with circuit depth
\begin{equation}
    d_{\rm all}^{\rm unit}=O\left((\log N)^3(\log\log N)^2 k\,(\log k)^2\right),
\end{equation}
in the all-to-all interaction geometry.
We note that throughout this Supplemental Material, logarithmic factors $\log k$ appearing in asymptotic bounds are understood as $\max\{1,\log k\}$, to avoid the trivial issue at $k=1$.
These results are presented as Theorem~3 in the main text and are proved in Sec.~\ref{s:efficient_const_asym} as Theorem~\ref{restatethm:efficient_design_with_asymmetric_gates} in the Supplemental Material.
Unlike the random-circuit architectures introduced in Sec.~\ref{ss:circuit_structure}, this result concerns an explicitly designed architecture optimized for generating $U(1)$-symmetric unitary designs.
In this construction, only gates between sites connected by the underlying geometry are allowed, and only gates with mutually disjoint supports can be applied simultaneously within a single layer.

\section{Auxiliary site-permutation and phase rotation groups} \label{s:auxiliary-permutation-phase}

In this section, we introduce several subgroups of the group of $U(1)$-symmetric unitaries $\mathcal{U}_{U(1)}$, which play an essential role in the derivation of our main results.
In Sec.~\ref{ss:mixing}, we introduce the mixing group, consisting of all permutations of the $N$ physical sites.
In Sec.~\ref{ss:RP}, we define the RP group, consisting of arbitrary phase rotations diagonal in the computational basis.
In Sec.~\ref{ss:2RP}, we introduce the 2RP group, generated by two-qubit diagonal phase rotations acting on arbitrary pairs of sites.
Finally, in Sec.~\ref{ss:RPmix_2RPmix}, we introduce the RPmix and 2RPmix groups, defined as the subgroups generated by the mixing group together with the RP and 2RP groups, respectively.

\subsection{Mixing group} \label{ss:mixing}

In this subsection, we introduce the mixing group, denoted by $\mathcal{G}_{\rm mix}$, which is the group of all permutations of $N$ sites.
We denote the site permutation corresponding to $\pi \in S_N$ by $U_\pi$, which acts as $U_\pi\ket{x_1,x_2,\dots,x_N} = \ket{x_{\pi^{-1}(1)},x_{\pi^{-1}(2)},\dots,x_{\pi^{-1}(N)}}$. 
Then, the $k$-th moment operator for the Haar measure on the mixing group is given by
\begin{equation}
P_{\mathrm{mix}}^{(k)} =\frac{1}{N!}\sum_{\pi\in S_N}U_{\pi}^{\otimes k,k}.
\end{equation}

We first characterize the action of this moment operator on a computational-basis state, whose $k$ ket copies and $k$ bra copies are labeled by $\bm x = (x^{1},\ldots,x^{k})$ and $\bm y = (y^{1},\ldots,y^{k})$, respectively, where $x^\alpha,y^\alpha\in\{0,1\}^N$ for every $\alpha\in[k]$.
For each physical site $i\in[N]$, we collect the occupations of the $k$ ket and bra copies into $x_i=\left(x_i^{1},\ldots,x_i^{k}\right)^{T}, y_i=\left(y_i^{1},\ldots,y_i^{k}\right)^{T} \in\{0,1\}^k$. 
We call $u_i=(x_i,y_i)\in\{0,1\}^{2k}$ the \emph{local type} at site $i$.
Then, since a site permutation changes only the positions of the local types and leaves their values unchanged, we have 
\begin{equation}
\label{eq:U_pi_comp_basis_action}
    U_{\pi}^{\otimes k,k} \ket{u_1,u_2,\dots,u_N}  = \ket{u_{\pi^{-1}(1)},u_{\pi^{-1}(2)},\dots,u_{\pi^{-1}(N)}}.
\end{equation}

This equation implies that the multiplicity of each local type is invariant under site-permutation. 
We therefore introduce the notion of \emph{type vector} as
\begin{equation}
\bm{m} = \left(m_{x,y}\right)_{x,y\in\{0,1\}^k},\qquad m_{x,y}\in\mathbb{Z}_{\geq 0},
\end{equation}
where $m_{x,y}$ denotes the number of physical sites whose local type is $(x,y)$. By definition, we have 
\begin{equation}
\label{eq:cond_site_num}
\sum_{x,y\in\{0,1\}^k}m_{x,y}=N.
\end{equation}
Equation~\eqref{eq:U_pi_comp_basis_action} implies that two computational-basis configurations are related by a site permutation if and only if they have the same type vector.
Therefore, by denoting the set of computational-basis configurations having type vector $\bm m$ by $\mathcal{T}_{\bm m}$, and the number of such configurations by
\begin{equation}
D_{\bm m}
\equiv
|\mathcal{T}_{\bm m}|
=
\frac{N!}
{\prod_{x,y\in\{0,1\}^k}m_{x,y}!},
\end{equation}
the action of the moment operator $P^{(k)}_{\rm mix}$ on the computational-basis state is given by
\begin{align}
    P_{\mathrm{mix}}^{(k)}  \ket{u_1,u_2,\dots,u_N} &=\frac{1}{N!}\sum_{\pi\in S_N}U_{\pi}^{\otimes k,k} \ket{u_1,u_2,\dots,u_N} \nonumber \\
    &= \frac{1}{N!} \displaystyle \prod_{x,y\in\{0,1\}^k}m_{x,y}!\sum_{(v_1,\dots,v_N) \in \mathcal{T}_{\bm m}} \ket{v_1,v_2,\dots,v_N} \nonumber \\
    &= \frac{1}{D_{\bm m}}\sum_{(v_1,\dots,v_N) \in \mathcal{T}_{\bm m}} \ket{v_1,v_2,\dots,v_N},
\end{align}
where we used the fact that each configuration in $\mathcal T_{\bm m}$ appears exactly $\prod_{x,y}m_{x,y}!$ times in the sum over $S_N$.
Thus, by defining the normalized orbit state by 
\begin{equation}
\label{eq:perm_inv_orbit_state}
|\Omega_{\bm m}\rangle \equiv \frac{1}{\sqrt{D_{\bm m}}} \sum_{\bm u\in\mathcal{T}_{\bm m}} |\bm u\rangle,
\end{equation}
we can say that the restriction of $P_{\mathrm{mix}}^{(k)}$ to the orbit subspace $\mathrm{span}\{\ket{\bm u}:\bm u \in \mathcal{T}_{\bm m}\}$ is the rank-one projector $\ket{\Omega_{\bm m}} \bra{\Omega_{\bm m}}$. 
Since distinct type vectors correspond to disjoint computational-basis orbits, the states $|\Omega_{\bm m}\rangle$ are mutually orthogonal, and the full moment operator is given by
\begin{equation}
P_{\mathrm{mix}}^{(k)}=\sum_{\bm m:\,\sum_{x,y}m_{x,y}=N}|\Omega_{\bm m}\rangle\langle\Omega_{\bm m}|.
\end{equation}

Each orbit state $|\Omega_{\bm m}\rangle$ belongs to a definite charge sector.
Specifically, a type vector $\bm m$ belongs to the charge sector $(\bm n,\overline{\bm n}) = ((n_1,\ldots,n_k)^T,(n_{\bar 1},\ldots,n_{\bar k})^T)$, when the following conditions are satisfied:
\begin{equation}
\label{eq:cond_part_num}
\bm n=\sum_{x,y\in\{0,1\}^k}m_{x,y}x,
\qquad
\overline{\bm n}=\sum_{x,y\in\{0,1\}^k}m_{x,y}y.
\end{equation}

\subsection{RP group}\label{ss:RP}

We introduce the RP group, denoted by $\mathcal{G}_{\rm RP}$, consisting of all phase rotations diagonal in the computational basis.
An element of $\mathcal{G}_{\rm RP}$ can be written as
\begin{equation}
V(\{\theta_z\}_{z}) \equiv \sum_{z\in\{0,1\}^N} e^{i\theta_z}\ket{z}\bra{z},
\end{equation}
where $\theta_z\in[0,2\pi)$ for every $z\in\{0,1\}^N$.
Its action on a computational-basis state of the $k$-th moment space is
\begin{equation}
V(\{\theta_z\}_{z})^{\otimes k,k}
\ket{x^1,\ldots,x^k,y^1,\ldots,y^k}
=
\exp\left[
i\sum_{\alpha=1}^k
\left(
\theta_{x^\alpha}-\theta_{y^\alpha}
\right)
\right]
\ket{x^1,\ldots,x^k,y^1,\ldots,y^k}.
\end{equation}

Averaging each phase $\theta_z$ independently and uniformly over $[0,2\pi)$ eliminates the basis state unless
\begin{equation}
\#\{\alpha\in[k]:x^\alpha=z\}
=
\#\{\alpha\in[k]:y^\alpha=z\}
\end{equation}
for every $z\in\{0,1\}^N$.
Equivalently, $(y^1,\ldots,y^k)$ must be a permutation of $(x^1,\ldots,x^k)$, i.e., there exists $\sigma\in S_k$ such that
\begin{equation}
\bm y=\sigma\bm x,
\qquad
(\sigma\bm x)^\alpha\equiv x^{\sigma^{-1}(\alpha)}.
\end{equation}
Therefore, the $k$-th moment operator of the Haar measure on $\mathcal{G}_{\rm RP}$ is
\begin{equation}
P_{\rm RP}^{(k)}
=
\sum_{\substack{\bm x,\bm y\\
\exists\,\sigma\in S_k:\,\bm y=\sigma\bm x}}
\ket{\bm x,\bm y}\bra{\bm x,\bm y}.
\end{equation}

For later use, it is useful to rewrite this condition in terms of the local types introduced in the previous subsection.
For each $\sigma\in S_k$, define
\begin{equation}
\label{eq:def_perm_comp_class}
F_\sigma
\equiv 
\left\{
(x,\sigma x):
x\in\{0,1\}^k
\right\}.
\end{equation}
The condition $\bm y=\sigma\bm x$ is equivalent to $u_i=(x_i,y_i)\in F_\sigma$ for every $i\in[N]$, with the same permutation $\sigma$ for all physical sites.
Therefore, a computational-basis state with type vector $\bm m$ survives the RP average if and only if 
\begin{equation}
    \operatorname{supp}(\bm m)\subseteq F_\sigma,
\end{equation}
for some $\sigma\in S_k$, where $\operatorname{supp}(\bm m) \equiv  \left\{(x,y):m_{x,y}>0\right\}.$

\subsection{2RP group}\label{ss:2RP}

In this subsection, we introduce the 2RP group, denoted by $\mathcal{G}_{\rm 2RP}$, generated by two-qubit diagonal phase rotations acting on arbitrary pairs of physical sites.
For a pair of distinct sites $i,j\in[N]$, define
\begin{equation}
R_{ij}(\{\theta_{pq}\}_{p,q})
=
\left(
\sum_{p,q\in\{0,1\}}
e^{i\theta_{pq}}
|pq\rangle\langle pq|_{ij}
\right)
\otimes
\mathbb{I}_{\overline{ij}},
\end{equation}
where the four phases $\theta_{00},\theta_{01},\theta_{10},\theta_{11}$ can be chosen arbitrarily.
The 2RP group is then defined by
\begin{equation}
\mathcal{G}_{\rm 2RP}
\equiv
\left\langle
R_{ij}(\{\theta_{pq}\}_{p,q})
:
1\leq i<j\leq N,\ 
\theta_{pq}\in[0,2\pi)
\right\rangle.
\end{equation}
Since every generator is diagonal in the computational basis, $\mathcal{G}_{\rm 2RP}$ is a subgroup of $\mathcal{G}_{\rm RP}$.

We next characterize the $k$-th moment space of $\mathcal{G}_{\rm 2RP}$.
Consider the phase average on a fixed pair of sites $(i,j)$.
For $z,w\in\{0,1\}^k$, define
\begin{equation}
\#_{pq}(z,w)
\equiv 
\#\left\{
\alpha\in[k]:
(z^\alpha,w^\alpha)=(p,q)
\right\}.
\end{equation}
Averaging the four phases of $R_{ij}$ independently and uniformly over $[0,2\pi)$ eliminates a computational-basis state unless
\begin{equation}
\label{eq:non_eliminate_cond_2RP}
\#_{pq}(x_i,x_j)
=
\#_{pq}(y_i,y_j)
\end{equation}
for every $p,q\in\{0,1\}$.
The four multiplicities on the ket side can be written as
\begin{equation}
\#_{11}(x_i,x_j)=x_i\cdot x_j,
\qquad
\#_{10}(x_i,x_j)=|x_i|-x_i\cdot x_j,
\qquad
\#_{01}(x_i,x_j)=|x_j|-x_i\cdot x_j,
\qquad
\#_{00}(x_i,x_j)=k-|x_i|-|x_j|+x_i\cdot x_j.
\end{equation}
Therefore, Eq.~\eqref{eq:non_eliminate_cond_2RP} is equivalent to the following conditions:
\begin{gather}
|x_i|=|y_i|,
\qquad
|x_j|=|y_j|,
\label{eq:sing_part_cond}\\
x_i\cdot x_j
=
y_i\cdot y_j.
\label{eq:two_part_cond}
\end{gather}

Here, Eq.~\eqref{eq:sing_part_cond} represents the single-site part of the phase-matching condition.
This condition restricts the allowed local types to
\begin{equation}
\mathcal{C}_k
\equiv 
\left\{
(x,y)\in\{0,1\}^k\times\{0,1\}^k:
|x|=|y|
\right\}.
\end{equation}
We refer to an element $(x,y)\in\mathcal{C}_k$ as a \emph{color}.
The number of possible colors is
\begin{equation}
|\mathcal{C}_k|
=
\sum_{r=0}^{k}\binom{k}{r}^2
=
\binom{2k}{k},
\end{equation}
where $r$ denotes the common Hamming weight of $x$ and $y$.

To characterize the genuinely two-site constraint in Eq.~\eqref{eq:two_part_cond}, we introduce the notion of a \emph{compatible class}.
\begin{definition}[Compatible class]
A subset $F\subseteq\mathcal{C}_k$ is called a \emph{compatible class} if
\begin{enumerate}
    \item every two colors $(x,y),(x',y')\in F$ satisfy
    \begin{equation}
    \label{eq:comp_cond_inner}
    x\cdot x'
    =
    y\cdot y',
    \end{equation}
    and
    \item $F$ is maximal with respect to this property, i.e., no color in $\mathcal{C}_k\setminus F$ can be added to $F$ while preserving Eq.~\eqref{eq:comp_cond_inner} for all pairs of colors.
\end{enumerate}
We denote the set of all compatible classes by
\begin{equation}
\mathcal{F}_k
\equiv 
\left\{
F\subseteq\mathcal{C}_k:
F\text{ is a compatible class}
\right\}.
\end{equation}
\end{definition}

A simple family of compatible classes is obtained from permutations of the $k$ copies, which is defined as $F_\sigma$ in Eq.~\eqref{eq:def_perm_comp_class}. Since permutations preserve inner products, the colors in $F_\sigma$ are mutually compatible. 
Moreover, $F_\sigma$ is maximal. Indeed, if a color $(x,y)\in\mathcal{C}_k$ is compatible with $(x,\sigma x)\in F_\sigma$, then $x\cdot x=y\cdot(\sigma x)=|x|$. Since $|y|=|\sigma x|=|x|$, this implies $y=\sigma x$. Hence no additional color can be added to $F_\sigma$.
We call these \emph{permutation compatible classes}, and denote their set by
\begin{equation}
\mathcal{F}_k^{\rm perm}
\equiv 
\left\{
F_\sigma:
\sigma\in S_k
\right\}
\subseteq
\mathcal{F}_k.
\end{equation}

For example, for $k=2$, there are six allowed colors,
\begin{equation}
\mathcal{C}_2
=
\left\{
0000,\,
1111,\,
1010,\,
0101,\,
1001,\,
0110
\right\},
\end{equation}
where each four-bit string represents the concatenated pair $(x,y)$.
In this case, there are only two compatible classes,
\begin{equation}
F_{\rm id}
=
\left\{
0000,\,
1111,\,
1010,\,
0101
\right\},
\qquad
F_{\rm swap}
=
\left\{
0000,\,
1111,\,
1001,\,
0110
\right\}.
\end{equation}
We can see, for instance, that the colors $(10,10)$ and $(10,01)$ are incompatible because $(10)\cdot(10)=1$ and $(10)\cdot(01)=0.$
This example also shows that distinct compatible classes need not be disjoint, since $F_{\rm id}$ and $F_{\rm swap}$ share the colors $0000$ and $1111$.
For $k\leq3$, all compatible classes are permutation compatible, whereas for $k\geq4$ there exist compatible classes that are not associated with permutations of the copies.
We prove these structural properties in Sec.~\ref{s:gap_doped_2RPmix_circuit}.

We now return to the moment space of the 2RP group.
A computational-basis state $\ket{\bm u}=\ket{u_1,\ldots,u_N}$ survives the 2RP average if and only if every local type $u_i$ is a color and every pair of colors $u_i,u_j$ satisfies Eq.~\eqref{eq:comp_cond_inner}.
Since any pairwise-compatible subset of the finite set $\mathcal{C}_k$ is contained in at least one compatible class, this condition is equivalent to the existence of $F\in\mathcal{F}_k$ such that $u_i\in F$ for every $i\in[N]$.
Therefore, the $k$-th moment operator of the Haar measure on $\mathcal{G}_{\rm 2RP}$ is
\begin{equation}
P_{\rm 2RP}^{(k)}
=
\sum_{\substack{
\bm u=(u_1,\ldots,u_N)\\
\exists\,F\in\mathcal{F}_k:\,
u_i\in F\ \forall i\in[N]
}}
|\bm u\rangle\langle\bm u|.
\end{equation}
Thus, we can see that the computational-basis state with type vector $\bm m$ survives the 2RP average if and only if
\begin{equation}
\operatorname{supp}(\bm m)
\subseteq
F
\end{equation}
for some $F\in\mathcal{F}_k$.

\subsection{RPmix and 2RPmix groups} \label{ss:RPmix_2RPmix}

In this subsection, we introduce the RPmix and 2RPmix groups, denoted by $\mathcal{G}_{\rm RPmix}$ and $\mathcal{G}_{\rm 2RPmix}$, respectively.
The RPmix group is generated by the mixing group and the RP group, i.e., $\mathcal{G}_{\rm RPmix} \equiv \langle \mathcal{G}_{\rm mix}, \mathcal{G}_{\rm RP}\rangle$.
Similarly, the 2RPmix group is generated by the mixing group and the 2RP group, i.e., $\mathcal{G}_{\rm 2RPmix} \equiv \langle \mathcal{G}_{\rm mix}, \mathcal{G}_{\rm 2RP}\rangle$.

We first consider the $k$-th moment operator of the RPmix ensemble.
Since conjugation by a site permutation preserves the Haar measure on the RP group, we have
\begin{equation}
    U_\pi^{\otimes k,k} P^{(k)}_{\rm RP} U_{\pi^{-1}}^{\otimes k,k} = P^{(k)}_{\rm RP}
\end{equation}
for any $\pi \in S_N$.
Using this equation, we have
\begin{align}
    P^{(k)}_{\rm mix} P^{(k)}_{\rm RP}
    &= \frac{1}{N!}\sum_{\pi \in S_N} U_\pi^{\otimes k,k} P^{(k)}_{\rm RP} \nonumber \\
    &= \frac{1}{N!}\sum_{\pi \in S_N} P^{(k)}_{\rm RP} U_{\pi}^{\otimes k,k} \nonumber \\
    &= P^{(k)}_{\rm RP} P^{(k)}_{\rm mix}, \label{eq:commute_RP_and_mix}
\end{align}
or equivalently, $[P^{(k)}_{\rm mix},P^{(k)}_{\rm RP}]=0.$
Since these two operators are orthogonal projectors, their product is the orthogonal projector onto the intersection of their ranges.
Moreover, a state is invariant under $\mathcal{G}_{\rm RPmix}$ if and only if it is invariant under both $\mathcal{G}_{\rm mix}$ and $\mathcal{G}_{\rm RP}$.
Hence,
\begin{equation}
P_{\rm RPmix}^{(k)}
=
P_{\rm mix}^{(k)}P_{\rm RP}^{(k)},\label{eq:product_RP_and_mix}
\end{equation}
with
\begin{equation}
\operatorname{Ran}P_{\rm RPmix}^{(k)}
=
\operatorname{Ran}P_{\rm mix}^{(k)}
\cap
\operatorname{Ran}P_{\rm RP}^{(k)}.
\end{equation}
Using the characterizations obtained in the previous subsections, we have
\begin{align}
\operatorname{Ran}P_{\rm mix}^{(k)}
&=
\operatorname{span}
\left\{
|\Omega_{\bm m}\rangle:
\sum_{x,y\in\{0,1\}^k}m_{x,y}=N
\right\}, \\
\operatorname{Ran}P_{\rm RP}^{(k)}
&=
\operatorname{span}
\left\{
|\bm u\rangle:
\exists\,\sigma\in S_k
\text{ such that }
u_i\in F_\sigma
\ \forall i\in[N]
\right\}.
\end{align}
Therefore, it follows that
\begin{equation}
P_{\rm RPmix}^{(k)}
=
\sum_{\substack{
\bm m:\,\sum_{x,y}m_{x,y}=N\\
\exists\,\sigma\in S_k:\,
\operatorname{supp}(\bm m)\subseteq F_\sigma
}}
|\Omega_{\bm m}\rangle
\langle\Omega_{\bm m}|.
\end{equation}

We next consider the 2RPmix group.
By the same argument as for the RPmix group, we obtain $\left[P_{\rm mix}^{(k)},P_{\rm 2RP}^{(k)}\right]=0$, which implies
\begin{equation}
P_{\rm 2RPmix}^{(k)}
=
P_{\rm mix}^{(k)}P_{\rm 2RP}^{(k)},
\end{equation}
and
\begin{equation}
\operatorname{Ran}P_{\rm 2RPmix}^{(k)}
=
\operatorname{Ran}P_{\rm mix}^{(k)}
\cap
\operatorname{Ran}P_{\rm 2RP}^{(k)}.
\end{equation}
Using the characterization of the 2RP moment space obtained in the previous subsection, we have
\begin{equation}
P_{\rm 2RPmix}^{(k)}
=
\sum_{\substack{
\bm m:\,\sum_{x,y}m_{x,y}=N\\
\exists\,F\in\mathcal{F}_k:\,
\operatorname{supp}(\bm m)\subseteq F
}}
|\Omega_{\bm m}\rangle
\langle\Omega_{\bm m}|.
\label{eq:P_2RPmix_def}
\end{equation}
Thus, the RPmix and 2RPmix moment spaces have the same orbit-state structure, with the only difference that the former allows only permutation compatible classes $F_\sigma\in\mathcal{F}_k^{\rm perm}$, whereas the latter allows arbitrary compatible classes $F\in\mathcal{F}_k$.

Finally, for later use, we briefly discuss the relationship between the normalized orbit states $\ket{\Omega_{\bm m}}$ and the $U(1)$-symmetric Haar moment space.
As introduced in Sec.~\ref{subsec:u1_haar_moment}, for each charge sector $(\bm n,\overline{\bm n})$, we have
\begin{equation}
    \operatorname{Ran}P_{U(1)\mathrm{Haar};\bm n,\overline{\bm n}}^{(k)}
    =
    \operatorname{span}\left\{\ket{\bm n;\sigma}:\sigma\in\Sigma(\bm n,\overline{\bm n})\right\}.
\end{equation}
By grouping the computational-basis states in Eq.~\eqref{eq:number_permutation_state} according to their type vectors, the number-permutation state can be rewritten as
\begin{equation}
    \ket{\bm n;\sigma}
    =
    \sum_{\substack{
        \bm m:\,
        \operatorname{supp}(\bm m)\subseteq F_\sigma\\
        \sum_{x\in\{0,1\}^k}m_{x,\sigma x}x=\bm n
    }}
    \sqrt{\frac{D_{\bm m}}{Z_{\bm n}}}\,
    \ket{\Omega_{\bm m}},
    \qquad
    \sigma\in\Sigma(\bm n,\overline{\bm n}).
    \label{eq:number_permutation_orbit_expansion}
\end{equation}
Therefore, if a type vector $\bm m$ satisfies
\begin{equation}
    \operatorname{supp}(\bm m)\nsubseteq F_\sigma
    \qquad
    \text{for every }\sigma\in S_k,
\end{equation}
then $\ket{\Omega_{\bm m}}$ is orthogonal to every number-permutation state.
Since the $U(1)$-symmetric Haar moment space is spanned by these states, it follows that
\begin{equation}
\label{eq:Omega_m_U(1)Haar_zero_intersection}
    P_{U(1)\mathrm{Haar}}^{(k)}\ket{\Omega_{\bm m}}=0.
\end{equation}

\section{Spectral gap upper bound for $U(1)$-symmetric random circuits} \label{s:gap_upper_bound}

In this section, we derive spectral-gap upper bounds for $U(1)$-symmetric random circuits.
In Sec.~\ref{ss:trans_mat_two_part_enc}, we first introduce a classical substochastic process describing two particles moving on a graph, whose principal decay rate is governed by the encounter of the two particles.
Then, in Sec.~\ref{ss:two_particle_encounter_mode}, we show that the transition matrix of this process appears exactly as a restriction of the moment operator of the $U(1)$-symmetric random circuit. This identifies the corresponding slow mode of the moment operator as the two-particle encounter mode and proves Theorem~\ref{restatethm:gap_upper_bound_two_particle_encounter}.
In Sec. \ref{ss:gap_upper_bound_robust}, we prove Theorem 2, establishing a geometry-independent spectral gap upper bound for circuits with general $\ell$-qubit gates.
Finally, in Sec.~\ref{ss:gap_upper_bound_parallel_fixed}, we extend the spectral-gap upper bounds obtained for single-edge and single-hyperedge circuits to parallel and fixed-architecture circuits, using architecture-comparison arguments and a direct argument for parallel $\ell$-local circuits.

\subsection{Transition matrix of two-particle encounter process} \label{ss:trans_mat_two_part_enc}

In this subsection, we introduce a classical substochastic process describing two distinguishable particles moving on a connected $N$-vertex graph $G=(V,E)$, together with its transition matrix.
The process is defined as follows:
\begin{itemize}
    \item[$\mathrm{(1)}$] At each step, an edge $e\in E$ is chosen uniformly at random.
    \item[$\mathrm{(2}$-$\mathrm{i)}$] If the selected edge does not connect the two particles, the two endpoint vertices of the edge are transposed with probability $1/2$, while the configuration is left unchanged with probability $1/2$.
    \item[$\mathrm{(2}$-$\mathrm{ii)}$] If the selected edge directly connects the two particles, the corresponding component is eliminated.
\end{itemize}

Denoting the positions of the two particles by $i$ and $j$, respectively, the configuration space is $\mathsf{X}_{G}^{\rm enc}\equiv\{(i,j)\in V\times V:i\neq j\}.$ We denote by $\ket{i,j}$ the configuration in which particle $1$ occupies vertex $i$ and particle $2$ occupies vertex $j$.
For an edge $e\in E$, let $\pi_e\in S_N$ denote the transposition of the two vertices connected by $e$.
The transition associated with the choice of $e$ is then represented by
\begin{equation}
    K_e^{\rm enc}\ket{i,j}=
    \begin{cases}
        0, & e=\{i,j\},\\[2mm]
        \dfrac{1}{2}\left(\ket{i,j}+\ket{\pi_e(i),\pi_e(j)}\right), & e\neq\{i,j\}.
    \end{cases}
\end{equation}
The transition matrix of one step of the full process is therefore
\begin{equation}
    K_G^{\rm enc}\equiv\frac{1}{|E|}\sum_{e\in E}K_e^{\rm enc}.
\end{equation}
Equivalently, its matrix elements are
\begin{equation}
    [K_G^{\rm enc}]_{(i',j'),(i,j)}
    =
    \frac{1}{2|E|}
    \sum_{\substack{e\in E\\ e\neq\{i,j\}}}
    \left[
        \delta_{(i',j'),(i,j)}
        +
        \delta_{(i',j'),(\pi_e(i),\pi_e(j))}
    \right].
    \label{eq:two_part_enc_trans_mat_element}
\end{equation}

All matrix elements of $K_G^{\rm enc}$ are nonnegative, and the sum of each column is
\begin{equation}
\label{eq:two_part_enc_process_column_sum}
    \sum_{(i',j')\in\mathsf{X}_{G}^{\rm enc}}
    [K_G^{\rm enc}]_{(i',j'),(i,j)}
    =
    1-\frac{\delta_{\{i,j\}\in E}}{|E|}
    \leq1,
\end{equation}
where $\delta_{\{i,j\}\in E}=1$ if $\{i,j\}\in E$ and $\delta_{\{i,j\}\in E}=0$ otherwise.
Thus, $K_G^{\rm enc}$ is a substochastic transition matrix.
The deficit of a column sum from unity represents the probability that the process is killed when the two particles encounter each other across the selected edge.

Since $K_G^{\rm enc}$ is substochastic, its spectral radius satisfies $\rho(K_G^{\rm enc})\leq1$.
We define the principal decay rate of the encounter process by
\begin{equation}
\label{eq:two_part_enc_decay_rate}
    \lambda_G^{\rm enc}\equiv1-\rho(K_G^{\rm enc}).
\end{equation}
Moreover, $K_G^{\rm enc}$ is a real symmetric matrix, and hence
\begin{equation}
    \|K_G^{\rm enc}\|_{\infty}
    =
    \rho(K_G^{\rm enc})
    =
    1-\lambda_G^{\rm enc}.
    \label{eq:two_part_enc_op_norm}
\end{equation}
We choose a normalized eigenvector $|v_G^{\rm enc}\rangle$ associated with the spectral radius,
\begin{equation}
    K_G^{\rm enc}|v_G^{\rm enc}\rangle
    =
    \left(1-\lambda_G^{\rm enc}\right)|v_G^{\rm enc}\rangle,
\end{equation}
which represents a slowest-decaying mode of the encounter process.

\subsection{Spectral gap upper bound from the two-particle encounter mode}
\label{ss:two_particle_encounter_mode}

In this subsection, we derive an upper bound on the spectral gap of the single-edge random circuit from the two-particle encounter mode.
Specifically, we prove the following theorem.

\begin{theorem} \label{restatethm:gap_upper_bound_two_particle_encounter}
For any connected $N$-vertex graph $G=(V,E)$, the second-moment spectral gap of the corresponding single-edge random circuit satisfies
\begin{equation}
\label{restateeq:gap_upper_bound_two_particle_encounter}
    \Delta_{\nu_G}^{(2)}
    \leq
    \lambda_G^{\rm enc},
\end{equation}
where $\lambda_G^{\rm enc}$ is the principal decay rate of the two-particle encounter process defined in Eq.~\eqref{eq:two_part_enc_decay_rate}.
\end{theorem}

\begin{proof}
Consider the type vector $\bm m$ specified by $m_{0000}=N-2$ and $m_{1010}=m_{0110}=1$, with all other multiplicities equal to zero.
Let $\mathcal T_{\bm m}$ denote the set of computational-basis configurations with this type vector, and let $\mathcal H_{\mathcal T_{\bm m}}$ be the subspace spanned by these configurations.
We denote the orthogonal projector onto this subspace by $\Pi_{\mathcal T_{\bm m}}$.

A direct evaluation of the second moment of a two-qubit $U(1)$-symmetric Haar-random gate shows that, for $v_1,v_2\in\{0000,1010,0110\}$, the matrix element vanishes whenever either $u_1$ or $u_2$ lies outside $\{0000,1010,0110\}$.
For $u_1,u_2,v_1,v_2\in\{0000,1010,0110\}$, we have
\begin{equation}
\label{eq:trans_2nd_mom_local}
    \bra{u_1,u_2}P^{(2)}_{a,b}\ket{v_1,v_2}
    =
    \begin{cases}
        0, & \{v_1,v_2\}=\{1010,0110\},\\[2mm]
        \dfrac{1}{2}\left(\delta_{u_1,v_1}\delta_{u_2,v_2}+\delta_{u_1,v_2}\delta_{u_2,v_1}\right), & \mathrm{otherwise}.
    \end{cases}
\end{equation}
Here, $\ket{v_1,v_2}$ denotes the two-site state with local types $v_1$ and $v_2$ at sites $a$ and $b$, respectively.
It follows from Eq.~\eqref{eq:trans_2nd_mom_local} that the moment operator
\begin{equation}
    M_{\nu_G}^{(2)}
    \equiv
    \frac{1}{|E|}
    \sum_{\{a,b\}\in E}
    P_{a,b}^{(2)}\otimes\mathbb I_{\overline{a,b}}
\end{equation}
preserves the subspace $\mathcal H_{\mathcal T_{\bm m}}$.

Since a computational-basis state in $\mathcal T_{\bm m}$ is uniquely determined by the positions of the local types $1010$ and $0110$, we relabel its basis states as follows.
Let $\bm 0\equiv0000$, $w_1\equiv1010$, and $w_2\equiv0110$, and define
\begin{equation}
    \ket{\psi_{i,j}}
    \equiv
    \ket{
        \bm 0,\ldots,\bm 0,
        \overset{i}{\overset{\wedge}{w_1}},
        \bm 0,\ldots,\bm 0,
        \overset{j}{\overset{\wedge}{w_2}},
        \bm 0,\ldots,\bm 0
    }.
\end{equation}
Then $\{\ket{\psi_{i,j}}:i,j\in V,\ i\neq j\}$ forms an orthonormal basis of $\mathcal H_{\mathcal T_{\bm m}}$.

For an edge $e=\{a,b\}\in E$, Eq.~\eqref{eq:trans_2nd_mom_local} gives
\begin{equation}
\label{eq:local_action_enc_mode}
    \left(P_{a,b}^{(2)}\otimes\mathbb I_{\overline{a,b}}\right)\ket{\psi_{i,j}}
    =
    \begin{cases}
        0, & e=\{i,j\},\\[2mm]
        \dfrac{1}{2}\left(\ket{\psi_{i,j}}+\ket{\psi_{\pi_e(i),\pi_e(j)}}\right), & e\neq\{i,j\},
    \end{cases}
\end{equation}
where $\pi_e\in S_N$ denotes the transposition of the two vertices connected by $e$.
Therefore, for every $i\neq j$ and $i'\neq j'$,
\begin{align}
    \bra{\psi_{i',j'}}M_{\nu_G}^{(2)}\ket{\psi_{i,j}}
    &=
    \frac{1}{2|E|}
    \sum_{\substack{e\in E\\e\neq\{i,j\}}}
    \left[
        \delta_{(i',j'),(i,j)}
        +
        \delta_{(i',j'),(\pi_e(i),\pi_e(j))}
    \right]
    \nonumber\\
    &=
    [K_G^{\rm enc}]_{(i',j'),(i,j)},
\end{align}
where $K_G^{\rm enc}$ is the substochastic transition matrix defined in Eq.~\eqref{eq:two_part_enc_trans_mat_element}.
Thus, under the basis correspondence $(i,j)\leftrightarrow\ket{\psi_{i,j}}$, the matrix representation of the restriction of $M_{\nu_G}^{(2)}$ to $\mathcal H_{\mathcal T_{\bm m}}$ is exactly $K_G^{\rm enc}$.
In particular,
\begin{equation}
\label{eq:mom_op_identity_enc_mode}
    \left\|
        \Pi_{\mathcal T_{\bm m}}
        M_{\nu_G}^{(2)}
        \Pi_{\mathcal T_{\bm m}}
    \right\|_{\infty}
    =
    \|K_G^{\rm enc}\|_{\infty}
    =
    1-\lambda_G^{\rm enc},
\end{equation}
where the second equality follows from Eq.~\eqref{eq:two_part_enc_op_norm}.

Using Eq.~\eqref{eq:mom_op_identity_enc_mode}, we obtain
\begin{align}
    1-\Delta_{\nu_G}^{(2)}
    &\equiv
    \left\|
        M_{\nu_G}^{(2)}
        -
        P_{U(1)\mathrm{Haar}}^{(2)}
    \right\|_{\infty}
    \nonumber\\
    &\geq
    \left\|
        \Pi_{\mathcal T_{\bm m}}
        \left(
            M_{\nu_G}^{(2)}
            -
            P_{U(1)\mathrm{Haar}}^{(2)}
        \right)
        \Pi_{\mathcal T_{\bm m}}
    \right\|_{\infty}
    \nonumber\\
    &=
    \left\|
        \Pi_{\mathcal T_{\bm m}}
        M_{\nu_G}^{(2)}
        \Pi_{\mathcal T_{\bm m}}
    \right\|_{\infty}
    \nonumber\\
    &=
    1-\lambda_G^{\rm enc}.
\end{align}
In the third line, we used the fact that $P_{U(1)\mathrm{Haar}}^{(2)}$ vanishes in the charge sector $(\bm n,\overline{\bm n})=((1,1)^T,(2,0)^T)$ containing $\mathcal H_{\mathcal T_{\bm m}}$.
Rearranging the above inequality proves Eq.~\eqref{restateeq:gap_upper_bound_two_particle_encounter}.
\end{proof}

\begin{rem}[Relationship to the Goldstone-mode ansatz in Ref.~\cite{hearth2025unitary}]
The two-particle encounter mode constructed in the proof of Theorem~\ref{restatethm:gap_upper_bound_two_particle_encounter} is supported on the type-vector subspace $\mathcal H_{\mathcal T_{\bm m}}$ for which $\operatorname{supp}(\bm m)\nsubseteq F_{\sigma}$ for every $\sigma\in S_2$.
In fact, the type vector used there is a minimal example of such an incompatible type vector.

By contrast, the Goldstone-mode ansatz of Ref.~\cite{hearth2025unitary} is obtained by applying diagonal number operators to number-permutation states and is therefore supported entirely on permutation-compatible type configurations.
Hence, denoting such a mode by $\ket{g}$, we have $\Pi_{\mathcal T_{\bm m}}\ket{g}=0$ for every type vector $\bm m$ satisfying $\operatorname{supp}(\bm m)\nsubseteq F_\sigma$ for all $\sigma\in S_k$.
Thus, the two-particle encounter mode constructed above is not captured by the Goldstone-mode ansatz.

\end{rem}

\subsection{Spectral gap upper bounds for general $\ell$-local random circuits}
\label{ss:gap_upper_bound_robust}

In this subsection, we derive spectral-gap upper bounds for random circuits composed of general $\ell$-qubit $U(1)$-symmetric Haar-random gates.
Such a circuit is specified by an $\ell$-uniform hypergraph $G=(V,E)$, where $E\subseteq\binom{V}{\ell}$ is the set of hyperedges consisting of $\ell$ vertices.
The moment operator of the corresponding single-hyperedge random circuit is
\begin{equation}
\label{eq:l_local_single_edge_def}
    M_{\nu_{G,\ell\mathrm{-loc}}}^{(k)}
    \equiv
    \frac{1}{|E|}
    \sum_{e\in E}
    P_e^{(k)}\otimes\mathbb I_{\overline e},
\end{equation}
where $e=\{i_1,\ldots,i_\ell\}$ consists of $\ell$ sites, and $P_e^{(k)}$ is the moment operator of the $U(1)$-symmetric Haar-random unitary ensemble acting on the sites in $e$.
For this general class of $U(1)$-symmetric random circuits, we obtain the following theorem.

\begin{theorem}
\label{restatethm:universal_gap_upper_bound}
Consider a charge sector specified by $\bm n=(n_1,n_2)^T$ and $\overline{\bm n}=(n_{\overline{1}},n_{\overline{2}})^T$ satisfying
\begin{gather}
    n_1+n_2=n_{\overline{1}}+n_{\overline{2}},
    \label{eq:slow_mode_cond1}\\
    n_1+n_2\leq N.
    \label{eq:slow_mode_cond2}
\end{gather}
Suppose in addition that at least one of the following four conditions is satisfied:
\begin{equation}
\label{eq:slow_mode_cond3}
\begin{split}
    n_1&\geq1,\quad n_2\geq1,\quad n_{\overline{1}}\geq2,\quad n_{\overline{2}}\geq0,\\
    n_1&\geq1,\quad n_2\geq1,\quad n_{\overline{1}}\geq0,\quad n_{\overline{2}}\geq2,\\
    n_1&\geq2,\quad n_2\geq0,\quad n_{\overline{1}}\geq1,\quad n_{\overline{2}}\geq1,\\
    n_1&\geq0,\quad n_2\geq2,\quad n_{\overline{1}}\geq1,\quad n_{\overline{2}}\geq1.
\end{split}
\end{equation}
Then there exists a normalized state $\ket{\psi_{\bm n,\overline{\bm n}}}\in\mathcal H_{\bm n,\overline{\bm n}}$ satisfying
\begin{equation}
\label{eq:universal_slow_mode_def}
\begin{split}
    P_{U(1)\mathrm{Haar}}^{(2)}\ket{\psi_{\bm n,\overline{\bm n}}}&=0,\\
    \left|\bra{\psi_{\bm n,\overline{\bm n}}}M_{\nu_{G,\ell\mathrm{-loc}}}^{(2)}\ket{\psi_{\bm n,\overline{\bm n}}}\right|
    &\geq
    1-\frac{1}{2}\left(\frac{\ell n_{\rm tot}}{N}\right)^2,
\end{split}
\end{equation}
where $n_{\rm tot}\equiv n_1+n_2=n_{\overline{1}}+n_{\overline{2}}$.
Here, the state $\ket{\psi_{\bm n,\overline{\bm n}}}$ can be chosen independently of the detailed geometry encoded by $G$.
Consequently, the spectral gap in this charge sector satisfies the geometry-independent upper bound
\begin{equation}
\label{eq:universal_slow_mode_gap_bound}
    \Delta_{\nu_{G,\ell\mathrm{-loc}};\bm n,\overline{\bm n}}^{(2)}
    \leq
    \frac{1}{2}\left(\frac{\ell n_{\rm tot}}{N}\right)^2.
\end{equation}
\end{theorem}

In particular, choosing $(\bm n,\overline{\bm n})=((1,1)^T,(2,0)^T)$ gives $n_{\rm tot}=2$, and hence Theorem~\ref{restatethm:universal_gap_upper_bound} immediately yields
\begin{equation}
    \Delta_{\nu_{G,\ell\mathrm{-loc}}}^{(2)}
    \leq
    2\left(\frac{\ell}{N}\right)^2,
\end{equation}
which recovers the upper bound for the $\ell$-local circuit given in the main text.

More generally, for any fixed $\ell=O(1)$, one may choose a sector with $n_{\rm tot}=O(1)$ satisfying Eqs.~\eqref{eq:slow_mode_cond1}--\eqref{eq:slow_mode_cond3}, which implies that the spectral gap is at most $O(N^{-2})$.
In particular, for three- and higher-dimensional lattices, expander graphs, and the all-to-all graph, this gives a mode whose decay is asymptotically slower than that associated with charge transport.

Equation~\eqref{eq:universal_slow_mode_gap_bound} further shows that such slow modes are not restricted to a particular low-particle-number sector, but appear more generally in sufficiently dilute charge sectors.
In particular, consider sectors satisfying Eqs.~\eqref{eq:slow_mode_cond1}--\eqref{eq:slow_mode_cond3} with $n_{\rm tot}=O(N^\beta)$.
For fixed $\ell=O(1)$, Eq.~\eqref{eq:universal_slow_mode_gap_bound} gives
\begin{equation}
    \Delta_{\nu_{G,\ell\mathrm{-loc}};\bm n,\overline{\bm n}}^{(2)}
    =
    O\left(N^{-2+2\beta}\right).
\end{equation}
For example, in the all-to-all circuit, this mode is asymptotically slower than the charge-transport mode when $\beta<1/2$.
Although Theorem~\ref{restatethm:universal_gap_upper_bound} is stated for dilute charge sectors, the same argument applies to dense sectors after exchanging $0\leftrightarrow1$.
Thus, an analogous slow mode and spectral-gap upper bound also arise near the fully occupied sectors.

\begin{proof}[Proof of Theorem~\ref{restatethm:universal_gap_upper_bound}]
We first explicitly construct a state $\ket{\psi_{\bm n,\overline{\bm n}}}$ satisfying Eq.~\eqref{eq:universal_slow_mode_def}.
We choose $\ket{\psi_{\bm n,\overline{\bm n}}}$ to be a site-permutation-invariant orbit state $\ket{\Omega_{\bm m}}$ defined in Eq.~\eqref{eq:perm_inv_orbit_state}.
We first consider the case
\begin{equation*}
    n_1\geq1,\quad n_2\geq1,\quad n_{\overline{1}}\geq2,\quad n_{\overline{2}}\geq0.
\end{equation*}
The other three cases in Eq.~\eqref{eq:slow_mode_cond3} follow from the same construction by relabeling the ket or bra copies and, for the last two cases, interchanging the ket and bra labels.

In this case, defining $r\equiv\max\{1,n_{\overline{1}}-n_2\},$ we choose the nonzero components of $\bm m$ as
\begin{equation}
\begin{split}
    m_{1010}&=r,\quad
    m_{0110}=n_{\overline{1}}-r,\quad
    m_{1001}=n_1-r,\\
    m_{0101}&=n_2-n_{\overline{1}}+r,\quad
    m_{0000}=N-n_{\rm tot},
\end{split}
\end{equation}
and set all other components to zero.
By Eq.~\eqref{eq:slow_mode_cond1}, $n_{\overline{1}}-n_2=n_1-n_{\overline{2}}\leq n_1,$ while $n_2\geq1$ and $n_{\overline{1}}\geq2$ imply $ n_{\overline{1}}-n_2\leq n_{\overline{1}}-1.$
Therefore,
\begin{equation*}
    1\leq r\leq\min\{n_1,n_{\overline{1}}-1\},
\end{equation*}
and all the above multiplicities are nonnegative.
Moreover, Eq.~\eqref{eq:slow_mode_cond2} guarantees $m_{0000}=N-n_{\rm tot}\geq0$.
One can directly verify that the resulting type vector belongs to the charge sector $(\bm n,\overline{\bm n})$ and satisfies
\begin{align}
    m_{1010}&\geq1,\qquad m_{0110}\geq1,
    \label{eq:slow_mode_type_vector_cond1}\\
    m_u&=0
    \qquad
    \left(\text{if }u\notin\{0000,1010,0101,1001,0110\}\right).
    \label{eq:slow_mode_type_vector_cond2}
\end{align}
For this type vector, we define
\begin{equation}
    \ket{\psi_{\bm n,\overline{\bm n}}}
    \equiv
    \ket{\Omega_{\bm m}}.
\end{equation}

We first show that
\begin{equation}
\label{eq:slow_mode_perp_Haar_moment}
    P_{U(1)\mathrm{Haar}}^{(2)}\ket{\Omega_{\bm m}}=0.
\end{equation}
For $k=2$, the only permutation compatible classes are $F_{\mathrm{id}}$ and $F_{\mathrm{swap}}$.
Since $m_{0110}\geq1$, we have $\operatorname{supp}(\bm m)\nsubseteq F_{\mathrm{id}}$, while $m_{1010}\geq1$ implies $\operatorname{supp}(\bm m)\nsubseteq F_{\mathrm{swap}}$.
Equation~\eqref{eq:Omega_m_U(1)Haar_zero_intersection} therefore directly gives Eq.~\eqref{eq:slow_mode_perp_Haar_moment}.

It remains to show that
\begin{equation}
    \bra{\Omega_{\bm m}}
    M_{\nu_{G,\ell\mathrm{-loc}}}^{(2)}
    \ket{\Omega_{\bm m}}
    \geq
    1-\frac{1}{2}\left(\frac{\ell n_{\rm tot}}{N}\right)^2.
\end{equation}
For each hyperedge $e=\{i_1^e,\ldots,i_\ell^e\}\in E$, choose a site permutation $\pi_e\in S_N$ satisfying $\pi_e(i_j^e)=j$ for every $j\in[\ell]$.
Then, by denoting $e_0\equiv \{1,2,\ldots,\ell\}$, we have
\begin{align}
    \bra{\Omega_{\bm m}}M_{\nu_{G,\ell\mathrm{-loc}}}^{(2)}\ket{\Omega_{\bm m}}
    &=
    \frac{1}{|E|}
    \sum_{e\in E}
    \bra{\Omega_{\bm m}}
    P_e^{(2)}\otimes\mathbb I_{\overline e}
    \ket{\Omega_{\bm m}}
    \nonumber\\
    &=
    \frac{1}{|E|}
    \sum_{e\in E}
    \bra{\Omega_{\bm m}}
    U_{\pi_e^{-1}}^{\otimes2,2}
    \left(
        P_{e_0}^{(2)}
        \otimes
        \mathbb I_{\overline{e_0}}
    \right)
    U_{\pi_e}^{\otimes2,2}
    \ket{\Omega_{\bm m}}
    \nonumber\\
    &=
    \bra{\Omega_{\bm m}}
    \left(
        P_{e_0}^{(2)}
        \otimes
        \mathbb I_{\overline{e_0}}
    \right)
    \ket{\Omega_{\bm m}},
    \label{eq:gap_UB_perm_lsites}
\end{align}
where in the last equality we used the site-permutation invariance $U_{\pi}^{\otimes2,2}\ket{\Omega_{\bm m}}=\ket{\Omega_{\bm m}}$ for every $\pi\in S_N$.

To evaluate the last expression in Eq.~\eqref{eq:gap_UB_perm_lsites}, we partition $\mathcal T_{\bm m}$ into disjoint subsets according to the configuration of local types on the sites in $e_0$.
In the following, we use $\bm 0$ to denote the local type $0000$.
We define
\begin{itemize}
    \item $\mathcal T_{\bm m}^{0}$ as the subset in which $u_i=\bm 0$ for every $i\in e_0$;
    \item $\mathcal T_{\bm m}^{1,u}$ for $u\neq\bm 0$ as the subset in which exactly one site in $e_0$ carries the local type $u$, while all other sites in $e_0$ carry $\bm 0$;
    \item $\mathcal T_{\bm m}^{\geq2}$ as the subset in which at least two sites in $e_0$ carry nonzero local types.
\end{itemize}

Let $\Pi_{\mathcal T_{\bm m}^{0}}$, $\Pi_{\mathcal T_{\bm m}^{1,u}}$, and $\Pi_{\mathcal T_{\bm m}^{\geq2}}$ denote the orthogonal projectors onto the spans of the corresponding computational-basis configurations.
Then,
\begin{align}
    \ket{\Omega_{\bm m}}
    &=
    \Pi_{\mathcal T_{\bm m}^{0}}\ket{\Omega_{\bm m}}
    +
    \sum_{\substack{u\neq\bm 0\\m_u\geq1}}
    \Pi_{\mathcal T_{\bm m}^{1,u}}\ket{\Omega_{\bm m}}
    +
    \Pi_{\mathcal T_{\bm m}^{\geq2}}\ket{\Omega_{\bm m}}
    \nonumber\\
    &=
    \frac{1}{\sqrt{D_{\bm m}}}
    \Bigg\{
    \left(\bigotimes_{i\in e_0}\ket{\bm 0}_i\right)
    \otimes
    \left(
        \sum_{\bm u'\in\mathcal T_{\bm m^{\ell,\bm 0}}}
        \ket{\bm u'}
    \right)
    \nonumber\\
    &\qquad\qquad
    +
    \sum_{\substack{u\neq\bm 0\\m_u\geq1}}
    \left[
        \sum_{j\in e_0}
        \left(
            \bigotimes_{\substack{i\in e_0\\i\neq j}}
            \ket{\bm 0}_i
        \right)
        \otimes
        \ket{u}_j
    \right]
    \otimes
    \left(
        \sum_{\bm u'\in\mathcal T_{\bm m^{\ell,u}}}
        \ket{\bm u'}
    \right)
    +
    \sum_{\bm u\in\mathcal T_{\bm m}^{\geq2}}
    \ket{\bm u}
    \Bigg\}.
\end{align}
Here, $\ket{u}_i$ denotes the local-type state $u$ at site $i$.
Whenever $\mathcal T_{\bm m}^{0}$ is nonempty, $\bm m^{\ell,\bm 0}$ denotes the type vector on the remaining $N-\ell$ sites obtained from $\bm m$ by replacing $m_{\bm 0}$ with $m_{\bm 0}-\ell$ and leaving all other multiplicities unchanged.
Similarly, whenever $\mathcal T_{\bm m}^{1,u}$ is nonempty, $\bm m^{\ell,u}$ is obtained by replacing $m_{\bm 0}$ with $m_{\bm 0}-(\ell-1)$ and $m_u$ with $m_u-1$, while leaving all other multiplicities unchanged.

The number of configurations in $\mathcal T_{\bm m}^{0}$ is
\begin{equation}
    \left|\mathcal T_{\bm m}^{0}\right|
    =
    \left|\mathcal T_{\bm m^{\ell,\bm 0}}\right|
    =
    \frac{(N-\ell)!}{(m_{\bm 0}-\ell)!\prod_{v\neq\bm 0}m_v!}
    =
    D_{\bm m}
    \frac{\binom{m_{\bm 0}}{\ell}}{\binom{N}{\ell}}.
\end{equation}
Similarly, for $u\neq\bm 0$,
\begin{equation}
    \left|\mathcal T_{\bm m}^{1,u}\right|
    =
    \ell\left|\mathcal T_{\bm m^{\ell,u}}\right|
    =
    \frac{\ell(N-\ell)!}{(m_{\bm 0}-\ell+1)!(m_u-1)!\prod_{v\neq\bm 0,u}m_v!}
    =
    D_{\bm m}
    \frac{m_u\binom{m_{\bm 0}}{\ell-1}}{\binom{N}{\ell}}.
\end{equation}
The factor $\ell$ accounts for the choice of the site in $e_0$ at which the unique nonzero local type $u$ is placed.
Consequently,
\begin{align}
    \left|\mathcal T_{\bm m}^{\geq2}\right|
    &=
    \left|\mathcal T_{\bm m}\right|
    -
    \left|\mathcal T_{\bm m}^{0}\right|
    -
    \sum_{u\neq\bm 0}
    \left|\mathcal T_{\bm m}^{1,u}\right|
    \nonumber\\
    &=
    D_{\bm m}
    \left[
        1
        -
        \frac{\binom{m_{\bm 0}}{\ell}}{\binom{N}{\ell}}
        -
        \sum_{u\neq\bm 0}
        \frac{m_u\binom{m_{\bm 0}}{\ell-1}}{\binom{N}{\ell}}
    \right].
\end{align}

For each nonzero local type $u$ appearing in Eq.~\eqref{eq:slow_mode_type_vector_cond2}, the state with a single $u$ on the sites in $e_0$, symmetrized over its position, belongs to the fixed space of the $\ell$-qubit $U(1)$-symmetric Haar moment projector $P_{e_0}^{(2)}$.
Indeed, for $u\in\{1010,0101,1001,0110\}$, this state is proportional to a number-permutation state associated with either the identity or the swap permutation.
Therefore,
\begin{equation}
\label{eq:slow_mode_l_loc_inner_0}
    \left(
        \bigotimes_{i\in e_0}\bra{\bm 0}_i
    \right)
    P_{e_0}^{(2)}
    \left(
        \bigotimes_{i\in e_0}\ket{\bm 0}_i
    \right)
    =
    1,
\end{equation}
and
\begin{equation}
\label{eq:slow_mode_l_loc_inner_u}
    \left[
        \sum_{j\in e_0}
        \left(
            \bigotimes_{\substack{i\in e_0\\i\neq j}}
            \bra{\bm 0}_i
        \right)
        \otimes
        \bra{u}_j
    \right]
    P_{e_0}^{(2)}
    \left[
        \sum_{j\in e_0}
        \left(
            \bigotimes_{\substack{i\in e_0\\i\neq j}}
            \ket{\bm 0}_i
        \right)
        \otimes
        \ket{u}_j
    \right]
    =
    \ell.
\end{equation}
The second equality follows because the state inside the brackets is fixed by $P_{e_0}^{(2)}$ and has squared norm $\ell$.

Using Eqs.~\eqref{eq:slow_mode_l_loc_inner_0} and \eqref{eq:slow_mode_l_loc_inner_u}, we obtain
\begin{align}
    \bra{\Omega_{\bm m}}
    \left(
        P_{e_0}^{(2)}\otimes\mathbb I_{\overline{e_0}}
    \right)
    \ket{\Omega_{\bm m}}
    &=
    \frac{\left|\mathcal T_{\bm m^{\ell,\bm 0}}\right|}{D_{\bm m}}
    +
    \sum_{\substack{u\neq\bm 0\\m_u\geq1}}
    \frac{\ell\left|\mathcal T_{\bm m^{\ell,u}}\right|}{D_{\bm m}}+
    \bra{\Omega_{\bm m}}
    \Pi_{\mathcal T_{\bm m}^{\geq2}}
    \left(
        P_{e_0}^{(2)}\otimes\mathbb I_{\overline{e_0}}
    \right)
    \Pi_{\mathcal T_{\bm m}^{\geq2}}
    \ket{\Omega_{\bm m}}
    \nonumber\\
    &\geq
    \frac{\left|\mathcal T_{\bm m}^{0}\right|}{D_{\bm m}}
    +
    \sum_{\substack{u\neq\bm 0\\m_u\geq1}}
    \frac{\left|\mathcal T_{\bm m}^{1,u}\right|}{D_{\bm m}}
    \nonumber\\
    &=
    1-\frac{\left|\mathcal T_{\bm m}^{\geq2}\right|}{D_{\bm m}}.
    \label{eq:slow_mode_local_expectation}
\end{align}
In the first equality, the cross terms between the three types of contributions vanish because their configurations on the remaining $N-\ell$ sites have disjoint type-vector supports.
The inequality follows from the positive semidefiniteness of the Haar moment projector $P_{e_0}^{(2)}$.

We next bound $\left|\mathcal T_{\bm m}^{\geq2}\right|$.
By Eq.~\eqref{eq:slow_mode_type_vector_cond2}, every nonzero local type appearing in $\bm m$ has Hamming weight two as a four-bit local type.
Therefore,
\begin{equation}
\label{eq:slow_mode_n_tot_cond}
    \sum_{u\neq\bm 0}m_u
    =
    \frac{1}{2}
    \left(
        n_1+n_2+n_{\overline{1}}+n_{\overline{2}}
    \right)
    =
    n_{\rm tot},
    \qquad
    m_{\bm 0}
    =
    N-\sum_{u\neq\bm 0}m_u
    =
    N-n_{\rm tot}.
\end{equation}
Substituting Eq.~\eqref{eq:slow_mode_n_tot_cond} into the expression for $\left|\mathcal T_{\bm m}^{\geq2}\right|$, we obtain
\begin{align}
    \left|\mathcal T_{\bm m}^{\geq2}\right|
    &=
    D_{\bm m}
    \left[
        1
        -
        \frac{
            \binom{N-n_{\rm tot}}{\ell}
            +
            n_{\rm tot}\binom{N-n_{\rm tot}}{\ell-1}
        }{
            \binom{N}{\ell}
        }
    \right]
    \nonumber\\
    &=
    \frac{D_{\bm m}}{\binom{N}{\ell}}
    \sum_{r=2}^{\min\{\ell,n_{\rm tot}\}}
    \binom{n_{\rm tot}}{r}
    \binom{N-n_{\rm tot}}{\ell-r}
    \nonumber\\
    &\leq
    \frac{D_{\bm m}}{\binom{N}{\ell}}
    \sum_{r=2}^{\min\{\ell,n_{\rm tot}\}}
    \binom{r}{2}
    \binom{n_{\rm tot}}{r}
    \binom{N-n_{\rm tot}}{\ell-r}
    \nonumber\\
    &=
    \frac{D_{\bm m}}{\binom{N}{\ell}}
    \binom{n_{\rm tot}}{2}
    \sum_{r=2}^{\min\{\ell,n_{\rm tot}\}}
    \binom{n_{\rm tot}-2}{r-2}
    \binom{N-n_{\rm tot}}{\ell-r}
    \nonumber\\
    &=
    D_{\bm m}
    \frac{
        \binom{n_{\rm tot}}{2}
        \binom{N-2}{\ell-2}
    }{
        \binom{N}{\ell}
    }
    \nonumber\\
    &=
    D_{\bm m}
    \frac{
        n_{\rm tot}(n_{\rm tot}-1)\ell(\ell-1)
    }{
        2N(N-1)
    }
    \nonumber\\
    &\leq
    \frac{D_{\bm m}}{2}
    \left(
        \frac{\ell n_{\rm tot}}{N}
    \right)^2.
    \label{eq:slow_mode_T_geq2_bound}
\end{align}
In the second and fifth lines, we used Vandermonde's identity
\begin{equation}
    \sum_r\binom{p}{r}\binom{q}{s-r}=\binom{p+q}{s},
\end{equation}
together with the convention that $\binom{p}{r}=0$ whenever $r<0$ or $r>p$.
In the fourth line, we used
\begin{equation*}
    \binom{r}{2}\binom{n_{\rm tot}}{r}
    =
    \binom{n_{\rm tot}}{2}\binom{n_{\rm tot}-2}{r-2}.
\end{equation*}
The final inequality follows from $\frac{\ell(\ell-1)}{N(N-1)}\leq \frac{\ell^2}{N^2},$ and $n_{\rm tot}(n_{\rm tot}-1)\leq n_{\rm tot}^2.$
Combining Eqs.~\eqref{eq:gap_UB_perm_lsites}, \eqref{eq:slow_mode_local_expectation}, and \eqref{eq:slow_mode_T_geq2_bound}, we obtain
\begin{equation}
    \bra{\Omega_{\bm m}}
    M_{\nu_{G,\ell\mathrm{-loc}}}^{(2)}
    \ket{\Omega_{\bm m}}
    \geq
    1-\frac{1}{2}
    \left(
        \frac{\ell n_{\rm tot}}{N}
    \right)^2.\label{eq:moment_Omega_m_upper_bound}
\end{equation}
Together with Eq.~\eqref{eq:slow_mode_perp_Haar_moment}, this proves Eq.~\eqref{eq:universal_slow_mode_def}.

Finally, since $\ket{\Omega_{\bm m}}$ is a normalized state in the charge sector $(\bm n,\overline{\bm n})$ and is annihilated by $P_{U(1)\mathrm{Haar}}^{(2)}$, we have
\begin{align}
    1-\Delta_{\nu_{G,\ell\mathrm{-loc}};\bm n,\overline{\bm n}}^{(2)}
    &=
    \left\|
        \Pi_{\bm n,\overline{\bm n}}
        \left(
            M_{\nu_{G,\ell\mathrm{-loc}}}^{(2)}
            -
            P_{U(1)\mathrm{Haar}}^{(2)}
        \right)
        \Pi_{\bm n,\overline{\bm n}}
    \right\|_\infty
    \nonumber\\
    &\geq
    \left|
        \bra{\Omega_{\bm m}}
        \left(
            M_{\nu_{G,\ell\mathrm{-loc}}}^{(2)}
            -
            P_{U(1)\mathrm{Haar}}^{(2)}
        \right)
        \ket{\Omega_{\bm m}}
    \right|
    \nonumber\\
    &=
    \bra{\Omega_{\bm m}}
    M_{\nu_{G,\ell\mathrm{-loc}}}^{(2)}
    \ket{\Omega_{\bm m}}
    \nonumber\\
    &\geq
    1-\frac{1}{2}
    \left(
        \frac{\ell n_{\rm tot}}{N}
    \right)^2.
\end{align}
Here, in the third line we used Eq.~\eqref{eq:slow_mode_perp_Haar_moment} together with the positive semidefiniteness of $M_{\nu_{G,\ell\mathrm{-loc}}}^{(2)}$.
This proves Eq.~\eqref{eq:universal_slow_mode_gap_bound}.

\end{proof}

\subsection{Spectral gap upper bounds for parallel and fixed-architecture circuits} \label{ss:gap_upper_bound_parallel_fixed}

In the previous subsections, we have focused on spectral gap upper bounds for single-edge and single-hyperedge circuits.
In this subsection, we briefly explain how these upper bounds can be directly translated into upper bounds for the corresponding parallel and fixed-architecture circuits through standard architecture comparison arguments.
The precise comparison statements and their proofs are given in Sec.~\ref{s:architecture_comparison_argument}, and here we summarize only their direct consequences.

First, combining Theorem~\ref{restatethm:gap_upper_bound_two_particle_encounter} with Lemma~\ref{lem:sing_edge_upper_bound_to_par_and_fix}, proved in Sec.~\ref{s:architecture_comparison_argument}, gives the following corollary.

\begin{corollary}\label{restatecor:gap_upper_bound_two_particle_encounter_parallel_fixed}
Let $\Gamma$ be a finite set of matchings whose associated graph $G_\Gamma=(V,E_\Gamma)$ is an $N$-vertex connected graph.
We denote by $g_\Gamma\equiv\max_{e\in E_\Gamma}|\{\gamma\in\Gamma:e\in\gamma\}|$ the maximum number of matchings in $\Gamma$ containing the same edge.
Then,
\begin{equation}
\label{restateeq:gap_upper_bound_two_particle_encounter_parallel}
    \Delta_{\nu_\Gamma}^{(2)}
    \leq
    \frac{g_\Gamma |E_\Gamma|}{|\Gamma|}\Delta_{\nu_{G_\Gamma}}^{(2)}
    \leq
    \frac{g_\Gamma |E_\Gamma|}{|\Gamma|}\lambda_{G_\Gamma}^{\rm enc}.
\end{equation}
Let $A=(\gamma_1,\ldots,\gamma_L)$ be a fixed circuit architecture whose associated graph $G_A=(V,E_A)$, with $E_A=\bigcup_{r=1}^{L}\gamma_r$, is connected.
We denote by $g_A\equiv\max_{e\in E_A}|\{r\in[L]:e\in\gamma_r\}|$ the maximum number of occurrences of the same edge within the architecture unit $A$.
Then,
\begin{equation}
\label{restateeq:gap_upper_bound_two_particle_encounter_fixed}
    \Delta_{\nu_A}^{(2)}
    \leq
    4g_A|E_A|\Delta_{\nu_{G_A}}^{(2)}
    \leq
    4g_A|E_A|\lambda_{G_A}^{\rm enc}.
\end{equation}
\end{corollary}

For an $\alpha$-dimensional lattice or a bounded-degree expander graph, one may choose a constant-size set of matchings such that $g_\Gamma=O(1)$, $|\Gamma|=O(1)$, and $|E_\Gamma|=\Theta(N)$.
For the all-to-all parallel circuit, if $\Gamma_{\rm all}$ is the set of all maximum matchings, double counting the pairs $(e,\gamma)$ with $e\in\gamma$ gives $g_{\Gamma_{\rm all}}|E_{\Gamma_{\rm all}}|/|\Gamma_{\rm all}|=\lfloor N/2\rfloor$.
Therefore, in all of these cases, Eq.~\eqref{restateeq:gap_upper_bound_two_particle_encounter_parallel} gives
\begin{equation}
    \Delta_{\nu_\Gamma}^{(2)}
    \leq
    O\left(N\lambda_{G_\Gamma}^{\rm enc}\right).
\end{equation}
For a fixed-architecture circuit on an $\alpha$-dimensional lattice or a bounded-degree expander graph, one may similarly take $g_A=O(1)$ and $|E_A|=\Theta(N)$.
Equation~\eqref{restateeq:gap_upper_bound_two_particle_encounter_fixed} therefore gives
\begin{equation}
    \Delta_{\nu_A}^{(2)}
    \leq
    O\left(N\lambda_{G_A}^{\rm enc}\right).
\end{equation}

\vspace{1em}

Furthermore, Theorem~\ref{restatethm:universal_gap_upper_bound} yields the following corollary.

\begin{corollary}
\label{cor:gap_upper_bound_l_local_parallel_fixed}
Let $\Gamma$ be any finite matching family whose associated $\ell$-uniform hypergraph $G_\Gamma=(V,E_\Gamma)$ is connected.
Then, we have
\begin{equation}
\label{eq:gap_upper_bound_l_local_parallel}
    \Delta_{\nu_{\Gamma,\ell\mathrm{-loc}}}^{(2)} \leq \frac{2\ell}{N}.
\end{equation}

Let $A=(\gamma_1,\ldots,\gamma_L)$ be a fixed circuit architecture whose associated $\ell$-uniform hypergraph $G_A=(V,E_A)$, with $E_A=\bigcup_{r=1}^{L}\gamma_r$, is connected.
We denote by $g_A\equiv\max_{e\in E_A}|\{r\in[L]:e\in\gamma_r\}|$ the maximum number of occurrences of the same hyperedge within the architecture unit $A$.
Then,
\begin{equation}
\label{eq:gap_upper_bound_l_local_fixed}
    \Delta_{\nu_{A,\ell\mathrm{-loc}}}^{(2)}
    \leq
    4g_A|E_A|
    \Delta_{\nu_{G_A,\ell\mathrm{-loc}}}^{(2)}
    \leq
    8g_A|E_A|
    \left(\frac{\ell}{N}\right)^2.
\end{equation}
\end{corollary}
\noindent
If the fixed architecture has depth $L=O(1)$, we have $g_A=O(1)$ and $|E_A|=O(N/\ell)$, and Eq.~\eqref{eq:gap_upper_bound_l_local_fixed} implies
\begin{equation}
    \Delta_{\nu_{A,\ell\mathrm{-loc}}}^{(2)}
    \leq
    O\left(\frac{\ell}{N}\right).
\end{equation}

\begin{proof}[Proof of Corollary~\ref{cor:gap_upper_bound_l_local_parallel_fixed}]
Equation~\eqref{eq:gap_upper_bound_l_local_fixed} immediately follows from Theorem~\ref{restatethm:universal_gap_upper_bound} and Lemma~\ref{lem:single_hyperedge_upper_bound_to_par_and_fix}.
On the other hand, we derive Eq.~\eqref{eq:gap_upper_bound_l_local_parallel} directly from Theorem~\ref{restatethm:universal_gap_upper_bound}, without using an architecture comparison argument.

First, consider the type vector $\bm m$ satisfying $m_{1010}=m_{0110}=1$ and $m_{0000}=N-2$.
From Eqs.~\eqref{eq:slow_mode_local_expectation} and \eqref{eq:slow_mode_T_geq2_bound}, we have
\begin{equation}
\label{eq:UB_l_loc_parallel_prf1}
    \bra{\Omega_{\bm m}}\left(P_{e}^{(2)}\otimes\mathbb I_{\overline{e}}\right)\ket{\Omega_{\bm m}}
    \geq
    1-2\left(\frac{\ell}{N}\right)^2,
\end{equation}
where $P_{e}^{(2)}$ denotes the second-moment projector of the $U(1)$-symmetric Haar ensemble on the hyperedge $e$.

Furthermore, using $P_{U(1)\mathrm{Haar}}^{(2)}\ket{\Omega_{\bm m}}=0$, we have
\begin{align}
    \Delta_{\nu_{\Gamma,\ell\mathrm{-loc}}}^{(2)}
    &\leq \bra{\Omega_{\bm m}}\left(\mathbb I-M_{\nu_{\Gamma,\ell\mathrm{-loc}}}^{(2)}\right)\ket{\Omega_{\bm m}} \nonumber \\
    &= \frac{1}{|\Gamma|}\sum_{\gamma\in\Gamma}
    \bra{\Omega_{\bm m}}\left(\mathbb I-P^{(2)}_{\gamma,\ell\mathrm{-loc}}\right)\ket{\Omega_{\bm m}}.
    \label{eq:UB_l_loc_parallel_prf2}
\end{align}
From the operator inequality
\begin{equation}
\label{eq:UB_l_loc_parallel_prf3}
    0\leq
    \mathbb I-P^{(2)}_{\gamma,\ell\mathrm{-loc}}
    =
    \mathbb I-\prod_{e\in\gamma}\left(P^{(2)}_{e}\otimes\mathbb I_{\overline{e}}\right)
    \leq
    \sum_{e\in\gamma}\left(\mathbb I-P^{(2)}_{e}\otimes\mathbb I_{\overline{e}}\right),
\end{equation}
we further obtain
\begin{align}
    \bra{\Omega_{\bm m}}\left(\mathbb I-P^{(2)}_{\gamma,\ell\mathrm{-loc}}\right)\ket{\Omega_{\bm m}}
    &\leq
    \bra{\Omega_{\bm m}}
    \sum_{e\in\gamma}\left(\mathbb I-P^{(2)}_{e}\otimes\mathbb I_{\overline{e}}\right)
    \ket{\Omega_{\bm m}} \nonumber \\
    &\leq
    2|\gamma|\left(\frac{\ell}{N}\right)^2,
    \label{eq:UB_l_loc_parallel_prf4}
\end{align}
where the first inequality follows from Eq.~\eqref{eq:UB_l_loc_parallel_prf3}, and the second follows from Eq.~\eqref{eq:UB_l_loc_parallel_prf1}.

Combining Eqs.~\eqref{eq:UB_l_loc_parallel_prf2} and \eqref{eq:UB_l_loc_parallel_prf4}, we obtain
\begin{equation}
    \Delta_{\nu_{\Gamma,\ell\mathrm{-loc}}}^{(2)}
    \leq
    \frac{1}{|\Gamma|}
    \sum_{\gamma\in\Gamma}
    2|\gamma|\left(\frac{\ell}{N}\right)^2
    \leq
    \frac{2\ell}{N},
\end{equation}
where the second inequality follows from $|\gamma|\leq\lfloor N/\ell\rfloor\leq N/\ell$ for every $\gamma\in\Gamma$.
This concludes the proof.
\end{proof}

\section{Two-particle encounter rate in representative graph geometries} \label{s:two_particle_encounter_rate}
In the previous section, we showed that the principal decay rate of the two-particle encounter process, denoted by $\lambda_{G}^{\rm enc}$, provides an upper bound on the spectral gap of the corresponding single-edge $U(1)$-symmetric random circuit.
In this section, we explicitly obtain its scaling in representative graph geometries, namely the $\alpha$-dimensional lattice graph, expander graph, and all-to-all interaction graph (i.e., complete graph).
Specifically, the goal of this section is to derive the following lemma:
\begin{lemma}\label{lem:enc_time_representative_graphs}
Let $G_{\alpha{\rm D}}$ be an $\alpha$-dimensional lattice graph with $\alpha=O(1)$, $G_{\rm exp}$ be an expander graph, and $G_{\rm all}$ be the complete graph on $N$ vertices.
Then, we have
\begin{align}
    &\lambda_{G_{1{\rm D}}}^{\rm enc} = \Theta(N^{-3}), \label{eq:enc_rate_1D}\\
    &\lambda_{G_{2{\rm D}}}^{\rm enc} = \Theta\!\left(N^{-2}(\log N)^{-1}\right), \label{eq:enc_rate_2D}\\
    &\lambda_{G_{\alpha{\rm D}}}^{\rm enc} = \Theta(N^{-2}), \qquad \mathrm{for}\ \alpha\geq 3, \label{eq:enc_rate_alphaD}\\
    &\lambda_{G_{\rm exp}}^{\rm enc} = \Theta(N^{-2}), \label{eq:enc_rate_exp} \\
    &\lambda_{G_{\rm all}}^{\rm enc} = \frac{2}{N(N-1)}, \label{eq:enc_rate_all}
\end{align}
where the lattice estimates hold for both open and periodic boundary conditions.
\end{lemma}

The key to deriving these results is the correspondence between the two-particle encounter process $K_G^{\rm enc}$ and the continuous-time random walks on $G$ of two independent particles absorbed upon their first meeting.
In Sec.~\ref{ss:encounter_rate_hitting_time}, we first explain this correspondence and bound $\lambda_G^{\rm enc}$ in terms of \emph{hitting times}.
In Sec.~\ref{ss:proof_lemma_encounter_rate}, we derive Lemma~\ref{lem:enc_time_representative_graphs} using these bounds.

\subsection{Relationship to meeting time and hitting time} \label{ss:encounter_rate_hitting_time}

In this subsection, we introduce the independent random walks of two particles.
Although this process differs from the discrete-time two-particle encounter process discussed in Sec.~\ref{ss:trans_mat_two_part_enc}, we show that they are related by a time rescaling and absorption upon collision.
We consider two independent continuous-time random walks on a graph $G=(V,E)$, whose Laplacian is given by
\begin{equation}
\label{eq:two_particle_indep_rand_walk}
\begin{split}
    &L_G^{(2)} \equiv L_G \otimes \mathbb I + \mathbb I \otimes L_G, \\
    &L_G \ket{i} \equiv d_i \ket{i} - \sum_{\{i,l\}\in E} \ket{l},
\end{split}
\end{equation}
where $\ket{i}$ denotes the state in which a particle is at site $i$, and $d_i$ is the degree of site $i$.
Then, the independent two-particle random walk is described by the master equation $\frac{d}{dt}\ket{f(t)}=-L_G^{(2)}\ket{f(t)}$, where the two-particle probability vector is denoted by $\ket{f(t)}\equiv\sum_{i_1,i_2}f(i_1,i_2;t)\ket{i_1}\ket{i_2}$.
From Eq.~\eqref{eq:two_particle_indep_rand_walk}, for $i\neq j$, we have
\begin{align}
    L_G^{(2)} \ket{i}\ket{j} &= (d_i+d_j)\ket{i}\ket{j} - \sum_{\{i,l\}\in E}\ket{l}\ket{j} - \sum_{\{j,l\}\in E}\ket{i}\ket{l} \nonumber \\
    &= 2|E|(\mathbb I-K_G^{\rm enc})\ket{i,j} - \delta_{\{i,j\}\in E}(\ket{i,i}+\ket{j,j}),
\end{align}
where $\delta_{\{i,j\}\in E}$ is one if $\{i,j\}\in E$ and zero otherwise.
This equation shows that the Laplacian $L_G^{(2)}$ corresponds to the two-particle encounter process when collisions are made absorbing.
To be more specific, if we define the projector $\Pi_{\rm dist}\equiv\mathbb I-\sum_{i=1}^N\ket{i}\ket{i}\bra{i}\bra{i}$, the Dirichlet Laplacian on its range exactly satisfies
\begin{equation}
    L_G^D \equiv \left.\Pi_{\rm dist}L_G^{(2)}\Pi_{\rm dist}\right|_{\operatorname{Ran}\Pi_{\rm dist}} = 2|E|(\mathbb I-K_G^{\rm enc}).
\end{equation}
Here, the identity and $K_G^{\rm enc}$ on the right-hand side act on the subspace of configurations with distinct particle positions.
Since $G$ is connected, $L_G^D$ is positive definite, and the principal decay rate of $K_G^{\rm enc}$ is given by $\lambda_G^{\rm enc}=\lambda_{\min}(L_G^D)/(2|E|)$.

To evaluate this quantity, we introduce the notion of \emph{meeting time} in the two-particle random walk.
The meeting time $T^{\rm meet}$ is the first time that the independent walks collide, and we denote its expectation for initial positions $i$ and $j$ by $t^{\rm meet}_{G;i,j}\equiv\mathbb E_{i,j}[T^{\rm meet}]$.
Then, it can be represented as
\begin{align}
    t^{\rm meet}_{G;i,j} &= \int_0^{\infty} ds\ \mathrm{Prob}_{i,j} [T^{\rm meet}\geq s] \nonumber\\
    &= \int_0^{\infty} ds\ \sum_{a\neq b}\bra{a,b}e^{-sL_G^D}\ket{i,j} \nonumber\\
    &= \sum_{a\neq b}\bra{a,b}\left(\int_0^{\infty} ds\ e^{-sL_G^D}\right)\ket{i,j} \nonumber\\
    &= \bra{\bm{1}}[L_G^D]^{-1}\ket{i,j},
\end{align}
where $\ket{\bm{1}}\equiv\sum_{a\neq b}\ket{a,b}$ is the unnormalized uniform vector on the distinct-position subspace.
We further define the maximum and average meeting times as
\begin{align}
    &t^{\rm meet}_{G;\mathrm{max}} \equiv \max_{i\neq j}t^{\rm meet}_{G;i,j} = \max_{i\neq j}\bra{\bm{1}}[L_G^D]^{-1}\ket{i,j}, \\
    &t^{\rm meet}_{G;\mathrm{ave}} \equiv \frac{1}{N(N-1)}\sum_{i\neq j}t^{\rm meet}_{G;i,j} = \frac{1}{N(N-1)}\bra{\bm{1}}[L_G^D]^{-1}\ket{\bm{1}}.
\end{align}
Then, by using these quantities, the principal decay rate $\lambda_G^{\rm enc}$ can be bounded as follows:
\begin{lemma}\label{lem:enc_rate_meeting_time_bound}
For any connected graph $G=(V,E)$, we have
\begin{equation}
\label{eq:enc_rate_meet_bounds}
    \frac{1}{2|E|t^{\rm meet}_{G;\mathrm{max}}}\leq\lambda_G^{\rm enc}\leq\frac{1}{2|E|t^{\rm meet}_{G;\mathrm{ave}}}.
\end{equation}
\end{lemma}

\begin{proof}
Since $2|E|\lambda_G^{\rm enc}=\lambda_{\min}(L_G^D)$, we can derive Eq.~\eqref{eq:enc_rate_meet_bounds} by bounding $\lambda_{\min}(L_G^D)$.
The lower bound follows from
\begin{equation}
    \frac{1}{\lambda_{\min}(L_G^D)}=\rho\!\left([L_G^D]^{-1}\right)\leq\max_{i\neq j}\sum_{a\neq b}\left([L_G^D]^{-1}\right)_{(a,b),(i,j)}=t^{\rm meet}_{G;\mathrm{max}},
\end{equation}
where we used the fact that $[L_G^D]^{-1}$ has nonnegative matrix elements, and that the spectral radius is bounded above by the maximum absolute column sum.
The upper bound is derived by using
\begin{equation}
    \frac{1}{\lambda_{\min}(L_G^D)}=\rho\!\left([L_G^D]^{-1}\right)\geq\frac{1}{N(N-1)}\bra{\bm{1}}[L_G^D]^{-1}\ket{\bm{1}}=t^{\rm meet}_{G;\mathrm{ave}},
\end{equation}
where the inequality follows from the variational characterization of the largest eigenvalue of the positive definite matrix $[L_G^D]^{-1}$.
\end{proof}

Lemma~\ref{lem:enc_rate_meeting_time_bound} provides a bound on the two-particle encounter rate in terms of the more accessible notion of meeting time for two independent random walks.
We next show that this meeting time can be bounded in terms of a yet more accessible quantity, the hitting time of a single-particle random walk described by the Laplacian $L_G$.
Specifically, we denote by $t^{\rm hit}_{G;i,j}$ the expected time for a particle initially at site $i$ to reach site $j$ for the first time.
By further defining the maximum and average hitting times as
\begin{align}
    t^{\rm hit}_{G;\mathrm{max}} &\equiv \max_{i,j}t^{\rm hit}_{G;i,j}, \\
    t^{\rm hit}_{G;\mathrm{ave}} &\equiv \frac{1}{N(N-1)}\sum_{i\neq j}t^{\rm hit}_{G;i,j},
\end{align}
the following relationship between the hitting time and the meeting time is known \cite{aldous2002reversible}:
\begin{lemma}\label{lem:meeting_time_hitting_time_relationship}
For any connected graph $G$, we have
\begin{equation}
\label{eq:max_meet_upper_bound_max_hit}
    t^{\rm meet}_{G;\mathrm{max}}\leq t^{\rm hit}_{G;\mathrm{max}}.
\end{equation}
For a vertex-transitive graph $G$, we have
\begin{equation}
\label{eq:ave_meet_lower_bound_max_hit}
    t^{\rm meet}_{G;\mathrm{ave}}\geq\frac{1}{4}t^{\rm hit}_{G;\mathrm{max}}.
\end{equation}
\end{lemma}
Here, a graph is vertex-transitive if, for any two vertices, there exists a graph automorphism mapping one to the other.
A representative example is the $\alpha$-dimensional torus, i.e., an $\alpha$-dimensional lattice with periodic boundary conditions.
While we do not provide the derivation of this lemma here, Eq.~\eqref{eq:max_meet_upper_bound_max_hit} directly follows from Proposition~14.5 of Ref.~\cite{aldous2002reversible}, and Eq.~\eqref{eq:ave_meet_lower_bound_max_hit} is derived by combining Proposition~14.5 and Lemma~7.2~$\mathrm{(ii)}$ of the same reference.

\subsection{Proof of Lemma~\ref{lem:enc_time_representative_graphs}} \label{ss:proof_lemma_encounter_rate}

In this subsection, we derive Lemma~\ref{lem:enc_time_representative_graphs} by combining Lemmas~\ref{lem:enc_rate_meeting_time_bound} and \ref{lem:meeting_time_hitting_time_relationship}.
For that purpose, we need the following fact about the maximum hitting time in various geometries:
\begin{lemma}\label{lem:hitting_time_for_various_geometries}
Let $G_{\alpha{\rm D}}$ be an $\alpha$-dimensional lattice graph with $\alpha=O(1)$, and $G_{\rm exp}$ be an expander graph.
Then, we have
\begin{align}
    &t^{\rm hit}_{G_{1\rm D};\mathrm{max}} = \Theta(N^2), \\
    &t^{\rm hit}_{G_{2\rm D};\mathrm{max}} = \Theta(N\log N), \\
    &t^{\rm hit}_{G_{\alpha\rm D};\mathrm{max}} = \Theta(N), \qquad \mathrm{for}\ \alpha\geq 3, \\
    &t^{\rm hit}_{G_{\rm exp};\mathrm{max}} = \Theta(N), \label{eq:max_hitting_time_expander}
\end{align}
where the lattice estimates hold for both open and periodic boundary conditions.
\end{lemma}
\noindent
These scalings directly follow by combining the results of previous works.
First, the maximum hitting time is characterized by the maximum effective resistance of the graph $G$, as follows from Proposition~10.7 of Ref.~\cite{levin2026markov}.
Using this relationship, the required estimates can be obtained for lattices with periodic boundary conditions (see Theorem~6.1 of Ref.~\cite{chandra} and Proposition~10.21 of Ref.~\cite{levin2026markov}), lattices with open boundary conditions (see Proposition~9.17 and the arguments in Sec.~10.6 of Ref.~\cite{levin2026markov}), and expander graphs (see Theorem~10.1 of Ref.~\cite{aldous2002reversible} for regular expanders), as shown in Lemma~\ref{lem:hitting_time_for_various_geometries}.

It is also useful to introduce the following universal upper bound on $\lambda_G^{\rm enc}$, which follows directly from $K_G^{\rm enc}$ without using the independent-random-walk comparison.
\begin{lemma} \label{lem:universal_upper_bound_encounter_rate}
For any $N$-vertex connected graph $G$, we have
\begin{equation}
\label{eq:robust_lambda_encount_upper_bound}
    \lambda_G^{\rm enc} \leq \frac{2}{N(N-1)}.
\end{equation}
\end{lemma}
\begin{proof}
This upper bound can be derived by considering the uniform superposition over all $N(N-1)$ ordered two-particle configurations, denoted by $\ket{\bm v_{\rm unif}}\equiv\frac{1}{\sqrt{N(N-1)}}\sum_{i\neq j}\ket{i,j}$.
For every edge $e$, only the two configurations in which the two particles occupy the two endpoints of $e$ are eliminated, while the uniform superposition is invariant under the transposition of the endpoints.
Therefore,
\begin{equation*}
    \bra{\bm v_{\rm unif}}K_e^{\rm enc}\ket{\bm v_{\rm unif}}=1-\frac{2}{N(N-1)},
\end{equation*}
and the same equality holds with $K_e^{\rm enc}$ replaced by $K_G^{\rm enc}$.
Since $K_G^{\rm enc}$ is symmetric, its spectral radius is at least this Rayleigh quotient, which gives Eq.~\eqref{eq:robust_lambda_encount_upper_bound}.
\end{proof}

By using these lemmas, we now give the proof of Lemma~\ref{lem:enc_time_representative_graphs}:
\begin{proof}[Proof of Lemma~\ref{lem:enc_time_representative_graphs}]
We first consider an $\alpha$-dimensional lattice with $\alpha=O(1)$.
For lattices with periodic boundary conditions, we can directly obtain Eqs.~\eqref{eq:enc_rate_1D}, \eqref{eq:enc_rate_2D}, and \eqref{eq:enc_rate_alphaD} by combining Lemmas~\ref{lem:enc_rate_meeting_time_bound}, \ref{lem:meeting_time_hitting_time_relationship}, and \ref{lem:hitting_time_for_various_geometries}, since they are vertex-transitive graphs and Eq.~\eqref{eq:ave_meet_lower_bound_max_hit} applies.
On the other hand, since lattices with open boundary conditions are not vertex-transitive, Eq.~\eqref{eq:ave_meet_lower_bound_max_hit} is not applicable to them, and we need a separate argument for the upper bound on $\lambda_G^{\rm enc}$.
For this purpose, we compare lattices with open and periodic boundary conditions, denoted by $G_{\alpha\rm D,OBC}$ and $G_{\alpha\rm D,PBC}$, respectively.
Specifically, since $G_{\alpha\rm D,OBC}$ is a subgraph of $G_{\alpha\rm D,PBC}$ and $\mathbb I-K_e^{\rm enc}$ is positive semidefinite for every edge $e$, we have the operator inequality $|E_{\alpha\rm D,OBC}|(\mathbb I-K^{\rm enc}_{G_{\alpha\rm D,OBC}})\leq|E_{\alpha\rm D,PBC}|(\mathbb I-K^{\rm enc}_{G_{\alpha\rm D,PBC}})$.
This leads to the upper bound $\lambda^{\rm enc}_{G_{\alpha\rm D,OBC}}\leq\frac{|E_{\alpha\rm D,PBC}|}{|E_{\alpha\rm D,OBC}|}\lambda^{\rm enc}_{G_{\alpha\rm D,PBC}}=\Theta(\lambda^{\rm enc}_{G_{\alpha\rm D,PBC}})$, since $\alpha=O(1)$.
The matching lower bound for open-boundary lattices follows from Eqs.~\eqref{eq:enc_rate_meet_bounds} and \eqref{eq:max_meet_upper_bound_max_hit}, together with the hitting-time estimates in Lemma~\ref{lem:hitting_time_for_various_geometries}.

We next discuss the encounter rate for the expander graph $\lambda_{G_{\rm exp}}^{\rm enc}$.
Its lower bound $\lambda_{G_{\rm exp}}^{\rm enc}\geq\Omega(N^{-2})$ is directly obtained by combining Lemma~\ref{lem:enc_rate_meeting_time_bound}, Eq.~\eqref{eq:max_meet_upper_bound_max_hit} of Lemma~\ref{lem:meeting_time_hitting_time_relationship}, and Eq.~\eqref{eq:max_hitting_time_expander} of Lemma~\ref{lem:hitting_time_for_various_geometries}.
Its matching upper bound also follows from Lemma~\ref{lem:universal_upper_bound_encounter_rate}, which gives Eq.~\eqref{eq:enc_rate_exp}.

Finally, we address the encounter rate for the complete graph.
By utilizing the site-permutation invariance of this graph, we can exactly obtain its principal decay rate, denoted by $\lambda_{G_{\rm all}}^{\rm enc}$.
Specifically, from Eq.~\eqref{eq:two_part_enc_process_column_sum}, we have $\sum_{(i',j')}[K_{G_{\rm all}}^{\rm enc}]_{(i',j'),(i,j)}=1-2/[N(N-1)]$ for all $(i,j)$, and therefore $K_{G_{\rm all}}^{\rm enc}$ has the uniform vector $\ket{\bm v_{\rm unif}}$ as an eigenvector with the corresponding eigenvalue $1-2/[N(N-1)]$.
Since $K_{G_{\rm all}}^{\rm enc}$ has nonnegative matrix elements, its spectral radius is bounded above by this common column sum, and hence equals $1-2/[N(N-1)]$.
This gives Eq.~\eqref{eq:enc_rate_all} and shows that the complete graph achieves equality in the universal upper bound given in Eq.~\eqref{eq:robust_lambda_encount_upper_bound}.
\end{proof}

\section{Circuit-architecture comparison arguments}\label{s:architecture_comparison_argument}

In this section, we present standard comparison arguments relating the spectral gaps of the single-edge, parallel, and fixed-architecture circuits.
The main tools are the detectability lemma \cite{aharonov2009detectability,anshu2016simple} and the quantum union bound \cite{gao2015quantum}, which we state first.

\begin{lemma}[Detectability lemma~\cite{anshu2016simple}]
\label{lem:detectability_lemma}
Let $H=\sum_{a=1}^{L}Q_a$ be a frustration-free Hamiltonian, where each $Q_a$ is an orthogonal projector.
Suppose that each $Q_a$ fails to commute with at most $d$ other projectors, with $d\geq1$.
Then, for any ordering of the projectors,
\begin{equation}
\label{eq:detectability_lemma}
    \max_{\substack{\ket{\psi^\perp}\perp\ker H\\ \|\psi^\perp\|=1}}
    \left\|\prod_{a=1}^{L}\left(\mathbb I-Q_a\right)\ket{\psi^\perp}\right\|^2
    \leq
    \frac{1}{1+d^{-2}\Delta(H)},
\end{equation}
where $\Delta(H)$ denotes the smallest nonzero eigenvalue of $H$.
\end{lemma}

\begin{lemma}[Quantum union bound~\cite{gao2015quantum}]
\label{lem:quantum_union_bound}
Let $H=\sum_{a=1}^{L}Q_a$ be a frustration-free Hamiltonian, where each $Q_a$ is an orthogonal projector.
Then, for any ordering of the projectors,
\begin{equation}
\label{eq:quantum_union_bound}
    \max_{\substack{\ket{\psi^\perp}\perp\ker H\\ \|\psi^\perp\|=1}}
    \left\|\prod_{a=1}^{L}\left(\mathbb I-Q_a\right)\ket{\psi^\perp}\right\|^2
    \geq
    1-4\Delta(H).
\end{equation}
\end{lemma}

We note that throughout this section, we restrict attention to moment orders $k$ for which
\begin{equation*}
    \bigcap_e \operatorname{Ran} (P_e^{(k)} \otimes \mathbb{I}_{\overline{e}})
    =
    \operatorname{Ran} P_{U(1)\mathrm{Haar}}^{(k)},
\end{equation*}
where the intersection is taken over all edges on which a two-qubit gate can be applied in the circuit.
For $U(1)$-symmetric random circuits composed of two-qubit gates, it is known that this condition is no longer satisfied for $k\geq 2(N-1)$~\cite{mitsuhashi2024unitary,mitsuhashi2024characterization,liu2024unitary}.
In this case, the common fixed space of the local moment operators strictly contains the $U(1)$-symmetric Haar moment space, and hence the spectral gap defined with respect to $P_{U(1)\mathrm{Haar}}^{(k)}$ vanishes.
We therefore restrict the discussion below to the regime in which the above equality holds, which includes the moment orders relevant to our main results.

\subsection{Upper bound based on the gap of single-edge circuit}\label{ss:comparison_argument_UB_with_single_edge}

We first upper-bound the spectral gaps of the parallel and fixed-architecture circuits in terms of the gap of the corresponding single-edge circuit.

\begin{lemma}
\label{lem:sing_edge_upper_bound_to_par_and_fix}
Let $\Gamma$ be a finite set of matchings whose associated graph $G_\Gamma=(V,E_\Gamma)$ is an $N$-vertex connected graph.
We denote by $g_\Gamma\equiv\max_{e\in E_\Gamma}|\{\gamma\in\Gamma:e\in\gamma\}|$ the maximum number of matchings in $\Gamma$ containing the same edge.
Then,
\begin{equation}
\label{eq:upper_bound_par_gap_with_sing_gap}
    \Delta_{\nu_\Gamma}^{(k)}
    \leq
    \frac{g_\Gamma |E_\Gamma|}{|\Gamma|}
    \Delta_{\nu_{G_\Gamma}}^{(k)}.
\end{equation}

Let $A=(\gamma_1,\ldots,\gamma_L)$ be a fixed circuit architecture whose associated graph $G_A=(V,E_A)$, with $E_A=\bigcup_{r=1}^{L}\gamma_r$, is connected.
We denote by $g_A\equiv\max_{e\in E_A}|\{r\in[L]:e\in\gamma_r\}|$ the maximum number of occurrences of the same edge within the architecture unit $A$.
Then,
\begin{equation}
\label{eq:upper_bound_fix_gap_with_sing_gap_general}
    \Delta_{\nu_A}^{(k)}
    \leq
    4g_A|E_A|\Delta_{\nu_{G_A}}^{(k)}.
\end{equation}
\end{lemma}

\begin{proof}
For an edge $e=\{i,j\}$, let $Q_e\equiv\mathbb I-P_{i,j}^{(k)}\otimes\mathbb I_{\overline{i,j}}$.
For the graph $G_\Gamma$ and the set of matchings $\Gamma$, define the frustration-free Hamiltonians
\begin{align}
    H_{G_\Gamma} &\equiv |E_\Gamma|\left(\mathbb I-M_{\nu_{G_\Gamma}}^{(k)}\right)=\sum_{e\in E_\Gamma}Q_e, \label{eq:H_G_def}\\
    H_\Gamma &\equiv |\Gamma|\left(\mathbb I-M_{\nu_\Gamma}^{(k)}\right)=\sum_{\gamma\in\Gamma}\left(\mathbb I-P_\gamma^{(k)}\right). \label{eq:H_Gamma_def}
\end{align}
Since these Hamiltonians have the same ground-state space, 
\begin{equation}
\label{eq:ground_state_subsp_U(1)Haar}
    \ker H_{G_\Gamma}
    =
    \ker H_\Gamma
    =
    \operatorname{Ran}P_{U(1)\mathrm{Haar}}^{(k)}.
\end{equation}
Their spectral gaps are therefore related to the corresponding moment spectral gaps by
\begin{equation}
\label{eq:spectral_gap_Hamiltonian_gap}
    \Delta(H_{G_\Gamma})
    =
    |E_\Gamma|\Delta_{\nu_{G_\Gamma}}^{(k)},
    \qquad
    \Delta(H_\Gamma)
    =
    |\Gamma|\Delta_{\nu_\Gamma}^{(k)}.
\end{equation}

For each matching $\gamma$, the projectors $\{Q_e:e\in\gamma\}$ commute because the edges in $\gamma$ are mutually disjoint.
Hence,
\begin{equation}
    0
    \leq
    \mathbb I-P_\gamma^{(k)}
    =
    \mathbb I-\prod_{e\in\gamma}\left(\mathbb I-Q_e\right)
    \leq
    \sum_{e\in\gamma}Q_e.
\end{equation}
Summing this inequality over $\gamma\in\Gamma$ and using the definition of $g_\Gamma$, we obtain
\begin{equation}
    H_\Gamma
    \leq
    \sum_{\gamma\in\Gamma}\sum_{e\in\gamma}Q_e
    \leq
    g_\Gamma\sum_{e\in E_\Gamma}Q_e
    =
    g_\Gamma H_{G_\Gamma}.
\end{equation}
Since $H_\Gamma$ and $H_{G_\Gamma}$ have the same kernel, the variational characterization of the smallest nonzero eigenvalue gives
\begin{equation}
    |\Gamma|\Delta_{\nu_\Gamma}^{(k)}
    =
    \Delta(H_\Gamma)
    \leq
    g_\Gamma\Delta(H_{G_\Gamma})
    =
    g_\Gamma|E_\Gamma|\Delta_{\nu_{G_\Gamma}}^{(k)}.
\end{equation}
This proves Eq.~\eqref{eq:upper_bound_par_gap_with_sing_gap}.

We next consider the fixed architecture $A=(\gamma_1,\ldots,\gamma_L)$.
Define
\begin{equation}
    J_A
    \equiv
    \sum_{r=1}^{L}\sum_{e\in\gamma_r}Q_e.
\end{equation}
Then, we have $\ker J_A=\ker H_{G_A}=\operatorname{Ran}P_{U(1)\mathrm{Haar}}^{(k)}$.
Moreover, $J_A\leq g_AH_{G_A}$, and hence
\begin{equation}
\label{eq:fixed_architecture_occurrence_gap_bound}
    \Delta(J_A)
    \leq
    g_A\Delta(H_{G_A})
    =
    g_A|E_A|\Delta_{\nu_{G_A}}^{(k)}.
\end{equation}
Using $M_{\nu_A}^{(k)}P_{U(1)\mathrm{Haar}}^{(k)}=P_{U(1)\mathrm{Haar}}^{(k)}M_{\nu_A}^{(k)}=P_{U(1)\mathrm{Haar}}^{(k)}$, together with the quantum union bound, we obtain
\begin{align}
    \left(1-\Delta_{\nu_A}^{(k)}\right)^2
    &=
    \max_{\substack{\ket{\psi^\perp}\perp\ker H_{G_A}\\ \|\psi^\perp\|=1}}
    \left\|
        \left[\prod_{e\in\gamma_L}\left(\mathbb I-Q_e\right)\right]
        \cdots
        \left[\prod_{e\in\gamma_1}\left(\mathbb I-Q_e\right)\right]
        \ket{\psi^\perp}
    \right\|^2
    \nonumber\\
    &\geq
    1-4\Delta(J_A).
\end{align}
Since $0\leq\Delta_{\nu_A}^{(k)}\leq1$, it follows that
\begin{align}
    \Delta_{\nu_A}^{(k)}
    &\leq
    1-\left(1-\Delta_{\nu_A}^{(k)}\right)^2
    \nonumber\\
    &\leq
    4\Delta(J_A)
    \nonumber\\
    &\leq
    4g_A|E_A|\Delta_{\nu_{G_A}}^{(k)},
\end{align}
where the last inequality follows from Eq.~\eqref{eq:fixed_architecture_occurrence_gap_bound}.
This proves Eq.~\eqref{eq:upper_bound_fix_gap_with_sing_gap_general}.
\end{proof}

The same comparison argument directly extends to $\ell$-local circuits on an $\ell$-uniform hypergraph.
\begin{lemma}
\label{lem:single_hyperedge_upper_bound_to_par_and_fix}
Let $\Gamma$ be a finite set of matchings of $\ell$-site hyperedges whose associated $\ell$-uniform hypergraph $G_\Gamma=(V,E_\Gamma)$ is connected.
We denote by $g_\Gamma\equiv\max_{e\in E_\Gamma}|\{\gamma\in\Gamma:e\in\gamma\}|$ the maximum number of matchings in $\Gamma$ containing the same hyperedge.
Then,
\begin{equation}
\label{eq:upper_bound_par_gap_with_single_hyperedge_gap}
    \Delta_{\nu_{\Gamma,\ell\mathrm{-loc}}}^{(k)}
    \leq
    \frac{g_\Gamma |E_\Gamma|}{|\Gamma|}
    \Delta_{\nu_{G_\Gamma,\ell\mathrm{-loc}}}^{(k)}.
\end{equation}
Let $A=(\gamma_1,\ldots,\gamma_L)$ be a fixed circuit architecture whose associated $\ell$-uniform hypergraph $G_A=(V,E_A)$, with $E_A=\bigcup_{r=1}^{L}\gamma_r$, is connected.
We denote by $g_A\equiv\max_{e\in E_A}|\{r\in[L]:e\in\gamma_r\}|$ the maximum number of occurrences of the same hyperedge within the architecture unit $A$.
Then,
\begin{equation}
\label{eq:upper_bound_fix_gap_with_single_hyperedge_gap}
    \Delta_{\nu_{A,\ell\mathrm{-loc}}}^{(k)}
    \leq
    4g_A|E_A|\Delta_{\nu_{G_A,\ell\mathrm{-loc}}}^{(k)}.
\end{equation}
\end{lemma}

The proof is identical to that of Lemma~\ref{lem:sing_edge_upper_bound_to_par_and_fix}, and we therefore omit it.
To see the correspondence explicitly, for a hyperedge $e=\{i_1,\ldots,i_\ell\}$, we replace the two-site Haar moment projector by the $\ell$-site Haar moment projector $P_e^{(k)}$ and define $Q_e\equiv\mathbb I-P_e^{(k)} \otimes \mathbb{I}_{\overline{e}}$.
Since the hyperedges belonging to the same matching are mutually disjoint, the corresponding projectors commute, and hence $P_\gamma^{(k)}=\prod_{e\in\gamma}P_e^{(k)} \otimes \mathbb{I}_{\overline{\gamma}}$ and $\mathbb I-P_\gamma^{(k)}\leq\sum_{e\in\gamma}Q_e$ hold exactly as in the $\ell=2$ case.
Consequently, the operator inequalities $H_\Gamma\leq g_\Gamma H_{G_\Gamma}$ and $J_A\leq g_AH_{G_A}$, which are the key ingredients of the proof, remain unchanged.

\subsection{Lower bound based on the gap of single-edge circuit}\label{ss:comparison_argument_LB_with_single_edge}

In this subsection, we lower bound the spectral gaps of parallel and fixed-architecture circuits in terms of those of the corresponding single-edge circuits.
\begin{lemma}
\label{lem:single_edge_to_parallel_gap_lower_bound}
Let $\Gamma$ be a finite set of matchings whose associated graph $G_\Gamma=(V,E_\Gamma)$ is an $N$-vertex connected graph.
Then,
\begin{equation}
    \Delta_{\nu_\Gamma}^{(k)}
    \geq
    \frac{|E_\Gamma|\Delta_{\nu_{G_\Gamma}}^{(k)}}{4|\Gamma|\left(4|\Gamma|^2+|E_\Gamma|\Delta_{\nu_{G_\Gamma}}^{(k)}\right)}.
\end{equation}

Let $A=(\gamma_1,\ldots,\gamma_L)$ be a fixed circuit architecture whose associated graph $G_A=(V,E_A)$, with $E_A=\bigcup_{r=1}^{L}\gamma_r$, is connected. Then, we have 
\begin{equation}
    \Delta_{\nu_A}^{(k)} \geq \frac{|E_A|\Delta_{\nu_{G_A}}^{(k)}}{8L^2+2|E_A|\Delta_{\nu_{G_A}}^{(k)}}.
\end{equation}

\end{lemma}

\begin{proof}
We first consider the parallel circuit.
As above, we use the Hamiltonians $H_{G_\Gamma}$ and $H_\Gamma$, and additionally define $J_\Gamma$ by
\begin{align*}
    H_{G_\Gamma} &\equiv |E_\Gamma|\left(\mathbb I-M_{\nu_{G_\Gamma}}^{(k)}\right)=\sum_{e\in E_\Gamma}Q_e, \\
    H_\Gamma &\equiv |\Gamma|\left(\mathbb I-M_{\nu_\Gamma}^{(k)}\right)=\sum_{\gamma\in\Gamma}Q_\gamma, \\
    J_\Gamma &\equiv \sum_{\gamma\in\Gamma}\sum_{e\in\gamma}Q_e, \label{eq:J_Gamma_def}
\end{align*}
where $Q_e\equiv\mathbb I-P_{e}^{(k)}\otimes\mathbb{I}_{\overline{e}}$ and $Q_\gamma \equiv \mathbb{I} - P_{\gamma}^{(k)}$.
These three Hamiltonians have the same ground-state space,
\begin{equation}
\label{eq:kernel_U(1)Haar_range_identity}
    \ker H_\Gamma
    =
    \ker J_\Gamma
    =
    \ker H_{G_\Gamma}
    =
    \operatorname{Ran}P_{U(1)\mathrm{Haar}}^{(k)}.
\end{equation}
We can therefore apply the quantum union bound to $H_\Gamma$ and the detectability lemma to $J_\Gamma$ to obtain
\begin{align}
    1-4\Delta(H_\Gamma)
    &\leq
    \max_{\substack{\ket{\psi^\perp}\perp\operatorname{Ran}P_{U(1)\mathrm{Haar}}^{(k)}\\ \|\psi^\perp\|=1}}
    \left\|
        \prod_{\gamma\in\Gamma}
        \left(\mathbb I-Q_\gamma\right)
        \ket{\psi^\perp}
    \right\|^2
    \nonumber\\
    &\leq
    \frac{1}{1+(2|\Gamma|)^{-2}\Delta(J_\Gamma)}
    \nonumber\\
    &\leq
    \frac{1}{1+(2|\Gamma|)^{-2}\Delta(H_{G_\Gamma})}.
\end{align}
Here, the first inequality follows from Lemma~\ref{lem:quantum_union_bound}, while the second follows from Lemma~\ref{lem:detectability_lemma}.
To apply the latter, we expand each layer projector as $\mathbb I-Q_\gamma=P_\gamma^{(k)}=\prod_{e\in\gamma}(\mathbb I-Q_e)$.
For any edge projector appearing in $J_\Gamma$, each matching contains at most two edge projectors that can fail to commute with it, so the number of noncommuting projectors is at most $2|\Gamma|$.
Finally, since every edge in $E_\Gamma$ appears in at least one matching in $\Gamma$, we have $J_\Gamma\geq H_{G_\Gamma}$.
Together with Eq.~\eqref{eq:kernel_U(1)Haar_range_identity}, this implies $\Delta(J_\Gamma)\geq\Delta(H_{G_\Gamma})$, which gives the last inequality.

Using $\Delta(H_\Gamma)=|\Gamma|\Delta_{\nu_\Gamma}^{(k)}$ and $\Delta(H_{G_\Gamma})=|E_\Gamma|\Delta_{\nu_{G_\Gamma}}^{(k)}$, we obtain
\begin{align}
    \Delta_{\nu_\Gamma}^{(k)}
    &= \frac{\Delta(H_\Gamma)}{|\Gamma|}
    \nonumber\\
    &\geq
    \frac{1}{4|\Gamma|}
    \left[
        1-
        \frac{1}{1+(2|\Gamma|)^{-2}|E_\Gamma|\Delta_{\nu_{G_\Gamma}}^{(k)}}
    \right]
    \nonumber\\
    &=
    \frac{|E_\Gamma|\Delta_{\nu_{G_\Gamma}}^{(k)}}{4|\Gamma|\left(4|\Gamma|^2+|E_\Gamma|\Delta_{\nu_{G_\Gamma}}^{(k)}\right)}.
\end{align}

Next, we show the result for the fixed-architecture circuit.
By introducing
\begin{align*}
    &J_A \equiv \sum_{r=1}^{L}\sum_{e\in\gamma_r}Q_e, \\
    &H_{G_A} \equiv |E_A| \left(\mathbb{I} - M_{\nu_{G_A}}^{(k)}\right) = \sum_{e \in E_A}Q_e,
\end{align*}
we have $J_A \geq H_{G_A}\geq 0$, and $\ker H_{G_A}=\ker J_A= \operatorname{Ran}P_{U(1)\mathrm{Haar}}^{(k)}$. Therefore, $\Delta(J_A) \geq \Delta(H_{G_A})\geq 0$ holds.

Then, by using the detectability lemma, we obtain
\begin{align}
    \left(1-\Delta_{\nu_A}^{(k)}\right)^2
    &=
    \max_{\substack{\ket{\psi^\perp}\perp\ker H_{G_A}\\ \|\psi^\perp\|=1}}
    \left\|
        \left[\prod_{e\in\gamma_L}\left(\mathbb I-Q_e\right)\right]
        \cdots
        \left[\prod_{e\in\gamma_1}\left(\mathbb I-Q_e\right)\right]
        \ket{\psi^\perp}
    \right\|^2
    \nonumber\\
    &\leq \frac{1}{1+(2L )^{-2}\Delta(J_A)} \nonumber \\
    &\leq \frac{1}{1+(2L)^{-2}\Delta(H_{G_A})},
\end{align}
where we used $\Delta(J_A) \geq \Delta(H_{G_A})\geq 0$ in the final line.
We also used the fact that, as in the parallel-circuit case, any edge projector appearing in $J_A$ can fail to commute with at most two edge projectors in each matching $\gamma_r$, so the number of noncommuting projectors is at most $2L$.
By utilizing this inequality, we have
\begin{align}
\Delta_{\nu_A}^{(k)} &\geq \frac{1}{2} \left[ 1-\left(1-\Delta_{\nu_A}^{(k)}\right)^2 \right] \nonumber \\
&\geq \frac{1}{2} \left[ 1-\frac{1}{1+(2L)^{-2}\Delta(H_{G_A})} \right] \nonumber \\
&=  \frac{|E_A|\Delta_{\nu_{G_A}}^{(k)}}{2(4L^2+|E_A|\Delta_{\nu_{G_A}}^{(k)})} ,
\end{align}
where we used $0\leq \Delta_{\nu_A}^{(k)}\le 1$ in the first line, and $\Delta(H_{G_A}) = |E_A|\Delta_{\nu_{G_A}}^{(k)}$ in the final line.
This concludes the proof.

\end{proof}

The argument above is not specific to moment operators of $U(1)$-symmetric unitaries.
It only uses the fact that the local operators associated with the edges are orthogonal projectors and that the corresponding Hamiltonians have the same kernel, as in Eq.~\eqref{eq:kernel_U(1)Haar_range_identity}.
Indeed, in Sec.~\ref{s:generate_2RPmix_group_with_circuit}, we apply the same comparison argument to circuits whose local gates are drawn from the 2RPmix group.

\subsection{Lower bound for the all-to-all interaction models}\label{ss:comparison_argument_all_to_all}

Finally, we derive spectral gap lower bounds for the single-edge and parallel circuits with all-to-all interactions.
For the natural all-to-all parallel circuit, the set of all maximum matchings contains a large number of matchings, and a direct application of the comparison bound in Sec.~\ref{ss:comparison_argument_LB_with_single_edge} does not yield a useful lower bound.
We therefore use permutation symmetry to obtain the tighter lower bounds.
Specifically, we show that the spectral gaps of the single-edge and parallel circuits with all-to-all interactions are no smaller than those of arbitrary single-edge and parallel circuits, respectively.

\begin{lemma}
\label{lem:all_to_all_matching_comparison}
Let $G_{\rm all}$ be the complete graph on $N$ sites.
Then, for any connected graph $G=(V,E)$ on the same set of sites,
\begin{equation}
\label{eq:all_to_all_single_edge_gap_comparison}
    \Delta_{\nu_{G_{\rm all}}}^{(k)}
    \geq
    \Delta_{\nu_G}^{(k)}.
\end{equation}
Similarly, let $\Gamma_{\rm all}$ denote the set of all maximum matchings on the $N$ sites.
Then, for any finite set of matchings $\Gamma$,
\begin{equation}
\label{eq:all_to_all_matching_comparison}
    \Delta_{\nu_{\Gamma_{\rm all}}}^{(k)}
    \geq
    \Delta_{\nu_\Gamma}^{(k)}.
\end{equation}
\end{lemma}

\begin{proof}
We first prove Eq.~\eqref{eq:all_to_all_single_edge_gap_comparison}.
For a site permutation $\pi\in S_N$, let $U_\pi$ denote the corresponding unitary permutation operator.
For an edge $e=\{i,j\}$, we write $\pi(e)\equiv\{\pi(i),\pi(j)\}$.
Since a site permutation maps a two-qubit $U(1)$-symmetric Haar moment projector to the same projector acting on the permuted edge, 
\begin{equation*}
    U_\pi^{\otimes k,k}P_{e}^{(k)} \otimes \mathbb{I}_{\overline{e}}\, U_{\pi^{-1}}^{\otimes k,k}=  P_{\pi(e)}^{(k)} \otimes \mathbb{I}_{\overline{\pi(e)}} . 
\end{equation*}
We now symmetrize the single-edge circuit over all site permutations:
\begin{align}
    \frac{1}{N!}\sum_{\pi\in S_N}U_\pi^{\otimes k,k}M_{\nu_G}^{(k)}U_{\pi^{-1}}^{\otimes k,k}
    &=
    \frac{1}{|E|}
    \sum_{e\in E}
    \left[
        \frac{1}{N!}
        \sum_{\pi\in S_N}
        P_{\pi(e)}^{(k)}\otimes \mathbb{I}_{\overline{\pi(e)}}
    \right]
    \nonumber\\
    &=
    M_{\nu_{G_{\rm all}}}^{(k)}.
\label{eq:Haar_single_edge_symmetrization}
\end{align}
Indeed, for any fixed edge $e$, a uniformly random site permutation maps it uniformly to one of the $\binom{N}{2}$ edges of the complete graph.

Moreover, the $U(1)$-symmetric Haar moment projector is invariant under arbitrary site permutations,
\begin{equation}
    U_\pi^{\otimes k,k}P_{U(1)\mathrm{Haar}}^{(k)}U_{\pi^{-1}}^{\otimes k,k}
    =
    P_{U(1)\mathrm{Haar}}^{(k)}.
\end{equation}
Therefore, Eq.~\eqref{eq:Haar_single_edge_symmetrization} gives
\begin{align}
    \left\|M_{\nu_{G_{\rm all}}}^{(k)}-P_{U(1)\mathrm{Haar}}^{(k)}\right\|_\infty
    &=
    \left\|
        \frac{1}{N!}
        \sum_{\pi\in S_N}
        U_\pi^{\otimes k,k}
        \left(
            M_{\nu_G}^{(k)}-P_{U(1)\mathrm{Haar}}^{(k)}
        \right)
        U_{\pi^{-1}}^{\otimes k,k}
    \right\|_\infty
    \nonumber\\
    &\leq
    \frac{1}{N!}
    \sum_{\pi\in S_N}
    \left\|
        U_\pi^{\otimes k,k}
        \left(
            M_{\nu_G}^{(k)}-P_{U(1)\mathrm{Haar}}^{(k)}
        \right)
        U_{\pi^{-1}}^{\otimes k,k}
    \right\|_\infty
    \nonumber\\
    &=
    \left\|M_{\nu_G}^{(k)}-P_{U(1)\mathrm{Haar}}^{(k)}\right\|_\infty.
\end{align}
The inequality follows from the triangle inequality, while the final equality follows from unitary invariance of the operator norm.
Equation~\eqref{eq:all_to_all_single_edge_gap_comparison} then follows immediately from the definition of the spectral gap.

We next prove Eq.~\eqref{eq:all_to_all_matching_comparison}.
For a matching $\gamma$, we write $\pi(\gamma)\equiv\{\pi(e):e\in\gamma\}$.
Since 
\begin{equation*}
    U_\pi^{\otimes k,k}P_\gamma^{(k)} \, U_{\pi^{-1}}^{\otimes k,k}=P_{\pi(\gamma)}^{(k)}, 
\end{equation*}
symmetrizing the parallel circuit over all site permutations gives
\begin{align}
    \frac{1}{N!}
    \sum_{\pi\in S_N}
    U_\pi^{\otimes k,k}
    M_{\nu_\Gamma}^{(k)}
    U_{\pi^{-1}}^{\otimes k,k}
    &=
    \frac{1}{|\Gamma|}
    \sum_{\gamma\in\Gamma}
    \left[
        \frac{1}{N!}
        \sum_{\pi\in S_N}
        P_{\pi(\gamma)}^{(k)}
    \right]
    \nonumber\\
    &\geq
    M_{\nu_{\Gamma_{\rm all}}}^{(k)}.
\label{eq:matching_symmetrization_bound}
\end{align}
Indeed, for any fixed $\gamma$, the permutation average in the square brackets is the uniform average over all matchings containing $|\gamma|$ edges.
Equivalently, one may first choose a maximum matching uniformly at random and then choose $|\gamma|$ of its edges uniformly at random.
Since the local projectors on disjoint edges commute, if a matching $\gamma$ is contained in a maximum matching $\gamma'$, then $P_\gamma^{(k)}\geq P_{\gamma'}^{(k)}$.
Averaging this operator inequality gives Eq.~\eqref{eq:matching_symmetrization_bound}.
In particular, when every $\gamma\in\Gamma$ is itself a maximum matching, the inequality in Eq.~\eqref{eq:matching_symmetrization_bound} becomes an equality.

The range of $P_{U(1)\mathrm{Haar}}^{(k)}$ is contained in the range of every matching projector $P_\gamma^{(k)}$.
Therefore, subtracting $P_{U(1)\mathrm{Haar}}^{(k)}$ from Eq.~\eqref{eq:matching_symmetrization_bound} preserves the operator ordering, and both sides are positive semidefinite.
It follows that
\begin{align}
    \left\|
        M_{\nu_{\Gamma_{\rm all}}}^{(k)}
        -
        P_{U(1)\mathrm{Haar}}^{(k)}
    \right\|_\infty
    &\leq
    \left\|
        \frac{1}{N!}
        \sum_{\pi\in S_N}
        U_\pi^{\otimes k,k}
        M_{\nu_\Gamma}^{(k)}
        U_{\pi^{-1}}^{\otimes k,k}
        -
        P_{U(1)\mathrm{Haar}}^{(k)}
    \right\|_\infty
    \nonumber\\
    &=
    \left\|
        \frac{1}{N!}
        \sum_{\pi\in S_N}
        U_\pi^{\otimes k,k}
        \left(
            M_{\nu_\Gamma}^{(k)}
            -
            P_{U(1)\mathrm{Haar}}^{(k)}
        \right)
        U_{\pi^{-1}}^{\otimes k,k}
    \right\|_\infty
    \nonumber\\
    &\leq
    \frac{1}{N!}
    \sum_{\pi\in S_N}
    \left\|
        U_\pi^{\otimes k,k}
        \left(
            M_{\nu_\Gamma}^{(k)}
            -
            P_{U(1)\mathrm{Haar}}^{(k)}
        \right)
        U_{\pi^{-1}}^{\otimes k,k}
    \right\|_\infty
    \nonumber\\
    &=
    \left\|
        M_{\nu_\Gamma}^{(k)}
        -
        P_{U(1)\mathrm{Haar}}^{(k)}
    \right\|_\infty.
\end{align}
Here, the second equality follows from the site-permutation invariance of $P_{U(1)\mathrm{Haar}}^{(k)}$, while the subsequent inequality follows from the triangle inequality and the final equality from unitary invariance of the operator norm.
Equation~\eqref{eq:all_to_all_matching_comparison} then follows from the definition of the spectral gap.
\end{proof}

The argument above is not specific to moment operators of $U(1)$-symmetric unitaries.
It only requires a site-permutation-covariant family of local orthogonal projectors, together with a target fixed-space projector that is invariant under site permutations and whose range is contained in the range of every local layer projector.

\section{Spectral gap lower bound for $U(1)$-symmetric random circuits} \label{s:lower_bound_spectral_gap}

In this section, we prove spectral gap lower bounds for $U(1)$-symmetric random circuits.
In Sec.~\ref{ss:lower_bound_spectral_gap_parallel}, we first derive the lower bound for parallel random circuits with restricted structure.
To make the overall proof strategy transparent, we defer the proofs of the key lemma and proposition used in this derivation.
The key lemma is proved in Sec.~\ref{s:generate_2RPmix_group_with_circuit}, while the key proposition is proved in Sec.~\ref{s:gap_doped_2RPmix_circuit}.
In Sec.~\ref{ss:lower_bound_spectral_gap_single_fixed}, by using the result obtained in Sec.~\ref{ss:lower_bound_spectral_gap_parallel}, 
we derive lower bounds for a broader range of circuit structures, using the architecture-comparison arguments introduced in Sec.~\ref{s:architecture_comparison_argument}.

\subsection{Lower bounds for parallel circuits with restricted structures} \label{ss:lower_bound_spectral_gap_parallel}

In this subsection, we first prove the spectral gap lower bound for parallel random circuits with restricted structure.
We show the lower bounds for more general circuit structures in the next subsection, by using the architecture comparison argument.
Specifically, we prove the following theorem.
\begin{theorem}\label{thm:gap_lower_bound_parallel}
Let $\Gamma$ be a set of matchings satisfying $|\gamma|=\lfloor N/2\rfloor$ for at least one $\gamma\in\Gamma$ and $|\Gamma|=O(1)$, whose associated graph $G_\Gamma$ is a bounded-degree connected $N$-vertex graph.
Then, for $2\leq k\leq O(\sqrt{\log N})$, we have
\begin{equation}
    \Delta_{\nu_\Gamma}^{(k)}\geq \Omega\left(N\lambda_{G_\Gamma}^{\rm enc}\right).
\end{equation}
\end{theorem}

For the proof of this theorem, the first essential ingredient is an auxiliary circuit that we call the \textit{doped 2RPmix circuit}.
One unit of this circuit consists of a layer of $\lfloor N/2\rfloor$ disjoint $U(1)$-symmetric Haar-random two-qubit gates sandwiched between two independent Haar-random unitaries drawn from the 2RPmix group on the entire $N$-qubit system.
We define the spectral gap of this doped 2RPmix circuit by
\begin{equation}
    \Delta_{\mathrm{dp2rpmix}}^{(k)} \equiv 1-\left\|P^{(k)}_{\mathrm{2RPmix}}P^{(k)}_{\gamma_0}P^{(k)}_{\mathrm{2RPmix}}-P^{(k)}_{U(1)\mathrm{Haar}}\right\|_{\infty},
\end{equation}
where $\gamma_0\equiv\left\{\{1,2\},\{3,4\},\dots,\{2\lfloor N/2\rfloor-1,2\lfloor N/2\rfloor\}\right\}$ is a fixed matching containing $\lfloor N/2\rfloor$ edges, and $P^{(k)}_{\gamma_0}$ is the corresponding projector.
For this spectral gap, we establish the following proposition.
\begin{proposition}[Doped 2RPmix circuit]\label{prop:gap_LB_dp_2RPmix_circuit}
There exist positive constants $C_1$ and $C_2$, independent of the system size $N$, such that
\begin{equation}
    \Delta_{\mathrm{dp2rpmix}}^{(k)} \geq C_1, \label{eq:dp2rpmix_circuit_gap_LB}
\end{equation}
for $k\leq C_2 \sqrt{\log N}$.
\end{proposition}
\noindent
The proof of this proposition is provided in Sec.~\ref{s:gap_doped_2RPmix_circuit}.

The second essential ingredient is to quantify how rapidly the full 2RPmix group can be generated using local two-qubit gates.
Specifically, we consider a local random circuit, which we call the \textit{local 2RPmix random circuit}, whose two-qubit generator on each selected edge is obtained by applying either the identity or SWAP with equal probability, followed by an independent two-qubit random-phase gate.
For two qubits, the RPmix and 2RPmix groups coincide, and this local ensemble is precisely their Haar measure.
Hence, its moment operator is an orthogonal projector, which we denote by $P_{i,j;\mathrm{2rpmix}}^{(k)}$ on sites $i$ and $j$.
The $k$-th moment operator of the parallel local 2RPmix random circuit associated with $\Gamma$ is then
\begin{equation}
    M_{\nu_{\Gamma}^{\mathrm{2rpmix}}}^{(k)} \equiv \frac{1}{|\Gamma|}\sum_{\gamma\in\Gamma}P_{\gamma;\mathrm{2rpmix}}^{(k)}, \qquad P_{\gamma;\mathrm{2rpmix}}^{(k)}\equiv\left[\bigotimes_{j=1}^{|\gamma|}P_{i_{2j-1}^\gamma,i_{2j}^\gamma;\mathrm{2rpmix}}^{(k)}\right]\otimes\mathbb I_{\overline{\gamma}}.
\end{equation}
Since $G_\Gamma$ is connected, these local generators generate the full 2RPmix group on the $N$-qubit system.
Accordingly, the target fixed-space projector is $P_{\mathrm{2RPmix}}^{(k)}$, and we define the spectral gap of this circuit by
\begin{equation}
\label{eq:loc_2RPmix_Gamma_gap_def}
    \Delta^{(k;\,\mathrm{2RPmix})}_{\nu_{\Gamma}^{\mathrm{2rpmix}}} \equiv 1-\left\|M^{(k)}_{\nu_{\Gamma}^{\mathrm{2rpmix}}}-P^{(k)}_{\mathrm{2RPmix}}\right\|_\infty,
\end{equation}
which quantifies the rate at which the moment operator approaches the 2RPmix moment space.

For this local 2RPmix random circuit, we establish the following lemma.
\begin{lemma}
\label{lem:2RPmix_parallel_spectral_gap_LB}
Let $\Gamma$ be a set of matchings, and let $G_{\Gamma}=(V,E_{\Gamma})$ be the corresponding connected graph.
Then, for $k\geq 2$,
\begin{equation}
\label{eq:parallel_single_edge_gap_comparison_simple}
    \Delta^{(k;\,\mathrm{2RPmix})}_{\nu_{\Gamma}^{\mathrm{2rpmix}}} \geq \frac{|E_\Gamma|}{20|\Gamma|^3} \lambda_{G_\Gamma}^{\rm enc}.
\end{equation}
\end{lemma}
\noindent
This lemma relates the local generation rate of the 2RPmix group directly to the two-particle encounter timescale.
In particular, for the constant-size families of matchings considered in Theorem~\ref{thm:gap_lower_bound_parallel}, connectedness implies $|E_\Gamma|\geq N-1$, and hence the lemma gives $\Delta^{(k;\,\mathrm{2RPmix})}_{\nu_{\Gamma}^{\mathrm{2rpmix}}}\geq\Omega(N\lambda_{G_\Gamma}^{\rm enc})$.
The proof of this lemma is provided in Sec.~\ref{s:generate_2RPmix_group_with_circuit}.

Combining Proposition~\ref{prop:gap_LB_dp_2RPmix_circuit} and Lemma~\ref{lem:2RPmix_parallel_spectral_gap_LB}, we can now prove Theorem~\ref{thm:gap_lower_bound_parallel}.

\begin{proof}[Proof of Theorem~\ref{thm:gap_lower_bound_parallel}]
Since $P_{\mathrm{2RPmix}}^{(k)}$, $P^{(k)}_{\gamma_0}$, and $P_{U(1)\mathrm{Haar}}^{(k)}$ are all orthogonal projectors whose ranges contain the $U(1)$-symmetric Haar moment space, we have
\begin{align}
    \left\|P_{\mathrm{2RPmix}}^{(k)}P^{(k)}_{\gamma_0}-P_{U(1)\mathrm{Haar}}^{(k)}\right\|_\infty^2
    &=\left\|\left(P_{\mathrm{2RPmix}}^{(k)}P^{(k)}_{\gamma_0}-P_{U(1)\mathrm{Haar}}^{(k)}\right)\left(P_{\mathrm{2RPmix}}^{(k)}P^{(k)}_{\gamma_0}-P_{U(1)\mathrm{Haar}}^{(k)}\right)^\dagger\right\|_\infty \nonumber \\
    &=\left\|P_{\mathrm{2RPmix}}^{(k)}P^{(k)}_{\gamma_0}P_{\mathrm{2RPmix}}^{(k)}-P_{U(1)\mathrm{Haar}}^{(k)}\right\|_\infty \nonumber \\
    &=1-\Delta_{\mathrm{dp2rpmix}}^{(k)}.
\end{align}
We can also bound
\begin{align}
    \left\|\frac{1}{2}\left(P_{\mathrm{2RPmix}}^{(k)}+P_{\gamma_0}^{(k)}\right)-P_{U(1)\mathrm{Haar}}^{(k)}\right\|_\infty^2
    &=\left\|\frac{1}{4}\left(P_{\mathrm{2RPmix}}^{(k)}+P_{\gamma_0}^{(k)}+P_{\mathrm{2RPmix}}^{(k)}P_{\gamma_0}^{(k)}+P_{\gamma_0}^{(k)}P_{\mathrm{2RPmix}}^{(k)}\right)-P_{U(1)\mathrm{Haar}}^{(k)}\right\|_\infty \nonumber \\
    &\leq \frac{1}{2}\left\|\frac{1}{2}\left(P_{\mathrm{2RPmix}}^{(k)}+P_{\gamma_0}^{(k)}\right)-P_{U(1)\mathrm{Haar}}^{(k)}\right\|_\infty
    +\frac{1}{2}\left\|P_{\mathrm{2RPmix}}^{(k)}P_{\gamma_0}^{(k)}-P_{U(1)\mathrm{Haar}}^{(k)}\right\|_\infty \nonumber \\
    &\leq \frac{1+\sqrt{1-\Delta_{\mathrm{dp2rpmix}}^{(k)}}}{2}.
\end{align}
Here, in the last inequality we used $\left\|\frac{1}{2}(P_{\mathrm{2RPmix}}^{(k)}+P_{\gamma_0}^{(k)})-P_{U(1)\mathrm{Haar}}^{(k)}\right\|_\infty\leq1$.
For later convenience, we define
\begin{equation}
    c_{\rm dp}^{(k)}\equiv1-\left[\frac{1+\sqrt{1-\Delta_{\mathrm{dp2rpmix}}^{(k)}}}{2}\right]^{1/2}.
\end{equation}
Proposition~\ref{prop:gap_LB_dp_2RPmix_circuit} implies $c_{\rm dp}^{(k)}=\Omega(1)$ for $k\leq O(\sqrt{\log N})$.

For $\gamma\in\Gamma$ satisfying $|\gamma|=\lfloor N/2\rfloor$, choose a site permutation $\pi_\gamma\in S_N$ such that $\pi_\gamma(\gamma_0)=\gamma$.
Since both $P_{\mathrm{2RPmix}}^{(k)}$ and $P_{U(1)\mathrm{Haar}}^{(k)}$ are invariant under arbitrary site permutations, we obtain
\begin{align}
    \left\|\frac{1}{2}\left(P_{\mathrm{2RPmix}}^{(k)}+P_{\gamma}^{(k)}\right)-P_{U(1)\mathrm{Haar}}^{(k)}\right\|_\infty
    &=\left\|U_{\pi_\gamma}^{\otimes k,k}\left[\frac{1}{2}\left(P_{\mathrm{2RPmix}}^{(k)}+P_{\gamma_0}^{(k)}\right)-P_{U(1)\mathrm{Haar}}^{(k)}\right]U_{\pi_\gamma^{-1}}^{\otimes k,k}\right\|_\infty \nonumber \\
    &\leq1-c_{\rm dp}^{(k)}.
\end{align}
Moreover, $\frac{1}{2}(P_{\mathrm{2RPmix}}^{(k)}+P_{\gamma}^{(k)})-P_{U(1)\mathrm{Haar}}^{(k)}$ is positive semidefinite and vanishes on $\operatorname{Ran}P_{U(1)\mathrm{Haar}}^{(k)}$.
Therefore,
\begin{equation}
    \frac{1}{2}\left(P_{\mathrm{2RPmix}}^{(k)}+P_{\gamma}^{(k)}\right)
    \leq\left(1-c_{\rm dp}^{(k)}\right)\mathbb I+c_{\rm dp}^{(k)}P_{U(1)\mathrm{Haar}}^{(k)}.
\end{equation}
Then, we have 
\begin{align}
\frac{1}{2}\left(P_{\mathrm{2RPmix}}^{(k)}+M_{\nu_{\Gamma}}^{(k)}\right) 
&\leq \left( 1- \frac{1}{|\Gamma|}\right)  \mathbb{I} +  \frac{1}{2|\Gamma|}\left(P_{\mathrm{2RPmix}}^{(k)}+P_{\gamma}^{(k)}\right) \nonumber \\
&\leq \left(1- \frac{c_{\rm dp}^{(k)}}{|\Gamma|} \right)\mathbb I+ \frac{c_{\rm dp}^{(k)}}{|\Gamma|}P_{U(1)\mathrm{Haar}}^{(k)}. \label{eq:last_gapLB_prf_1}
\end{align}

On the other hand, Eq.~\eqref{eq:loc_2RPmix_Gamma_gap_def} gives
\begin{align}
    M_{\nu_{\Gamma}}^{(k)} &\leq M_{\nu_{\Gamma}^{\mathrm{2rpmix}}}^{(k)} \nonumber \\
    &\leq\left(1-\Delta_{\nu_{\Gamma}^{\mathrm{2rpmix}}}^{(k;\,\mathrm{2RPmix})}\right)\mathbb I+\Delta_{\nu_{\Gamma}^{\mathrm{2rpmix}}}^{(k;\,\mathrm{2RPmix})}P_{\mathrm{2RPmix}}^{(k)},
    \label{eq:last_gapLB_prf_2}
\end{align}
where the first inequality follows from $P_{\gamma}^{(k)}\leq P_{\gamma;\mathrm{2rpmix}}^{(k)}$ for every $\gamma\in\Gamma$.
For brevity, we set $\delta_\Gamma^{(k)}\equiv\Delta_{\nu_{\Gamma}^{\mathrm{2rpmix}}}^{(k;\,\mathrm{2RPmix})}$.
Combining Eqs.~\eqref{eq:last_gapLB_prf_1} and \eqref{eq:last_gapLB_prf_2}, we obtain
\begin{align}
    M_{\nu_{\Gamma}}^{(k)}
    &\leq\frac{1}{1+\delta_\Gamma^{(k)}}M_{\nu_{\Gamma}^{\mathrm{2rpmix}}}^{(k)}
    +\frac{\delta_\Gamma^{(k)}}{1+\delta_\Gamma^{(k)}}M_{\nu_{\Gamma}}^{(k)} \nonumber \\
    &\leq\frac{1-\delta_\Gamma^{(k)}}{1+\delta_\Gamma^{(k)}}\mathbb I
    +\frac{\delta_\Gamma^{(k)}}{1+\delta_\Gamma^{(k)}}\left(P_{\mathrm{2RPmix}}^{(k)}+M_{\nu_{\Gamma}}^{(k)}\right) \nonumber \\
    &\leq\left[1-\frac{\delta_\Gamma^{(k)}}{1+\delta_\Gamma^{(k)}} \frac{2c_{\rm dp}^{(k)}}{|\Gamma|}\right]\mathbb I
    +\frac{\delta_\Gamma^{(k)}}{1+\delta_\Gamma^{(k)}} \frac{2c_{\rm dp}^{(k)}}{|\Gamma|}P_{U(1)\mathrm{Haar}}^{(k)} \nonumber \\
    &\leq\left[1-\frac{c_{\rm dp}^{(k)}\delta_\Gamma^{(k)}}{|\Gamma|}\right]\mathbb I
    + \frac{c_{\rm dp}^{(k)}\delta_\Gamma^{(k)}}{|\Gamma|}P_{U(1)\mathrm{Haar}}^{(k)},
\end{align}
where in the last inequality we used $0\leq\delta_\Gamma^{(k)}\leq1$.

Since $M_{\nu_{\Gamma}}^{(k)}-P_{U(1)\mathrm{Haar}}^{(k)}$ is positive semidefinite, it follows that
\begin{align}
    \left\|M_{\nu_{\Gamma}}^{(k)}-P_{U(1)\mathrm{Haar}}^{(k)}\right\|_\infty
    &\leq1- \frac{c_{\rm dp}^{(k)}\Delta_{\nu_{\Gamma}^{\mathrm{2rpmix}}}^{(k;\,\mathrm{2RPmix})}}{|\Gamma|}  \nonumber \\
    &\leq1-c_{\rm dp}^{(k)}\frac{|E_\Gamma|}{20 |\Gamma|^4} \lambda_{G_\Gamma}^{\rm enc},
\end{align}
where the second inequality follows from Lemma~\ref{lem:2RPmix_parallel_spectral_gap_LB}.
Since $c_{\rm dp}^{(k)}=\Omega(1)$, $|\Gamma|=O(1)$, and connectedness of $G_\Gamma$ implies $|E_\Gamma|\geq N-1$, we conclude that
\begin{equation}
    \Delta_{\nu_{\Gamma}}^{(k)}\geq c_{\rm dp}^{(k)}\frac{|E_\Gamma|}{20 |\Gamma|^4} \lambda_{G_\Gamma}^{\rm enc}=\Omega\left(N\lambda_{G_\Gamma}^{\rm enc}\right).
\end{equation}
\end{proof}

\subsection{Lower bounds for random circuits with a broad range of structures} \label{ss:lower_bound_spectral_gap_single_fixed}

In this subsection, we derive lower bounds for single-edge and fixed-architecture circuits, as well as more general parallel circuits than addressed in Sec.~\ref{ss:lower_bound_spectral_gap_parallel}.
Specifically, we prove the following corollary:
\begin{corollary}\label{restatecor:gap_lower_bound_sing_fix}
Let $G=(V,E)$ be a bounded-degree graph that admits a perfect or near-perfect matching, let $\Gamma$ be a constant-size matching family (i.e., $|\Gamma|=O(1)$) whose associated graph $G_\Gamma$ admits a perfect or near-perfect matching, and let $A$ be a fixed architecture of a constant depth $d_A=O(1)$, whose associated graph $G_A$ admits a perfect or near-perfect matching.
Then, for $2\leq k \leq O(\sqrt{\log N})$, we have
\begin{align}
    &\Delta_{\nu_{G}}^{(k)} \geq \Omega\left(\lambda_{G}^{\rm enc}\right), \label{eq:lower_bound_single_edge_bounded_degree} \\
    &\Delta_{\nu_{\Gamma}}^{(k)} \geq \Omega\left(N \lambda_{G_\Gamma}^{\rm enc}\right), \label{eq:lower_bound_parallel_bounded_degree} \\
    &\Delta_{\nu_{A}}^{(k)} \geq \Omega\left(N \lambda_{G_A}^{\rm enc}\right). \label{eq:lower_bound_fixed_bounded_degree} 
\end{align}
\end{corollary}

\begin{proof}
We first consider the single-edge circuit associated with $G$. 
Since $G$ admits a perfect or near-perfect matching, we can construct the matching family $\Upsilon$ from $G$ such that $\Upsilon$ contains at least one (near-)perfect matching and its size is bounded by a constant $|\Upsilon|\leq d_G+2=O(1)$. Here, $d_G$ is the maximum degree of $G$.
We can prepare such a matching family by first constructing the constant-size matching family $\Upsilon'$ by using Vizing's theorem~\cite{vizing1964estimate} and then adding a (near-)perfect matching to $\Upsilon'$. The resulting matching family $\Upsilon$ also becomes constant size. Note that we here allow the overlap of edges in $\Upsilon$, which can be upper bounded as $g_\Upsilon \leq 2 |\Upsilon|$.

By using such a matching family $\Upsilon$, we can derive 
\begin{align}
    \Delta_{\nu_{G}}^{(k)} &\geq \frac{|\Upsilon|}{g_\Upsilon |E_\Upsilon|}\Delta_{\nu_\Upsilon}^{(k)}  \nonumber \\
    &\geq \frac{|\Upsilon|}{g_\Upsilon |E_\Upsilon|}  \frac{c_{\rm dp}^{(k)}|E_\Upsilon|}{20 |\Upsilon|^4} \lambda_{G }^{\rm enc} \nonumber \\
    &\geq \frac{c_{\rm dp}^{(k)}}{40(d_G+2)^4} \lambda_{G }^{\rm enc} = \Omega\left(\lambda_{G}^{\rm enc}\right), \label{eq:gap_lower_bound_sing_in_the_proof}
\end{align}
where we used Lemma~\ref{lem:sing_edge_upper_bound_to_par_and_fix} in the first line, Theorem~\ref{thm:gap_lower_bound_parallel} in the second line, and the inequalities $g_\Upsilon \leq 2 |\Upsilon|$ and $|\Upsilon| \leq d_G +2$ in the final line. This leads to Eq.~\eqref{eq:lower_bound_single_edge_bounded_degree}.

We next consider the parallel circuit defined by the matching family $\Gamma$.
Since the associated graph $G_{\Gamma}$ has at least one (near-)perfect matching, we have
\begin{align}
\Delta_{\nu_{\Gamma}}^{(k)} &\geq \frac{|E_\Gamma|}{16|\Gamma|^3+4|\Gamma||E_\Gamma|\Delta_{\nu_{G_\Gamma}}^{(k)}} \Delta_{\nu_{G_\Gamma}}^{(k)} \nonumber \\
&\geq \frac{|E_\Gamma|}{16|\Gamma|^3+4|\Gamma|^2/(N-1)} \Delta_{\nu_{G_\Gamma}}^{(k)}\nonumber \\
&\geq \frac{|E_\Gamma|}{20|\Gamma|^3} \Delta_{\nu_{G_\Gamma}}^{(k)} \nonumber \\
&\geq \frac{|E_\Gamma|}{20|\Gamma|^3}  \frac{c_{\rm dp}^{(k)}}{40(|\Gamma|+2)^4}\lambda_{G_\Gamma}^{\rm enc}= \Omega\left(N \lambda_{G_\Gamma}^{\rm enc}\right),
\end{align}
where the first line is derived by using Lemma~\ref{lem:single_edge_to_parallel_gap_lower_bound}, the second line follows from inequalities $|E_\Gamma| \leq \frac{N}{2} |\Gamma|$ and $\Delta_{\nu_{G_\Gamma}}^{(k)} \leq \Delta_{\nu_{G_\Gamma}}^{(2)}\leq  \frac{2}{N(N-1)}$ for $k\geq 2$ for any $G_\Gamma$, which is derived from Theorem~\ref{restatethm:gap_upper_bound_two_particle_encounter}, and the final line is derived from Eq.~\eqref{eq:gap_lower_bound_sing_in_the_proof} and the bound on the maximum degree $d_{G_\Gamma} \leq |\Gamma|$.

Finally, we discuss the case of the fixed-architecture circuit $\nu_A$. 
Since the associated graph $G_{A}$ has at least one (near-)perfect matching, we have
\begin{align}
\Delta_{\nu_{A}}^{(k)} &\geq \frac{|E_A|\Delta_{\nu_{G_A}}^{(k)}}{8d_A^2+2|E_A|\Delta_{\nu_{G_A}}^{(k)}}\nonumber \\
&\geq \frac{|E_A|\Delta_{\nu_{G_A}}^{(k)}}{8d_A^2+2d_A/(N-1) }\nonumber \\
&\geq \frac{|E_A|}{10d_A^2} \Delta_{\nu_{G_A}}^{(k)}\nonumber \\
&\geq \frac{|E_A|}{10d_A^2} \frac{c_{\rm dp}^{(k)}}{40(d_A+2)^4}\lambda_{G_A}^{\rm enc}=\Omega\left(N \lambda_{G_A}^{\rm enc}\right),
\end{align}
where the first line is derived by using Lemma~\ref{lem:single_edge_to_parallel_gap_lower_bound}, the second line follows from inequalities $|E_A| \leq \frac{N}{2} d_A $ and $ \Delta_{\nu_{G_A}}^{(k)} \leq\Delta_{\nu_{G_A}}^{(2)} \leq \frac{2}{N(N-1)}$ for $k\geq 2$ for any graph $G_A$ from Theorem~\ref{restatethm:gap_upper_bound_two_particle_encounter}, and the final line is derived from Eq.~\eqref{eq:gap_lower_bound_sing_in_the_proof} and the bound on the maximum degree $d_{G_A} \leq d_A$.
\end{proof}

We note that Corollary~\ref{restatecor:gap_lower_bound_sing_fix} does not cover the all-to-all single-edge circuit $\nu_{G_{\rm all}}$ and parallel circuit $\nu_{\Gamma_{\rm all}}$ since the complete graph is not bounded-degree, and the matching family $\Gamma_{\rm all}$ is not constant size.
However, by using Lemma~\ref{lem:all_to_all_matching_comparison}, which provides a comparison argument for the all-to-all model, we can derive, for $2\leq k\leq O(\sqrt{\log N})$,
\begin{align}
    &\Delta_{\nu_{G_{\rm all}}}^{(k)} \geq \Delta_{\nu_{G_{3\rm D}}}^{(k)} = \Omega (N^{-2}) = \Omega \left(\lambda_{G_{\rm all}}^{\rm enc}\right ), \\
    &\Delta_{\nu_{\Gamma_{\rm all}}}^{(k)} \geq \Delta_{\nu_{\Gamma_{3\rm D}}}^{(k)} = \Omega (N^{-1}) = \Omega \left(N\lambda_{G_{\rm all}}^{\rm enc}\right ), 
\end{align}
where $G_{3D}$ and $\Gamma_{3\rm D}$ are the $3$-dimensional lattice graph and the corresponding matching family.
The final inequality follows from the fact that $\lambda_{G_{\rm all}}^{\rm enc} = \Theta (N^{-2})$.

\section{Generation of the 2RPmix group with local gates}
\label{s:generate_2RPmix_group_with_circuit}

In this section, we quantify the rate at which the global 2RPmix group is generated by local 2RPmix gates, which provides a key ingredient in the proof of Theorem~\ref{thm:gap_lower_bound_parallel}.
In Sec.~\ref{ss:classical_color_exchange}, we first introduce and analyze classical stochastic and substochastic dynamics on the type-vector sectors.
In the compatible sectors, the dynamics describes the exchange of local types on the graph, whereas in the incompatible sectors probability is lost when an incompatible pair is selected. Using purely classical arguments, we show that the slowest nonstationary behavior of both types of sectors is controlled by the two-particle encounter process introduced in Sec.~\ref{ss:trans_mat_two_part_enc}.
Then, in Sec.~\ref{ss:local_2RPmix_circuit_spectral_gap}, we show that the restriction of the moment operator of the single-edge local 2RPmix random circuit to each type-vector sector exactly coincides with the corresponding classical transition matrix.
This blockwise identification determines the spectral gap of the single-edge local 2RPmix random circuit in terms of the two-particle encounter decay rate.
Finally, in Sec.~\ref{ss:2RPmix_parallel_generation}, we extend this result to parallel 2RPmix random circuits and derive Lemma~\ref{lem:2RPmix_parallel_spectral_gap_LB}, which is used in Theorem~\ref{thm:gap_lower_bound_parallel}.

\subsection{Classical color-exchange and killing dynamics}
\label{ss:classical_color_exchange}

In this subsection, we analyze classical dynamics on the configurations associated with a fixed type vector $\bm m$ in the $k$-th moment space.
As introduced in Sec.~\ref{ss:mixing}, we denote by $\mathcal T_{\bm m}$ the set of computational-basis configurations having the type vector $\bm m$.
The multiplicity of each local type is fixed within $\mathcal T_{\bm m}$, while different elements of $\mathcal T_{\bm m}$ correspond to different spatial configurations of these local types.
We write each configuration as
\begin{equation}
    \bm u\equiv(u_1,u_2,\ldots,u_N),\qquad \sum_{i=1}^{N}\delta_{u_i,v}=m_v\quad\text{for every }v\in\{0,1\}^{2k},
\end{equation}
and denote the corresponding basis state by $\ket{\bm u}$.

We also recall the compatibility relation introduced for the 2RP group.
For two local types $u=(x,y)$ and $v=(x',y')$, we say that $u$ and $v$ are compatible if and only if
\begin{gather}
    |x|=|y|,\qquad |x'|=|y'|,\\
    x\cdot x'=y\cdot y'.
\end{gather}
Here, $|x|$ denotes the Hamming weight of $x\in\{0,1\}^k$, and $x\cdot x'$ denotes the standard inner product.
These are precisely the conditions under which the pair of local types survives the two-qubit random-phase average.
When $u$ and $v$ are compatible, we write $u\sim v$, and otherwise we write $u\not\sim v$.
In particular, a local type $u=(x,y)\notin\mathcal C_k$ is incompatible with every local type, including itself, because it does not satisfy $|x|=|y|$.

We now define the classical dynamics on a connected graph $G=(V,E)$ as follows:
\begin{itemize}
    \item[$\mathrm{(1)}$] At each step, an edge $e\in E$ is chosen uniformly at random.
    \item[$\mathrm{(2}$-$\mathrm{i)}$] If the local types at the two endpoints of the selected edge are compatible, they are exchanged with probability $1/2$, while the configuration is left unchanged with probability $1/2$.
    \item[$\mathrm{(2}$-$\mathrm{ii)}$] If the local types at the two endpoints are incompatible, the corresponding component is eliminated.
\end{itemize}
For a selected edge $e=\{i,j\}$, the corresponding transition matrix is defined by
\begin{equation}
    K_{\{i,j\}}^{\bm m}\ket{\bm u}
    =
    \begin{cases}
        0, & u_i\not\sim u_j,\\[2mm]
        \dfrac{1}{2}\left(\ket{\bm u}+\ket{\pi_{(i,j)}(\bm u)}\right), & u_i\sim u_j,
    \end{cases}
\end{equation}
where $\pi_{(i,j)}$ denotes the transposition of sites $i$ and $j$.
The transition matrix of one step of the full process is therefore
\begin{equation}
    [K_G^{\bm m}]_{\bm v,\bm u}
    \equiv
    \frac{1}{|E|}\sum_{e\in E}[K_e^{\bm m}]_{\bm v,\bm u}
    =
    \frac{1}{2|E|}
    \sum_{\substack{\{i,j\}\in E\\u_i\sim u_j}}
    \left(
        \delta_{\bm v,\bm u}
        +
        \delta_{\bm v,\pi_{(i,j)}(\bm u)}
    \right).
    \label{eq:K_m^G_each_element}
\end{equation}

All matrix elements of $K_G^{\bm m}$ are nonnegative, and the sum of the column corresponding to $\bm u$ is
\begin{equation}
\label{eq:column_sum_multi_color_trans_mat}
    \sum_{\bm v\in\mathcal T_{\bm m}}[K_G^{\bm m}]_{\bm v,\bm u}
    =
    1-\frac{|\{\{i,j\}\in E:u_i\not\sim u_j\}|}{|E|}
    \leq 1.
\end{equation}
Thus, probability mass is lost only when the selected edge has incompatible local types at its endpoints.
Moreover, each $K_e^{\bm m}$ is real symmetric, and hence so is $K_G^{\bm m}$.
Its operator norm therefore coincides with its spectral radius,
\begin{equation}
    \left\|K_G^{\bm m}\right\|_{\infty}
    =
    \rho\left(K_G^{\bm m}\right).
\end{equation}

Depending on the type vector $\bm m$, this dynamics exhibits two qualitatively different behaviors.
First, suppose that all local types appearing in $\bm m$ are mutually compatible, or equivalently,
\begin{equation}
\label{eq:color_exchange_compatible_condition}
    \operatorname{supp}(\bm m)\subseteq F \qquad\text{for some }F\in\mathcal F_k.
\end{equation}
In this case, every pair of local types appearing in any configuration $\bm u\in\mathcal T_{\bm m}$ is compatible.
Equation~\eqref{eq:column_sum_multi_color_trans_mat} therefore equals unity for every column, and $K_G^{\bm m}$ is a stochastic transition matrix describing a colored interchange process.
Since $G$ is connected, the transpositions associated with its edges generate all site permutations, and hence the process is irreducible on $\mathcal T_{\bm m}$.
Its unique steady state is therefore the uniform distribution on $\mathcal T_{\bm m}$, whose normalized eigenvector is the orbit state $\ket{\Omega_{\bm m}}$.

On the other hand, suppose that
\begin{equation}
\label{eq:color_exchange_incompatible_condition}
    \operatorname{supp}(\bm m)\not\subseteq F
    \qquad\text{for every }F\in\mathcal F_k.
\end{equation}
Then $\operatorname{supp}(\bm m)$ contains at least one pair of incompatible local types, and $K_G^{\bm m}$ is substochastic.
In particular, Eq.~\eqref{eq:column_sum_multi_color_trans_mat} is strictly smaller than unity for configurations in which an incompatible pair occupies the endpoints of an edge.
Since $G$ is connected, from every configuration there exists a finite sequence of edge selections and exchanges that either eliminates the component along the way or brings an incompatible pair to the endpoints of a selected edge.
Consequently, the process has no nonzero stationary component and $\rho(K_G^{\bm m})<1$.
Thus, rather than converging to a normalized steady distribution, the total probability mass eventually decays to zero, with principal decay rate $1-\rho(K_G^{\bm m})$.
In what follows, we analyze the relaxation and decay rates of these two cases in turn.

\vspace{1em}

We first consider a compatible type vector satisfying Eq.~\eqref{eq:color_exchange_compatible_condition}.
As explained above, $K_G^{\bm m}$ describes a colored interchange process on the connected graph $G$, with steady-state eigenvector $\ket{\Omega_{\bm m}}$.

The simplest example of such a colored interchange process is the dynamics associated with the $k=1$ type vector $\bm m_{\rm sing}$ specified by $(m_{\rm sing})_{00}=N-1$ and $(m_{\rm sing})_{11}=1$, with all other multiplicities equal to zero.
A configuration in this sector is completely specified by the position of the single nontrivial color $11$, and hence $K_G^{\bm m_{\rm sing}}$ is the transition matrix of the single-particle random walk on $G$.
We define its spectral gap by
\begin{equation}
\label{eq:single_particle_transport_gap}
    \lambda_G^{\rm tr}
    \equiv
    1-
    \rho\left(
        K_G^{\bm m_{\rm sing}}
        -
        \ket{\Omega_{\bm m_{\rm sing}}}\bra{\Omega_{\bm m_{\rm sing}}}
    \right),
\end{equation}
where the superscript $\rm tr$ stands for single-particle transport.

It is well known that this single-particle random walk controls the relaxation rate of the general interchange process on the same graph.
This result is widely known as Aldous' spectral-gap theorem~\cite{caputo2010proof}:

\begin{lemma}[Aldous' spectral-gap theorem]
\label{lem:Aldous_thm}
For any connected graph $G$, the spectral gap of the interchange process is equal to that of the corresponding single-particle random walk, with the same edge-selection normalization.
\end{lemma}

The colored interchange process described by $K_G^{\bm m}$ is obtained from the interchange process by identifying particles carrying the same color.
Therefore, Aldous' spectral-gap theorem implies that, for every compatible type vector $\bm m$,
\begin{equation}
\label{eq:Aldous_application_Km}
    \rho\left(
        K_G^{\bm m}
        -
        \ket{\Omega_{\bm m}}\bra{\Omega_{\bm m}}
    \right)
    \leq
    1-\lambda_G^{\rm tr}.
\end{equation}
For any nontrivial color composition, the bound is in fact saturated by a single-particle mode, although only the inequality above will be needed here.
Since $K_G^{\bm m}-\ket{\Omega_{\bm m}}\bra{\Omega_{\bm m}}$ is real symmetric, Eq.~\eqref{eq:Aldous_application_Km} is equivalently written as
\begin{equation}
    \left\|
        K_G^{\bm m}
        -
        \ket{\Omega_{\bm m}}\bra{\Omega_{\bm m}}
    \right\|_{\infty}
    =
    \rho\left(
        K_G^{\bm m}
        -
        \ket{\Omega_{\bm m}}\bra{\Omega_{\bm m}}
    \right)
    \leq
    1-\lambda_G^{\rm tr}.
\end{equation}

\vspace{1em}

Next, we consider type vectors containing mutually incompatible local types.
In this case, $K_G^{\bm m}$ describes a substochastic process that evolves by exchanging local types until an edge whose endpoints carry incompatible local types is selected, at which point the corresponding component is eliminated.
We show that the decay of any such process is at least as fast as that of the two-particle encounter process described by $K_G^{\rm enc}$ in Sec.~\ref{ss:trans_mat_two_part_enc}.

\begin{lemma}
\label{lem:multi_color_substoc_decay_rate}
Let $G=(V,E)$ be a connected graph with $N\geq2$ vertices and let $k\geq2$.
For any type vector $\bm m$ satisfying $\operatorname{supp}(\bm m)\not\subseteq F$ for every $F\in\mathcal F_k$, we have
\begin{equation}
\label{eq:multi_color_substoc_decay_rate}
    \left\|K_G^{\bm m}\right\|_{\infty}
    \leq
    1-\lambda_G^{\rm enc},
\end{equation}
where $\lambda_G^{\rm enc}$ is the principal decay rate of the two-particle encounter process defined in Eq.~\eqref{eq:two_part_enc_decay_rate}.
\end{lemma}

The bound in Lemma~\ref{lem:multi_color_substoc_decay_rate} is attained by a particular type-vector sector.
Specifically, for $k=2$, consider the type vector $\bm m_{\rm enc}$ defined by $(m_{\rm enc})_{0000}=N-2$, $(m_{\rm enc})_{1010}=1$, and $(m_{\rm enc})_{0110}=1$, with all other multiplicities equal to zero.
The two colors $1010$ and $0110$ are incompatible with each other, whereas $0000$ is compatible with both.
Hence, a configuration in $\mathcal T_{\bm m_{\rm enc}}$ is completely specified by the positions of the colors $1010$ and $0110$, and these two colors undergo the ordinary exchange dynamics until the selected edge directly connects them.
Under the natural basis correspondence, we have $K_G^{\bm m_{\rm enc}}=K_G^{\rm enc}$, and therefore
\begin{equation}
    \left\|K_G^{\bm m_{\rm enc}}\right\|_{\infty}=1-\lambda_G^{\rm enc},
\end{equation}
is satisfied. This means that Eq.~\eqref{eq:multi_color_substoc_decay_rate} is tight for some type vector.

\begin{proof}[Proof of Lemma~\ref{lem:multi_color_substoc_decay_rate}]
For a matrix $A$, let $c(A)\equiv\max_j\sum_i|A_{ij}|$ denote its maximum absolute column sum.
Since $K_G^{\bm m}$ is nonnegative, so is $(K_G^{\bm m})^t$, and hence, for every $t\geq1$,
\begin{equation}
\label{eq:Km_column_sum}
    c\left((K_G^{\bm m})^t\right)
    =
    \max_{\bm v\in\mathcal T_{\bm m}}
    \sum_{\bm u\in\mathcal T_{\bm m}}
    \left[(K_G^{\bm m})^t\right]_{\bm u,\bm v}.
\end{equation}
Since $c(\cdot)$ is a matrix norm, the Gelfand formula gives
\begin{equation}
\label{eq:Km_Gelfand_formula}
    \rho\left(K_G^{\bm m}\right)
    =
    \lim_{t\to\infty}
    \left[
        c\left((K_G^{\bm m})^t\right)
    \right]^{1/t}.
\end{equation}
We therefore analyze the column sums of the $t$-step transition matrix.

For this purpose, we resolve the $t$-step evolution into histories of the selected edges and the identity/SWAP choices.
Let
\begin{equation}
    \bm X_t
    =
    \left\{
        (e_s,\sigma_s)
    \right\}_{s=1}^t,
    \qquad
    e_s\in E,
    \quad
    \sigma_s\in\{\mathrm{id},\mathrm{swap}\},
\end{equation}
denote such a history.
Each history occurs with probability $ \Pr(\bm X_t)=\left(\frac{1}{2|E|}\right)^t.$
For an initial configuration $\bm v\in\mathcal T_{\bm m}$, let $b_{\rm surv}(\bm v;\bm X_t)\in\{0,1\}$ indicate whether the process starting from $\bm v$ survives the entire history $\bm X_t$.
Thus, $b_{\rm surv}(\bm v;\bm X_t)=1$ if no selected edge connects incompatible local types during the first $t$ steps, and $b_{\rm surv}(\bm v;\bm X_t)=0$ otherwise.
It follows that
\begin{equation}
\label{eq:survival_history_expansion}
    \sum_{\bm u\in\mathcal T_{\bm m}}
    \left[(K_G^{\bm m})^t\right]_{\bm u,\bm v}
    =
    \sum_{\bm X_t}
    \Pr(\bm X_t)
    b_{\rm surv}(\bm v;\bm X_t).
\end{equation}

Using Eq.~\eqref{eq:survival_history_expansion}, we now show that the two-particle encounter process $K_G^{\rm enc}$ decays no faster than the process associated with any incompatible type vector $\bm m$.
For any type vector $\bm m$ satisfying Eq.~\eqref{eq:color_exchange_incompatible_condition}, there exists at least one pair of incompatible local types.
For a fixed initial configuration $\bm v\in\mathcal T_{\bm m}$, choose two occurrences of such incompatible local types and denote their positions by $i$ and $j$.
We compare the original process with the two-particle encounter process in which the two distinguishable particles are initially located at the same sites $i$ and $j$.

For a fixed history $\bm X_t$, as long as the original process survives, the two marked local types undergo exactly the same sequence of site transpositions as the two particles in the encounter process.
Therefore, if the encounter process is eliminated at some step, then either the original process has already been eliminated, or the same selected edge connects the two marked incompatible local types and eliminates the original process at that step.
Hence, for every history $\bm X_t$,
\begin{equation}
\label{eq:survival_history_comparison}
    b_{\rm surv}(\bm v;\bm X_t)
    \leq
    b_{\rm surv}^{\rm enc}((i,j);\bm X_t),
\end{equation}
where $b_{\rm surv}^{\rm enc}((i,j);\bm X_t)\in\{0,1\}$ denotes the corresponding survival indicator for the two-particle encounter process initialized at $(i,j)$.

Using Eqs.~\eqref{eq:survival_history_expansion} and \eqref{eq:survival_history_comparison}, we obtain
\begin{align}
    \sum_{\bm u\in\mathcal T_{\bm m}}
    \left[(K_G^{\bm m})^t\right]_{\bm u,\bm v}
    &=
    \sum_{\bm X_t}
    \Pr(\bm X_t)
    b_{\rm surv}(\bm v;\bm X_t)
    \nonumber\\
    &\leq
    \sum_{\bm X_t}
    \Pr(\bm X_t)
    b_{\rm surv}^{\rm enc}((i,j);\bm X_t)
    \nonumber\\
    &=
    \sum_{\substack{i',j'\in V\\i'\neq j'}}
    \left[(K_G^{\rm enc})^t\right]_{(i',j'),(i,j)}
    \nonumber\\
    &\leq
    c\left((K_G^{\rm enc})^t\right).
\end{align}
Since this holds for every $\bm v\in\mathcal T_{\bm m}$, we have
\begin{equation}
    c\left((K_G^{\bm m})^t\right)
    \leq
    c\left((K_G^{\rm enc})^t\right).
\end{equation}
Applying the Gelfand formula to both matrices gives
\begin{equation}
    \rho\left(K_G^{\bm m}\right)
    \leq
    \rho\left(K_G^{\rm enc}\right)
    =
    1-\lambda_G^{\rm enc}.
\end{equation}
Finally, since $K_G^{\bm m}$ is real symmetric, $\|K_G^{\bm m}\|_{\infty}=\rho(K_G^{\bm m})$.
This proves Eq.~\eqref{eq:multi_color_substoc_decay_rate}.
\end{proof}

\vspace{1em}

Finally, we compare the single-particle transport rate $\lambda_G^{\rm tr}$ with the two-particle encounter decay rate $\lambda_G^{\rm enc}$.

\begin{lemma}
\label{lem:one_two_particle_comparison}
For any connected graph $G$, we have
\begin{equation}
\label{eq:one_two_particle_comparison}
    \lambda_G^{\rm tr}\geq\lambda_G^{\rm enc}.
\end{equation}
\end{lemma}

\begin{proof}
Recall the single-particle type vector $\bm m_{\rm sing}$ introduced above.
A configuration in this sector is completely specified by the position of the single nontrivial color $11$.
We denote by $\ket{x}$ the configuration in which the color $11$ occupies vertex $x\in V$, while all other vertices carry the color $00$.
Let $\ket{\psi}$ be a nontrivial eigenvector of $K_G^{\bm m_{\rm sing}}$ with eigenvalue $1-\lambda_G^{\rm tr}$, written as
\begin{equation}
    \ket{\psi}
    =
    \sum_{x\in V}\psi(x)\ket{x},
    \qquad
    K_G^{\bm m_{\rm sing}}\ket{\psi}
    =
    (1-\lambda_G^{\rm tr})\ket{\psi}.
\end{equation}
In components, this eigenvalue equation reads
\begin{equation}
\label{eq:single_particle_eigenvalue_equation}
    \psi(x)
    -
    \frac{1}{2|E|}
    \sum_{\{x,z\}\in E}
    \bigl[\psi(x)-\psi(z)\bigr]
    =
    (1-\lambda_G^{\rm tr})\psi(x).
\end{equation}

We now construct an eigenvector of the two-particle encounter matrix $K_G^{\rm enc}$ from $\ket{\psi}$.
Recall that the configuration space of the encounter process consists of ordered pairs $(x,y)$ with $x,y\in V$ and $x\neq y$.
We denote by $\ket{x,y}$ the configuration in which particle $1$ occupies vertex $x$ and particle $2$ occupies vertex $y$.
Define
\begin{equation}
    \ket{\Psi}
    =
    \sum_{\substack{x,y\in V\\x\neq y}}
    \Psi(x,y)\ket{x,y},
    \qquad
    \Psi(x,y)
    \equiv 
    \psi(x)-\psi(y).
\end{equation}
There is no configuration $\ket{x,x}$ in the state space of the encounter process.
For notational convenience in the calculation below, however, we extend the scalar function $\Psi(x,y)$ to the diagonal by defining $\Psi(x,x)\equiv 0$.
This is only an auxiliary convention and does not introduce any additional basis state.

With this convention, the action of $K_G^{\rm enc}$ can be written uniformly for every $x\neq y$ as
\begin{align}
\label{eq:two_particle_transition_compact}
    \bra{x,y}K_G^{\rm enc}\ket{\Psi}
    &=
    \Psi(x,y)
    -
    \frac{1}{2|E|}
    \left[
        \sum_{\{x,z\}\in E}
        \bigl(\Psi(x,y)-\Psi(z,y)\bigr)
        +
        \sum_{\{y,z\}\in E}
        \bigl(\Psi(x,y)-\Psi(x,z)\bigr)
    \right].
\end{align}
When $\{x,y\}\in E$, the terms involving $\Psi(y,y)=\Psi(x,x)=0$ account precisely for the elimination of the component when the selected edge directly connects the two particles.

Substituting $\Psi(x,y)=\psi(x)-\psi(y)$ into Eq.~\eqref{eq:two_particle_transition_compact} and using Eq.~\eqref{eq:single_particle_eigenvalue_equation}, we obtain
\begin{align}
    \bra{x,y}K_G^{\rm enc}\ket{\Psi}
    &=
    \left[
        \psi(x)
        -
        \frac{1}{2|E|}
        \sum_{\{x,z\}\in E}
        \bigl(\psi(x)-\psi(z)\bigr)
    \right]
    -
    \left[
        \psi(y)
        -
        \frac{1}{2|E|}
        \sum_{\{y,z\}\in E}
        \bigl(\psi(y)-\psi(z)\bigr)
    \right]
    \nonumber\\
    &=
    (1-\lambda_G^{\rm tr})
    \bigl[\psi(x)-\psi(y)\bigr]
    =
    (1-\lambda_G^{\rm tr})\Psi(x,y).
\end{align}
Thus, $\ket{\Psi}$ is an eigenvector of $K_G^{\rm enc}$ with eigenvalue $1-\lambda_G^{\rm tr}$.
Since $\ket{\psi}$ is a nontrivial single-particle eigenvector, it is not constant, and hence $\ket{\Psi}\neq0$.
Therefore,
\begin{equation}
    1-\lambda_G^{\rm tr}
    \leq
    \rho\left(K_G^{\rm enc}\right)
    =
    1-\lambda_G^{\rm enc},
\end{equation}
which is equivalent to Eq.~\eqref{eq:one_two_particle_comparison}.
\end{proof}

\subsection{Generation in single-edge circuits}
\label{ss:local_2RPmix_circuit_spectral_gap}

In this subsection, we derive a spectral gap lower bound for the single-edge local 2RPmix random circuit introduced in Sec.~\ref{ss:lower_bound_spectral_gap_parallel}.
In this circuit, on each selected edge, either the identity or SWAP is applied with equal probability, followed by an independent two-qubit random-phase gate.
For a connected graph $G=(V,E)$, the corresponding moment operator is given by
\begin{equation}
\label{eq:loc_2RPmix_G_def}
    M^{(k)}_{\nu_G^{\mathrm{2rpmix}}}\equiv\frac{1}{|E|}\sum_{\{i,j\}\in E}P^{(k)}_{i,j;\mathrm{2rpmix}}\otimes\mathbb I_{\overline{i,j}}.
\end{equation}
Since the local 2RPmix gates preserve the type vector, we have $[M^{(k)}_{\nu_G^{\mathrm{2rpmix}}},\Pi_{\mathcal T_{\bm m}}]=0$, where $\Pi_{\mathcal T_{\bm m}}\equiv\sum_{\bm u\in\mathcal T_{\bm m}}\ket{\bm u}\bra{\bm u}$ is the orthogonal projector onto the subspace spanned by the configurations in $\mathcal T_{\bm m}$.
Therefore, the moment operator is block diagonal with respect to the type vector,
\begin{equation}
    M^{(k)}_{\nu_G^{\mathrm{2rpmix}}}
    =
    \bigoplus_{\bm m}M^{(k)}_{\nu_G^{\mathrm{2rpmix}};\bm m},
    \qquad
    M^{(k)}_{\nu_G^{\mathrm{2rpmix}};\bm m}
    \equiv
    \Pi_{\mathcal T_{\bm m}}M^{(k)}_{\nu_G^{\mathrm{2rpmix}}}\Pi_{\mathcal T_{\bm m}}.
\end{equation}

Since $G$ is connected, the transpositions associated with its edges generate all permutations of the $N$ sites.
Together with the two-site random-phase gates, these local generators generate the full 2RPmix group.
Hence, the common fixed space of the local moment operators coincides with the fixed space of the global 2RPmix moment operator $P^{(k)}_{\mathrm{2RPmix}}$.
We therefore define the spectral gap of the single-edge local 2RPmix circuit by
\begin{equation}
\label{eq:loc_2RPmix_G_gap_def}
    \Delta^{(k;\,\mathrm{2RPmix})}_{\nu_G^{\mathrm{2rpmix}}}
    \equiv
    1-\left\|M^{(k)}_{\nu_G^{\mathrm{2rpmix}}}-P^{(k)}_{\mathrm{2RPmix}}\right\|_\infty.
\end{equation}

For this spectral gap, we obtain the following relationship.
\begin{lemma}
\label{lem:2RPmix_random_edge_gap}
Let $G=(V,E)$ be a connected graph and $k\geq2$.
Then,
\begin{equation}
\label{eq:2RPmix_random_edge_gap}
    \Delta^{(k;\,\mathrm{2RPmix})}_{\nu_G^{\mathrm{2rpmix}}} =\lambda_G^{\rm enc}. 
\end{equation}
\end{lemma}

\begin{proof}
The derivation of $\Delta^{(k;\,\mathrm{2RPmix})}_{\nu_G^{\mathrm{2rpmix}}} \leq \lambda_G^{\rm enc}$ is immediate: consider $k=2$ and the type vector $\bm m$ such that $m_{1010}=1$, $m_{0110}=1$, and $m_{0000}=N-2$ are satisfied. Then, the moment operator for that subspace $M^{(2)}_{\nu_G^{\mathrm{2rpmix}};\bm m}$ exactly coincides with $K_G^{\rm enc}$, yielding $\Delta^{(k;\,\mathrm{2RPmix})}_{\nu_G^{\mathrm{2rpmix}}} \leq \Delta^{(2;\,\mathrm{2RPmix})}_{\nu_G^{\mathrm{2rpmix}}}\leq \lambda_G^{\rm enc}$ for $k\geq 2$.

We next prove the inequality in the opposite direction $\Delta^{(k;\,\mathrm{2RPmix})}_{\nu_G^{\mathrm{2rpmix}}} \geq \lambda_G^{\rm enc}$.
For this purpose, we first identify each type-vector block $M^{(k)}_{\nu_G^{\mathrm{2rpmix}};\bm m}$ with the classical transition matrix $K_G^{\bm m}$ introduced in Sec.~\ref{ss:classical_color_exchange}.
Consider the action of the single-edge moment operator $P^{(k)}_{i,j;\mathrm{2rpmix}}\otimes\mathbb I_{\overline{i,j}}$ on a basis state $\ket{\bm u}$ with $\bm u=(u_1,u_2,\ldots,u_N)\in\mathcal T_{\bm m}$.
A direct evaluation gives
\begin{equation}
    \left(P^{(k)}_{i,j;\mathrm{2rpmix}}\otimes\mathbb I_{\overline{i,j}}\right)\ket{\bm u}
    =
    \begin{cases}
        0, & u_i\not\sim u_j,\\[2mm]
        \dfrac{1}{2}\left(\ket{\bm u}+\ket{\pi_{(i,j)}(\bm u)}\right), & u_i\sim u_j,
    \end{cases}
\end{equation}
where $\pi_{(i,j)}$ denotes the transposition of sites $i$ and $j$.
Averaging uniformly over the edges therefore gives
\begin{equation}
    \bra{\bm v}M^{(k)}_{\nu_G^{\mathrm{2rpmix}};\bm m}\ket{\bm u}
    =
    \frac{1}{2|E|}
    \sum_{\substack{\{i,j\}\in E\\u_i\sim u_j}}
    \left(
        \delta_{\bm v,\bm u}
        +
        \delta_{\bm v,\pi_{(i,j)}(\bm u)}
    \right).
\end{equation}
This is exactly the transition matrix defined for $K_G^{\bm m}$ in Sec.~\ref{ss:classical_color_exchange}.
Hence, on every type-vector sector, we have
\begin{equation}
\label{eq:2RPmix_block_classical_identification}
    M^{(k)}_{\nu_G^{\mathrm{2rpmix}};\bm m} =K_G^{\bm m}.
\end{equation}

We next identify the corresponding blocks of the global 2RPmix projector.
From Eq.~\eqref{eq:P_2RPmix_def},
\begin{equation*}
    P_{\mathrm{2RPmix}}^{(k)}
    =
    \sum_{\substack{\bm m:\, \exists F\in\mathcal F_k, \\ \operatorname{supp}(\bm m)\subseteq F}}
    \ket{\Omega_{\bm m}}\bra{\Omega_{\bm m}}.
\end{equation*}
Therefore, for an incompatible type vector satisfying Eq.~\eqref{eq:color_exchange_incompatible_condition},
\begin{equation}
    \Pi_{\mathcal T_{\bm m}}P_{\mathrm{2RPmix}}^{(k)}\Pi_{\mathcal T_{\bm m}}
    =
    0,
\end{equation}
whereas for a compatible type vector satisfying Eq.~\eqref{eq:color_exchange_compatible_condition},
\begin{equation}
    \Pi_{\mathcal T_{\bm m}}P_{\mathrm{2RPmix}}^{(k)}\Pi_{\mathcal T_{\bm m}}
    =
    \ket{\Omega_{\bm m}}\bra{\Omega_{\bm m}}.
\end{equation}

Combining these block decompositions with Eq.~\eqref{eq:2RPmix_block_classical_identification}, we obtain
\begin{align}
    \left\|M^{(k)}_{\nu_G^{\mathrm{2rpmix}}}-P^{(k)}_{\mathrm{2RPmix}}\right\|_\infty
    &=
    \max_{\bm m}
    \left\|
        \Pi_{\mathcal T_{\bm m}}
        \left(
            M^{(k)}_{\nu_G^{\mathrm{2rpmix}}}
            -
            P^{(k)}_{\mathrm{2RPmix}}
        \right)
        \Pi_{\mathcal T_{\bm m}}
    \right\|_\infty
    \nonumber\\
    &=
    \max\left\{
        \max_{\substack{\bm m:\, \exists F\in\mathcal F_k,\ \\\operatorname{supp}(\bm m)\subseteq F}}
        \left\|
            K_G^{\bm m}
            -
            \ket{\Omega_{\bm m}}\bra{\Omega_{\bm m}}
        \right\|_\infty,
        \max_{\substack{\bm m:\,  \forall F\in\mathcal F_k, \\ \operatorname{supp}(\bm m)\not\subseteq F}}
        \left\|K_G^{\bm m}\right\|_\infty
    \right\}
    \nonumber\\
    &\leq
    \max\left\{
        1-\lambda_G^{\rm tr},
        1-\lambda_G^{\rm enc}
    \right\}
    \nonumber\\
    &=
    1-\lambda_G^{\rm enc}.
\end{align}
Here, the inequality follows from Lemma~\ref{lem:multi_color_substoc_decay_rate}, Eq.~\eqref{eq:Aldous_application_Km}, and the fact that $K_G^{\bm m}$ is a real symmetric matrix.
In the final equality, we used Lemma~\ref{lem:one_two_particle_comparison}, which gives $\lambda_G^{\rm tr}\geq\lambda_G^{\rm enc}$.
By the definition in Eq.~\eqref{eq:loc_2RPmix_G_gap_def}, this proves Eq.~\eqref{eq:2RPmix_random_edge_gap}.
\end{proof}

\subsection{Generation in parallel circuits} \label{ss:2RPmix_parallel_generation}

We now consider the spectral gap for the parallel 2RPmix random circuit. 
As introduced in Sec.~\ref{ss:lower_bound_spectral_gap_parallel}, the moment operator for the parallel circuit whose matching family is $\Gamma$ is 
\begin{equation*}
    M_{\nu_{\Gamma}^{\mathrm{2rpmix}}}^{(k)} \equiv \frac{1}{|\Gamma|}\sum_{\gamma\in\Gamma}P_{\gamma;\mathrm{2rpmix}}^{(k)}, \qquad P_{\gamma;\mathrm{2rpmix}}^{(k)}\equiv\left[\bigotimes_{j=1}^{|\gamma|}P_{i_{2j-1}^\gamma,i_{2j}^\gamma;\mathrm{2rpmix}}^{(k)}\right]\otimes\mathbb I_{\overline{\gamma}},
\end{equation*}
and its spectral gap is given by
\begin{equation*}
    \Delta^{(k;\,\mathrm{2RPmix})}_{\nu_{\Gamma}^{\mathrm{2rpmix}}} \equiv 1-\left\|M^{(k)}_{\nu_{\Gamma}^{\mathrm{2rpmix}}}-P^{(k)}_{\mathrm{2RPmix}}\right\|_\infty.
\end{equation*}
By using Lemma \ref{lem:2RPmix_random_edge_gap} and the circuit architecture comparison argument given in Lemma~\ref{lem:single_edge_to_parallel_gap_lower_bound}, we can now derive Lemma~\ref{lem:2RPmix_parallel_spectral_gap_LB} as follows:

\begin{proof}[Proof of Lemma~\ref{lem:2RPmix_parallel_spectral_gap_LB}]
By using the preceding lemmas, we can obtain
\begin{align}
    \Delta^{(k;\,\mathrm{2RPmix})}_{\nu_{\Gamma}^{\mathrm{2rpmix}}} &\geq \frac{|E_\Gamma|}{16|\Gamma|^3+4|\Gamma||E_\Gamma|\,\Delta_{\nu_{G_\Gamma}^{\mathrm{2rpmix}}}^{(k;\,\mathrm{2RPmix})}} \ \Delta_{\nu_{G_\Gamma}^{\mathrm{2rpmix}}}^{(k;\,\mathrm{2RPmix})} \nonumber \\
    &= \frac{|E_\Gamma|}{16|\Gamma|^3+4|\Gamma||E_\Gamma|\lambda_{G_\Gamma}^{\rm enc}} \lambda_{G_\Gamma}^{\rm enc} \nonumber \\
    &\geq \frac{|E_\Gamma|}{16|\Gamma|^3+4|\Gamma|^2/(N-1) } \lambda_{G_\Gamma}^{\rm enc} \nonumber \\
    &\geq \frac{|E_\Gamma|}{20|\Gamma|^3} \lambda_{G_\Gamma}^{\rm enc},
\end{align}
where the architecture comparison argument (Lemma~\ref{lem:single_edge_to_parallel_gap_lower_bound}) is used in the first line, Lemma~\ref{lem:2RPmix_random_edge_gap} is used in the second line, the inequalities $|E_\Gamma| \leq \frac{N}{2} |\Gamma|$ and $\lambda_{G_\Gamma}^{\rm enc} \leq \frac{2}{N(N-1)}$ for any graph $G_\Gamma$ (see Eq.~\eqref{eq:robust_lambda_encount_upper_bound}) are used in the third line, and the inequalities $N\geq 2$ and $|\Gamma|\geq 1$ are used in the final line.
\end{proof}

\section{Constant gap lower bound for doped 2RPmix circuit} \label{s:gap_doped_2RPmix_circuit}

In this section, we prove Proposition~\ref{prop:gap_LB_dp_2RPmix_circuit}, which gives a constant spectral gap lower bound for the doped 2RPmix circuit.
This proposition is a crucial ingredient in the proof of Theorem~\ref{thm:gap_lower_bound_parallel}.
In Sec.~\ref{ss:dp2RP_setup_strategy}, we present the overall proof strategy.
In Sec.~\ref{subsec:classical_dynamics}, for later use, we introduce a classical stochastic process whose transition matrix corresponds to the matrix representation of the moment operator.
In Secs.~\ref{ss:Properties of compatible classes} and \ref{ss:compatible_class_projectors}, we introduce the additional properties of the 2RPmix moment space.
In Sec.~\ref{subsec:u1_haar_interior_approximation}, we introduce an approximate form of the moment operator of the $U(1)$-symmetric Haar-random unitary ensemble.
In Secs.~\ref{ss:bound_diagonal_block_mom} and \ref{ss:bound_off_diagonal_block_mom}, we derive operator-norm bounds for the diagonal and off-diagonal blocks of the moment operator, respectively, using the results of the preceding subsections.
Finally, in Sec.~\ref{ss:proof_of_doped_2RPmix_gap}, we prove Proposition~\ref{prop:gap_LB_dp_2RPmix_circuit}, mainly using the results of Secs.~\ref{subsec:u1_haar_interior_approximation}, \ref{ss:bound_diagonal_block_mom}, and \ref{ss:bound_off_diagonal_block_mom}.

\subsection{Setup and proof strategy} \label{ss:dp2RP_setup_strategy}

In this section, we derive Proposition~\ref{prop:gap_LB_dp_2RPmix_circuit}, which gives a lower bound on the spectral gap of the doped 2RPmix circuit.
The spectral gap for doped 2RPmix circuit is defined as
\begin{equation*}
    \Delta_{\mathrm{dp2rpmix}}^{(k)}
    \equiv
    1-\left\|P_{\mathrm{2RPmix}}^{(k)}P_{\gamma_0}^{(k)}P_{\mathrm{2RPmix}}^{(k)}-P_{U(1)\mathrm{Haar}}^{(k)}\right\|_{\infty},
\end{equation*}
where $P_{\gamma_0}^{(k)}$ is the moment operator associated with $\lfloor N/2\rfloor$ independent $U(1)$-symmetric Haar-random two-qubit gates acting on the fixed matching $\gamma_0$. 

In order to derive Proposition~\ref{prop:gap_LB_dp_2RPmix_circuit}, it is convenient to replace this Haar-random interleaving layer by a simpler auxiliary ensemble.
Specifically, we define the \(V\)-doped 2RPmix circuit by deterministically applying the following Hadamard-type gate to every pair in $\gamma_0$:
\begin{equation}
\label{eq:Hadamard_type_local_unitary_def}
    V_{2i-1,2i}
    \equiv
    \begin{pmatrix}
        1 & 0 & 0 & 0 \\
        0 & \frac{1}{\sqrt{2}} & \frac{1}{\sqrt{2}} & 0 \\
        0 & \frac{1}{\sqrt{2}} & -\frac{1}{\sqrt{2}} & 0 \\
        0 & 0 & 0 & 1
    \end{pmatrix},
\end{equation}
where the matrix is represented in the ordered basis $\{\ket{11},\ket{10},\ket{01},\ket{00}\}$.
For odd $N$, the remaining unpaired qubit is acted on trivially.
The moment operator of this auxiliary interleaving ensemble is given by
\begin{equation}
    M_{\mathrm{dope}}^{(k)}
    \equiv
    V^{\otimes k,k}
    \equiv
    \left[
        \bigotimes_{1\leq i\leq \lfloor N/2\rfloor}
        V_{2i-1,2i}
    \right]^{\otimes k,k},
\end{equation}
and the spectral gap for the $V$-doped 2RPmix circuit is defined as
\begin{equation}
    \Delta_{V\mathrm{dp2rpmix}}^{(k)}
    \equiv
    1-
    \left\|P_{\mathrm{2RPmix}}^{(k)}M_{\mathrm{dope}}^{(k)}P_{\mathrm{2RPmix}}^{(k)}-P_{U(1)\mathrm{Haar}}^{(k)}\right\|_{\infty}.
\end{equation}
Then, by utilizing the auxiliary moment operator 
\begin{equation}
    Q_{\mathrm{dope}}^{(k)} \equiv \frac{1}{2}\left(\mathbb{I}+M_{\mathrm{dope}}^{(k)}\right),
\end{equation}
we can derive the following lemma.
\begin{lemma}\label{lem:gap_Vdp2RPmix_circuit_to_dp2RPmix_circuit}
Let $\Delta_{V\mathrm{dp2rpmix}}^{(k)}$ and $\Delta_{\mathrm{dp2rpmix}}^{(k)}$ be the spectral gaps of the $V$-doped 2RPmix circuit and doped 2RPmix circuit in an $N$-qubit system. Then, we have
\begin{equation}
\label{eq:Lower_bound_Delta_dp2rpmix_with_Delta_Vdp2rpmix}
    \Delta_{\mathrm{dp2rpmix}}^{(k)} \geq \frac{1}{2} \Delta_{V\mathrm{dp2rpmix}}^{(k)}.
\end{equation}
\end{lemma}

\begin{proof}
Since each $V_{2i-1,2i}$ is Hermitian and satisfies $V_{2i-1,2i}^2=\mathbb{I}$, we have $(M_{\mathrm{dope}}^{(k)})^\dagger=M_{\mathrm{dope}}^{(k)}$ and $(M_{\mathrm{dope}}^{(k)})^2=\mathbb{I}$.
Hence $Q_{\mathrm{dope}}^{(k)}$ is an orthogonal projector.
Moreover, $V_{2i-1,2i}$ is a $U(1)$-symmetric two-qubit unitary and therefore belongs to the group over which the corresponding local Haar moment projector is defined.
The left and right invariance of the Haar measure thus give
\begin{equation}
    M_{\mathrm{dope}}^{(k)}P_{\gamma_0}^{(k)}
    =
    P_{\gamma_0}^{(k)}M_{\mathrm{dope}}^{(k)}
    =
    P_{\gamma_0}^{(k)}.
\end{equation}
Consequently, we have $Q_{\mathrm{dope}}^{(k)}P_{\gamma_0}^{(k)}=P_{\gamma_0}^{(k)}Q_{\mathrm{dope}}^{(k)}=P_{\gamma_0}^{(k)}$, and hence
\begin{equation}
    Q_{\mathrm{dope}}^{(k)}\geq P_{\gamma_0}^{(k)}.
\end{equation}

Since $\operatorname{Ran}P_{U(1)\mathrm{Haar}}^{(k)}$ is contained in the ranges of both $P_{\mathrm{2RPmix}}^{(k)}$ and $P_{\gamma_0}^{(k)}$, the operators obtained after subtracting $P_{U(1)\mathrm{Haar}}^{(k)}$ below are positive semidefinite.
Using the above operator inequality, followed by the triangle inequality, we obtain
\begin{align}
    \left\|P_{\mathrm{2RPmix}}^{(k)}P_{\gamma_0}^{(k)}P_{\mathrm{2RPmix}}^{(k)}-P_{U(1)\mathrm{Haar}}^{(k)}\right\|_{\infty}
    &\leq
    \left\|P_{\mathrm{2RPmix}}^{(k)}Q_{\mathrm{dope}}^{(k)}P_{\mathrm{2RPmix}}^{(k)}-P_{U(1)\mathrm{Haar}}^{(k)}\right\|_{\infty}
    \nonumber \\
    &\leq
    \frac{1}{2}
    \left\|P_{\mathrm{2RPmix}}^{(k)}-P_{U(1)\mathrm{Haar}}^{(k)}\right\|_{\infty} +
    \frac{1}{2}
    \left\|P_{\mathrm{2RPmix}}^{(k)}M_{\mathrm{dope}}^{(k)}P_{\mathrm{2RPmix}}^{(k)}-P_{U(1)\mathrm{Haar}}^{(k)}\right\|_{\infty}
    \nonumber \\
    &\leq
    \frac{1}{2}
    +
    \frac{1}{2}
    \left\|P_{\mathrm{2RPmix}}^{(k)}M_{\mathrm{dope}}^{(k)}P_{\mathrm{2RPmix}}^{(k)}-P_{U(1)\mathrm{Haar}}^{(k)}\right\|_{\infty}.
\end{align}
This concludes the proof.
\end{proof}
\noindent
By Lemma~\ref{lem:gap_Vdp2RPmix_circuit_to_dp2RPmix_circuit}, it suffices to derive a constant lower bound on $\Delta_{V\mathrm{dp2rpmix}}^{(k)}$.

In this section, we mainly analyze this spectral gap separately in each charge sector $(\bm n,\overline{\bm n})$.
For this purpose, we define the spectral gap in each charge sector by
\begin{equation}
    \Delta_{V\mathrm{dp2rpmix};\bm n,\overline{\bm n}}^{(k)}
    \equiv
    1-
    \left\|\Pi_{\bm n,\overline{\bm n}}
    \left(P_{\mathrm{2RPmix}}^{(k)}M_{\mathrm{dope}}^{(k)}P_{\mathrm{2RPmix}}^{(k)}-P_{U(1)\mathrm{Haar}}^{(k)}\right)
    \Pi_{\bm n,\overline{\bm n}}\right\|_{\infty}.
\end{equation}
In this charge-sector-wise analysis, the trivial charge sectors with particle number $n=0,N$ require special treatment in some cases.
Specifically, we use the following fact several times in the proof.
\begin{rem}[Reduction of trivial charge sectors] \label{rem:reduction_boundary_charge_sector}
Suppose $\Sigma(\bm n,\overline{\bm n})\neq\varnothing$, i.e., $\overline{\bm n}$ is a permutation of $\bm n$.
If $n_i=0$ or $n_i=N$ for some ket copy $i$, then there is a corresponding bra copy with the same particle number.
Since the charge sectors with particle number $0$ or $N$ are one-dimensional, such a ket-bra copy pair contributes only trivially to the moment operators and can be factored out.
Repeating this reduction for all such copies reduces the analysis in the sector $(\bm n,\overline{\bm n})$ to that in a sector $(\bm n',\overline{\bm n}')$ of moment order $k'\leq k$ satisfying $1\leq n_i'\leq N-1$ for every $i\in[k']$.
Therefore, for sectors satisfying $\Sigma(\bm n,\overline{\bm n})\neq\varnothing$, it suffices to derive the gap lower bound for sectors satisfying $1\leq n_i\leq N-1$ for every $i\in[k]$.
\end{rem}

We now outline the proof strategy.
We represent $M_{\mathrm{dope}}^{(k)}$ and $P_{U(1)\mathrm{Haar}}^{(k)}$ in the orbit-state basis $\{\ket{\Omega_{\bm m}}\}$ spanning the 2RPmix moment space, and refine this space into mutually orthogonal sectors according to the compatible-class membership vectors.
We then bound the diagonal blocks and the off-diagonal blocks of $M_{\mathrm{dope}}^{(k)}$ separately.
The off-diagonal contribution of $P_{U(1)\mathrm{Haar}}^{(k)}$ is controlled independently using the approximate orthogonality of the number-permutation states.
Finally, combining these estimates by the triangle inequality gives a uniform contraction strictly below one, which proves Proposition~\ref{prop:gap_LB_dp_2RPmix_circuit}.
For later use, we here define the matrix representation of $M_{\mathrm{dope}}^{(k)}$ in the orbit-state basis as 
\begin{equation}
    T_{\bm m; \bm m'}^{\rm dp} \equiv \bra{\Omega_{\bm m}} M_{\mathrm{dope}}^{(k)}\ket{\Omega_{\bm m'}}.
\end{equation}

\subsection{Classical dynamics of ket type vectors}
\label{subsec:classical_dynamics}

In this subsection, we introduce a classical stochastic process that will be used in the proof of Proposition~\ref{prop:gap_LB_dp_2RPmix_circuit}.
Specifically, we consider a $k$-copy classical process in which the $l$-th copy is an $N$-bit configuration with particle number $n_l$.
The corresponding state space is
\begin{equation}
    \mathcal X_{\bm n}
    \equiv
    \prod_{l=1}^{k}\mathcal X_{n_l},
    \qquad
    \mathcal X_n
    \equiv
    \{x\in\{0,1\}^N:|x|=n\}.
\end{equation}
Copies with $n_l=0$ or $N$ are trivial, since their state spaces are one-dimensional.

At each step, we first choose a uniformly random maximum matching $\gamma$ of the $N$ sites, common to all $k$ copies.
Thus, every $\gamma$ contains $\lfloor N/2\rfloor$ edges, and for odd $N$ one site is left unmatched.
For every edge of $\gamma$ and every copy independently, we then apply either the identity or SWAP with probability $1/2$.
For a fixed matching $\gamma$, the transition probability from $\bm y=(y^1,\ldots,y^k)$ to $\bm x=(x^1,\ldots,x^k)$ is
\begin{align}
    W_{\gamma}(\bm x,\bm y)
    &=
    \prod_{l=1}^{k}w_{\gamma}^l(x^l,y^l), \\
    w_{\gamma}^l(x^l,y^l)
    &=
    \prod_{\{i,j\}\in\gamma}
    w_{\rm 2bit}(x_i^l,x_j^l;y_i^l,y_j^l)
    \prod_{i\notin V(\gamma)}
    \delta_{x_i^l,y_i^l},
\end{align}
where $V(\gamma)$ denotes the set of sites covered by $\gamma$, and
\begin{equation}
    w_{\rm 2bit}(a,b;c,d)
    \equiv 
    \delta_{a+b,c+d}
    \left(\frac12\right)^{|c-d|}.
\end{equation}
For even $N$, the last product is empty and is understood to be equal to one, while for odd $N$, exactly one site lies outside $V(\gamma)$.

Let $\Gamma_{\rm all}$ denote the set of all maximum matchings on the $N$ sites.
The transition matrix of the full $k$-copy process is
\begin{equation}
\label{eq:W^k_def}
    W^{(k)}
    \equiv
    \frac{1}{|\Gamma_{\rm all}|}
    \sum_{\gamma\in\Gamma_{\rm all}}
    W_{\gamma}.
\end{equation}
Since every update preserves the particle number in each copy, $W^{(k)}$ is block diagonal with respect to $\bm n=(n_1,\ldots,n_k)^T$.
We denote its restriction to the particle-number sector $\bm n$ by $W_{\bm n}^{(k)}$.
Similarly, we denote the restrictions of $W_{\gamma}$ and $w_{\gamma}^l$ by $W_{\gamma,\bm n}$ and $w_{\gamma,n_l}^l$, respectively.
In particular,
\begin{equation}
    W_{\gamma,\bm n}
    =
    \bigotimes_{l=1}^{k}w_{\gamma,n_l}^l,
    \qquad
    W_{\bm n}^{(k)}
    =
    \frac{1}{|\Gamma_{\rm all}|}
    \sum_{\gamma\in\Gamma_{\rm all}}
    W_{\gamma,\bm n}.
\end{equation}

Each $W_{\gamma,\bm n}$ is symmetric and stochastic, and hence so is $W_{\bm n}^{(k)}$.
Moreover, the process is irreducible on $\mathcal X_{\bm n}$: any transposition of two sites in any one copy can occur with nonzero probability, while all updates in the other copies are chosen to be identities.
It follows that $W_{\bm n}^{(k)}$ has the unique steady state
\begin{equation}
    \ket{v_{\bm n}^{\rm ss}}
    \equiv
    \frac{1}{\sqrt{\prod_{l=1}^{k}D_{n_l}}}
    \sum_{\bm x\in\mathcal X_{\bm n}}
    \ket{\bm x},
\end{equation}
where $D_n\equiv|\mathcal X_n|=\binom{N}{n}$.

For later convenience, we also introduce an equivalent representation of $W_{\bm n}^{(k)}$.
A uniformly random maximum matching can be generated by applying a uniformly random site permutation to any fixed maximum matching.
Therefore,
\begin{equation}
\label{eq:W^k_perm}
    W_{\bm n}^{(k)}(\bm x,\bm y)
    =
    \frac{1}{N!}
    \sum_{\pi\in S_N}
    W_{\gamma_0,\bm n}(\pi\bm x,\pi\bm y),
\end{equation}
where $\gamma_0=\{\{1,2\},\{3,4\},\ldots,\{2\lfloor N/2\rfloor-1,2\lfloor N/2\rfloor\}\}$ is the fixed maximum matching introduced above.

We next derive a constant lower bound on the spectral gap of $W_{\bm n}^{(k)}$, which is defined as
\begin{equation}
    \Delta(W_{\bm n}^{(k)}) \equiv 1 - \rho(W_{\bm n}^{(k)} -  \ket{v_{\bm n}^{\rm ss}} \bra{v_{\bm n}^{\rm ss}})=  1 - \|W_{\bm n}^{(k)} -  \ket{v_{\bm n}^{\rm ss}} \bra{v_{\bm n}^{\rm ss}}\|_{\infty},
\end{equation}
where  $\rho(\cdot)$ represents the spectral radius, and the second equality follows from the Hermiticity of the operator.
To derive a lower bound, we define the single-copy transition matrix
\begin{equation}
\label{eq:single_copy_W_def}
    w^l
    \equiv
    \frac{1}{|\Gamma_{\rm all}|}
    \sum_{\gamma\in\Gamma_{\rm all}}
    w_{\gamma}^l,
\end{equation}
and denote its restriction to the $n$-particle sector by
\begin{equation}
    w_n^l
    \equiv
    \frac{1}{|\Gamma_{\rm all}|}
    \sum_{\gamma\in\Gamma_{\rm all}}
    w_{\gamma,n}^l.
\end{equation}
For particle number $n$, the normalized uniform steady state is
\begin{equation}
    \ket{u_n^{\rm ss}}
    \equiv
    \frac{1}{\sqrt{D_n}}
    \sum_{x\in\mathcal X_n}
    \ket{x},
\end{equation}
and we denote the corresponding projector by $P_n^{\rm ss}\equiv\ket{u_n^{\rm ss}}\bra{u_n^{\rm ss}}$.
Then, the spectral gap for this single-copy transition matrix is given by $\Delta(w_n^l)\equiv 1- \rho (w_n^l-P_n^{\rm ss}) = 1- \|w_n^l-P_n^{\rm ss}\|_{\infty}$.
For $n=0$ or $N$, the state space is one-dimensional and $w_n^l=P_n^{\rm ss}=\mathbb I_n$, so that $\Delta(w_n^l)=1$.

We first relate the spectral gap of the $k$-copy process to those of the single-copy processes.

\begin{lemma}
\label{lem:kcopy_gap}
The spectral gap of the $k$-copy process in the particle-number sector $\bm n$ satisfies
\begin{equation}
\label{eq:kcopy_gap}
    \Delta(W_{\bm n}^{(k)})
    =
    \min_{1\leq l\leq k}\Delta(w_{n_l}^l).
\end{equation}
\end{lemma}

\begin{proof}
For a fixed matching $\gamma$, $w_{\gamma,n_l}^l$ is an orthogonal projector and satisfies $w_{\gamma,n_l}^l\ket{u_{n_l}^{\rm ss}}=\ket{u_{n_l}^{\rm ss}}$.
We may therefore write
\begin{equation}
\label{eq:Wgamma_decomp}
    w_{\gamma,n_l}^l
    =
    P_{n_l}^{\rm ss}
    +
    \widetilde w_{\gamma,n_l}^l,
    \qquad
    0\leq\widetilde w_{\gamma,n_l}^l\leq\mathbb I_{n_l}-P_{n_l}^{\rm ss},
\end{equation}
where $\mathbb I_{n_l}$ denotes the identity on the $n_l$-particle sector.
The projector onto the uniform $k$-copy state in the sector $\bm n$ is
\begin{equation}
    P_{\bm n}^{\rm ss}
    \equiv
    \ket{v_{\bm n}^{\rm ss}}\bra{v_{\bm n}^{\rm ss}}
    =
    \bigotimes_{l=1}^{k}P_{n_l}^{\rm ss}.
\end{equation}
Using Eq.~\eqref{eq:Wgamma_decomp}, the Hilbert space decomposes into mutually orthogonal sectors according to the set of copies lying in the orthogonal complement of their uniform states.
Consequently,
\begin{align}
    1-\Delta(W_{\bm n}^{(k)})
    &=
    \left\|
        W_{\bm n}^{(k)}-P_{\bm n}^{\rm ss}
    \right\|_{\infty}
    \nonumber\\
    &=
    \left\|
        \frac{1}{|\Gamma_{\rm all}|}
        \sum_{\gamma\in\Gamma_{\rm all}}
        \left[
            \bigotimes_{l=1}^{k}
            \left(
                P_{n_l}^{\rm ss}
                +
                \widetilde w_{\gamma,n_l}^l
            \right)
            -
            \bigotimes_{l=1}^{k}
            P_{n_l}^{\rm ss}
        \right]
    \right\|_{\infty}
    \nonumber\\
    &=
    \max_{\varnothing\neq\mathcal S\subseteq[k]}
    \left\|
        \frac{1}{|\Gamma_{\rm all}|}
        \sum_{\gamma\in\Gamma_{\rm all}}
        \bigotimes_{l\in\mathcal S}
        \widetilde w_{\gamma,n_l}^l
    \right\|_{\infty},
    \label{eq:kcopy_gap_decomp}
\end{align}
where the one-dimensional factors $P_{n_l}^{\rm ss}$ for $l\notin\mathcal S$ have been suppressed.

For any $j\in\mathcal S$, Eq.~\eqref{eq:Wgamma_decomp} gives
\begin{equation}
    0
    \leq
    \bigotimes_{l\in\mathcal S}\widetilde w_{\gamma,n_l}^l
    \leq
    \widetilde w_{\gamma,n_j}^j\otimes\mathbb I,
\end{equation}
and therefore
\begin{align}
    \left\|
        \frac{1}{|\Gamma_{\rm all}|}
        \sum_{\gamma\in\Gamma_{\rm all}}
        \bigotimes_{l\in\mathcal S}
        \widetilde w_{\gamma,n_l}^l
    \right\|_{\infty}
    &\leq
    \left\|
        \frac{1}{|\Gamma_{\rm all}|}
        \sum_{\gamma\in\Gamma_{\rm all}}
        \widetilde w_{\gamma,n_j}^j
    \right\|_{\infty}
    \nonumber\\
    &=
    1-\Delta(w_{n_j}^j).
\end{align}
Since the singleton sets $\mathcal S=\{j\}$ are also included in Eq.~\eqref{eq:kcopy_gap_decomp}, we obtain
\begin{equation}
    1-\Delta(W_{\bm n}^{(k)})
    =
    \max_{1\leq j\leq k}
    \left(1-\Delta(w_{n_j}^j)\right),
\end{equation}
which proves Eq.~\eqref{eq:kcopy_gap}.
\end{proof}

We now establish the required constant lower bound on the single-copy spectral gap.
\begin{lemma}
\label{lem:Bernstein}
For every $N\geq2$, $l\in[k]$, and $0<n<N$, the single-copy process satisfies
\begin{equation}
\label{eq:single_copy_gap}
    \Delta(w_n^l)\geq\frac12.
\end{equation}
\end{lemma}

\begin{proof}
We first consider even $N$.
In this case, $w_n^l$ is the $n$-particle version of the $p=1/2$ random involution walk analyzed in Ref.~\cite{bernstein2018random}.
For $N\geq6$, the two-row spectrum obtained there gives
\begin{equation}
    \Delta(w_n^l)
    =
    \frac{N}{2(N-1)}
    \geq
    \frac12.
\end{equation}
The cases $N=2,4$ can be verified directly.

We next consider odd $N$. The cases $N=3,5$ can be verified directly, so we assume $N\geq7$ below.
For each site $i\in[N]$, let $w_n^{l,(i)}$ denote the transition matrix in the $n$-particle sector conditioned on $i$ being the unmatched site.
Since the unmatched site is uniformly distributed,
\begin{equation}
    w_n^l=\frac{1}{N}\sum_{i=1}^{N}w_n^{l,(i)}.
\end{equation}
For fixed $i$, the occupation of site $i$ is unchanged, and $w_n^{l,(i)}$ is block diagonal with respect to whether site $i$ is occupied.
Each block is the corresponding single-copy process on the remaining $N-1$ sites.
Since $N-1\geq6$ is even, the even-$N$ result above implies that all eigenvalues orthogonal to the steady state in each nontrivial block are at most $(N-3)/[2(N-2)]$.
The same bound trivially holds when a block is one-dimensional.
Therefore,
\begin{align}
    w_n^{l,(i)}
    &\leq \frac{N-3}{2(N-2)}\mathbb{I}
    +\frac{N-1}{2(N-2)}
    \left(
        \ket{u_{n,\overline{i}}^{\rm ss}}\bra{u_{n,\overline{i}}^{\rm ss}}\otimes\ket{0_i}\bra{0_i}
        +
        \ket{u_{n-1,\overline{i}}^{\rm ss}}\bra{u_{n-1,\overline{i}}^{\rm ss}}\otimes\ket{1_i}\bra{1_i}
    \right)
    \nonumber\\
    &=
    \frac{N-3}{2(N-2)}\mathbb{I}
    +\frac{N-1}{2(N-2)}
    \left(
        \ket{u_n^{\rm ss}}\bra{u_n^{\rm ss}}
        +
        \ket{\phi_i}\bra{\phi_i}
    \right),
\end{align}
where $\ket{u_{n,\overline{i}}^{\rm ss}}$ denotes the uniform steady state in the $n$-particle sector of the $N-1$ sites excluding site $i$.
Here, $\ket{u_n^{\rm ss}}$ is the steady-state vector on the full $N$-qubit system, and $\ket{\phi_i}$ is the normalized vector orthogonal to it defined by
\begin{align}
    \ket{u_n^{\rm ss}}
    &=
    \sqrt{\frac{N-n}{N}}\ket{u_{n,\overline{i}}^{\rm ss}}\ket{0_i}
    +
    \sqrt{\frac{n}{N}}\ket{u_{n-1,\overline{i}}^{\rm ss}}\ket{1_i}
    =
    \frac{1}{\sqrt{D_n}}\sum_{x\in\mathcal X_n}\ket{x},
    \\
    \ket{\phi_i}
    &\equiv
    \sqrt{\frac{N-n}{N}}\ket{u_{n-1,\overline{i}}^{\rm ss}}\ket{1_i}
    -
    \sqrt{\frac{n}{N}}\ket{u_{n,\overline{i}}^{\rm ss}}\ket{0_i}
    =
    \sqrt{\frac{N^2}{D_n n(N-n)}}
    \sum_{x\in\mathcal X_n}
    \left(x_i-\frac{n}{N}\right)\ket{x}.
\end{align}
Using this representation, we obtain
\begin{align}
    w_n^l
    &=\frac{1}{N}\sum_{i=1}^{N}w_n^{l,(i)}
    \nonumber\\
    &\leq
    \frac{1}{N}
    \left[
        \sum_{i=1}^{N}
        \frac{N-3}{2(N-2)}\mathbb{I}
        +
        \frac{N-1}{2(N-2)}
        \left(
            \ket{u_n^{\rm ss}}\bra{u_n^{\rm ss}}
            +
            \ket{\phi_i}\bra{\phi_i}
        \right)
    \right]
    \nonumber\\
    &=
    \frac{N-3}{2(N-2)}\mathbb{I}
    +
    \frac{N-1}{2(N-2)}
    \ket{u_n^{\rm ss}}\bra{u_n^{\rm ss}}
    +
    \frac{N-1}{2N(N-2)}
    \sum_{i=1}^{N}\ket{\phi_i}\bra{\phi_i}.
\end{align}
Furthermore,
\begin{align}
    \sum_{i=1}^{N}\ket{\phi_i}
    &=
    \sqrt{\frac{N^2}{D_n n(N-n)}}
    \sum_{i=1}^{N}\sum_{x\in\mathcal X_n}
    \left(x_i-\frac{n}{N}\right)\ket{x}
    =0,
    \\
    \langle\phi_i|\phi_j\rangle
    &=
    -\frac{1}{N-1},
    \qquad
    i,j\in[N],\quad i\neq j.
\end{align}
Here, the second relation follows from site-permutation symmetry: the off-diagonal inner product is independent of $i\neq j$, and taking the squared norm of $\sum_i\ket{\phi_i}=0$ determines its value.

We define $\mathcal H_\phi\equiv\operatorname{span}\{\ket{\phi_1},\ket{\phi_2},\ldots,\ket{\phi_N}\}$ and denote the orthogonal projector onto this subspace by $\Pi_{\mathcal H_\phi}$.
Then,
\begin{equation}
    \sum_{i=1}^{N}\ket{\phi_i}\bra{\phi_i}
    =
    \frac{N}{N-1}\Pi_{\mathcal H_\phi}.
\end{equation}
Indeed, for any $j\in[N]$,
\begin{equation}
    \left(
        \sum_{i=1}^{N}\ket{\phi_i}\bra{\phi_i}
    \right)\ket{\phi_j}
    =
    \ket{\phi_j}
    -
    \frac{1}{N-1}\sum_{i\neq j}\ket{\phi_i}
    =
    \frac{N}{N-1}\ket{\phi_j},
\end{equation}
while the operator vanishes on any vector orthogonal to $\mathcal H_\phi$.

Thus,
\begin{align}
    w_n^l
    &\leq
    \frac{N-3}{2(N-2)}\mathbb{I}
    +
    \frac{N-1}{2(N-2)}
    \ket{u_n^{\rm ss}}\bra{u_n^{\rm ss}}
    +
    \frac{N-1}{2N(N-2)}
    \sum_{i=1}^{N}\ket{\phi_i}\bra{\phi_i}
    \nonumber\\
    &=
    \frac{N-3}{2(N-2)}\mathbb{I}
    +
    \frac{N-1}{2(N-2)}
    \ket{u_n^{\rm ss}}\bra{u_n^{\rm ss}}
    +
    \frac{1}{2(N-2)}
    \Pi_{\mathcal H_\phi}.
\end{align}
Since every $\ket{\phi_i}$ is orthogonal to $\ket{u_n^{\rm ss}}$, we have $\Pi_{\mathcal H_\phi}\leq\mathbb{I}-\ket{u_n^{\rm ss}}\bra{u_n^{\rm ss}}$.
Therefore, we have
\begin{equation}
    0\leq w_n^l - \ket{u_n^{\rm ss}}\bra{u_n^{\rm ss}}
    \leq \frac{N-3}{2(N-2)} (\mathbb{I} - \ket{u_n^{\rm ss}}\bra{u_n^{\rm ss}}) +\frac{1}{2(N-2)}\Pi_{\mathcal H_\phi}
    \leq \frac12 (\mathbb{I} - \ket{u_n^{\rm ss}}\bra{u_n^{\rm ss}}) .
\end{equation}
Hence all nontrivial eigenvalues of $w_n^l$ lie in $[0,1/2]$, and therefore
\begin{equation}
    \Delta(w_n^l)\geq\frac12.
\end{equation}
This concludes the proof.

\end{proof}

Combining Lemmas~\ref{lem:kcopy_gap} and \ref{lem:Bernstein}, and recalling that $\Delta(w_n^l)=1$ for $n=0$ or $N$, we obtain
\begin{equation}
\label{eq:kcopy_constant_gap}
    \Delta(W_{\bm n}^{(k)})
    \geq
    \frac12
\end{equation}
for every particle-number sector $\bm n$.

Finally, we coarse-grain the $k$-copy process $W_{\bm n}^{(k)}$ by the ket type vector.
To distinguish this object from the joint ket--bra type vector $\bm m$ used for the quantum moment space, we denote the ket type vector by $\bm r$.
For $\bm x\in\mathcal X_{\bm n}$, define
\begin{equation}
\label{eq:type_map_Phi}
    \Phi:
    \mathcal X_{\bm n}
    \longrightarrow
    \mathbb Z_{\geq0}^{2^k},
    \qquad
    [\Phi(\bm x)]_a
    \equiv
    \left|\left\{i\in[N]:(x_i^1,\ldots,x_i^k)^T=a\right\}\right|,
    \qquad
    a\in\{0,1\}^k.
\end{equation}
For a ket type vector $\bm r$, we denote the corresponding set of configurations by
\begin{equation}
    \mathcal T_{\bm r}
    \equiv
    \{\bm x\in\mathcal X_{\bm n}:\Phi(\bm x)=\bm r\},
\end{equation}
whose cardinality is
\begin{equation}
    D_{\bm r}
    \equiv
    |\mathcal T_{\bm r}|
    =
    \frac{N!}{\prod_{a\in\{0,1\}^k}r_a!}.
\end{equation}
The possible ket type vectors in the particle-number sector $\bm n$ form the set
\begin{equation}
    \mathcal L_{\bm n}
    \equiv
    \left\{
        \bm r\in\mathbb Z_{\geq0}^{2^k}
        :
        \sum_{a\in\{0,1\}^k}r_a=N,
        \quad
        \sum_{a\in\{0,1\}^k}r_a a=\bm n
    \right\}.
\end{equation}

For any $\bm y,\bm y'\in\mathcal T_{\bm r}$, there exists $\pi\in S_N$ such that $\bm y'=\pi\bm y$.
Since $W_{\bm n}^{(k)}$ is invariant under simultaneous site permutations, for every $\bm r'\in\mathcal L_{\bm n}$,
\begin{align}
    \sum_{\bm x\in\mathcal T_{\bm r'}}
    W_{\bm n}^{(k)}(\bm x,\bm y)
    &=
    \sum_{\bm x\in\mathcal T_{\bm r'}}
    W_{\bm n}^{(k)}(\pi\bm x,\pi\bm y)
    \nonumber\\
    &=
    \sum_{\bm x\in\mathcal T_{\bm r'}}
    W_{\bm n}^{(k)}(\bm x,\bm y').
    \label{eq:strong_lumping}
\end{align}
Thus, the coarse-grained process is Markovian, with transition matrix
\begin{equation}
\label{eq:Wphi_def}
    W_{\bm n}^{\Phi}(\bm r',\bm r)
    \equiv
    \sum_{\bm x\in\mathcal T_{\bm r'}}
    W_{\bm n}^{(k)}(\bm x,\bm y),
    \qquad
    \bm y\in\mathcal T_{\bm r}.
\end{equation}
Equation~\eqref{eq:strong_lumping} is the standard strong-lumping condition~\cite{levin2026markov}.

The coarse-grained process has the following properties.

\begin{lemma}
\label{lem:classical_coarse_process}
The transition matrix $W_{\bm n}^{\Phi}$ satisfies the following.
\begin{enumerate}
    \item[$\mathrm{(i)}$]
    Its unique stationary probability vector is
    \begin{equation}
    \label{eq:Wphi_ss}
        \ket{v_{\bm n,\Phi}^{\rm ss}}
        =
        \frac{1}{\prod_{l=1}^{k}D_{n_l}}
        \sum_{\bm r\in\mathcal L_{\bm n}}
        D_{\bm r}\ket{\bm r},
    \end{equation}
    where $\ket{\bm r}$ denotes the canonical basis vector of the coarse-grained state space associated with $\bm r\in\mathcal L_{\bm n}$.

    \item[$\mathrm{(ii)}$]
    Defining the spectral gap of this transition matrix by $\Delta(W_{\bm n}^{\Phi})\equiv1-\rho\left(W_{\bm n}^{\Phi}-\ket{v_{\bm n,\Phi}^{\rm ss}}\bra{\bm 1}\right)$, where $\bra{\bm 1}\equiv\sum_{\bm r\in\mathcal L_{\bm n}}\bra{\bm r}$, we have
    \begin{equation}
    \label{eq:Wphi_gap}
        \Delta(W_{\bm n}^{\Phi})
        \geq
        \Delta(W_{\bm n}^{(k)})
        \geq
        \frac12.
    \end{equation}

    \item[$\mathrm{(iii)}$]
    For the fixed maximum matching $\gamma_0$,
    \begin{equation}
    \label{eq:Wphi_fixed_matching}
        W_{\bm n}^{\Phi}(\bm r',\bm r)
        =
        \frac{1}{D_{\bm r}}
        \sum_{\bm x\in\mathcal T_{\bm r'}}
        \sum_{\bm y\in\mathcal T_{\bm r}}
        W_{\gamma_0,\bm n}(\bm x,\bm y).
    \end{equation}
\end{enumerate}
\end{lemma}

\begin{proof}
\emph{$\mathrm{(i)}$ and $\mathrm{(ii)}$}
The unique stationary distribution of $W_{\bm n}^{(k)}$ is uniform on $\mathcal X_{\bm n}$.
Under the coarse-graining map $\Phi$, the set $\mathcal T_{\bm r}$ contains $D_{\bm r}$ configurations, and hence the induced stationary weight of $\bm r$ is $D_{\bm r}/\prod_{l=1}^{k}D_{n_l}$.
This proves Eq.~\eqref{eq:Wphi_ss}, and the stationary probability vector is unique because the coarse-grained process inherits irreducibility from $W_{\bm n}^{(k)}$.

By the strong-lumping property, every eigenvalue of $W_{\bm n}^{\Phi}$ is also an eigenvalue of $W_{\bm n}^{(k)}$. 
Moreover, $W_{\bm n}^{(k)}$ is positive semidefinite, since it is an average of the orthogonal projectors $W_{\gamma,\bm n}$. 
Hence both spectra are contained in $[0,1]$, and therefore
\begin{equation}
    \Delta(W_{\bm n}^{\Phi})
    \geq
    \Delta(W_{\bm n}^{(k)}).
\end{equation}
Combining this with Eq.~\eqref{eq:kcopy_constant_gap} proves Eq.~\eqref{eq:Wphi_gap}.

\emph{$\mathrm{(iii)}$}
Fix $\bm y\in\mathcal T_{\bm r}$.
Using Eqs.~\eqref{eq:W^k_perm} and \eqref{eq:Wphi_def}, we obtain
\begin{align}
    W_{\bm n}^{\Phi}(\bm r',\bm r)
    &=
    \frac{1}{N!}
    \sum_{\pi\in S_N}
    \sum_{\bm x\in\mathcal T_{\bm r'}}
    W_{\gamma_0,\bm n}(\pi\bm x,\pi\bm y)
    \nonumber\\
    &=
    \frac{1}{N!}
    \sum_{\pi\in S_N}
    \sum_{\bm x\in\mathcal T_{\bm r'}}
    W_{\gamma_0,\bm n}(\bm x,\pi\bm y)
    \nonumber\\
    &=
    \frac{1}{D_{\bm r}}
    \sum_{\bm y'\in\mathcal T_{\bm r}}
    \sum_{\bm x\in\mathcal T_{\bm r'}}
    W_{\gamma_0,\bm n}(\bm x,\bm y').
\end{align}
Here, we used the fact that $\pi\bm y$ runs over every element of $\mathcal T_{\bm r}$ exactly $N!/D_{\bm r}$ times.
This proves Eq.~\eqref{eq:Wphi_fixed_matching}.
\end{proof}

\subsection{Properties of compatible classes} \label{ss:Properties of compatible classes}

In this subsection, we show several important properties of the compatible classes introduced to characterize the 2RPmix group.
In particular, it is useful to represent each compatible class in terms of a partially defined isometry on the Boolean cube.
We first summarize this representation and derive an upper bound on the number of compatible classes.

\begin{lemma}[Representation and number of compatible classes]
\label{lem:compatible_class_representation}
For every compatible class $F\in\mathcal{F}_k$, the following properties hold.
\begin{enumerate}
    \item[$\mathrm{(i)}$]
    There exist a subset $\mathcal{I}_F\subseteq\{0,1\}^k$ and a unique injective map $f_F:\mathcal{I}_F\to\{0,1\}^k$ such that
    \begin{equation}
        F
        =
        \left\{
        (x,f_F(x)):
        x\in\mathcal{I}_F
        \right\}.
    \end{equation}
    The map $f_F$ preserves inner products on $\mathcal{I}_F$, i.e., $f_F(x)\cdot f_F(x')=x\cdot x'$ for all $x,x'\in\mathcal{I}_F$.
    Moreover, $|f_F(x)|=|x|$ for every $x\in\mathcal{I}_F$.

    \item[$\mathrm{(ii)}$]
    The all-zero and all-one vectors $\bm 0,\bm 1\in\{0,1\}^k$ belong to $\mathcal{I}_F$ for every $F$, and
    \begin{equation}
        f_F(\bm 0)=\bm 0,
        \qquad
        f_F(\bm 1)=\bm 1.
    \end{equation}

    \item[$\mathrm{(iii)}$]
    The map $f_F$ extends uniquely to a linear isometry on $\operatorname{span}_{\mathbb R}\mathcal{I}_F$.
    Then, we can choose a $k$-dimensional orthogonal matrix $Q_F\in \operatorname{Orth}(k)$ satisfying $Q_Fx=f_F(x)$ for every $x\in\mathcal{I}_F$.
    Such an orthogonal matrix is unique whenever $\operatorname{span}_{\mathbb R}\mathcal{I}_F=\mathbb R^k$.

    \item[$\mathrm{(iv)}$]
    Let $F,G\in\mathcal{F}_k$, and let $Q_F,Q_G\in \operatorname{Orth}(k)$ be orthogonal matrices chosen as in $\mathrm{(iii)}$.
    Then, for every $x\in\mathcal{I}_F$,
    \begin{equation}
        (Q_F-Q_G)x=\bm 0
        \quad\Longleftrightarrow\quad
        (x,f_F(x))\in F\cap G.
    \end{equation}
    In particular, this equivalence holds independently of the choices of the orthogonal extensions $Q_F$ and $Q_G$.
    
    \item[$\mathrm{(v)}$]
    The total number of compatible classes satisfies
    \begin{equation}
        |\mathcal{F}_k|
        \leq
        \sum_{r=1}^{k}2^{2kr}
        \leq
        k\,2^{2k^2}.
    \end{equation}
\end{enumerate}
\end{lemma}

\begin{proof}
\emph{$\mathrm{(i)}$}
Suppose that $(x,y),(x,y')\in F$ have the same ket component $x$.
Their compatibility gives $x\cdot x=y\cdot y'$.
Since both colors belong to $\mathcal{C}_k$, we also have $|y|=|y'|=|x|$, and hence $y\cdot y'=|y|=|y'|$.
For binary vectors, this is possible only if $y=y'$.
Therefore, each ket component appearing in $F$ is associated with a unique bra component.

We may thus define $\mathcal{I}_F\equiv\{x\in\{0,1\}^k:\exists y\ {\rm such\ that}\ (x,y)\in F\}$ and write $F=\{(x,f_F(x)):x\in\mathcal{I}_F\}$ with a unique map $f_F$.
The compatibility condition immediately gives $f_F(x)\cdot f_F(x')=x\cdot x'$ for all $x,x'\in\mathcal{I}_F$.
Since $(x,f_F(x))\in\mathcal{C}_k$, we also have $|f_F(x)|=|x|$.

It remains to note that $f_F$ is injective.
If $f_F(x)=f_F(x')=y$, then compatibility gives
\begin{equation*}
    x\cdot x'
    =
    y\cdot y
    =
    |y|
    =
    |x|
    =
    |x'|.
\end{equation*}
For binary vectors, this implies $x=x'$.

\emph{$\mathrm{(ii)}$}
The colors $(\bm 0,\bm 0)$ and $(\bm 1,\bm 1)$ are compatible with every color $(x,y)\in\mathcal{C}_k$.
Compatibility with $(\bm 0,\bm 0)$ is immediate, while $\bm 1\cdot x=|x|=|y|=\bm 1\cdot y$.
Since $F$ is maximal, both colors must belong to $F$.
Therefore, we have $\bm 0,\bm 1\in\mathcal{I}_F$.
Since $|f_F(\bm 0)|=0$ and $|f_F(\bm 1)|=k$, we obtain $f_F(\bm 0)=\bm 0$ and $f_F(\bm 1)=\bm 1$, respectively.

\emph{$\mathrm{(iii)}$}
Let $V_F\equiv\operatorname{span}_{\mathbb R}\mathcal{I}_F$ and $r\equiv\dim V_F$.
Choose linearly independent vectors
$x_1,\ldots,x_r\in\mathcal{I}_F$ forming a basis of $V_F$.
Since $f_F$ preserves inner products,
$f_F(x_1),\ldots,f_F(x_r)$ have the same Gram matrix as
$x_1,\ldots,x_r$ and are therefore linearly independent.
Thus, the assignment $x_i\mapsto f_F(x_i)$ uniquely defines a linear isometry on $V_F$.

For any $x\in\mathcal{I}_F$, this linear isometry maps $x$ to $f_F(x)$.
Indeed, writing $x=\sum_i c_i x_i$, we have
\begin{equation*}
    f_F(x)\cdot\sum_i c_i f_F(x_i)
    =
    \sum_i c_i\,x\cdot x_i
    =
    x\cdot x
    =
    |x|,
\end{equation*}
while both $f_F(x)$ and $\sum_i c_i f_F(x_i)$ have squared norm $|x|$. Therefore, $f_F(x) = \sum_i c_i f_F(x_i)$ must be satisfied.

Finally, extending orthonormal bases of $V_F$ and its image to orthonormal bases of $\mathbb R^k$ gives an orthogonal extension $Q_F\in\operatorname{Orth}(k)$.
If $V_F=\mathbb R^k$, no further extension is required, and hence $Q_F$ is unique.

\emph{$\mathrm{(iv)}$}
Let $x\in\mathcal{I}_F$.
If $(x,f_F(x))\in F\cap G$, then $x\in\mathcal{I}_G$ and
$f_F(x)=f_G(x)$.
Therefore, we have $Q_Fx=f_F(x)=f_G(x)=Q_Gx,$ which gives $(Q_F-Q_G)x=\bm 0$.

Conversely, suppose that $(Q_F-Q_G)x=\bm 0$.
Since $x\in\mathcal{I}_F$, we have $Q_Fx=f_F(x)$, and hence $Q_Gx=f_F(x)$.
For every $z\in\mathcal{I}_G$, orthogonality of $Q_G$ gives
\begin{equation*}
    x\cdot z
    =
    (Q_Gx)\cdot(Q_Gz)
    =
    f_F(x)\cdot f_G(z).
\end{equation*}
Thus, the color $(x,f_F(x))$ is compatible with every color in $G$.
By maximality of $G$, we have $(x,f_F(x))\in G$.
Since $(x,f_F(x))\in F$ by definition, it follows that
$(x,f_F(x))\in F\cap G$.

\emph{$\mathrm{(v)}$}
Fix $F\in\mathcal{F}_k$ and let $r=\dim V_F$.
Choose an ordered basis $x_1,\ldots,x_r\in\mathcal{I}_F$.
Then the $r$ pairs $\left(x_1,f_F(x_1)\right),\ldots,\left(x_r,f_F(x_r)\right)$ uniquely determine $F$.
Indeed, for every $z\in\mathcal{I}_F$, writing $z=\sum_i c_i x_i$, we have $f_F(z)=\sum_i c_i f_F(x_i)$.
Thus, any color $(x,y)\in\mathcal{C}_k$ that is compatible with all $\left(x_i,f_F(x_i)\right)$ is compatible with every $(z,f_F(z))\in F$.
By maximality, such a color must already belong to $F$.

Each $x_i$ and $f_F(x_i)$ has at most $2^k$ possible values.
Hence, for fixed $r$, there are at most $2^{2kr}$ possible ordered collections of basis pairs.
Summing over $1\leq r\leq k$ yields
\begin{equation}
    |\mathcal{F}_k|
    \leq
    \sum_{r=1}^{k}2^{2kr}
    \leq
    k\,2^{2k^2}.
\end{equation}
\end{proof}

We next distinguish the compatible classes generated by permutations of the $k$ copies from genuinely non-permutation compatible classes.

\begin{lemma}[Permutation and non-permutation compatible classes]
\label{lem:nonpermutation_compatible_class}
The following properties hold.
\begin{enumerate}
    \item[$\mathrm{(i)}$]
    For every permutation $\sigma\in S_k$, define
    \begin{equation}
        F_\sigma
        \equiv
        \left\{
        (x,\sigma x):
        x\in\{0,1\}^k
        \right\}.
    \end{equation}
    Then $F_\sigma\in\mathcal{F}_k$.
    We denote the set of these permutation compatible classes by
    \begin{equation}
        \mathcal{F}_k^{\rm perm}
        \equiv
        \left\{
        F_\sigma:
        \sigma\in S_k
        \right\}
        \subseteq\mathcal{F}_k.
    \end{equation}

    \item[$\mathrm{(ii)}$]
    Let $F\in\mathcal{F}_k\setminus\mathcal{F}_k^{\rm perm}$.
    Then there exists at least one coordinate $j\in[k]$ for which there is no $l\in[k]$ satisfying $e_j\in\mathcal{I}_F$ and $f_F(e_j)=e_l$.
    Here $e_j\in\{0,1\}^k$ denotes the unit vector on coordinate $j$, i.e., $(e_j)_i=\delta_{ij}$.
    We call such a coordinate $j$ a \emph{non-permutation coordinate} of $F$.

    For every non-permutation coordinate $j$, no two elements of $\mathcal{I}_F$ can differ only in their $j$-th bit.
    In particular, for any $x\in\{0,1\}^k$ with $x_j=0$, either $x\notin \mathcal{I}_F$ or $x+e_j\notin\mathcal{I}_F$ is satisfied.

    \item[$\mathrm{(iii)}$]
    For $k\leq3$, all compatible classes are permutation compatible classes, $\mathcal{F}_k=\mathcal{F}_k^{\rm perm}$.
    In contrast, for every $k\geq4$, $\mathcal{F}_k^{\rm perm}\subsetneq\mathcal{F}_k$.
\end{enumerate}
\end{lemma}

\begin{proof}
\emph{$\mathrm{(i)}$}
A permutation $\sigma$ preserves inner products, so $(\sigma x)\cdot(\sigma x')=x\cdot x'$ for all $x,x'\in\{0,1\}^k$.
Hence the colors in $F_\sigma$ are pairwise compatible.
Moreover, $F_\sigma$ contains one color for every possible ket vector $x\in\{0,1\}^k$.
By the uniqueness of the bra component established in Lemma~\ref{lem:compatible_class_representation}, no further color can be added.
Thus $F_\sigma$ is a compatible class.

\emph{$\mathrm{(ii)}$}
Suppose, to the contrary, that every coordinate $j\in[k]$ satisfies $e_j\in\mathcal{I}_F$ and $f_F(e_j)=e_{l(j)}$ for some $l(j)\in[k]$.
Inner-product preservation implies that the indices $l(1),\ldots,l(k)$ are all distinct, and hence define a permutation $\sigma\in S_k$.
Since $e_1,\ldots,e_k\in\mathcal{I}_F$, we have $V_F=\mathbb R^k$, and the corresponding orthogonal matrix $Q_F$ satisfies $Q_Fe_j=\sigma e_j$ for every $j\in[k]$.
Therefore, $f_F(x)=\sigma x$ for every $x\in\mathcal{I}_F$, so $F\subseteq F_\sigma$.
Since $F$ is maximal, this implies $F=F_\sigma$, contradicting $F\notin\mathcal{F}_k^{\rm perm}$.
Thus at least one non-permutation coordinate exists.

Now let $j$ be a non-permutation coordinate and suppose that $x,x+e_j\in\mathcal{I}_F$ for some $x$ with $x_j=0$.
Since $f_F$ preserves Hamming weights and inner products, it also preserves Hamming distance.
Thus $f_F(x)$ and $f_F(x+e_j)$ differ in exactly one bit.
Moreover, $|f_F(x+e_j)|=|x+e_j|=|x|+1=|f_F(x)|+1$, so there exists $l\in[k]$ such that
\begin{equation}
    f_F(x+e_j)=f_F(x)+e_l.
\end{equation}
For any $z\in\mathcal{I}_F$, inner-product preservation then gives
\begin{equation*}
    e_j\cdot z
    =
    (x+e_j)\cdot z-x\cdot z
    =
    f_F(x+e_j)\cdot f_F(z)-f_F(x)\cdot f_F(z)
    =
    e_l\cdot f_F(z).
\end{equation*}
Therefore, the color $(e_j,e_l)$ is compatible with every color in $F$.
By maximality, $(e_j,e_l)\in F$, implying $e_j\in\mathcal{I}_F$ and $f_F(e_j)=e_l$.
This contradicts the assumption that $j$ is a non-permutation coordinate.
Hence either $x\notin\mathcal{I}_F$ or $x+e_j\notin\mathcal{I}_F$ must be satisfied.

\emph{$\mathrm{(iii)}$}
We first note that $\mathcal{I}_F$ is closed under taking complements.
Indeed, for any $x,z\in\mathcal{I}_F$,
\begin{equation*}
    (\bm 1-x)\cdot z
    =
    |z|-x\cdot z
    =
    |f_F(z)|-f_F(x)\cdot f_F(z)
    =
    (\bm 1-f_F(x))\cdot f_F(z).
\end{equation*}
Thus $(\bm 1-x,\bm 1-f_F(x))$ is compatible with every color in $F$.
By maximality, $\bm 1-x\in\mathcal{I}_F$ and
$f_F(\bm 1-x)=\bm 1-f_F(x)$.

For $k=1$, the claim is immediate.

For $k=2$, maximality implies that $\mathcal{I}_F$ contains a vector other than $\bm 0$ and $\bm 1$, since otherwise a weight-one color such as $(e_1,e_1)$ could be added.
Such a vector has Hamming weight one, and complement closure gives the other weight-one vector.
Their images are distinct weight-one vectors, and hence determine a permutation $\sigma\in S_2$.
Therefore $F\subseteq F_\sigma$, and maximality gives $F=F_\sigma$.

For $k=3$, the same argument shows that $\mathcal{I}_F$ contains at least one weight-one vector $e_j$, with $f_F(e_j)=e_l$.
It must contain at least two distinct weight-one vectors.
Indeed, otherwise complement closure implies that the only nontrivial vectors in $\mathcal{I}_F$ are $e_j$ and $\bm 1-e_j$.
Then, for any $j'\neq j$ and $l'\neq l$, the color $(e_{j'},e_{l'})$ is compatible with every color in $F$, contradicting maximality.

Thus there exist distinct $j_1,j_2$ and distinct $l_1,l_2$ such that
$f_F(e_{j_a})=e_{l_a}$ for $a=1,2$.
Let $j_3$ and $l_3$ be the remaining coordinates.
For every $x\in\mathcal{I}_F$, inner-product preservation gives
$f_F(x)_{l_a}=x_{j_a}$ for $a=1,2$, while Hamming-weight preservation gives
$f_F(x)_{l_3}=x_{j_3}$.
Hence $f_F(x)=\sigma x$ for the permutation $\sigma$ satisfying
$\sigma(j_a)=l_a$ for $a=1,2,3$.
Therefore $F\subseteq F_\sigma$, and maximality gives $F=F_\sigma$.
This proves $\mathcal{F}_k=\mathcal{F}_k^{\rm perm}$ for $k\leq3$.

For $k\geq4$, non-permutation compatible classes exist.
For example, for $k=4$, the orthogonal matrix
\begin{equation}
    Q_{\mathrm P}
    \equiv
    \frac{1}{2}J_4-\mathbb I_4,
\end{equation}
where $J_4$ denotes the $4\times4$ all-ones matrix, induces a non-permutation compatible class with the corresponding domain $\mathcal I_{\mathrm P} =\{x\in\{0,1\}^4:|x|\in\{0,2,4\}\}$. Indeed, $Q_{\mathrm P}x$ is a binary vector for $x\in \mathcal I_{\mathrm P}$, while $Q_{\mathrm P}$ is not a permutation matrix.
This construction extends to every $k>4$ by $Q_{\mathrm P}\oplus\mathbb I_{k-4}$.
\end{proof}

\subsection{2RPmix moment space in a fixed charge sector}
\label{ss:compatible_class_projectors}

We now characterize the fixed space of the 2RPmix moment operator within a fixed charge sector $(\bm n,\overline{\bm n})$.
The key simplification is that, once a compatible class $F$ is fixed, the bra local type is uniquely determined by the ket local type, as shown in Lemma~\ref{lem:compatible_class_representation}. 
Hence, within each compatible class, the orbit states can be labeled solely by the ket type vectors introduced in Eq.~\eqref{eq:type_map_Phi}.

Let $F\in\mathcal F_k$, and let $\mathcal I_F$, $f_F$, and $Q_F$ denote the corresponding domain, class map, and a fixed orthogonal extension introduced in Lemma~\ref{lem:compatible_class_representation}.
For a ket type vector $\bm r\in\mathcal L_{\bm n}$ satisfying $\operatorname{supp}(\bm r)\subseteq\mathcal I_F$, we denote by $\ket{\Omega_{\bm r;F}}$ the orbit state $\ket{\Omega_{\bm m}}$ whose type vector is specified by
\begin{equation}
    m_{x,f_F(x)}=r_x
    \quad\text{for }x\in\mathcal I_F,
    \qquad
    m_{x,y}=0
    \quad\text{otherwise}. \label{eq:full_type_vector_from_ket_type_vector_fixed_F}
\end{equation}
The corresponding orbit state is given by
\begin{equation}
    \ket{\Omega_{\bm r;F}} \equiv \frac{1}{\sqrt{D_{\bm r}}} \sum_{\bm x \in \mathcal{T}_{\bm r}} \ket{(x_1,f_F(x_1)), (x_2,f_F(x_2)),\dots , (x_N,f_F(x_N))}.
\end{equation}
Thus, $\bm r$ specifies the ket local-type multiplicities, while the corresponding bra local types are fixed by $f_F$.

For each compatible class $F$, we define
\begin{equation}
\label{eq:charge_class_projector}
    \Pi_{\bm n;F}
    \equiv 
    \sum_{\substack{
    \bm r\in\mathcal L_{\bm n}\\
    \operatorname{supp}(\bm r)\subseteq\mathcal I_F
    }}
    \ket{\Omega_{\bm r;F}}
    \bra{\Omega_{\bm r;F}}.
\end{equation}
This is the orthogonal projector onto the span of the 2RPmix orbit states compatible with $F$ and having ket particle-number vector $\bm n$.
The projector may vanish if no $\bm r\in\mathcal L_{\bm n}$ satisfies $\operatorname{supp}(\bm r)\subseteq\mathcal I_F$.
Although only the ket particle-number vector is specified in Eq.~\eqref{eq:charge_class_projector}, the bra particle-number vector is uniquely determined by the compatible class.
Indeed,
\begin{equation}
\label{eq:compatible_class_charge_relation}
    \overline{\bm n}=\sum_{x\in\mathcal I_F}r_x f_F(x)=Q_F\sum_{x\in\mathcal I_F}r_x x=Q_F\bm n.
\end{equation}
To characterize the allowed compatible classes in the charge sector $(\bm n,\overline{\bm n})$, we define
\begin{equation}
\label{eq:charge_compatible_classes}
    \mathcal F_k(\bm n,\overline{\bm n})
    \equiv
    \left\{
        F\in\mathcal F_k :
        \Pi_{\bm n;F}\neq0,\ 
        Q_F\bm n=\overline{\bm n}
    \right\}.
\end{equation}
By definition, for $F\in\mathcal F_k(\bm n,\overline{\bm n})$, we have $\bm n\in\operatorname{span}_{\mathbb R}\mathcal I_F$ and $\overline{\bm n}=Q_F\bm n$.

For a permutation compatible class $F_\sigma$, we have $\mathcal I_{F_\sigma}=\{0,1\}^k$ and $Q_{F_\sigma}=\sigma$.
Hence the support constraint in Eq.~\eqref{eq:charge_class_projector} is absent, and it is convenient to write
\begin{equation}
    \Pi_{\bm n;\sigma}
    \equiv
    \Pi_{\bm n;F_\sigma}
    =
    \sum_{\bm r\in\mathcal L_{\bm n}}
    \ket{\Omega_{\bm r;F_\sigma}}
    \bra{\Omega_{\bm r;F_\sigma}}.
\end{equation}
Such a compatible class contributes to the charge sector $(\bm n,\overline{\bm n})$ only when $\overline{\bm n}=\sigma\bm n$, or equivalently, $\sigma \in \Sigma (\bm n,\overline{\bm n})$.
Therefore, if $\overline{\bm n}$ is not a permutation of $\bm n$, no permutation compatible class contributes to this charge sector.

For a non-permutation compatible class, by contrast, the additional constraint $\operatorname{supp}(\bm r)\subseteq\mathcal I_F$ is generally nontrivial.
In particular, $\Pi_{\bm n;F}=0$ whenever $\bm n\notin\operatorname{span}_{\mathbb R}\mathcal I_F$.
The converse need not hold: even when $\bm n\in\operatorname{span}_{\mathbb R}\mathcal I_F$, there may be no $\bm r\in\mathcal L_{\bm n}$ whose support is contained in $\mathcal I_F$.

\vspace{1em}

We are now ready to consider the restriction of the 2RPmix moment projector to this charge sector, defined by
\begin{equation}
\label{eq:2rpmix_fixed_charge_projector}
    P^{(k)}_{\mathrm{2RPmix};\bm n,\overline{\bm n}}
    \equiv 
    \Pi_{\bm n,\overline{\bm n}}
    P^{(k)}_{\mathrm{2RPmix}}
    \Pi_{\bm n,\overline{\bm n}}.
\end{equation}
Its fixed space is characterized as follows.

\begin{lemma}[2RPmix fixed space in a fixed charge sector]
\label{lem:compatible_class_charge_sector}
For every charge sector $(\bm n,\overline{\bm n})$,
\begin{equation}
\label{eq:2rpmix_fixed_charge_range}
    \operatorname{Ran}
    P^{(k)}_{\mathrm{2RPmix};\bm n,\overline{\bm n}}
    =
    \operatorname{span}
    \left(
    \bigcup_{F\in\mathcal F_k(\bm n,\overline{\bm n})}
    \operatorname{Ran}\Pi_{\bm n;F}
    \right).
\end{equation}

\end{lemma}
\noindent 
Equation~\eqref{eq:2rpmix_fixed_charge_range} shows that, within a fixed charge sector $(\bm n,\overline{\bm n})$, the 2RPmix fixed space is spanned by the compatible-class subspaces associated with $F\in\mathcal F_k(\bm n,\overline{\bm n})$.
We note that distinct compatible classes can contain the same orbit state, and hence the projectors $\Pi_{\bm n;F}$ are in general not mutually orthogonal.

\begin{proof}
By Eq.~\eqref{eq:compatible_class_charge_relation}, every orbit state contributing to $\Pi_{\bm n;F}$ belongs to the charge sector $(\bm n,Q_F\bm n)$.
Hence, for $F\in\mathcal F_k(\bm n,\overline{\bm n})$, every such orbit state belongs to the charge sector $(\bm n,\overline{\bm n})$.
Moreover, since its type vector is supported on $F$, it is fixed by $P^{(k)}_{\mathrm{2RPmix}}$.
Therefore,
\begin{equation}
    \operatorname{Ran}\Pi_{\bm n;F}
    \subseteq
    \operatorname{Ran}P^{(k)}_{\mathrm{2RPmix};\bm n,\overline{\bm n}}.
\end{equation}

Conversely, consider an orbit state $\ket{\Omega_{\bm m}}$ in the range of $P^{(k)}_{\mathrm{2RPmix};\bm n,\overline{\bm n}}$.
From the characterization of the 2RPmix moment space, there exists $F\in\mathcal F_k$ such that $\operatorname{supp}(\bm m)\subseteq F$.
Writing $F=\{(x,f_F(x)):x\in\mathcal I_F\}$ and setting $r_x=m_{x,f_F(x)}$, we obtain a ket type vector $\bm r\in\mathcal L_{\bm n}$ satisfying $\operatorname{supp}(\bm r)\subseteq\mathcal I_F$.
In particular, $\bm n\in\operatorname{span}_{\mathbb R}\mathcal I_F$, while Eq.~\eqref{eq:compatible_class_charge_relation} gives $Q_F\bm n=\overline{\bm n}$.
Thus $F\in\mathcal F_k(\bm n,\overline{\bm n})$ and $\ket{\Omega_{\bm m}}\in\operatorname{Ran}\Pi_{\bm n;F}$.
This proves Eq.~\eqref{eq:2rpmix_fixed_charge_range}.
\end{proof}

For later use, we further prove the following lemma.

\begin{lemma}
\label{lem:active_nonperm_coordinate}
Let $(\bm n,\overline{\bm n})$ be a charge sector satisfying
$\Sigma(\bm n,\overline{\bm n})=\varnothing$.
Then, for every compatible class
$F\in\mathcal F_k(\bm n,\overline{\bm n})$,
there exists a non-permutation coordinate $j\in[k]$ such that
\begin{equation}
    1\leq n_j\leq N-1.
\end{equation}
\end{lemma}

\begin{proof}
Suppose, toward a contradiction, that every coordinate $j$ satisfying
$1\leq n_j\leq N-1$ is a permutation coordinate of $F$.
Let
\begin{equation*}
    A\equiv\{j\in[k]:1\leq n_j\leq N-1\},
    \qquad
    B\equiv\{j\in[k]:n_j=N\}.
\end{equation*}
For every $j\in A$, there exists $\tau(j)\in[k]$ such that $Q_Fe_j=e_{\tau(j)}$.
Furthermore, denoting $\bm 1_B\equiv\sum_{j\in B}e_j$, we have
\begin{equation}
    \overline{\bm n}
    =
    Q_F\bm n
    =
    \sum_{j\in A}n_je_{\tau(j)}
    +
    NQ_F\bm 1_B.
\end{equation}
By orthogonality, $Q_F\bm 1_B$ is orthogonal to $e_{\tau(j)}$ for every $j\in A$, and hence is supported outside $\tau(A)$.
Moreover, since every component of $\overline{\bm n}$ lies between $0$ and $N$, every component of $Q_F\bm 1_B$ lies between $0$ and $1$.
We also have
\begin{align*}
    \|Q_F\bm 1_B\|^2 &= |B|, \\
    \bm 1^TQ_F\bm 1_B
    &= (Q_F^T\bm 1)^T\bm 1_B
    = \bm 1^T\bm 1_B
    = |B|,
\end{align*}
where the first equality follows from the orthogonality of $Q_F$, while the second follows from $Q_F^T\bm 1=\bm 1$, which in turn follows from $Q_F\bm 1=\bm 1$ and the orthogonality of $Q_F$.
These equalities imply that every component of $Q_F\bm 1_B$ is either $0$ or $1$, with exactly $|B|$ components equal to $1$.

Therefore, the entries of $\overline{\bm n}$ consist of the entries $n_j$ with $j\in A$, together with $|B|$ entries equal to $N$ and the remaining entries equal to $0$.
Thus $\overline{\bm n}$ is a permutation of $\bm n$, implying $\Sigma(\bm n,\overline{\bm n})\neq\varnothing$, a contradiction.
\end{proof}

\subsection{Approximate form of the Haar projector in interior charge sectors}
\label{subsec:u1_haar_interior_approximation}

In this subsection, we derive a simple approximation of the Haar projector for a fixed charge sector by analyzing the Gram matrix of the normalized number-permutation states.
If $\Sigma(\bm n,\overline{\bm n})=\varnothing$, then $P_{U(1)\mathrm{Haar};\bm n,\overline{\bm n}}^{(k)}=0$, and hence there is nothing to approximate. We therefore assume $\Sigma(\bm n,\overline{\bm n})\neq\varnothing$ below.

Here, we only consider interior charge sectors satisfying $1\leq n_\alpha\leq N-1$ for every $\alpha\in[k]$.
If some $n_\alpha$ is equal to $0$ or $N$, the corresponding particle-number subspace is one-dimensional.
Since $\Sigma(\bm n,\overline{\bm n})\neq\varnothing$, the same trivial factors appear on the bra side, and removing these factors reduces the Haar-projector analysis to a lower-order moment on the active copies.
Thus, it is sufficient to derive the estimates below for interior sectors; the corresponding bounds for boundary sectors are obtained by applying them to the reduced active sector. We further assume $k\leq N$, since it is sufficient for our purpose.

As explained in Sec.~\ref{subsec:u1_haar_moment}, the $U(1)$-symmetric Haar moment space in the charge sector $(\bm n,\overline{\bm n})$ is
\begin{equation}
    \operatorname{Ran}
    P_{U(1)\mathrm{Haar};\bm n,\overline{\bm n}}^{(k)}
    =
    \operatorname{span}
    \left\{
        \ket{\bm n;\sigma}:
        \sigma\in\Sigma(\bm n,\overline{\bm n})
    \right\}.
    \label{eq:u1_haar_fixed_space}
\end{equation}
To evaluate the overlaps among these states, let $q_1,\ldots,q_p$ be the distinct values appearing among $n_1,\ldots,n_k$, and define
\begin{equation}
    J_l
    \equiv
    \{\alpha\in[k]:n_\alpha=q_l\},
    \qquad
    \alpha_l
    \equiv
    |J_l|,
    \qquad
    \sum_{l=1}^{p}\alpha_l=k.
    \label{eq:particle_multiplicities}
\end{equation}
Since $\Sigma(\bm n,\overline{\bm n})\neq\varnothing$, the same values $q_l$ appear in $\overline{\bm n}$ with the same multiplicities $\alpha_l$.
Choose and fix permutations $\eta,\overline{\eta}\in S_k$ such that
\begin{equation}
    \eta\bm n
    =
    \overline{\eta}\,\overline{\bm n}
    =
    \bigl(
        \underbrace{q_1,\ldots,q_1}_{\alpha_1},
        \underbrace{q_2,\ldots,q_2}_{\alpha_2},
        \ldots,
        \underbrace{q_p,\ldots,q_p}_{\alpha_p}
    \bigr)^T.
    \label{eq:particle_number_ordering}
\end{equation}
For notational simplicity, we suppress the dependence of $q_l,p,J_l,\alpha_l,\eta$, and $\overline{\eta}$ on $(\bm n,\overline{\bm n})$.
With this ordering, every $\sigma\in\Sigma(\bm n,\overline{\bm n})$ can be written uniquely as
\begin{equation}
    \sigma
    =
    \overline{\eta}^{-1}\tau\eta,
    \qquad
    \tau
    \in
    S_{\alpha_1}\times\cdots\times S_{\alpha_p},
    \label{eq:sigma_block_decomposition}
\end{equation}
where we write $\tau=(\tau_1,\ldots,\tau_p)$ with $\tau_l\in S_{\alpha_l}$.

For $\sigma=\overline{\eta}^{-1}\tau\eta$ and $\sigma'=\overline{\eta}^{-1}\tau'\eta$, the normalization of the number-permutation states gives
\begin{equation}
    \langle
        \bm n;\sigma
        |
        \bm n;\sigma'
    \rangle
    =
    \prod_{l=1}^{p}
    D_{q_l}^{
        \#_{\rm cyc}(\tau_l^{-1}\tau_l')-\alpha_l
    }
    =
    \prod_{l=1}^{p}
    G^{(q_l,\alpha_l)}_{\tau_l,\tau_l'},
    \label{eq:u1_haar_gram_factorization}
\end{equation}
where $\#_{\rm cyc}(\tau)$ denotes the number of cycles of $\tau$, including one-cycles, and
\begin{equation}
    G^{(q_l,\alpha_l)}_{\tau_l,\tau_l'}
    \equiv
    D_{q_l}^{
        \#_{\rm cyc}(\tau_l^{-1}\tau_l')-\alpha_l
    }.
    \label{eq:sector_gram_matrix}
\end{equation}
Thus, under the identification in Eq.~\eqref{eq:sigma_block_decomposition}, the Gram matrix
\begin{equation}
    (G_{\bm n,\overline{\bm n}})_{\sigma,\sigma'}
    \equiv
    \langle\bm n;\sigma|\bm n;\sigma'\rangle
    \label{eq:u1_haar_gram}
\end{equation}
factorizes into the permutation Gram matrices $G^{(q_l,\alpha_l)}$.

For an interior charge $q_l$, we have $D_{q_l}\geq N$, while $\alpha_l\leq k\leq N$.
The standard permutation Gram matrix on $\alpha_l$ copies of a $D_{q_l}$-dimensional space is therefore nonsingular.
Hence the number-permutation states in Eq.~\eqref{eq:u1_haar_fixed_space} are linearly independent, and the Haar projector can be written as
\begin{equation}
    P_{U(1)\mathrm{Haar};\bm n,\overline{\bm n}}^{(k)}
    =
    \sum_{\sigma,\sigma'\in\Sigma(\bm n,\overline{\bm n})}
    (G_{\bm n,\overline{\bm n}}^{-1})_{\sigma,\sigma'}
    \ket{\bm n;\sigma}
    \bra{\bm n;\sigma'}.
    \label{eq:u1_haar_weingarten_general}
\end{equation}
This inverse-Gram representation is the usual Weingarten representation, written here for the normalized number-permutation states.

Using Eq.~\eqref{eq:u1_haar_gram_factorization}, we obtain the following approximate-orthogonality bound.
Similar estimates for systems without symmetry were obtained in Refs.~\cite{harrow2023approximate,schuster2024random}.

\begin{lemma}[Approximate orthogonality of number-permutation states]
\label{lem:u1_haar_approximate_orthogonality}
Let $\bm n,\overline{\bm n}\in\{1,\ldots,N-1\}^k$ satisfy $\Sigma(\bm n,\overline{\bm n})\neq\varnothing$, and assume $k\leq N$.
Define
\begin{equation}
    \widetilde P_{\bm n,\overline{\bm n}}^{(k)}
    \equiv
    \sum_{\sigma\in\Sigma(\bm n,\overline{\bm n})}
    \ket{\bm n;\sigma}
    \bra{\bm n;\sigma}.
    \label{eq:leading_haar_projector}
\end{equation}
Then, we have
\begin{equation}
    \left\|
        P_{U(1)\mathrm{Haar};\bm n,\overline{\bm n}}^{(k)}
        -
        \widetilde P_{\bm n,\overline{\bm n}}^{(k)}
    \right\|_\infty
    \leq
    \exp\left(\frac{k(k-1)}{2N}\right)-1.
    \label{eq:u1_haar_small_error}
\end{equation}
\end{lemma}

\begin{proof}
Let
$\mathcal V
\equiv
\operatorname{span}
\{\ket{\bm n;\sigma}:\sigma\in\Sigma(\bm n,\overline{\bm n})\}$.
Both $P_{U(1)\mathrm{Haar};\bm n,\overline{\bm n}}^{(k)}$ and $\widetilde P_{\bm n,\overline{\bm n}}^{(k)}$ vanish on $\mathcal V^\perp$, while the former acts as the identity on $\mathcal V$.

For $\bm u=(u_\sigma)_{\sigma\in\Sigma(\bm n,\overline{\bm n})}$, write
\begin{equation}
    \ket{\Psi_{\bm u}}
    \equiv
    \sum_{\sigma\in\Sigma(\bm n,\overline{\bm n})}
    u_\sigma
    \ket{\bm n;\sigma}.
\end{equation}
Then, we have
\begin{align}
    \widetilde P_{\bm n,\overline{\bm n}}^{(k)}
    \ket{\Psi_{\bm u}}
    &=
    \sum_{\sigma,\sigma'\in\Sigma(\bm n,\overline{\bm n})}
    u_{\sigma'}
    \ket{\bm n;\sigma}
    \langle\bm n;\sigma|\bm n;\sigma'\rangle
    \nonumber\\
    &=
    \sum_{\sigma\in\Sigma(\bm n,\overline{\bm n})}
    (G_{\bm n,\overline{\bm n}}\bm u)_\sigma
    \ket{\bm n;\sigma}
    =
    \ket{\Psi_{G_{\bm n,\overline{\bm n}}\bm u}}.
    \label{eq:gram_action}
\end{align}
Since the number-permutation states are linearly independent, the map $\bm u\mapsto\ket{\Psi_{\bm u}}$ is an isomorphism onto $\mathcal V$.
Equation~\eqref{eq:gram_action} therefore shows that
$\widetilde P_{\bm n,\overline{\bm n}}^{(k)}$ restricted to $\mathcal V$
is similar to $G_{\bm n,\overline{\bm n}}$.
Since $P_{U(1)\mathrm{Haar};\bm n,\overline{\bm n}}^{(k)}$ acts as the identity on $\mathcal V$, while both operators vanish on $\mathcal V^\perp$, we obtain
\begin{equation}
    \left\|
        P_{U(1)\mathrm{Haar};\bm n,\overline{\bm n}}^{(k)}
        -
        \widetilde P_{\bm n,\overline{\bm n}}^{(k)}
    \right\|_{\infty}
    =
    \rho\left(
        P_{U(1)\mathrm{Haar};\bm n,\overline{\bm n}}^{(k)}
        -
        \widetilde P_{\bm n,\overline{\bm n}}^{(k)}
    \right)
    =
    \rho\left(
        I-G_{\bm n,\overline{\bm n}}
    \right)
    =
    \left\|
        I-G_{\bm n,\overline{\bm n}}
    \right\|_{\infty},
    \label{eq:haar_error_gram}
\end{equation}
where $\rho(A)$ denotes the spectral radius of $A$.
Here, the first and last equalities follow from the Hermiticity of the corresponding operators.

For any fixed $\sigma=\overline{\eta}^{-1}\tau\eta$, Eq.~\eqref{eq:u1_haar_gram_factorization} gives
\begin{align}
    \sum_{\sigma'\in\Sigma(\bm n,\overline{\bm n})}
    (G_{\bm n,\overline{\bm n}})_{\sigma,\sigma'}
    &=
    \prod_{l=1}^{p}
    D_{q_l}^{-\alpha_l}
    \sum_{\tau_l'\in S_{\alpha_l}}
    D_{q_l}^{\#_{\rm cyc}(\tau_l^{-1}\tau_l')}
    \nonumber\\
    &=
    \prod_{l=1}^{p}
    \prod_{j=1}^{\alpha_l-1}
    \left(
        1+\frac{j}{D_{q_l}}
    \right),
    \label{eq:normalized_gram_row_sum}
\end{align}
where in the second line we used the standard cycle-counting identity
\begin{equation}
    \sum_{\tau\in S_\alpha }
    D^{\#_{\rm cyc}(\tau)}
    =
    D(D+1)\cdots(D+\alpha-1).
\end{equation}

The matrix $G_{\bm n,\overline{\bm n}}-I$ is real symmetric, has nonnegative entries, and has the same row sum for every row.
Therefore, by using Gershgorin's theorem, we obtain
\begin{align}
    \left\|
        G_{\bm n,\overline{\bm n}}-I
    \right\|_\infty
    &\leq
    \prod_{l=1}^{p}
    \prod_{j=1}^{\alpha_l-1}
    \left(
        1+\frac{j}{D_{q_l}}
    \right)-1
    \nonumber\\
    &\leq
    \exp\left[
        \sum_{l=1}^{p}
        \sum_{j=1}^{\alpha_l-1}
        \frac{j}{D_{q_l}}
    \right]-1
    \nonumber\\
    &=
    \exp\left[
        \frac{1}{2}
        \sum_{l=1}^{p}
        \frac{\alpha_l(\alpha_l-1)}
             {D_{q_l}}
    \right]-1
    \nonumber\\
    &\leq
    \exp\left(\frac{k(k-1)}{2N}\right)-1,
\end{align}
where the final inequality follows from $D_{q_l}\geq N$ and
$\sum_{l=1}^{p}\alpha_l(\alpha_l-1)\leq k(k-1)$.
\end{proof}

By using this lemma, we can also derive the corresponding approximation after restricting to a permutation compatible class:
\begin{lemma}
\label{lem:approx_Haar_sector_sigma_n}
Under the assumptions of Lemma~\ref{lem:u1_haar_approximate_orthogonality}, for any $\sigma\in\Sigma(\bm n,\overline{\bm n})$,
\begin{equation}
    \left\|
        \Pi_{\bm n;\sigma}
        \left(
            P_{U(1)\mathrm{Haar};\bm n,\overline{\bm n}}^{(k)}
            -
            \ket{\bm n;\sigma}
            \bra{\bm n;\sigma}
        \right)
        \Pi_{\bm n;\sigma}
    \right\|_\infty
    \leq
    2\left[
        \exp\left(\frac{k(k-1)}{2N}\right)-1
    \right].
    \label{eq:u1_haar_fixed_class_error}
\end{equation}
\end{lemma}

\begin{proof}
By Eq.~\eqref{eq:number_permutation_orbit_expansion}, $\Pi_{\bm n;\sigma}$ selects from $\ket{\bm n;\tau}$ precisely the computational-basis configurations that are also compatible with $F_\sigma$.
Equivalently, these are the tuples $(x^{1},\ldots,x^{k})$ for which $\tau x_i=\sigma x_i$ at every site $i$.
Since every computational-basis configuration in a number-permutation state has the same positive amplitude, the squared norm of the projected state is equal to the overlap of the two normalized number-permutation states:
\begin{equation}
    \left\|
        \Pi_{\bm n;\sigma}
        \ket{\bm n;\tau}
    \right\|^2
    =
    \langle\bm n;\sigma|\bm n;\tau\rangle.
    \label{eq:compatible_projection_gram_overlap}
\end{equation}
Therefore,
\begin{align}
    &\left\|
        \Pi_{\bm n;\sigma}
        \left(
            P_{U(1)\mathrm{Haar};\bm n,\overline{\bm n}}^{(k)}
            -
            \ket{\bm n;\sigma}
            \bra{\bm n;\sigma}
        \right)
        \Pi_{\bm n;\sigma}
    \right\|_\infty \nonumber \\
    &\qquad \leq
    \left\|
        \Pi_{\bm n;\sigma}
        \left(
            P_{U(1)\mathrm{Haar};\bm n,\overline{\bm n}}^{(k)}
            -
            \widetilde P_{\bm n,\overline{\bm n}}^{(k)}
        \right)
        \Pi_{\bm n;\sigma}
    \right\|_\infty
    +
    \sum_{\substack{
        \tau\in\Sigma(\bm n,\overline{\bm n})\\
        \tau\neq\sigma
    }}
    \left\|
        \Pi_{\bm n;\sigma}
        \ket{\bm n;\tau}
        \bra{\bm n;\tau}
        \Pi_{\bm n;\sigma}
    \right\|_\infty
    \nonumber\\
    &\qquad \leq
    \left\|
        P_{U(1)\mathrm{Haar};\bm n,\overline{\bm n}}^{(k)}
        -
        \widetilde P_{\bm n,\overline{\bm n}}^{(k)}
    \right\|_\infty
    +
    \sum_{\substack{
        \tau\in\Sigma(\bm n,\overline{\bm n})\\
        \tau\neq\sigma
    }}
    \langle\bm n;\sigma|\bm n;\tau\rangle
    \nonumber\\
    &\qquad \leq
    2\left[
        \exp\left(\frac{k(k-1)}{2N}\right)-1
    \right],
    \label{eq:fixed_class_error_decomposition}
\end{align}
where the first inequality follows from the triangle inequality, and the final inequality follows from Lemma~\ref{lem:u1_haar_approximate_orthogonality}, Eq.~\eqref{eq:normalized_gram_row_sum}, and $ (G_{\bm n,\overline{\bm n}})_{\sigma,\tau} \equiv \langle\bm n;\sigma|\bm n;\tau\rangle$.
\end{proof}

For later use, we also show that a number-permutation state has only a small overlap with the subspace associated with any distinct compatible class.

\begin{lemma}
\label{lem:distinct_class_overlap_number_permutation}
Let $\bm n,\overline{\bm n}\in\{1,\ldots,N-1\}^k$ satisfy $\Sigma(\bm n,\overline{\bm n})\neq\varnothing$.
Then, for any $\sigma\in\Sigma(\bm n,\overline{\bm n})$ and any $F\in\mathcal F_k(\bm n,\overline{\bm n})$ with $F\neq F_\sigma$, we have 
\begin{equation}
    \left\|
        \Pi_{\bm n;F}
        \ket{\bm n;\sigma}
    \right\|^2
    \leq
    \frac{1}{N}.
    \label{eq:nonperm_number_perm_overlap}
\end{equation}
\end{lemma}

\begin{proof}
We first consider $F\in\mathcal F_k^{\rm perm}$.
Since $F\in\mathcal F_k(\bm n,\overline{\bm n})$, we can write $F=F_\tau$ for some $\tau\in\Sigma(\bm n,\overline{\bm n})$.
Since $F\neq F_\sigma$, we have $\tau\neq\sigma$.
By Eq.~\eqref{eq:compatible_projection_gram_overlap},
\begin{equation*}
    \left\|
        \Pi_{\bm n;\tau}
        \ket{\bm n;\sigma}
    \right\|^2
    =
    \langle\bm n;\tau|\bm n;\sigma\rangle.
\end{equation*}
Since $\tau\neq\sigma$, their relative permutation is nontrivial in at least one of the equal-charge blocks appearing in Eq.~\eqref{eq:u1_haar_gram_factorization}.
The number of cycles in that block is therefore smaller than its size by at least one.
Hence,
\begin{equation*}
    \langle\bm n;\tau|\bm n;\sigma\rangle
    \leq
    \frac{1}{D_{q_l}}
    \leq
    \frac{1}{N}
\end{equation*}
for some interior charge $q_l$.
 
We next consider $F\in\mathcal F_k(\bm n,\overline{\bm n})\setminus\mathcal F_k^{\rm perm}$, and let $j\in[k]$ be a non-permutation coordinate of $F$.
Recall that
\begin{equation*}
    \ket{\bm n;\sigma}
    =
    \frac{1}{\sqrt{Z_{\bm n}}}
    \sum_{\substack{
        x^{1},\ldots,x^{k}\in\{0,1\}^N\\
        |x^{l}|=n_l\ \forall l\in[k]
    }}
    \ket{\bm x,\sigma \bm x}.
\end{equation*}
Therefore, $\Pi_{\bm n;F}$ keeps precisely those terms for which the local type $(x_i,\sigma x_i)$ belongs to $F$ at every site $i\in[N]$, where $x_i=(x_i^1,\ldots,x_i^k)^T$.
Since all terms in $\ket{\bm n;\sigma}$ have the same amplitude,
$\|\Pi_{\bm n;F}\ket{\bm n;\sigma}\|^2$ is the fraction of such tuples satisfying this condition.
In particular, since $F=\{(x,f_F(x)):x\in\mathcal I_F\}$, every surviving tuple satisfies $x_i\in\mathcal I_F$ for all $i\in[N]$.

By Lemma~\ref{lem:nonpermutation_compatible_class}$\mathrm{(ii)}$, no two elements of $\mathcal I_F$ can differ only in their $j$-th coordinate.
Hence, once the $k-1$ strings $x^{l}$ with $l\neq j$ are fixed, the bit $x_i^j$ is uniquely determined at every site $i$ whenever an admissible choice exists.
Therefore, for each fixed choice of the other $k-1$ strings, there is at most one admissible string $x^{j}$.
There are $\prod_{l\neq j}D_{n_l}$ possible choices of the other strings, whereas the total number of tuples in $\ket{\bm n;\sigma}$ is $Z_{\bm n}=\prod_{l=1}^{k}D_{n_l}$.
Thus,
\begin{equation}
    \left\|
        \Pi_{\bm n;F}
        \ket{\bm n;\sigma}
    \right\|^2
    \leq
    \frac{\prod_{l\neq j}D_{n_l}}
         {\prod_{l=1}^{k}D_{n_l}}
    =
    \frac{1}{D_{n_j}}.
\end{equation}
Finally, since $1\leq n_j\leq N-1$, we have $D_{n_j}=\binom{N}{n_j}\geq N$, which proves the claim.
\end{proof}

\subsection{Bound on diagonal blocks of moment operator} \label{ss:bound_diagonal_block_mom}

In this subsection, we analyze the diagonal blocks of the moment operator $M_{\rm dope}^{(k)}$.
Specifically, we derive the following upper bounds on their operator norms.

\begin{lemma}
\label{lem:bound_each_block}
For a fixed charge sector $(\bm n,\overline{\bm n})$, we have the following bounds:
\begin{enumerate}
    \item[$\mathrm{(i)}$]
    For every $\sigma\in\Sigma(\bm n,\overline{\bm n})$,
    \begin{equation}
        \left\|
        \Pi_{\bm n;\sigma}
        \left(
        M_{\rm dope}^{(k)}
        -
        \ket{\bm n;\sigma}\bra{\bm n;\sigma}
        \right)
        \Pi_{\bm n;\sigma}
        \right\|_\infty
        \leq
        \frac{1}{2}.
    \end{equation}

    \item[$\mathrm{(ii)}$]
    For every compatible class $F\in\mathcal F_k(\bm n,\overline{\bm n})$ having a non-permutation coordinate $j\in[k]$ such that $1\leq n_j\leq N-1$,
    \begin{equation}
        \left\|
        \Pi_{\bm n;F}M_{\rm dope}^{(k)}\Pi_{\bm n;F}
        \right\|_\infty
        \leq
        \frac{1}{2}+\frac{1}{2(N-1)}.
    \end{equation}
\end{enumerate}
\end{lemma}

\begin{proof}
We first establish the relation between the matrix elements of $M_{\rm dope}^{(k)}$ within a compatible class and the transition matrix $W_{\bm n}^{\Phi}$ of the classical dynamics introduced in Sec.~\ref{subsec:classical_dynamics}.

As explained in Eq.~\eqref{eq:full_type_vector_from_ket_type_vector_fixed_F}, once we fix a compatible class $F\in\mathcal F_k(\bm n,\overline{\bm n})$, the orbit state $\ket{\Omega_{\bm m}}$ is completely specified by the ket type vector and can be denoted by $\ket{\Omega_{\bm r;F}}$.
Then, for $\bm r,\bm r'\in\mathcal L_{\bm n}$ satisfying $\operatorname{supp}(\bm r)\subseteq\mathcal I_F$ and $\operatorname{supp}(\bm r')\subseteq\mathcal I_F$, we first consider even $N$ and obtain
\begin{align}
    \bra{\Omega_{\bm r;F}}M_{\rm dope}^{(k)}\ket{\Omega_{\bm r';F}}
    &=
    \frac{1}{\sqrt{D_{\bm r}D_{\bm r'}}}
    \sum_{\bm x\in\mathcal T_{\bm r}}
    \sum_{\bm y\in\mathcal T_{\bm r'}}
    \left(\bigotimes_{i=1}^{N}\bra{x_i,Q_Fx_i}\right)
    \bigotimes_{j=1}^{N/2}V_{2j-1,2j}^{\otimes k,k}
    \left(\bigotimes_{i=1}^{N}\ket{y_i,Q_Fy_i}\right)
    \nonumber\\
    &=
    \frac{1}{\sqrt{D_{\bm r}D_{\bm r'}}}
    \sum_{\bm x\in\mathcal T_{\bm r}}
    \sum_{\bm y\in\mathcal T_{\bm r'}}
    \prod_{j=1}^{N/2}
    \bra{x_{2j-1}}\bra{x_{2j}}V_{2j-1,2j}^{\otimes k}\ket{y_{2j-1}}\ket{y_{2j}}
    \nonumber\\
    &\hspace{5em}\times
    \bra{Q_Fx_{2j-1}}\bra{Q_Fx_{2j}}V_{2j-1,2j}^{\otimes k}\ket{Q_Fy_{2j-1}}\ket{Q_Fy_{2j}}
    \nonumber\\
    &=
    \frac{1}{\sqrt{D_{\bm r}D_{\bm r'}}}
    \sum_{\bm x\in\mathcal T_{\bm r}}
    \sum_{\bm y\in\mathcal T_{\bm r'}}
    \prod_{j=1}^{N/2}
    \left|
    \bra{x_{2j-1}}\bra{x_{2j}}V_{2j-1,2j}^{\otimes k}\ket{y_{2j-1}}\ket{y_{2j}}
    \right|^2
    \nonumber\\
    &=
    \frac{1}{\sqrt{D_{\bm r}D_{\bm r'}}}
    \sum_{\bm x\in\mathcal T_{\bm r}}
    \sum_{\bm y\in\mathcal T_{\bm r'}}
    \prod_{j=1}^{N/2}\prod_{l=1}^{k}
    w_{\rm 2bit}(x_{2j-1}^{l},x_{2j}^{l};y_{2j-1}^{l},y_{2j}^{l})
    \nonumber\\
    &=
    \frac{1}{\sqrt{D_{\bm r}D_{\bm r'}}}
    \sum_{\bm x\in\mathcal T_{\bm r}}
    \sum_{\bm y\in\mathcal T_{\bm r'}}
    \prod_{l=1}^{k}
    w_{\gamma_0,n_l}^{l}(x^{l};y^{l})
    \nonumber\\
    &=
    \frac{1}{\sqrt{D_{\bm r}D_{\bm r'}}}
    \sum_{\bm x\in\mathcal T_{\bm r}}
    \sum_{\bm y\in\mathcal T_{\bm r'}}
    W_{\gamma_0,\bm n}(\bm x,\bm y)
    \nonumber\\
    &=
    \sqrt{\frac{D_{\bm r'}}{D_{\bm r}}}\,
    W_{\bm n}^{\Phi}(\bm r,\bm r').
    \label{eq:Mdope_Wcl_similarity}
\end{align}
Here, $x_i=(x_i^1,\ldots,x_i^k)^T$ and $y_i=(y_i^1,\ldots,y_i^k)^T$ denote the ket local types at site $i\in[N]$, while $w_{\rm 2bit}$, $w_{\gamma_0,n_l}^{l}$, $W_{\gamma_0,\bm n}$, and $W_{\bm n}^{\Phi}$ are the transition probabilities introduced in Sec.~\ref{subsec:classical_dynamics}.
In deriving the third line of Eq.~\eqref{eq:Mdope_Wcl_similarity}, we used the invariance of the two-site matrix element under $Q_F$.
From the definition of the doped gate, Eq.~\eqref{eq:Hadamard_type_local_unitary_def}, the matrix element for $x,x',y,y'\in\mathcal I_F$ is given by 
\begin{equation}
\label{eq:Hadamard_trans_prob}
    \bra{x}\bra{x'}V_{2j-1,2j}^{\otimes k}\ket{y}\ket{y'}
    =
    \delta_{x+x',y+y'}\,
    2^{-\lvert y-y'\rvert/2}
    (-1)^{(x'+y)\cdot y'}.
\end{equation}
Since $Q_F$ is orthogonal and $Q_Fx,Q_Fx',Q_Fy,Q_Fy'\in\{0,1\}^k$, we have
\begin{gather*}
    \delta_{Q_Fx+Q_Fx',Q_Fy+Q_Fy'} = \delta_{x+x',y+y'},\\
    \lvert Q_Fy-Q_Fy'\rvert = \lvert y-y'\rvert,\\
    (Q_Fx'+Q_Fy)\cdot Q_Fy' = (x'+y)\cdot y'.
\end{gather*}
Here, the second equality follows because orthogonality preserves squared Euclidean distance, which equals Hamming distance for binary vectors.
Therefore,
\begin{equation}
    \bra{Q_Fx}\bra{Q_Fx'}V_{2j-1,2j}^{\otimes k}\ket{Q_Fy}\ket{Q_Fy'}
    =
    \bra{x}\bra{x'}V_{2j-1,2j}^{\otimes k}\ket{y}\ket{y'}.
\end{equation}

For odd $N$, the unmatched site contributes an additional factor $\prod_{l=1}^{k}\delta_{x_N^l,y_N^l}$ to Eq.~\eqref{eq:Mdope_Wcl_similarity}, in accordance with the definition of $W_{\gamma_0,\bm n}$.
Thus, the final relation in Eq.~\eqref{eq:Mdope_Wcl_similarity} holds for odd $N$ as well.

\vspace{1em}

We now derive $\mathrm{(i)}$ and $\mathrm{(ii)}$ separately using Eq.~\eqref{eq:Mdope_Wcl_similarity}.

\noindent
\emph{$\mathrm{(i)}$}
For $\sigma\in\Sigma(\bm n,\overline{\bm n})$, any ket type vector $\bm r\in\mathcal L_{\bm n}$ is allowed because $\mathcal I_{F_\sigma}=\{0,1\}^k$.
Defining the diagonal matrix by $\mathcal R_{\bm r,\bm r}\equiv D_{\bm r}^{-1/2}$, we therefore have
\begin{equation}
    \Pi_{\bm n;\sigma}M_{\rm dope}^{(k)}\Pi_{\bm n;\sigma}
    =
    \mathcal R W_{\bm n}^{\Phi}\mathcal R^{-1},
\end{equation}
where the orbit-state basis $\{\ket{\Omega_{\bm r;F_\sigma}}\}_{\bm r\in\mathcal L_{\bm n}}$ corresponds to the canonical basis $\{\ket{\bm r}\}_{\bm r\in\mathcal L_{\bm n}}$ of the coarse-grained dynamics.

Under this basis correspondence, Eq.~\eqref{eq:number_permutation_orbit_expansion} and Lemma~\ref{lem:classical_coarse_process}$\mathrm{(i)}$ give
\begin{equation}
    \ket{\bm n;\sigma}
    =
    \sum_{\bm r\in\mathcal L_{\bm n}}
    \sqrt{\frac{D_{\bm r}}{Z_{\bm n}}}\,
    \ket{\Omega_{\bm r;F_\sigma}}
    =
    \sqrt{Z_{\bm n}}\,
    \mathcal R\ket{v_{\bm n,\Phi}^{\rm ss}}.
\end{equation}
We also have
\begin{equation}
    \ket{\bm n;\sigma}
    =
    \frac{1}{\sqrt{Z_{\bm n}}}\,
    \mathcal R^{-1}\ket{\bm 1},
\end{equation}
where $\ket{\bm 1}\equiv\sum_{\bm r\in\mathcal L_{\bm n}}\ket{\bm r}$.
Combining these two expressions, we obtain
\begin{equation}
    \ket{\bm n;\sigma}\bra{\bm n;\sigma}
    =
    \mathcal R
    \ket{v_{\bm n,\Phi}^{\rm ss}}
    \bra{\bm 1}
    \mathcal R^{-1}.
\end{equation}
Hence, the matrix representation of $\Pi_{\bm n;\sigma}(M_{\rm dope}^{(k)}-\ket{\bm n;\sigma}\bra{\bm n;\sigma})\Pi_{\bm n;\sigma}$ is identical to that of $\mathcal R(W_{\bm n}^{\Phi}-\ket{v_{\bm n,\Phi}^{\rm ss}}\bra{\bm 1})\mathcal R^{-1}$, and therefore
\begin{align}
    \left\|
    \Pi_{\bm n;\sigma}
    \left(
    M_{\rm dope}^{(k)}
    -
    \ket{\bm n;\sigma}\bra{\bm n;\sigma}
    \right)
    \Pi_{\bm n;\sigma}
    \right\|_\infty
    &=
    \rho\left(
    \Pi_{\bm n;\sigma}
    \left(
    M_{\rm dope}^{(k)}
    -
    \ket{\bm n;\sigma}\bra{\bm n;\sigma}
    \right)
    \Pi_{\bm n;\sigma}
    \right)
    \nonumber\\
    &=
    \rho\left(
    \mathcal R
    \left(
    W_{\bm n}^{\Phi}
    -
    \ket{v_{\bm n,\Phi}^{\rm ss}}\bra{\bm 1}
    \right)
    \mathcal R^{-1}
    \right)
    \nonumber\\
    &=
    \rho\left(
    W_{\bm n}^{\Phi}
    -
    \ket{v_{\bm n,\Phi}^{\rm ss}}\bra{\bm 1}
    \right)
    \nonumber\\
    &\leq
    \frac{1}{2}.
\end{align}
Here, the first equality follows from the Hermiticity of the operator, the third equality follows because a similarity transformation preserves the spectral radius, and the final inequality follows from Lemma~\ref{lem:classical_coarse_process}$\mathrm{(ii)}$.

\vspace{1em}

\noindent
\emph{$\mathrm{(ii)}$}
Let $F\in\mathcal F_k(\bm n,\overline{\bm n})$ be a compatible class having a non-permutation coordinate $j\in[k]$ satisfying $1\leq n_j\leq N-1$.
The retained ket type vectors form the subset
\begin{equation}
    \mathcal L_{\bm n}(F)
    \equiv
    \left\{
    \bm r\in\mathcal L_{\bm n}:
    \operatorname{supp}(\bm r)\subseteq\mathcal I_F
    \right\}.
\end{equation}
Define the corresponding projector in the classical ket type-vector space by
\begin{equation}
    \Pi_{\mathcal L_{\bm n}(F)}
    \equiv
    \sum_{\bm r\in\mathcal L_{\bm n}(F)}
    \ket{\bm r}\bra{\bm r}.
\end{equation}
Equation~\eqref{eq:Mdope_Wcl_similarity} then gives
\begin{equation}
\label{eq:non_perm_diag_correspondence}
    \Pi_{\bm n;F}M_{\rm dope}^{(k)}\Pi_{\bm n;F}
    =
    \mathcal R
    \left(
    \Pi_{\mathcal L_{\bm n}(F)}
    W_{\bm n}^{\Phi}
    \Pi_{\mathcal L_{\bm n}(F)}
    \right)
    \mathcal R^{-1},
\end{equation}
where the orbit-state basis $\{\ket{\Omega_{\bm r;F}}\}_{\bm r\in\mathcal L_{\bm n}(F)}$ corresponds to the canonical basis $\{\ket{\bm r}\}_{\bm r\in\mathcal L_{\bm n}(F)}$.
The matrix $\Pi_{\mathcal L_{\bm n}(F)}W_{\bm n}^{\Phi}\Pi_{\mathcal L_{\bm n}(F)}$ is substochastic, since all of its entries are nonnegative and each of its column sums is at most one.

From Eq.~\eqref{eq:non_perm_diag_correspondence}, we further obtain
\begin{align}
    \left\|
    \Pi_{\bm n;F}M_{\rm dope}^{(k)}\Pi_{\bm n;F}
    \right\|_\infty
    &=
    \rho\left(
    \Pi_{\bm n;F}M_{\rm dope}^{(k)}\Pi_{\bm n;F}
    \right)
    \nonumber\\
    &=
    \rho\left(
    \mathcal R
    \left(
    \Pi_{\mathcal L_{\bm n}(F)}
    W_{\bm n}^{\Phi}
    \Pi_{\mathcal L_{\bm n}(F)}
    \right)
    \mathcal R^{-1}
    \right)
    \nonumber\\
    &=
    \rho\left(
    \Pi_{\mathcal L_{\bm n}(F)}
    W_{\bm n}^{\Phi}
    \Pi_{\mathcal L_{\bm n}(F)}
    \right)
    \nonumber\\
    &\leq
    \max_{\bm r\in\mathcal L_{\bm n}(F)}
    \sum_{\bm r'\in\mathcal L_{\bm n}(F)}
    W_{\bm n}^{\Phi}(\bm r',\bm r).
    \label{eq:killed_chain_bound}
\end{align}
Here, the first equality follows from the Hermiticity of the operator, while in the last line we used $\rho(A)\leq\max_j\sum_i|A_{ij}|$.
The right-hand side of Eq.~\eqref{eq:killed_chain_bound} is the one-step survival probability of the substochastic process killed whenever the support of the ket type vector leaves $\mathcal I_F$.

We now bound this survival probability.
Fix a non-permutation coordinate $j\in[k]$ satisfying $1\leq n_j\leq N-1$, and suppose that a maximum matching contains $a$ pairs whose $j$-th bits are different.
We first show that, conditioned on this matching, the survival probability is at most $2^{-a}$.

Consider one such matched pair, whose two input ket local types $z,w\in\mathcal I_F$ satisfy $z^j\neq w^j$, and fix the output choices for all copies other than $j$.
For the $j$-th copy, the local update has two equiprobable possibilities, corresponding to keeping or exchanging the two different bits.
The two resulting local types at either endpoint differ only in their $j$-th coordinate.
Since $j$ is a non-permutation coordinate, Lemma~\ref{lem:nonpermutation_compatible_class}$\mathrm{(ii)}$ implies that no two elements of $\mathcal I_F$ can differ only in their $j$-th coordinate.
Hence, at most one of the two possibilities can have both output local types in $\mathcal I_F$.
Since the updates on distinct matched pairs are independent, the conditional survival probability is at most $2^{-a}$.

It remains to average this bound over the matching.
For every configuration in the particle-number sector $\bm n$, there are $n_j$ sites whose $j$-th bit is $1$ and $N-n_j$ sites whose $j$-th bit is $0$.
Moreover, the average over the permutation orbit with the fixed matching $\gamma_0$ is equivalent to a uniformly random maximum matching of these $N$ sites.
Denoting by $\operatorname{Pr}_{N,n_j}(a)$ the probability that such a matching contains exactly $a$ pairs connecting different $j$-th bits, Lemma~\ref{lem:crossing_pairs_maximum_matching}, which is proved later, gives
\begin{align}
    \sum_{\bm r'\in\mathcal L_{\bm n}(F)}
    W_{\bm n}^{\Phi}(\bm r',\bm r)
    &\leq
    \sum_{a}
    \operatorname{Pr}_{N,n_j}(a)\,2^{-a}
    \nonumber\\
    &\leq
    \frac{1}{2}
    +
    \frac{1}{2(N-1)}.
    \label{eq:survival_crossing_bound}
\end{align}
Combining Eqs.~\eqref{eq:killed_chain_bound} and \eqref{eq:survival_crossing_bound} proves $\mathrm{(ii)}$.
\end{proof}

\begin{lemma}
\label{lem:crossing_pairs_maximum_matching}
Consider $N$ sites labeled by bits, with $n$ sites carrying $1$ and $N-n$ sites carrying $0$, where $1\leq n\leq N-1$.
Let $a$ denote the number of edges connecting different bits in a uniformly random maximum matching of the $N$ sites.
Then,
\begin{equation}
    \mathbb E\left[2^{-a}\right]
    \leq
    \frac{1}{2}
    +
    \frac{1}{2(N-1)}.
\end{equation}
\end{lemma}

\begin{proof}
Since $2^{-a}\leq1/2$ whenever $a\geq1$, we have
\begin{equation}
    \mathbb E\left[2^{-a}\right]
    \leq
    \Pr[a=0]
    +
    \frac{1}{2}\Pr[a\geq1]
    =
    \frac{1}{2}
    +
    \frac{1}{2}\Pr[a=0].
\end{equation}
It therefore suffices to show that $\Pr[a=0]\leq1/(N-1)$.

Suppose first that $N$ is even.
If $n$ is odd, every perfect matching contains at least one edge connecting different bits, and hence $\Pr[a=0]=0$.
If $n$ is even, then
\begin{equation}
    \Pr[a=0]
    =
    \frac{(n-1)!!(N-n-1)!!}{(N-1)!!}.
\end{equation}
For even $n\in\{2,4,\ldots,N-2\}$, denote the right-hand side by $p_N(n)$.
Since $\frac{p_N(n+2)}{p_N(n)} =\frac{n+1}{N-n-1}$ and $p_N(n)=p_N(N-n)$, the maximum is attained at $n=2$ or $n=N-2$.
Therefore,
\begin{equation}
    \Pr[a=0] \leq p_N(2) = p_N(N-2)  = \frac{1}{N-1}.
\end{equation}

Suppose next that $N$ is odd.
Then, exactly one of the two bit classes has odd cardinality; denote this cardinality by $o$.
For $a=0$, the unmatched site must belong to this odd class, and all remaining sites must be paired within their respective bit classes.
Hence,
\begin{equation}
    \Pr[a=0]
    =
    \frac{o!!(N-o-1)!!}{N(N-2)!!}.
\end{equation}
Denoting the right-hand side by $q_N(o)$ for odd $o\in\{1,3,\ldots,N-2\}$, we have $\frac{q_N(o+2)}{q_N(o)}=\frac{o+2}{N-o-1}$, which means the maximum is attained at $o=1$ or $o=N-2$.
Therefore, we obtain
\begin{equation}
    \Pr[a=0]
    \leq q_N(1)=q_N(N-2)=
    \frac{1}{N}
    \leq
    \frac{1}{N-1}.
\end{equation}
The claimed bound follows.
\end{proof}

\subsection{Bound on off-diagonal blocks of moment operator}\label{ss:bound_off_diagonal_block_mom}

In this subsection, we consider the moment operator restricted to a fixed charge sector $(\bm n,\overline{\bm n})$, and two distinct compatible classes $F,G\in\mathcal F_k(\bm n,\overline{\bm n})$.
For notational simplicity, throughout this subsection we write $\Pi_F\equiv\Pi_{\bm n;F}$ and $\Pi_G\equiv\Pi_{\bm n;G}$, suppressing the dependence on $\bm n$ and $\overline{\bm n}$.

Since both $\Pi_F$ and $\Pi_G$ are diagonal in the orbit-state basis, they commute.
We define the orthogonal projector onto their intersection by
\begin{equation}
    \Pi_{F,G}\equiv\Pi_F\Pi_G,
\end{equation}
and further define
\begin{equation}
    \Pi_F^\circ\equiv\Pi_F-\Pi_{F,G},
    \qquad
    \Pi_G^\circ\equiv\Pi_G-\Pi_{F,G}.
\end{equation}
The orbit-state basis associated with $\Pi_{F,G}$ consists precisely of the orbit states that belong to both compatible-class subspaces, namely, those whose local colors are all contained in $F\cap G$.
Thus, the span of the orbit states associated with $F$ or $G$ is orthogonally decomposed into the ranges of $\Pi_{F,G}$, $\Pi_F^\circ$, and $\Pi_G^\circ$.

With respect to this decomposition, the restriction of $T^{\mathrm{dp}}$ can be written as
\begin{equation}
\label{eq:Tdp_FG_block}
    T^{\mathrm{dp}}_{F\cup G}
    =
    \begin{pmatrix}
        T_{FG,FG} & T_{FG,F^\circ} & T_{FG,G^\circ} \\
        T_{F^\circ,FG} & T_{F^\circ,F^\circ} & T_{F^\circ,G^\circ} \\
        T_{G^\circ,FG} & T_{G^\circ,F^\circ} & T_{G^\circ,G^\circ}
    \end{pmatrix}.
\end{equation}
Here, each block denotes the corresponding projection of $T^{\mathrm{dp}}$; for example, $T_{FG,FG}\equiv\Pi_{F,G}T^{\mathrm{dp}}\Pi_{F,G}$ and $T_{FG,F^\circ}\equiv\Pi_{F,G}T^{\mathrm{dp}}\Pi_F^\circ$.

In this subsection, we derive an upper bound on the off-diagonal blocks of this matrix representation.
In particular, we consider
\begin{equation}
\label{eq:def_T_offabs_FG}
    T^{\mathrm{dp},FG}_{\rm off,abs}
    \equiv
    \begin{pmatrix}
        0 & |T_{FG,F^\circ}| & |T_{FG,G^\circ}| \\
        |T_{F^\circ,FG}| & 0 & |T_{F^\circ,G^\circ}| \\
        |T_{G^\circ,FG}| & |T_{G^\circ,F^\circ}| & 0
    \end{pmatrix},
\end{equation}
where, for a matrix $A$, we denote by $|A|$ the matrix obtained by taking the absolute value of each matrix element in the orbit-state basis.
We then have the following lemma.

\begin{lemma}
\label{lem:T_offabs_FG_upper_bound}
Let $F,G\in\mathcal F_k(\bm n,\overline{\bm n})$ with $F\neq G$.
For $N\geq3$,
\begin{equation}
    \left\|T^{\mathrm{dp},FG}_{\rm off,abs}\right\|_\infty
    \leq 3\sqrt{\frac{2}{N-1}}.
\end{equation}
\end{lemma}

\begin{proof}
As discussed in Eq.~\eqref{eq:full_type_vector_from_ket_type_vector_fixed_F}, once we fix a compatible class $F\in\mathcal F_k(\bm n,\overline{\bm n})$, the orbit state $\ket{\Omega_{\bm m}}$ is completely specified by the ket type vector and can be denoted by $\ket{\Omega_{\bm r;F}}$.
Therefore, for $\bm r,\bm r'\in\mathcal L_{\bm n}$ satisfying $\operatorname{supp}(\bm r)\subseteq\mathcal I_F$ and $\operatorname{supp}(\bm r')\subseteq\mathcal I_G$, we first consider even $N$ and obtain
\begin{align}
    &T_{\bm r,F;\bm r',G}^{\mathrm{dp}}
    \equiv
    \bra{\Omega_{\bm r;F}}M_{\rm dope}^{(k)}\ket{\Omega_{\bm r';G}}
    \nonumber\\
    &=
    \frac{1}{\sqrt{D_{\bm r}D_{\bm r'}}}
    \sum_{\bm x\in\mathcal T_{\bm r}}
    \sum_{\bm y\in\mathcal T_{\bm r'}}
    \left(\bigotimes_{i=1}^{N}\bra{x_i,Q_Fx_i}\right)
    \left(\bigotimes_{j=1}^{N/2}V_{2j-1,2j}^{\otimes k,k}\right)
    \left(\bigotimes_{i=1}^{N}\ket{y_i,Q_Gy_i}\right)
    \nonumber\\
    &=
    \frac{1}{\sqrt{D_{\bm r}D_{\bm r'}}}
    \sum_{\bm x\in\mathcal T_{\bm r}}
    \sum_{\bm y\in\mathcal T_{\bm r'}}
    \prod_{j=1}^{N/2}
    \bra{x_{2j-1},Q_Fx_{2j-1}}
    \bra{x_{2j},Q_Fx_{2j}}
    V_{2j-1,2j}^{\otimes k,k}
    \ket{y_{2j-1},Q_Gy_{2j-1}}
    \ket{y_{2j},Q_Gy_{2j}}
    \nonumber\\
    &=
    \frac{1}{\sqrt{D_{\bm r}D_{\bm r'}}}
    \sum_{\bm x\in\mathcal T_{\bm r}}
    \sum_{\bm y\in\mathcal T_{\bm r'}}
    \prod_{j=1}^{N/2}
    \left[
        \left|
            \bra{x_{2j-1}}\bra{x_{2j}}
            V_{2j-1,2j}^{\otimes k}
            \ket{y_{2j-1}}\ket{y_{2j}}
        \right|^2
        \delta_{(Q_F-Q_G)y_{2j-1},(Q_G-Q_F)y_{2j}}
        (-1)^{x_{2j}\cdot y_{2j}+Q_Fx_{2j}\cdot Q_Gy_{2j}}
    \right].
    \label{eq:T^dp_off_w_delta_func}
\end{align}
In the last equality, we substituted the explicit transition amplitude generated by the doped gate $V_{2j-1,2j}$ given in Eq.~\eqref{eq:Hadamard_trans_prob}.
In particular, the matrix element $\bra{x}\bra{x'}V_{2j-1,2j}^{\otimes k}\ket{y}\ket{y'}$ is nonzero only if the particle number is conserved in every copy, namely, $x+x'=y+y'$.
Thus, for each pair of sites $\{2j-1,2j\}$, a nonzero contribution requires
\begin{equation*}
    x_{2j-1}+x_{2j}=y_{2j-1}+y_{2j},
    \qquad
    Q_F(x_{2j-1}+x_{2j})=Q_G(y_{2j-1}+y_{2j}).
\end{equation*}
Combining these two conditions gives
\begin{equation}
    (Q_F-Q_G)y_{2j-1}
    =
    (Q_G-Q_F)y_{2j},
\end{equation}
which yields the delta-function constraint in Eq.~\eqref{eq:T^dp_off_w_delta_func}.

Taking the absolute value and using the triangle inequality, we obtain from Eq.~\eqref{eq:T^dp_off_w_delta_func}
\begin{equation}
\label{eq:T^dp_abs_UB_for_even_N}
    \left|T_{\bm r,F;\bm r',G}^{\mathrm{dp}}\right|
    \leq
    \sqrt{\frac{D_{\bm r'}}{D_{\bm r}}}
    \left[
        \sum_{\bm x\in\mathcal T_{\bm r}}
        \sum_{\bm y\in\mathcal T_{\bm r'}}
        \frac{W_{\gamma_0,\bm n}(\bm x,\bm y)}{D_{\bm r'}}
        \prod_{j=1}^{N/2}
        \delta_{(Q_F-Q_G)y_{2j-1},(Q_G-Q_F)y_{2j}}
    \right].
\end{equation}
Here, $W_{\gamma_0,\bm n}$ is the transition matrix of the classical stochastic process introduced in Sec.~\ref{subsec:classical_dynamics}.

Defining $\mathcal L_{\bm n}(F)\equiv\{\bm r\in\mathcal L_{\bm n}:\operatorname{supp}(\bm r)\subseteq\mathcal I_F\}$, we obtain the following bound on the weighted column sum:
\begin{align}
    \sum_{\bm r\in\mathcal L_{\bm n}(F)}
    \sqrt{\frac{D_{\bm r}}{D_{\bm r'}}}
    \left|T_{\bm r,F;\bm r',G}^{\mathrm{dp}}\right|
    &\leq
    \sum_{\bm r\in\mathcal L_{\bm n}(F)}
    \sum_{\bm x\in\mathcal T_{\bm r}}
    \sum_{\bm y\in\mathcal T_{\bm r'}}
    W_{\gamma_0,\bm n}(\bm x,\bm y)
    \frac{1}{D_{\bm r'}}
    \prod_{j=1}^{N/2}
    \delta_{(Q_F-Q_G)y_{2j-1},(Q_G-Q_F)y_{2j}}
    \nonumber\\
    &\leq
    \sum_{\bm r\in\mathcal L_{\bm n}}
    \sum_{\bm x\in\mathcal T_{\bm r}}
    \sum_{\bm y\in\mathcal T_{\bm r'}}
    W_{\gamma_0,\bm n}(\bm x,\bm y)
    \frac{1}{D_{\bm r'}}
    \prod_{j=1}^{N/2}
    \delta_{(Q_F-Q_G)y_{2j-1},(Q_G-Q_F)y_{2j}}
    \nonumber\\
    &=
    \sum_{\bm x\in\mathcal X_{\bm n}}
    \sum_{\bm y\in\mathcal T_{\bm r'}}
    W_{\gamma_0,\bm n}(\bm x,\bm y)
    \frac{1}{D_{\bm r'}}
    \prod_{j=1}^{N/2}
    \delta_{(Q_F-Q_G)y_{2j-1},(Q_G-Q_F)y_{2j}}
    \nonumber\\
    &=
    \sum_{\bm y\in\mathcal T_{\bm r'}}
    \left[
        \sum_{\bm x\in\mathcal X_{\bm n}}
        W_{\gamma_0,\bm n}(\bm x,\bm y)
    \right]
    \frac{1}{D_{\bm r'}}
    \prod_{j=1}^{N/2}
    \delta_{(Q_F-Q_G)y_{2j-1},(Q_G-Q_F)y_{2j}}
    \nonumber\\
    &=
    \sum_{\bm y\in\mathcal T_{\bm r'}}
    \frac{1}{D_{\bm r'}}
    \prod_{j=1}^{N/2}
    \delta_{(Q_F-Q_G)y_{2j-1},(Q_G-Q_F)y_{2j}}.
    \label{eq:T_off_col_sum_bound_for_even_N}
\end{align}
The second inequality follows from the nonnegativity of each matrix element and $\mathcal L_{\bm n}(F)\subseteq\mathcal L_{\bm n}$, and the last equality follows from the fact that every column sum of $W_{\gamma_0,\bm n}$ is unity.

For odd $N$, the same calculation applies to the $(N-1)/2$ matched pairs.
The only additional factor comes from the unmatched site $N$, on which $M_{\rm dope}^{(k)}$ acts as the identity.
At this site,
\begin{equation*}
    \braket{x_N,Q_Fx_N|y_N,Q_Gy_N}
    =
    \delta_{x_N,y_N}\delta_{Q_Fx_N,Q_Gy_N},
\end{equation*}
and hence, after using $x_N=y_N$, the additional constraint is $(Q_F-Q_G)y_N=\bm0$.
We therefore obtain
\begin{equation}
\label{eq:T_off_col_sum_bound_for_odd_N}
    \sum_{\bm r\in\mathcal L_{\bm n}(F)}
    \sqrt{\frac{D_{\bm r}}{D_{\bm r'}}}
    \left|T_{\bm r,F;\bm r',G}^{\mathrm{dp}}\right|
    \leq
    \sum_{\bm y\in\mathcal T_{\bm r'}}
    \frac{1}{D_{\bm r'}}
    \left[
        \prod_{j=1}^{(N-1)/2}
        \delta_{(Q_F-Q_G)y_{2j-1},(Q_G-Q_F)y_{2j}}
    \right]
    \delta_{(Q_F-Q_G)y_N,\bm0}.
\end{equation}

We now interpret the right-hand sides of Eqs.~\eqref{eq:T_off_col_sum_bound_for_even_N} and \eqref{eq:T_off_col_sum_bound_for_odd_N} as a defect-pairing problem.
For each ket color $y$, define its defect color with respect to the two compatible classes $F$ and $G$ by
\begin{equation}
    \bm d_{FG}(y)\equiv(Q_F-Q_G)y.
\end{equation}
Then the constraints in Eqs.~\eqref{eq:T_off_col_sum_bound_for_even_N} and \eqref{eq:T_off_col_sum_bound_for_odd_N} can be written as
\begin{align}
    N\ {\rm even}:&\qquad
    \bm d_{FG}(y_{2j-1})+\bm d_{FG}(y_{2j})=\bm0,
    \qquad
    1\leq j\leq\frac{N}{2},
    \nonumber\\
    N\ {\rm odd}:&\qquad
    \bm d_{FG}(y_{2j-1})+\bm d_{FG}(y_{2j})=\bm0,
    \qquad
    1\leq j\leq\frac{N-1}{2},
    \qquad
    \bm d_{FG}(y_N)=\bm0.
    \label{eq:defect_matching_constraints}
\end{align}
Hence a site with defect $\bm d\neq\bm0$ must be paired with a site with the opposite defect $-\bm d$, whereas a zero-defect site must be paired with another zero-defect site.
For odd $N$, the unmatched site must additionally have zero defect.

For a fixed ket type vector $\bm r'$, the multiplicity of each defect color is fixed.
We may therefore regard the sites as carrying a neutral defect $\bm0$ and pairs of nonzero defects $\pm\bm d_1,\ldots,\pm\bm d_s$.
By Lemma~\ref{lem:compatible_class_representation}$\mathrm{(iv)}$, for a local ket type $y$ associated with the compatible class $G$,
\begin{equation}
    \bm d_{FG}(y)=\bm0
    \quad\Longleftrightarrow\quad
    Q_Fy=Q_Gy,
\end{equation}
which is precisely the condition that the corresponding local color belongs to $F\cap G$.
Thus, if the input orbit state belongs to the range of $\Pi_{F,G}$, every site has zero defect, whereas if it belongs to the range of $\Pi_G^\circ$, at least one site has a nonzero defect.

Since $\mathcal T_{\bm r'}$ is the permutation orbit of a fixed multiset of ket types, the uniform average over $\bm y\in\mathcal T_{\bm r'}$ is equivalent to choosing a uniformly random maximum matching of the sites with these fixed defect multiplicities.
For even $N$, this is a perfect matching, whereas for odd $N$ one site is left unmatched.
The products of delta functions in Eqs.~\eqref{eq:T_off_col_sum_bound_for_even_N} and \eqref{eq:T_off_col_sum_bound_for_odd_N} are equal to one exactly when every matched pair has defects summing to zero and, for odd $N$, the unmatched site has zero defect.

As shown in Lemma~\ref{lem:defect_pairing_bound} below, if at least one nonzero defect is present, the probability of satisfying these constraints is at most
\begin{equation}
\label{eq:def_epsilon_N}
    \varepsilon_N \equiv \frac{2}{N-1}.
\end{equation}
Combining these bounds with Eqs.~\eqref{eq:T_off_col_sum_bound_for_even_N} and \eqref{eq:T_off_col_sum_bound_for_odd_N}, we obtain
\begin{equation}
\label{eq:off_col_bound_eN}
    \sum_{\bm r\in\mathcal L_{\bm n}(F)} \sqrt{\frac{D_{\bm r}}{D_{\bm r'}}}\left|T_{\bm r,F;\bm r',G}^{\mathrm{dp}}\right|\leq \varepsilon_N
\end{equation}
for every input basis state belonging to the range of $\Pi_G^\circ$.
The corresponding bound with $F$ and $G$ exchanged holds for every input basis state belonging to the range of $\Pi_F^\circ$.

We next transfer these weighted column-sum bounds by a diagonal similarity transformation.
We use the same diagonal matrix as in the previous subsection, $\mathcal R_{\bm r,\bm r}\equiv D_{\bm r}^{-1/2}$, restricted to each of the three orbit-state blocks, and define
\begin{equation}
    W
    \equiv
    \mathcal R^{-1}T^{\mathrm{dp},FG}_{\rm off,abs}\mathcal R
    =
    \begin{pmatrix}
        0 & W_{FG,F^\circ} & W_{FG,G^\circ} \\
        W_{F^\circ,FG} & 0 & W_{F^\circ,G^\circ} \\
        W_{G^\circ,FG} & W_{G^\circ,F^\circ} & 0
    \end{pmatrix}.
\end{equation}
For a matrix $A$, let
\begin{equation}
    c(A)\equiv\max_j\sum_i|A_{ij}|
\end{equation}
denote its maximum column sum.
Equation~\eqref{eq:off_col_bound_eN} and its counterpart with $F$ and $G$ exchanged then imply
\begin{equation}
\label{eq:W_col_sum_bound_eN}
    c(W_{FG,F^\circ})\leq\varepsilon_N,\qquad
    c(W_{FG,G^\circ})\leq\varepsilon_N,\qquad
    c(W_{F^\circ,G^\circ})\leq\varepsilon_N,\qquad
    c(W_{G^\circ,F^\circ})\leq\varepsilon_N.
\end{equation}
Indeed, in each of these blocks the input belongs to the range of $\Pi_F^\circ$ or $\Pi_G^\circ$, so that Eq.~\eqref{eq:off_col_bound_eN} or its $F\leftrightarrow G$ counterpart applies.

By contrast, for $W_{F^\circ,FG}$ and $W_{G^\circ,FG}$, the input belongs to the range of $\Pi_{F,G}$, where all defects may vanish.
Hence Eq.~\eqref{eq:off_col_bound_eN} does not apply.
Instead, Eqs.~\eqref{eq:T_off_col_sum_bound_for_even_N} and \eqref{eq:T_off_col_sum_bound_for_odd_N}, together with $|\mathcal T_{\bm r'}|=D_{\bm r'}$, give the trivial bounds
\begin{equation}
\label{eq:W_col_sum_bound_1}
    c(W_{F^\circ,FG})\leq1,
    \qquad
    c(W_{G^\circ,FG})\leq1.
\end{equation}

We now convert these column-sum bounds into operator-norm bounds.
Since $T^{\mathrm{dp}}$ is Hermitian, we have $|T_{G^\circ,F^\circ}|^\dagger=|T_{F^\circ,G^\circ}|$, and therefore
\begin{align}
    \left\||T_{G^\circ,F^\circ}|\right\|_\infty^2
    &=
    \rho\left(
        |T_{G^\circ,F^\circ}|
        |T_{G^\circ,F^\circ}|^\dagger
    \right)
    \nonumber\\
    &=
    \rho\left(
        |T_{G^\circ,F^\circ}|
        |T_{F^\circ,G^\circ}|
    \right)
    \nonumber\\
    &=
    \rho\left(
        W_{G^\circ,F^\circ}
        W_{F^\circ,G^\circ}
    \right)
    \nonumber\\
    &\leq
    c\left(
        W_{G^\circ,F^\circ}
        W_{F^\circ,G^\circ}
    \right)
    \nonumber\\
    &\leq
    c(W_{G^\circ,F^\circ})
    c(W_{F^\circ,G^\circ})
    \nonumber\\
    &\leq
    \varepsilon_N^2.
    \label{eq:T_off_bound_Fcirc_Gcirc}
\end{align}
Here, $\rho(A)$ denotes the spectral radius of $A$.
The first equality follows from $\|A\|_\infty^2=\|AA^\dagger\|_\infty=\rho(AA^\dagger)$.
The third equality follows because the two matrices are similar, since
\begin{equation*}
    \mathcal R^{-1}
    |T_{G^\circ,F^\circ}|
    |T_{F^\circ,G^\circ}|
    \mathcal R
    =
    W_{G^\circ,F^\circ}W_{F^\circ,G^\circ}.
\end{equation*}
The fourth line follows from the standard bound $\rho(A)\leq c(A)$, while the fifth follows from the submultiplicativity of the maximum column sum.
The final line follows directly from Eq.~\eqref{eq:W_col_sum_bound_eN}.

In the same way, we obtain
\begin{align}
    \left\||T_{FG,F^\circ}|\right\|_\infty^2
    &=
    \rho\left(
        |T_{FG,F^\circ}|
        |T_{FG,F^\circ}|^\dagger
    \right)
    \nonumber\\
    &=
    \rho\left(
        |T_{FG,F^\circ}|
        |T_{F^\circ,FG}|
    \right)
    \nonumber\\
    &=
    \rho\left(
        W_{FG,F^\circ}
        W_{F^\circ,FG}
    \right)
    \nonumber\\
    &\leq
    c\left(
        W_{FG,F^\circ}
        W_{F^\circ,FG}
    \right)
    \nonumber\\
    &\leq
    c(W_{FG,F^\circ})
    c(W_{F^\circ,FG})
    \nonumber\\
    &\leq
    \varepsilon_N.
    \label{eq:T_off_bound_FG_Fcirc}
\end{align}
The argument is identical to that for Eq.~\eqref{eq:T_off_bound_Fcirc_Gcirc}, except that in the last line we use $c(W_{FG,F^\circ})\leq\varepsilon_N$ from Eq.~\eqref{eq:W_col_sum_bound_eN} and $c(W_{F^\circ,FG})\leq1$ from Eq.~\eqref{eq:W_col_sum_bound_1}.
By the same argument,
\begin{equation}
    \left\||T_{FG,G^\circ}|\right\|_\infty^2
    \leq
    \varepsilon_N.
    \label{eq:T_off_bound_FG_Gcirc}
\end{equation}

Using these bounds, we can now estimate the operator norm of $T^{\mathrm{dp},FG}_{\rm off,abs}$ as
\begin{align}
    \left\|T^{\mathrm{dp},FG}_{\rm off,abs}\right\|_\infty
    &\leq
    \left\|
    \begin{pmatrix}
        0 & |T_{FG,F^\circ}| & 0 \\
        |T_{F^\circ,FG}| & 0 & 0 \\
        0 & 0 & 0
    \end{pmatrix}
    \right\|_\infty
    +
    \left\|
    \begin{pmatrix}
        0 & 0 & |T_{FG,G^\circ}| \\
        0 & 0 & 0 \\
        |T_{G^\circ,FG}| & 0 & 0
    \end{pmatrix}
    \right\|_\infty
    +
    \left\|
    \begin{pmatrix}
        0 & 0 & 0 \\
        0 & 0 & |T_{F^\circ,G^\circ}| \\
        0 & |T_{G^\circ,F^\circ}| & 0
    \end{pmatrix}
    \right\|_\infty
    \nonumber\\
    &\leq
    2\sqrt{\varepsilon_N}+\varepsilon_N
    \nonumber\\
    &\leq
    3\sqrt{\varepsilon_N}\nonumber \\
    &= 3\sqrt{\frac{2}{N-1}}.
\end{align}
In the second line, we used Eqs.~\eqref{eq:T_off_bound_Fcirc_Gcirc}, \eqref{eq:T_off_bound_FG_Fcirc}, and \eqref{eq:T_off_bound_FG_Gcirc}, together with
\begin{equation}
    \left\|
    \begin{pmatrix}
        0 & A \\
        A^\dagger & 0
    \end{pmatrix}
    \right\|_\infty^2
    =
    \left\|
    \begin{pmatrix}
        AA^\dagger & 0 \\
        0 & A^\dagger A
    \end{pmatrix}
    \right\|_\infty
    =
    \|A\|_\infty^2.
\end{equation}
In the third line, we used that $\varepsilon_N=2/(N-1) \leq 1$ since we now assume $N\geq 3$.
This completes the proof.

\end{proof}

We now prove the upper bound on the defect-pairing probability used in the proof above.

\begin{lemma}[Defect-pairing bound]
\label{lem:defect_pairing_bound}
Let $N\geq2$, and consider $N$ labeled particles carrying colors
\begin{equation}
    \bm0,\ \pm\bm d_1,\ldots,\pm\bm d_s,
\end{equation}
where $\bm d_a\neq\bm0$ and $\bm d_a\neq\pm\bm d_b$ for $a\neq b$.
Let $n_0$ be the number of particles with color $\bm0$, and let $n_a^+$ and $n_a^-$ be the numbers of particles with colors $\bm d_a$ and $-\bm d_a$, respectively.
Assume that $n_0<N$.

Choose a maximum matching of the $N$ particles uniformly at random.
If $N$ is even, this is a perfect matching, whereas if $N$ is odd, one particle is left unmatched.
Then the probability that every matched pair has colors whose sum is zero and, for odd $N$, the unmatched particle has color $\bm0$, is at most
\begin{equation}
    \varepsilon_N\equiv\frac{2}{N-1}.
\end{equation}
\end{lemma}

\begin{proof}
The desired matching is impossible unless $n_a^+=n_a^-$ for every $a$.
Suppose therefore that
\begin{equation}
    n_a^+=n_a^-=r_a,
    \qquad
    R\equiv\sum_{a=1}^{s}r_a\geq1.
\end{equation}

We first consider even $N$.
In this case, $n_0=N-2R$, with $1\leq R\leq N/2$.
For each $a$, the $r_a$ particles of color $\bm d_a$ must be matched bijectively with the $r_a$ particles of color $-\bm d_a$, while the remaining $N-2R$ neutral particles must be matched among themselves.
Hence the probability is
\begin{equation*}
    \frac{
        \left(\prod_{a=1}^{s}r_a!\right)
        (N-2R-1)!!
    }{
        (N-1)!!
    }
    \leq
    \frac{
        R!(N-2R-1)!!
    }{
        (N-1)!!
    },
\end{equation*}
where we used $\prod_a r_a!\leq R!$ and the convention $(-1)!!=1$.
Defining
\begin{equation*}
    \alpha_N^{\rm even}(R)
    \equiv
    \frac{R!(N-2R-1)!!}{(N-1)!!},
\end{equation*}
we have
\begin{equation*}
    \frac{\alpha_N^{\rm even}(R+1)}{\alpha_N^{\rm even}(R)}
    =
    \frac{R+1}{N-2R-1}.
\end{equation*}
Thus $\alpha_N^{\rm even}(R)$ first decreases and then increases, so its maximum over $1\leq R\leq N/2$ is attained at one of the two endpoints.
Therefore,
\begin{equation*}
    \max_{1\leq R\leq N/2}\alpha_N^{\rm even}(R)
    =
    \max\left\{
        \frac{1}{N-1},
        \frac{(N/2)!}{(N-1)!!}
    \right\}.
\end{equation*}
The second term satisfies
\begin{equation*}
    \frac{(N/2)!}{(N-1)!!} \leq  \frac{2}{N-1},
\end{equation*}
where we used $(N/2)!\leq 2(N-3)!!$. Since $1/(N-1)\leq 2/(N-1)$, we conclude that the desired probability is at most $2/(N-1)$ for even $N$.

We next consider odd $N$.
In this case, the unmatched particle must be neutral.
Consequently, $n_0=N-2R\geq1$, with $1\leq R\leq(N-1)/2$.
The total number of maximum matchings of the $N$ labeled particles is $N!!$.
For a favorable matching, the nonzero defects can be paired in $\prod_a r_a!$ ways.
Among the $n_0=N-2R$ neutral particles, one is chosen as the unmatched particle and the remaining $n_0-1$ particles are paired among themselves.
Thus, the number of favorable maximum matchings is
\begin{equation*}
    \left(\prod_{a=1}^{s}r_a!\right)
    (N-2R)(N-2R-2)!!
    =
    \left(\prod_{a=1}^{s}r_a!\right)
    (N-2R)!!.
\end{equation*}
It follows that the probability is at most
\begin{equation*}
    \alpha_N^{\rm odd}(R)
    \equiv
    \frac{R!(N-2R)!!}{N!!}.
\end{equation*}
Moreover,
\begin{equation*}
    \frac{\alpha_N^{\rm odd}(R+1)}{\alpha_N^{\rm odd}(R)}
    =
    \frac{R+1}{N-2R}.
\end{equation*}
Hence $\alpha_N^{\rm odd}(R)$ again first decreases and then increases, so its maximum is attained at one of the endpoints.
At these endpoints,
\begin{equation*}
    \alpha_N^{\rm odd}(1)=\frac{1}{N},
    \qquad
    \alpha_N^{\rm odd}\left(\frac{N-1}{2}\right)
    =
    \frac{((N-1)/2)!}{N!!}.
\end{equation*}
Writing $N=2m+1$, we have
\begin{equation*} 
    \frac{((N-1)/2)!}{N!!} 
    =
    \frac{1}{N}\frac{((N-1)/2)!}{(N-2)!!} 
    \leq
    \frac{1}{N},
\end{equation*}
where we used $((N-1)/2)!\leq(N-2)!!$.
Therefore,
\begin{equation*}
    \max_{1\leq R\leq(N-1)/2}\alpha_N^{\rm odd}(R)
    =
    \frac{1}{N}
    \leq
    \frac{2}{N-1}
    =
    \varepsilon_N.
\end{equation*}
This completes the proof.
\end{proof}

\subsection{Proof of Proposition~\ref{prop:gap_LB_dp_2RPmix_circuit}}
\label{ss:proof_of_doped_2RPmix_gap}

In this subsection, we provide the proof of Proposition~\ref{prop:gap_LB_dp_2RPmix_circuit} by using the lemmas derived in the previous subsections.
We first introduce the disjoint decomposition of the 2RPmix moment subspace needed for this purpose and then proceed to the proof.

The set of all type vectors whose corresponding orbit states span the 2RPmix subspace is denoted by
\begin{equation}
    \mathcal M_{\bm n,\overline{\bm n}}
    \equiv
    \left\{
        \bm m:
        \ket{\Omega_{\bm m}}
        \in
        \operatorname{Ran}
        P_{\mathrm{2RPmix};\bm n,\overline{\bm n}}^{(k)}
    \right\}.
\end{equation}
For each $\bm m\in\mathcal M_{\bm n,\overline{\bm n}}$, define its compatible-class membership vector $\bm v(\bm m)$ by
\begin{equation}
    v_F(\bm m)
    \equiv
    \begin{cases}
        1, & \operatorname{supp}(\bm m)\subseteq F,\\
        0, & \operatorname{supp}(\bm m)\not\subseteq F,
    \end{cases}
    \qquad
    F\in\mathcal F_k(\bm n,\overline{\bm n}),
\end{equation}
where the dimension of the vector $\bm v(\bm m)$ is $|\mathcal F_k(\bm n,\overline{\bm n})|$.
We further denote the set of membership vectors that actually occur in this charge sector by
\begin{equation}
    \mathcal B_k(\bm n,\overline{\bm n})
    \equiv
    \left\{
        \bm v(\bm m):
        \bm m\in\mathcal M_{\bm n,\overline{\bm n}}
    \right\}.
\end{equation}
Then, for each $\bm v\in\mathcal B_k(\bm n,\overline{\bm n})$, we introduce the corresponding orthogonal projector as
\begin{equation}
    \Pi_{\bm v}
    \equiv
    \sum_{\substack{
        \bm m\in\mathcal M_{\bm n,\overline{\bm n}}\\
        \bm v(\bm m)=\bm v
    }}
    \ket{\Omega_{\bm m}}\bra{\Omega_{\bm m}}.
\end{equation}

Using these membership projectors, we obtain the disjoint decomposition of the 2RPmix moment space as
\begin{equation}
\label{eq:2RPmix_moment_op_disjoint_decomp}
    P_{\mathrm{2RPmix};\bm n,\overline{\bm n}}^{(k)}
    =
    \sum_{\bm v\in\mathcal B_k(\bm n,\overline{\bm n})}
    \Pi_{\bm v}.
\end{equation}
The projector onto a compatible class $F\in\mathcal F_k(\bm n,\overline{\bm n})$ is also given by
\begin{equation}
    \Pi_{\bm n;F}
    =
    \sum_{\substack{
        \bm v\in\mathcal B_k(\bm n,\overline{\bm n})\\
        v_F=1
    }}
    \Pi_{\bm v}.
\end{equation}
Furthermore, we introduce the following maps that decompose an operator into diagonal and off-diagonal parts with respect to the membership sectors:
\begin{align}
    \mathfrak{F}_{\rm diag}(A)
    &\equiv
    \sum_{\bm v\in\mathcal B_k(\bm n,\overline{\bm n})}
    \Pi_{\bm v}A\Pi_{\bm v},
    \\
    \mathfrak{F}_{\rm off}(A)
    &\equiv
    P_{\mathrm{2RPmix};\bm n,\overline{\bm n}}^{(k)}
    A
    P_{\mathrm{2RPmix};\bm n,\overline{\bm n}}^{(k)}
    -
    \mathfrak{F}_{\rm diag}(A)
    =
    \sum_{\substack{
        \bm v,\bm v'\in\mathcal B_k(\bm n,\overline{\bm n})\\
        \bm v\neq\bm v'
    }}
    \Pi_{\bm v}A\Pi_{\bm v'}.
\end{align}
Since the projectors $\{\Pi_{\bm v}\}_{\bm v}$ are mutually orthogonal, we have
\begin{align}
    \left\|\mathfrak{F}_{\rm diag}(A)\right\|_\infty
    &=
    \max_{\bm v\in\mathcal B_k(\bm n,\overline{\bm n})}
    \left\|\Pi_{\bm v}A\Pi_{\bm v}\right\|_\infty
    \leq
    \|A\|_\infty,
    \nonumber\\
    \left\|\mathfrak{F}_{\rm off}(A)\right\|_\infty
    &\leq
    \left\|
        P_{\mathrm{2RPmix};\bm n,\overline{\bm n}}^{(k)}
        A
        P_{\mathrm{2RPmix};\bm n,\overline{\bm n}}^{(k)}
    \right\|_\infty
    +
    \left\|\mathfrak{F}_{\rm diag}(A)\right\|_\infty
    \leq
    2\|A\|_\infty.
    \label{eq:opineq_F_off}
\end{align}

By using these notions, we can derive Proposition~\ref{prop:gap_LB_dp_2RPmix_circuit} as follows.

\begin{proof}[Proof of Proposition~\ref{prop:gap_LB_dp_2RPmix_circuit}]

We consider two cases, $\mathrm{(i)}$ $\Sigma(\bm n,\overline{\bm n})=\varnothing$ and $\mathrm{(ii)}$ $\Sigma(\bm n,\overline{\bm n})\neq\varnothing$, separately.

First, we assume $\Sigma(\bm n,\overline{\bm n})=\varnothing$.
For such a charge sector, we have
\begin{equation}
    1-\Delta_{V\mathrm{dp2rpmix};\bm n,\overline{\bm n}}^{(k)}
    =
    \left\|
        P_{\mathrm{2RPmix};\bm n,\overline{\bm n}}^{(k)}
        M_{\mathrm{dope}}^{(k)}
        P_{\mathrm{2RPmix};\bm n,\overline{\bm n}}^{(k)}
    \right\|_\infty,
\end{equation}
since $P_{U(1)\mathrm{Haar}}^{(k)}\Pi_{\bm n,\overline{\bm n}}=0$ in this case.
Then, by using Eq.~\eqref{eq:2RPmix_moment_op_disjoint_decomp}, we obtain
\begin{align}
    1-\Delta_{V\mathrm{dp2rpmix};\bm n,\overline{\bm n}}^{(k)}
    &=
    \left\|
        \sum_{\bm v,\bm v'\in\mathcal B_k(\bm n,\overline{\bm n})}
        \Pi_{\bm v}M_{\mathrm{dope}}^{(k)}\Pi_{\bm v'}
    \right\|_\infty
    \nonumber\\
    &\leq
    \left\|
        \mathfrak{F}_{\rm diag}\left(M_{\mathrm{dope}}^{(k)}\right)
    \right\|_\infty
    +
    \left\|
        \mathfrak{F}_{\rm off}\left(M_{\mathrm{dope}}^{(k)}\right)
    \right\|_\infty.
    \label{eq:diag_offdiag_decomposition}
\end{align}

We first bound the diagonal contribution.
For any $\bm v\in\mathcal B_k(\bm n,\overline{\bm n})$, choose a type vector $\bm m\in\mathcal M_{\bm n,\overline{\bm n}}$ satisfying $\bm v(\bm m)=\bm v$.
There exists $F\in\mathcal F_k(\bm n,\overline{\bm n})$ such that $\operatorname{supp}(\bm m)\subseteq F$, and hence $v_F=1$.
Since we now assume $\Sigma(\bm n,\overline{\bm n})=\varnothing$, such $F$ has to be a non-permutation compatible class.
Therefore, we obtain
\begin{align}
    \left\|
        \Pi_{\bm v}M_{\mathrm{dope}}^{(k)}\Pi_{\bm v}
    \right\|_\infty
    &\leq
    \left\|
        \Pi_{\bm n;F}M_{\mathrm{dope}}^{(k)}\Pi_{\bm n;F}
    \right\|_\infty
    \nonumber\\
    &\leq
    \frac{1}{2}+\frac{1}{2(N-1)},
    \label{eq:diag_nonperm_bound}
\end{align}
where the first inequality follows from $0\leq\Pi_{\bm v}\leq\Pi_{\bm n;F}$, since the left-hand side is a compression of the operator on the right-hand side.
The second inequality follows from Lemmas~\ref{lem:active_nonperm_coordinate} and~\ref{lem:bound_each_block}$\mathrm{(ii)}$.
This inequality upper bounds the first term of Eq.~\eqref{eq:diag_offdiag_decomposition}.

We next upper bound the off-diagonal contribution.
For this purpose, we first derive the following element-wise inequality:
\begin{equation}
\label{eq:element_wise_inequality_dp2rp_circuit}
    \left|
        \bra{\Omega_{\bm m}}
        \mathfrak{F}_{\rm off}\left(M_{\mathrm{dope}}^{(k)}\right)
        \ket{\Omega_{\bm m'}}
    \right|
    \leq
    \left[
        \sum_{\substack{
            \{F,G\}\subseteq\mathcal F_k(\bm n,\overline{\bm n})\\
            F\neq G
        }}
        T_{\rm off,abs}^{\mathrm{dp},FG}
    \right]_{\bm m,\bm m'}.
\end{equation}
Here, each $T_{\rm off,abs}^{\mathrm{dp},FG}$ is understood as its zero extension to $\operatorname{Ran}P_{\mathrm{2RPmix};\bm n,\overline{\bm n}}^{(k)}$.
Namely, its matrix elements are set to zero whenever its row or column orbit state lies outside the $F,G$ block on which it is originally defined.
Thus, all these matrices are represented in the common orbit-state basis $\{\ket{\Omega_{\bm m}}:\bm m\in\mathcal M_{\bm n,\overline{\bm n}}\}$, and $[\cdot]_{\bm m,\bm m'}$ denotes the corresponding matrix element.

If $\bm v(\bm m)=\bm v(\bm m')$, the left-hand side vanishes, and the inequality immediately follows from the nonnegativity of the right-hand side.
If $\bm v(\bm m)\neq\bm v(\bm m')$, we have $\bra{\Omega_{\bm m}}\mathfrak{F}_{\rm off}\left(M_{\mathrm{dope}}^{(k)}\right)\ket{\Omega_{\bm m'}}=T^{\rm dp}_{\bm m;\bm m'}$.
In this case, there exists $F\in\mathcal F_k(\bm n,\overline{\bm n})$ such that either $v_F(\bm m)=1$, $v_F(\bm m')=0$ or $v_F(\bm m)=0$, $v_F(\bm m')=1$.
Without loss of generality, we assume $v_F(\bm m)=1$ and $v_F(\bm m')=0$.
Then, by taking any $G\in\mathcal F_k(\bm n,\overline{\bm n})$ such that $v_G(\bm m')=1$, we have $\left[T_{\rm off,abs}^{\mathrm{dp},FG}\right]_{\bm m,\bm m'}=\left|T^{\rm dp}_{\bm m;\bm m'}\right|$.
Therefore,
\begin{equation*}
    \left|
        \bra{\Omega_{\bm m}}
        \mathfrak{F}_{\rm off}\left(M_{\mathrm{dope}}^{(k)}\right)
        \ket{\Omega_{\bm m'}}
    \right|
    =
    \left|T^{\rm dp}_{\bm m;\bm m'}\right|
    =
    \left[T_{\rm off,abs}^{\mathrm{dp},FG}\right]_{\bm m,\bm m'}
    \leq
    \left[
        \sum_{\substack{
            \{F,G\}\subseteq\mathcal F_k(\bm n,\overline{\bm n})\\
            F\neq G
        }}
        T_{\rm off,abs}^{\mathrm{dp},FG}
    \right]_{\bm m,\bm m'},
\end{equation*}
which shows Eq.~\eqref{eq:element_wise_inequality_dp2rp_circuit}.

From this element-wise inequality, we obtain the operator-norm bound.
Indeed, suppose that matrices $A$ and $B$, with $B$ entrywise nonnegative, satisfy $|A_{\bm m,\bm m'}|\leq B_{\bm m,\bm m'}$ for all $\bm m,\bm m'$.
For any unit vectors $\ket{x}$ and $\ket{y}$, let $\ket{|x|}$ and $\ket{|y|}$ denote the vectors obtained by taking the absolute values of their coefficients.
Then, $\|\ket{|x|}\|=\|\ket{|y|}\|=1$, and we have $|\bra{x}A\ket{y}|\leq \bra{|x|}B\ket{|y|}\leq\|B\|_\infty$.
Therefore, by taking the supremum over $\ket{x}$ and $\ket{y}$, Eq.~\eqref{eq:element_wise_inequality_dp2rp_circuit} gives
\begin{equation}
    \left\|
        \mathfrak{F}_{\rm off}\left(M_{\mathrm{dope}}^{(k)}\right)
    \right\|_\infty
    \leq
    \left\|
        \sum_{\substack{
            \{F,G\}\subseteq\mathcal F_k(\bm n,\overline{\bm n})\\
            F\neq G
        }}
        T_{\rm off,abs}^{\mathrm{dp},FG}
    \right\|_\infty.
    \label{eq:operator_norm_inequality_F_off_upper_bound}
\end{equation}
By using this inequality, we have
\begin{align}
    \left\|
        \mathfrak{F}_{\rm off}\left(M_{\mathrm{dope}}^{(k)}\right)
    \right\|_\infty
    &\leq
    \sum_{\substack{
        \{F,G\}\subseteq\mathcal F_k(\bm n,\overline{\bm n})\\
        F\neq G
    }}
    \left\|T_{\rm off,abs}^{\mathrm{dp},FG}\right\|_\infty
    \nonumber\\
    &\leq
    \frac{
        |\mathcal F_k(\bm n,\overline{\bm n})|
        \bigl(|\mathcal F_k(\bm n,\overline{\bm n})|-1\bigr)
    }{2}
    \cdot
    \frac{3\sqrt{2}}{\sqrt{N-1}}
    \nonumber\\
    &\leq
    \frac{3k^2 2^{4k^2}}{\sqrt{N-1}}
    \nonumber\\
    &\leq
    \exp\left[O(k^2)-\frac{1}{2}\log(N-1)\right],
    \label{eq:offdiag_dope_final_bound}
\end{align}
where in the third line we used Lemma~\ref{lem:compatible_class_representation}$\mathrm{(v)}$, which gives $|\mathcal F_k(\bm n,\overline{\bm n})|\leq|\mathcal F_k|\leq k\,2^{2k^2}$.

Combining Eqs.~\eqref{eq:diag_nonperm_bound} and~\eqref{eq:offdiag_dope_final_bound}, we have
\begin{align}
    1-\Delta_{V\mathrm{dp2rpmix};\bm n,\overline{\bm n}}^{(k)}
    &\leq
    \frac{1}{2}
    +\frac{1}{2(N-1)}
    +\frac{3k^2 2^{4k^2}}{\sqrt{N-1}}
    \nonumber\\
    &\leq
    \frac{1}{2}
    +
    \exp\left[O(k^2)-\frac{1}{2}\log(N-1)\right].
\end{align}

We next consider the case that $\Sigma(\bm n,\overline{\bm n})\neq\varnothing$ holds.
After removing the trivial copies with $n_j\in\{0,N\}$ as discussed in Sec.~\ref{ss:dp2RP_setup_strategy}, we may assume $1\leq n_j\leq N-1$ for all $j\in[k]$.
In such a charge sector, we decompose the moment operators as
\begin{align}
    1-\Delta_{V\mathrm{dp2rpmix};\bm n,\overline{\bm n}}^{(k)}
    &\equiv
    \left\|
        P_{\mathrm{2RPmix};\bm n,\overline{\bm n}}^{(k)}
        \left(
            M_{\mathrm{dope}}^{(k)}
            -
            P_{U(1)\mathrm{Haar}}^{(k)}
        \right)
        P_{\mathrm{2RPmix};\bm n,\overline{\bm n}}^{(k)}
    \right\|_\infty
    \nonumber\\
    &=
    \left\|
        \sum_{\bm v,\bm v'\in\mathcal B_k(\bm n,\overline{\bm n})}
        \Pi_{\bm v}
        \left(
            M_{\mathrm{dope}}^{(k)}
            -
            P_{U(1)\mathrm{Haar}}^{(k)}
        \right)
        \Pi_{\bm v'}
    \right\|_\infty
    \nonumber\\
    &\leq
    \left\|
        \mathfrak{F}_{\rm diag}
        \left(
            M_{\mathrm{dope}}^{(k)}
            -
            P_{U(1)\mathrm{Haar}}^{(k)}
        \right)
    \right\|_\infty
    +
    \left\|
        \mathfrak{F}_{\rm off}
        \left(M_{\mathrm{dope}}^{(k)}\right)
    \right\|_\infty
    +
    \left\|
        \mathfrak{F}_{\rm off}
        \left(P_{U(1)\mathrm{Haar}}^{(k)}\right)
    \right\|_\infty.
    \label{eq:diag_offdiag_decomposition_Sigma}
\end{align}

First, we consider the diagonal contribution.
If $\bm v\in\mathcal B_k(\bm n,\overline{\bm n})$ satisfies $v_{F_\sigma}=1$ for some $\sigma\in\Sigma(\bm n,\overline{\bm n})$, we can upper bound the operator norm as
\begin{align}
    &\left\|
        \Pi_{\bm v}
        \left(
            M_{\mathrm{dope}}^{(k)}
            -
            P_{U(1)\mathrm{Haar}}^{(k)}
        \right)
        \Pi_{\bm v}
    \right\|_\infty
    \leq
    \left\|
        \Pi_{\bm n;\sigma}
        \left(
            M_{\mathrm{dope}}^{(k)}
            -
            P_{U(1)\mathrm{Haar}}^{(k)}
        \right)
        \Pi_{\bm n;\sigma}
    \right\|_\infty
    \nonumber\\
    &\leq
    \left\|
        \Pi_{\bm n;\sigma}
        \left(
            M_{\mathrm{dope}}^{(k)}
            -
            \ket{\bm n;\sigma}\bra{\bm n;\sigma}
        \right)
        \Pi_{\bm n;\sigma}
    \right\|_\infty
    +
    \left\|
        \Pi_{\bm n;\sigma}
        \left(
            P_{U(1)\mathrm{Haar}}^{(k)}
            -
            \ket{\bm n;\sigma}\bra{\bm n;\sigma}
        \right)
        \Pi_{\bm n;\sigma}
    \right\|_\infty
    \nonumber\\
    &\leq
    \frac{1}{2}
    +
    2\left[
        \exp\left(\frac{k(k-1)}{2N}\right)-1
    \right],
    \label{eq:diag_perm_bound}
\end{align}
where the first inequality follows from $0\leq\Pi_{\bm v}\leq\Pi_{\bm n;\sigma}$, and the final inequality follows from Lemmas~\ref{lem:approx_Haar_sector_sigma_n} and~\ref{lem:bound_each_block}$\mathrm{(i)}$.

If $v_{F_\sigma}=0$ for all $\sigma\in\Sigma(\bm n,\overline{\bm n})$, this membership sector is orthogonal to the Haar moment space.
Indeed, $\operatorname{Ran}P_{U(1)\mathrm{Haar};\bm n,\overline{\bm n}}^{(k)}$ is spanned by $\{\ket{\bm n;\sigma}:\sigma\in\Sigma(\bm n,\overline{\bm n})\}$, and every orbit state appearing in $\ket{\bm n;\sigma}$ has its type-vector support contained in $F_\sigma$ and hence belongs to a membership sector with $v_{F_\sigma}=1$.
Therefore, $\Pi_{\bm v}P_{U(1)\mathrm{Haar}}^{(k)}=0$.
For such $\bm v$, there exists $F\in\mathcal F_k(\bm n,\overline{\bm n})\setminus\mathcal F_k^{\rm perm}$ satisfying $v_F=1$, and we have
\begin{align}
    \left\|
        \Pi_{\bm v}
        \left(
            M_{\mathrm{dope}}^{(k)}
            -
            P_{U(1)\mathrm{Haar}}^{(k)}
        \right)
        \Pi_{\bm v}
    \right\|_\infty
    &=
    \left\|
        \Pi_{\bm v}
        M_{\mathrm{dope}}^{(k)}
        \Pi_{\bm v}
    \right\|_\infty
    \nonumber\\
    &\leq
    \left\|
        \Pi_{\bm n;F}
        M_{\mathrm{dope}}^{(k)}
        \Pi_{\bm n;F}
    \right\|_\infty
    \nonumber\\
    &\leq
    \frac{1}{2}
    +
    \frac{1}{2(N-1)}.
    \label{eq:diag_perm_bound2}
\end{align}
These inequalities give the upper bounds on the operator norms of the diagonal blocks.

Regarding the off-diagonal block of the moment operator $\mathfrak{F}_{\rm off}\left(M_{\mathrm{dope}}^{(k)}\right)$, we can derive the following upper bound in the same way as Eqs.~\eqref{eq:element_wise_inequality_dp2rp_circuit}, \eqref{eq:operator_norm_inequality_F_off_upper_bound}, and~\eqref{eq:offdiag_dope_final_bound}:
\begin{equation}
\label{eq:offdiag_dope_final_bound_Sigma_with_perm}
    \left\|
        \mathfrak{F}_{\rm off}
        \left(M_{\mathrm{dope}}^{(k)}\right)
    \right\|_\infty
    \leq
    \frac{3k^2 2^{4k^2}}{\sqrt{N-1}}.
\end{equation}

Finally, we bound the off-diagonal contribution from the Haar moment operator.
We have
\begin{align}
    \left\|
        \mathfrak{F}_{\rm off}
        \left(P_{U(1)\mathrm{Haar}}^{(k)}\right)
    \right\|_\infty
    &\leq
    \left\|
        \mathfrak{F}_{\rm off}
        \left(
            P_{U(1)\mathrm{Haar}}^{(k)}
            -
            \sum_{\sigma\in\Sigma(\bm n,\overline{\bm n})}
            \ket{\bm n;\sigma}\bra{\bm n;\sigma}
        \right)
    \right\|_\infty +
    \sum_{\sigma\in\Sigma(\bm n,\overline{\bm n})}
    \left\|
        \mathfrak{F}_{\rm off}
        \left(
            \ket{\bm n;\sigma}\bra{\bm n;\sigma}
        \right)
    \right\|_\infty
    \nonumber\\
    &\leq
    2\left[
        \exp\left(\frac{k(k-1)}{2N}\right)-1
    \right]
    +
    \sum_{\sigma\in\Sigma(\bm n,\overline{\bm n})}
    \left\|
        \mathfrak{F}_{\rm off}
        \left(
            \ket{\bm n;\sigma}\bra{\bm n;\sigma}
        \right)
    \right\|_\infty,
    \label{eq:Haar_offdiag_reduce_rankone}
\end{align}
where the first inequality follows from the triangle inequality, and the second inequality follows from Lemma~\ref{lem:u1_haar_approximate_orthogonality} and Eq.~\eqref{eq:opineq_F_off}.

We now derive an operator-norm upper bound for $\mathfrak{F}_{\rm off}\left(\ket{\bm n;\sigma}\bra{\bm n;\sigma}\right)$.
For this purpose, we first consider the membership sector $\bm v$ satisfying $v_{F_\sigma}=1$ and $v_F=0$ for every $F\in\mathcal F_k(\bm n,\overline{\bm n})$ with $F\neq F_\sigma$. This membership sector is nonempty in the parameter regime considered here. Indeed, Lemma~\ref{lem:distinct_class_overlap_number_permutation} and the union bound imply that the total weight of \(\ket{\bm n;\sigma}\) in the union of the other compatible classes is at most \((|\mathcal F_k(\bm n,\overline{\bm n})|-1)/N<1\).
By using this $\bm v$, we decompose the state $\ket{\bm n;\sigma}$ as
\begin{equation}
    \ket{\bm n;\sigma}
    =
    \ket{\psi_\sigma}
    +
    \ket{\psi_{\overline{\sigma}}},
    \qquad
    \ket{\psi_\sigma}
    \equiv
    \Pi_{\bm v}\ket{\bm n;\sigma},
    \quad
    \ket{\psi_{\overline{\sigma}}}
    \equiv
    (\mathbb I-\Pi_{\bm v})\ket{\bm n;\sigma},
\end{equation}
and denote $q_\sigma\equiv\langle\psi_{\overline{\sigma}}|\psi_{\overline{\sigma}}\rangle$.
Then, from Lemma~\ref{lem:distinct_class_overlap_number_permutation}, we have
\begin{equation}
    q_\sigma
    \leq
    \sum_{\substack{
        F\in\mathcal F_k(\bm n,\overline{\bm n})\\
        F\neq F_\sigma
    }}
    \left\|
        \Pi_{\bm n;F}
        \ket{\bm n;\sigma}
    \right\|^2
    \leq
    \frac{
        |\mathcal F_k(\bm n,\overline{\bm n})|-1
    }{N}.
    \label{eq:psi_sigma_leakage}
\end{equation}
Furthermore, since
\begin{equation}
    \mathfrak{F}_{\rm off}
    \left(
        \ket{\bm n;\sigma}\bra{\bm n;\sigma}
    \right)
    =
    \ket{\psi_\sigma}\bra{\psi_{\overline{\sigma}}}
    +
    \ket{\psi_{\overline{\sigma}}}\bra{\psi_\sigma}
    +
    \mathfrak{F}_{\rm off}
    \left(
        \ket{\psi_{\overline{\sigma}}}
        \bra{\psi_{\overline{\sigma}}}
    \right),
\end{equation}
we can upper bound its operator norm as
\begin{align}
    \left\|
        \mathfrak{F}_{\rm off}
        \left(
            \ket{\bm n;\sigma}\bra{\bm n;\sigma}
        \right)
    \right\|_\infty
    &\leq
    \left\|
        \ket{\psi_\sigma}\bra{\psi_{\overline{\sigma}}}
    \right\|_\infty
    +
    \left\|
        \ket{\psi_{\overline{\sigma}}}\bra{\psi_\sigma}
    \right\|_\infty
    +
    \left\|
        \mathfrak{F}_{\rm off}
        \left(
            \ket{\psi_{\overline{\sigma}}}
            \bra{\psi_{\overline{\sigma}}}
        \right)
    \right\|_\infty
    \nonumber\\
    &\leq
    2\sqrt{q_\sigma}
    +
    2q_\sigma
    \nonumber\\
    &\leq
    4\sqrt{q_\sigma},
\end{align}
where we used Eq.~\eqref{eq:opineq_F_off} in the second inequality.

Therefore, using $|\Sigma(\bm n,\overline{\bm n})|\leq k!$ and $|\mathcal F_k(\bm n,\overline{\bm n})|\leq|\mathcal F_k|\leq k\,2^{2k^2}$, we obtain
\begin{align}
    \left\|
        \mathfrak{F}_{\rm off}
        \left(P_{U(1)\mathrm{Haar}}^{(k)}\right)
    \right\|_\infty
    &\leq
    2\left[
        \exp\left(\frac{k(k-1)}{2N}\right)-1
    \right]
    +
    4k!
    \sqrt{
        \frac{
            |\mathcal F_k(\bm n,\overline{\bm n})|-1
        }{N}
    }
    \nonumber\\
    &\leq
    2\left[
        \exp\left(\frac{k(k-1)}{2N}\right)-1
    \right]
    +
    \frac{4k!\sqrt{k}\,2^{k^2}}{\sqrt N}
    \nonumber\\
    &\leq
    \exp\left[
        O(k^2)-\frac{1}{2}\log N
    \right].
    \label{eq:offdiag_U(1)Haar_final_bound}
\end{align}

Combining Eqs.~\eqref{eq:diag_offdiag_decomposition_Sigma}, \eqref{eq:diag_perm_bound}, \eqref{eq:diag_perm_bound2}, \eqref{eq:offdiag_dope_final_bound_Sigma_with_perm}, and~\eqref{eq:offdiag_U(1)Haar_final_bound}, we finally obtain
\begin{align}
    1-\Delta_{V\mathrm{dp2rpmix};\bm n,\overline{\bm n}}^{(k)}
    &\leq
    \frac{1}{2}
    +
    \frac{1}{2(N-1)}
    +
    4\left[
        \exp\left(\frac{k(k-1)}{2N}\right)-1
    \right]
    +
    \frac{4k!\sqrt{k}\,2^{k^2}}{\sqrt N}
    +
    \frac{3k^2 2^{4k^2}}{\sqrt{N-1}}
    \nonumber\\
    &\leq 
    \frac{1}{2}
    +
    \exp\left[
        O(k^2)-\frac{1}{2}\log(N-1)
    \right].
    \label{eq:dp2RPmix_sector_final_bound}
\end{align}

Since the moment operators are block diagonal with respect to the charge sectors, the bounds derived in the two cases imply
\begin{equation}
    1-\Delta_{V\mathrm{dp2rpmix}}^{(k)}
    =
    \max_{\bm n,\overline{\bm n}}
    \left(
        1-\Delta_{V\mathrm{dp2rpmix};\bm n,\overline{\bm n}}^{(k)}
    \right)
    \leq
    \frac{1}{2}
    +
    \exp\left[
        O(k^2)-\frac{1}{2}\log(N-1)
    \right].
\end{equation}
Hence, $\Delta_{V\mathrm{dp2rpmix}}^{(k)}$ is bounded below by a positive constant independent of $N$ and $k$, for $k\leq O(\sqrt{\log N})$.
By further utilizing Lemma~\ref{lem:gap_Vdp2RPmix_circuit_to_dp2RPmix_circuit}, we finally obtain
\begin{equation}
    \Delta_{\mathrm{dp2rpmix}}^{(k)} \geq  \frac{1}{2}\Delta_{V\mathrm{dp2rpmix}}^{(k)} \geq \frac{1}{4} - \exp\left[ O(k^2)-\frac{1}{2}\log(N-1)\right],
\end{equation}
which completes the proof.

\end{proof}

\section{Spectral gap for the first moment $k=1$} \label{s:gap_scaling_k=1}

In this section, we determine the spectral gap $\Delta_{\nu}^{(1)}$ governing the formation rate of a $U(1)$-symmetric unitary $1$-design.
In this case, the spectral gap analysis becomes considerably simpler than in the $k\geq 2$ case.
Combined with the results in the previous sections, this section completes our characterization of the spectral gap scaling for $U(1)$-symmetric unitary $k$-design formation in the range $1\leq k\leq O(\sqrt{\log N})$.

Specifically, we show that the spectral gap $\Delta_{\nu_G}^{(1)}$ exactly coincides with the relaxation rate of the single-particle transport process on the same graph.
\begin{lemma}\label{lem:spectral_gap_1_design}
For any connected graph $G=(V,E)$ with $N\equiv |V|\geq 2$, we have
\begin{equation}
    \Delta_{\nu_G}^{(1)}=\lambda_G^{\rm tr}.
\end{equation}
\end{lemma}
\noindent
This lemma shows that $k=1$ is a special case in which the single-particle transport determines the spectral gap, in contrast to the $k\geq 2$ case, in which the two-particle encounter mode governs the spectral gap.
The proof of this lemma can be given using the notions introduced in Sec.~\ref{ss:classical_color_exchange}.

\begin{proof}[Proof of Lemma~\ref{lem:spectral_gap_1_design}]
We first show that for $k=1$, the moment projector of the $U(1)$-symmetric Haar measure coincides with that of the Haar measure on the 2RPmix group.
Indeed, the set of colors and the set of compatible classes are $\mathcal C_1=\{00,11\}$ and $\mathcal F_1=\{F_{\rm id}\}$, respectively, with $F_{\rm id}=\{00,11\}$.
A type vector satisfying $\operatorname{supp}(\bm m)\subseteq F_{\rm id}$ is therefore specified by $m_{00}=N-n$, $m_{11}=n$, and $m_u=0$ for $u\in\{01,10\}$.
The corresponding orbit state satisfies $\ket{\Omega_{\bm m}}=\ket{(n);\mathrm{id}}$, which implies
\begin{equation}
    P_{\rm 2RPmix}^{(1)}=P_{U(1)\mathrm{Haar}}^{(1)}.
\end{equation}
The same argument applied to a two-qubit system gives $P_{i,j;\mathrm{2rpmix}}^{(1)}=P_{i,j}^{(1)}$.
Hence, $M_{\nu_G^{\mathrm{2rpmix}}}^{(1)}=M_{\nu_G}^{(1)}$, and it is sufficient to evaluate
\begin{equation}
    \Delta_{\nu_G^{\mathrm{2rpmix}}}^{(1;\,\mathrm{2RPmix})}
    \equiv
    1-\left\|M_{\nu_G^{\mathrm{2rpmix}}}^{(1)}-P_{\rm 2RPmix}^{(1)}\right\|_{\infty}.
\end{equation}

As shown in the proof of Lemma~\ref{lem:2RPmix_random_edge_gap}, the moment operator $M_{\nu_G^{\mathrm{2rpmix}}}^{(1)}$ is block diagonal with respect to the type vectors $\bm m$, and each block exactly coincides with the classical stochastic or substochastic transition matrix $K_G^{\bm m}$.
The inequality $\Delta_{\nu_G^{\mathrm{2rpmix}}}^{(1;\,\mathrm{2RPmix})}\leq\lambda_G^{\rm tr}$ is then immediate, since the single-particle transport process is embedded in the moment operator as the sector $\bm m_{\rm sing}$ satisfying $(m_{\rm sing})_{00}=N-1$ and $(m_{\rm sing})_{11}=1$.

It remains to show that $\Delta_{\nu_G^{\mathrm{2rpmix}}}^{(1;\,\mathrm{2RPmix})}\geq\lambda_G^{\rm tr}$.
For the compatible type vectors satisfying $m_{00}=N-n$, $m_{11}=n$, and $m_u=0$ for $u\in\{01,10\}$, Aldous' spectral-gap theorem gives
\begin{equation}
    \rho\left(K_G^{\bm m}-\ket{\Omega_{\bm m}}\bra{\Omega_{\bm m}}\right)
    \leq
    1-\lambda_G^{\rm tr}.
\end{equation}

On the other hand, consider an incompatible type vector satisfying $m_{10}>0$ or $m_{01}>0$.
Since the local types $10$ and $01$ are incompatible with every local type, the corresponding component is eliminated whenever an edge having at least one such local type at its endpoint is selected.
Since $K_G^{\bm m}$ is real symmetric and nonnegative, $\|K_G^{\bm m}\|_\infty=\rho(K_G^{\bm m})$, and Gershgorin's theorem together with Eq.~\eqref{eq:K_m^G_each_element} gives
\begin{align}
\|K_G^{\bm m}\|_{\infty}
&\leq \max_{\bm u\in\mathcal T_{\bm m}}\sum_{\bm v\in\mathcal T_{\bm m}}[K_G^{\bm m}]_{\bm v,\bm u} \nonumber\\
&= \max_{\bm u\in\mathcal T_{\bm m}}
\left(
1-\frac{|\{\{i,j\}\in E:u_i\in\{01,10\}\text{ or }u_j\in\{01,10\}\}|}{|E|}
\right) \nonumber\\
&\leq 1-\frac{d_G^{\rm min}}{|E|},
\end{align}
where $d_G^{\rm min}\geq 1$ denotes the minimum degree of $G$.
The last inequality follows because every configuration contains at least one vertex carrying a local type $01$ or $10$, and the number of edges incident to this vertex is at least $d_G^{\rm min}$.

Finally, we show that $\lambda_G^{\rm tr}\leq d_G^{\rm min}/|E|$ for an arbitrary connected graph $G$.
For this purpose, we consider the subspace for the single-particle type vector $\bm m_{\rm sing}$, and we denote by $\ket{x}$ the configuration in which the color $11$ occupies vertex $x \in V$, while all other vertices carry the color $00$.
Then, we define the test vector
\begin{equation}
    \ket{\psi}\equiv\sum_{x\in V}\psi(x)\ket{x},
    \qquad
    \psi(x_{\rm min})=1-\frac{1}{N},
    \quad
    \psi(x)=-\frac{1}{N}\quad (x\neq x_{\rm min}),
\end{equation}
where $x_{\rm min}$ is a vertex of $G$ whose degree is $d_G^{\rm min}$.
Using the component form of $K_G^{\bm m_{\rm sing}}$ appearing in Eq.~\eqref{eq:single_particle_eigenvalue_equation}, we have
\begin{align}
\bra{\psi}\left(\mathbb I-K_G^{\bm m_{\rm sing}}\right)\ket{\psi}
&=\frac{1}{2|E|}\sum_{x\in V}\psi(x)\sum_{\{x,y\}\in E}\bigl[\psi(x)-\psi(y)\bigr] \nonumber\\
&=\frac{1}{2|E|}\sum_{\{x,y\}\in E}\bigl[\psi(x)-\psi(y)\bigr]^2 \nonumber\\
&=\frac{d_G^{\rm min}}{2|E|}.
\end{align}
Here, the second line follows from the identity $\psi(x)[\psi(x)-\psi(y)]+\psi(y)[\psi(y)-\psi(x)]=[\psi(x)-\psi(y)]^2$, while the third line follows from the fact that $\psi(x)=\psi(y)$ for any $x,y\neq x_{\rm min}$, whereas $\psi(x_{\rm min})-\psi(x)=1$ for any $x\neq x_{\rm min}$.

Furthermore, $\langle\Omega_{\bm m_{\rm sing}}|\psi\rangle=0$ and $\langle\psi|\psi\rangle=(N-1)/N$.
Therefore, the variational characterization of $\lambda_G^{\rm tr}$ gives
\begin{align}
    \lambda_G^{\rm tr}
    &\leq
    \frac{\bra{\psi}\left(\mathbb I-K_G^{\bm m_{\rm sing}}\right)\ket{\psi}}
    {\braket{\psi|\psi}}
    =\frac{d_G^{\rm min}}{2|E|}\frac{N}{N-1}
    \leq\frac{d_G^{\rm min}}{|E|},
\end{align}
where the final inequality follows from $N\geq 2$.
Hence, every incompatible block satisfies $\|K_G^{\bm m}\|_\infty\leq 1-\lambda_G^{\rm tr}$.
Together with the bound for the compatible blocks, this yields $\Delta_{\nu_G^{\mathrm{2rpmix}}}^{(1;\,\mathrm{2RPmix})}\geq\lambda_G^{\rm tr}$, which, combined with the opposite inequality shown above, concludes the proof.
\end{proof}

By further using the architecture-comparison arguments, for any constant-size matching family $\Gamma$ and any constant-depth fixed architecture $A$ whose associated graphs are connected, we have
\begin{align}
    &\Delta_{\nu_{\Gamma}}^{(1)}=\Theta\left(N\lambda_{G_{\Gamma}}^{\rm tr}\right), \label{eq:gap_1_design_par}\\
    &\Delta_{\nu_{A}}^{(1)}=\Theta\left(N\lambda_{G_{A}}^{\rm tr}\right).
\end{align}
The first relation also holds for the all-to-all parallel circuit $\Gamma_{\rm all}$, by combining Lemma~\ref{lem:all_to_all_matching_comparison} with the comparison to a constant-size matching family on a bounded-degree expander graph.

\section{Efficient construction of $U(1)$-symmetric unitary designs with asymmetric local gates} \label{s:efficient_const_asym}

In this section, we prove the efficient-construction result stated as Theorem~3 in the main text.
The basic idea is to combine a constant-gap auxiliary circuit based on the RPmix group with efficient local constructions of its mixing and random-phase components.
Although the local gates used to implement the random-phase component need not individually preserve the $U(1)$ symmetry, the resulting circuit unit is always $U(1)$-symmetric.

\subsection{Basic strategy for the construction}

In this subsection, we explain the basic strategy for constructing a $U(1)$-symmetric unitary design with asymmetric local gates.

\begin{theorem}
\label{restatethm:efficient_design_with_asymmetric_gates}
For $k\leq O(\log N/\log \log N)$, a constant-gap circuit unit can be implemented on an $\alpha$-dimensional lattice with fixed $\alpha$ with circuit depth
\begin{equation}
    d_{\alpha\mathrm D}^{\rm unit} =O(N^{1/\alpha}k^3 (\log k)^6 ).
\end{equation}
In the all-to-all interaction model, the required depth is
\begin{equation}
    d_{\rm all}^{\rm unit}
    =
    O\left((\log N)^3(\log\log N)^2 k\, (\log k)^2 \right).
\end{equation}
\end{theorem}
\noindent
Here, as in the main text, we call a circuit unit satisfying $\Delta_{\nu_{\rm unit}}^{(k)}\geq\Omega(1)$ a constant-gap unit.
We also recall that logarithmic factors $\log k$ appearing in asymptotic bounds are understood as $\max\{1,\log k\}$, to avoid the trivial issue at $k=1$.

The proof of this theorem consists essentially of two steps.
The first step is to establish a constant spectral gap for the doped RPmix circuit.
The doped RPmix circuit is obtained from the doped 2RPmix circuit by replacing the Haar-random unitaries drawn from the 2RPmix group with Haar-random unitaries drawn from the RPmix group.
Its spectral gap is defined by
\begin{equation}
    \Delta_{\mathrm{dprpmix}}^{(k)}
    \equiv
    1-
    \left\|
        P^{(k)}_{\mathrm{RPmix}}
        P^{(k)}_{\gamma_0}
        P^{(k)}_{\mathrm{RPmix}}
        -
        P^{(k)}_{U(1)\mathrm{Haar}}
    \right\|_{\infty},
\end{equation}
where
$\gamma_0\equiv\left\{\{1,2\},\{3,4\},\ldots,\{2\lfloor N/2\rfloor-1,2\lfloor N/2\rfloor\}\right\}$
is the fixed maximum matching containing $\lfloor N/2\rfloor$ edges.
In the lattice case, we label the sites along a nearest-neighbor Hamiltonian path so that every edge of $\gamma_0$ is geometrically local.

For the proof, we also use the auxiliary $V$-doped RPmix circuit, defined analogously to the $V$-doped 2RPmix circuit.
Its spectral gap is
\begin{equation}
    \Delta_{V\mathrm{dprpmix}}^{(k)}
    \equiv
    1-
    \left\|
        P_{\mathrm{RPmix}}^{(k)}
        M_{\mathrm{dope}}^{(k)}
        P_{\mathrm{RPmix}}^{(k)}
        -
        P_{U(1)\mathrm{Haar}}^{(k)}
    \right\|_\infty,
\end{equation}
where $M_{\mathrm{dope}}^{(k)}\equiv V^{\otimes k,k}$ for the Hadamard-type unitary $V$ introduced in Eq.~\eqref{eq:Hadamard_type_local_unitary_def}.
We then prove the following proposition.

\begin{proposition}[Doped RPmix circuit]
\label{prop:gap_LB_dp_RPmix_circuit}
There exist positive constants $C_3$ and $C_4$, independent of the system size $N$ and the moment order $k$, such that
\begin{equation}
    \Delta_{\mathrm{dprpmix}}^{(k)}
    \geq
    C_3,
\end{equation}
for $k\leq C_4 (\log N/\log \log N)$.
\end{proposition}
\noindent
The proof is analogous to that of Proposition~\ref{prop:gap_LB_dp_2RPmix_circuit} and is given in Sec.~\ref{ss:gap_lower_bound_dpRPmix}.

\vspace{1em}
The second step is to evaluate how efficiently the global RPmix group can be generated using local gates.
For this purpose, we generate the RP group and the mixing group separately.

To generate the mixing group on an $N$-qubit system, we use routing via matchings~\cite{alon1994routing}, in which each step consists of SWAP gates on mutually disjoint allowed edges.
These results imply the following lemma:
\begin{lemma}\label{lem:mixing_generation_via_routing}
On an $\alpha$-dimensional lattice with $\alpha=O(1)$ and $\Theta(N^{1/\alpha})$ sites along each spatial dimension, any site permutation $\pi \in S_N$ can be implemented exactly using nearest-neighbor SWAP gates with circuit depth
\begin{equation}
\label{eq:mix_routing_alphaD_lattice}
    d^{\rm mixunit}_{\alpha\mathrm D} = O(N^{1/\alpha}).
\end{equation}
In the all-to-all interaction model, any site permutation $\pi \in S_N$ can be implemented exactly using SWAP gates with circuit depth
\begin{equation}
\label{eq:mix_routing_all_to_all}
    d^{\rm mixunit}_{\rm all} \leq 2.
\end{equation}
\end{lemma}
\noindent
The lattice bound in Eq.~\eqref{eq:mix_routing_alphaD_lattice} follows from the Cartesian-product routing bound in Theorem~4 of Ref.~\cite{alon1994routing}; see also Lemma~24 of Ref.~\cite{folkertsma2026arts}. Equation~\eqref{eq:mix_routing_all_to_all} follows from Theorem~2 of Ref.~\cite{alon1994routing}.

By sampling $\pi$ uniformly from $S_N$ and implementing the corresponding site permutation using this lemma, we exactly sample the Haar measure on the mixing group within the depth bounds in Eqs.~\eqref{eq:mix_routing_alphaD_lattice} and \eqref{eq:mix_routing_all_to_all}, respectively.
The resulting moment operator is exactly $P_{\rm mix}^{(k)}$ for every $k$.
This routing-based implementation differs from a random walk on site permutations generated by repeated local random SWAPs.
In the former, we first sample the target permutation and then construct a coordinated sequence of SWAP gates tailored to it, whereas in the latter, local SWAPs are sampled according to a fixed stochastic update rule and the distribution over site permutations approaches uniformity through repeated updates.
On lattices, this distinction allows the routing construction to generate uniformly random site permutations in a depth parametrically smaller than the diffusive charge-transport timescale.
We therefore adopt the routing construction for the present purpose of generating symmetric unitary designs in shallow circuits.

The generation of the RP group requires a different construction and is discussed in detail in Sec.~\ref{subsec:generation_random_phase}.
For the present subsection, we only need the following consequence.

\begin{lemma}
\label{lem:RP_construction_with_local_gates}
For $ k\leq O(\log N/\log\log N)$ and any positive constant $\varepsilon$, one can construct a circuit unit whose overall unitaries belong to the RP group and whose moment operator satisfies
\begin{equation}
     \left\|
        M_{\nu^{\rm rpunit}}^{(k)}
        -
        P^{(k)}_{\rm RP}
     \right\|_{\infty}
     \leq
     \varepsilon,
\end{equation}
with the following circuit depths in an $\alpha$-dimensional lattice with fixed $\alpha$ and in the all-to-all interaction geometry, respectively:
\begin{align}
    d^{\rm rpunit}_{\alpha\mathrm D}
    &=
    O\left(N^{1/\alpha} k^3 (\log k)^6\right),
    \\
    d^{\rm rpunit}_{\rm all}
    &=
    O\left((\log N)^3(\log\log N)^2 k\,(\log k)^2\right).
\end{align}
\end{lemma}

By combining Proposition~\ref{prop:gap_LB_dp_RPmix_circuit} and Lemmas~\ref{lem:mixing_generation_via_routing} and~\ref{lem:RP_construction_with_local_gates}, we can now prove Theorem~\ref{restatethm:efficient_design_with_asymmetric_gates}.

\begin{proof}[Proof of Theorem~\ref{restatethm:efficient_design_with_asymmetric_gates}]

Proposition~\ref{prop:gap_LB_dp_RPmix_circuit} gives
\begin{equation}
    \left\|
        P_{\mathrm{RPmix}}^{(k)}
        P_{\gamma_0}^{(k)}
        P_{\mathrm{RPmix}}^{(k)}
        -
        P_{U(1)\mathrm{Haar}}^{(k)}
    \right\|_\infty
    \leq
    1-C_3.
\end{equation}
Using the projector relations
$P_{\mathrm{RPmix}}^{(k)}P_{U(1)\mathrm{Haar}}^{(k)}
=
P_{\gamma_0}^{(k)}P_{U(1)\mathrm{Haar}}^{(k)}
=
P_{U(1)\mathrm{Haar}}^{(k)}$,
we therefore obtain
\begin{equation}
    \left\|
        P_{\mathrm{RPmix}}^{(k)}
        P_{\gamma_0}^{(k)}
        -
        P_{U(1)\mathrm{Haar}}^{(k)}
    \right\|_\infty
    =
    \left\|
        P_{\mathrm{RPmix}}^{(k)}
        P_{\gamma_0}^{(k)}
        P_{\mathrm{RPmix}}^{(k)}
        -
        P_{U(1)\mathrm{Haar}}^{(k)}
    \right\|_\infty^{1/2}
    \leq
    \sqrt{1-C_3}
    \equiv
    1-c,
\end{equation}
where $c>0$ is independent of $N$ and $k$.

We next consider the construction of a circuit unit $\nu_{\rm unit}$ by independently sampling its RP, mixing, and Haar-gate components.
Its corresponding moment operator is given by
\begin{equation*}
    M_{\nu_{\rm unit}}^{(k)}
    \equiv
    M_{\nu^{\rm rpunit}}^{(k)}
    P_{\rm mix}^{(k)}
    P_{\gamma_0}^{(k)}.
\end{equation*}
Here, the Haar measure on the mixing group can be sampled exactly within the depth bounds given in Lemma~\ref{lem:mixing_generation_via_routing}.
Since the RP ensemble consists of $U(1)$-symmetric unitaries as complete operations, we have
\begin{equation*}
    M_{\nu^{\rm rpunit}}^{(k)}
    P_{U(1)\mathrm{Haar}}^{(k)}
    =
    P_{U(1)\mathrm{Haar}}^{(k)}.
\end{equation*}
Therefore, we can derive
\begin{align}
    \left\|
        M_{\nu_{\rm unit}}^{(k)}
        -
        P_{U(1)\mathrm{Haar}}^{(k)}
    \right\|_\infty
    &=
    \left\|
        M_{\nu^{\rm rpunit}}^{(k)}
        \left(
            P_{\rm mix}^{(k)}P_{\gamma_0}^{(k)}
            -
            P_{U(1)\mathrm{Haar}}^{(k)}
        \right)
    \right\|_\infty
    \nonumber\\
    &\leq
    \left\|P_{\rm RP}^{(k)} \left(P_{\rm mix}^{(k)}P_{\gamma_0}^{(k)}-P_{U(1)\mathrm{Haar}}^{(k)}\right) \right\|_\infty +
    \left\| \left(M_{\nu^{\rm rpunit}}^{(k)}-P_{\mathrm{RP}}^{(k)}\right) \left(P_{\rm mix}^{(k)}P_{\gamma_0}^{(k)}- P_{U(1)\mathrm{Haar}}^{(k)}\right) \right\|_\infty
    \nonumber\\
    &\leq 1-c+\left\| M_{\nu^{\rm rpunit}}^{(k)}-P_{\mathrm{RP}}^{(k)} \right\|_\infty.
\end{align}
Here, we used $P_{\rm RP}^{(k)}P_{\rm mix}^{(k)}=P_{\rm RPmix}^{(k)}$ and $\|P_{\rm mix}^{(k)}P_{\gamma_0}^{(k)}-P_{U(1)\mathrm{Haar}}^{(k)}\|_\infty\leq1$, since the latter operator is a product of the orthogonal projectors $P_{\rm mix}^{(k)}-P_{U(1)\mathrm{Haar}}^{(k)}$ and $P_{\gamma_0}^{(k)}-P_{U(1)\mathrm{Haar}}^{(k)}$.
Therefore, by taking $\nu^{\rm rpunit}$ so that $\left\| M_{\nu^{\rm rpunit}}^{(k)}-P_{\mathrm{RP}}^{(k)} \right\|_\infty \leq \frac{c}{2}$ is satisfied, we have
\begin{equation*}
    \Delta_{\nu_{\rm unit}}^{(k)}
    \geq
    \frac{c}{2}
    =
    \Omega(1),
\end{equation*}
and hence the constructed circuit unit has a constant spectral gap.

It remains to evaluate its circuit depth.
The circuit unit consists of the RP construction, the mixing construction, and one layer of $\lfloor N/2\rfloor$ independent $U(1)$-symmetric Haar-random two-qubit gates on $\gamma_0$.
By Lemma~\ref{lem:mixing_generation_via_routing}, the mixing part requires depth $O(N^{1/\alpha})$ on an $\alpha$-dimensional lattice and depth at most $2$ in the all-to-all interaction model.
By Lemma~\ref{lem:RP_construction_with_local_gates}, the RP part requires depth $O(N^{1/\alpha}k^3 (\log k)^6)$ on an $\alpha$-dimensional lattice and depth $O((\log N)^3(\log\log N)^2k\,(\log k)^2)$ in the all-to-all interaction model.

Therefore, for $2\leq k\leq O(\log N/\log\log N)$ and fixed $\alpha$, the mixing and Haar-gate depths are absorbed into the stated upper bounds for the RP part, giving
\begin{align}
    &d_{\alpha\mathrm D}^{\rm unit} =O(N^{1/\alpha}k^3 (\log k)^6 ), \\
    &d_{\rm all}^{\rm unit}=O\left((\log N)^3(\log\log N)^2k\,(\log k)^2\right),
\end{align}
which completes the proof.
\end{proof}

\subsection{Generation of the RP group}
\label{subsec:generation_random_phase}

In this subsection, we prove Lemma~\ref{lem:RP_construction_with_local_gates} by providing an explicit construction of a unitary ensemble whose $k$-th moment is close to that of the random phase unitary ensemble.
Let $\mathbb{B}_N\equiv\{0,1\}^N$ and $D\equiv|\mathbb{B}_N|=2^N$.
Recall that the random phase unitary ensemble is the ensemble of diagonal unitaries $V(\bm\phi)=\sum_{z\in\mathbb{B}_N}e^{i\phi_z}\ket{z}\bra{z}$, where the phases $\{\phi_z\}_{z\in\mathbb{B}_N}$ are independently and uniformly distributed over $[0,2\pi)$.
Its $k$-th moment operator is denoted by $P_{\rm RP}^{(k)}$.

We now construct an ensemble whose moment operator is close to $P_{\rm RP}^{(k)}$ using a sufficiently random permutation of the computational basis.
For $\bm\theta=(\theta_1,\ldots,\theta_N)$, define
\begin{equation}
    R(\bm\theta)\equiv\bigotimes_{j=1}^{N}\left(\ket{0}\bra{0}+e^{i\theta_j}\ket{1}\bra{1}\right),
\end{equation}
where $\theta_1,\ldots,\theta_N$ are independently and uniformly distributed over $[0,2\pi)$.
For a permutation $\zeta$ of $\mathbb{B}_N$, let $P_{\zeta}$ denote the corresponding permutation unitary, $P_{\zeta}\ket{z}=\ket{\zeta(z)}$.
Given a distribution $\eta$ over permutations of $\mathbb{B}_N$, we define $\nu_{\eta}^{\rm rp}$ as the ensemble of unitaries
\begin{equation}
\label{eq:RP_implement_conjugation}
    U(\zeta,\bm\theta)
    \equiv
    P_{\zeta^{-1}}R(\bm\theta)P_{\zeta},
    \qquad
    \zeta\sim\eta.
\end{equation}
In particular, each $U(\zeta,\bm\theta)$ is diagonal in the computational basis and satisfies $U(\zeta,\bm\theta)\ket{z}=e^{i\bm\theta\cdot\zeta(z)}\ket{z}$.
Notice that the phases $\theta_j$ are sampled independently for all $j\in[N]$.

To construct the circuit unit in Lemma~\ref{lem:RP_construction_with_local_gates}, we choose $\eta$ to be an approximately $2k$-wise independent permutation ensemble. The definition of $t$-wise independence is given as follows:

\begin{definition}[Exact and approximate $t$-wise independence]
Let $\eta$ be a distribution over permutations of $\mathbb{B}_N$.
For $t\leq D$, let $\mathbb{B}_N^{\underline{t}}$ denote the set of ordered $t$-tuples of distinct elements of $\mathbb{B}_N$, and let $\mathsf{U}_t$ denote the uniform distribution on $\mathbb{B}_N^{\underline{t}}$.
The distribution $\eta$ is \emph{exactly $t$-wise independent} if, for any distinct $z^1,\ldots,z^t\in\mathbb{B}_N$, the distribution of $(\zeta(z^1),\ldots,\zeta(z^t))$ for $\zeta\sim\eta$ coincides with $\mathsf{U}_t$:
\begin{equation}
    \Pr_{\zeta\sim\eta}\left[(\zeta(z^1),\ldots,\zeta(z^t))=\cdot\right]
    =
    \mathsf{U}_t.
\end{equation}
It is \emph{$\delta$-approximate $t$-wise independent} if, for any distinct $z^1,\ldots,z^t\in\mathbb{B}_N$, this distribution has total variation distance at most $\delta$ from $\mathsf{U}_t$:
\begin{equation}
    d_{\rm TV}\left(
        \Pr_{\zeta\sim\eta}\left[(\zeta(z^1),\ldots,\zeta(z^t))=\cdot\right],
        \mathsf{U}_t
    \right)
    \leq
    \delta,
\end{equation}
where $d_{\rm TV}(\mu,\nu)\equiv\frac{1}{2}\sum_{\omega}|\mu(\omega)-\nu(\omega)|$.
\end{definition}
\noindent
The corresponding conditions for any $s<t$ follow by marginalization.
The reason that we require $2k$-wise rather than $k$-wise independence below is that a basis element of the $k$-th moment space can involve up to $2k$ distinct computational-basis states.

Using this notion of $t$-wise independence, we obtain the following lemma.

\begin{lemma}
\label{lem:RP_ensemble_gap}
Assume $2k\leq 2^N$.
If $\eta_{\rm ex}$ is an exactly $2k$-wise independent permutation ensemble, then
\begin{equation}
    \left\|
        M_{\nu_{\eta_{\rm ex}}^{\rm rp}}^{(k)}
        -
        P_{\rm RP}^{(k)}
    \right\|_\infty
    \leq
    \frac{1}{2^N-2k+1}.
\end{equation}
More generally, if $\eta$ is a $\delta$-approximate $2k$-wise independent permutation ensemble, then
\begin{equation}
    \left\|
        M_{\nu_{\eta}^{\rm rp}}^{(k)}
        -
        P_{\rm RP}^{(k)}
    \right\|_\infty
    \leq
    \frac{1}{2^N-2k+1}
    +
    \delta.
\end{equation}
\end{lemma}

\begin{proof}
Every unitary in $\nu_{\eta}^{\rm rp}$ is diagonal in the computational basis.
For $\bm x=(x^1,\ldots,x^k)$ and $\bm y=(y^1,\ldots,y^k)$ with $x^j,y^j\in\mathbb{B}_N$, we write the corresponding basis state of the $k$-th moment space as
\begin{equation}
    \ket{\bm x,\bm y}
    \equiv
    \ket{x^1,\ldots,x^k,y^1,\ldots,y^k}.
\end{equation}
Using $U(\zeta,\bm\theta)\ket{z}=e^{i\bm\theta\cdot\zeta(z)}\ket{z}$, the moment operator can first be written as
\begin{equation}
    M_{\nu_{\eta}^{\rm rp}}^{(k)}
    =
    \sum_{\bm x,\bm y}
    \underset{\substack{
        \bm\theta\sim\mathrm{Unif}([0,2\pi)^N)\\
        \zeta\sim\eta
    }}{\mathbb{E}}
    \left[
        \exp\left(
            i\bm\theta\cdot
            \sum_{j=1}^k
            \bigl(\zeta(x^j)-\zeta(y^j)\bigr)
        \right)
    \right]
    \ket{\bm x,\bm y}\bra{\bm x,\bm y}.
\end{equation}

For each $z\in\mathbb{B}_N$, we now define the multiplicity imbalance
\begin{equation}
    q_z^{\bm x,\bm y}
    \equiv
    \#\{1\leq j\leq k:x^j=z\}
    -
    \#\{1\leq j\leq k:y^j=z\}.
\end{equation}
Thus $q^{\bm x,\bm y}=(q_z^{\bm x,\bm y})_{z\in\mathbb{B}_N}$ is an integer-valued vector satisfying $\sum_zq_z^{\bm x,\bm y}=0$ and $|\operatorname{supp}(q^{\bm x,\bm y})|\leq2k$.
By construction,
\begin{equation}
    \sum_{j=1}^k
    \bigl(\zeta(x^j)-\zeta(y^j)\bigr)
    =
    \sum_{z\in\mathbb{B}_N}
    q_z^{\bm x,\bm y}\zeta(z).
\end{equation}
Substituting this expression into the moment operator and averaging independently over $\theta_1,\ldots,\theta_N$, we obtain
\begin{align}
    M_{\nu_{\eta}^{\rm rp}}^{(k)}
    &=
    \sum_{\bm x,\bm y}
    \underset{\substack{
        \bm\theta\sim\mathrm{Unif}([0,2\pi)^N)\\
        \zeta\sim\eta
    }}{\mathbb{E}}
    \left[
        \exp\left(
            i\bm\theta\cdot
            \sum_{z\in\mathbb{B}_N}
            q_z^{\bm x,\bm y}\zeta(z)
        \right)
    \right]
    \ket{\bm x,\bm y}\bra{\bm x,\bm y}
    \nonumber\\
    &=
    \sum_{\bm x,\bm y}
    \Pr_{\zeta\sim\eta}
    \left[
        \sum_{z\in\mathbb{B}_N}
        q_z^{\bm x,\bm y}\zeta(z)
        =
        \bm0
    \right]
    \ket{\bm x,\bm y}\bra{\bm x,\bm y}.
\end{align}
Here the second equality follows from Fourier orthogonality of the independently uniform phases.

If $q^{\bm x,\bm y}=\bm0$, the multisets $\{x^1,\ldots,x^k\}$ and $\{y^1,\ldots,y^k\}$ coincide, and the corresponding eigenvalue is equal to one for every permutation ensemble $\eta$.
These basis states span precisely the fixed space of the full random phase moment operator $P_{\rm RP}^{(k)}$.
We therefore define
\begin{equation}
    \mathcal{Q}_k
    \equiv
    \left\{
        q^{\bm x,\bm y}:
        x^j,y^j\in\mathbb{B}_N\ \text{for all }j\in[k],\
        q^{\bm x,\bm y}\neq\bm0
    \right\},
\end{equation}
and obtain
\begin{equation}
    \left\|
        M_{\nu_{\eta}^{\rm rp}}^{(k)}
        -
        P_{\rm RP}^{(k)}
    \right\|_\infty
    =
    \max_{q\in\mathcal{Q}_k}
    \Pr_{\zeta\sim\eta}
    \left[
        \sum_{z\in\mathbb{B}_N}
        q_z\zeta(z)
        =
        \bm0
    \right].
\end{equation}

We first bound this value for the exactly $2k$-wise independent ensemble $\eta_{\rm ex}$.
Fix $q\in\mathcal{Q}_k$, let $s\equiv|\operatorname{supp}(q)|\leq2k$, and write $\operatorname{supp}(q)=\{z^1,\ldots,z^s\}$.
Since all $z^j$ are distinct, exact $2k$-wise independence implies that $(\zeta(z^1),\ldots,\zeta(z^s))$ is distributed according to $\mathsf{U}_s$.
Fix arbitrary distinct values $v^2,\ldots,v^s\in\mathbb{B}_N$ and condition on $\zeta(z^j)=v^j$ for $2\leq j\leq s$.
Under this conditioning, $\zeta(z^1)$ is uniformly distributed over the remaining $D-s+1$ elements of $\mathbb{B}_N$.
Since $q_{z^1}\neq0$, the zero-sum constraint uniquely determines the only possible value of $\zeta(z^1)$ as
\begin{equation}
    \zeta(z^1)
    =
    -
    \sum_{j=2}^s
    \frac{q_{z^j}}{q_{z^1}}v^j.
\end{equation}
If the right-hand side does not belong to $\mathbb{B}_N$, or coincides with one of $v^2,\ldots,v^s$, the conditional probability is zero.
Otherwise, the required value is one of the $D-s+1$ remaining possibilities, and hence
\begin{equation}
    \Pr_{\zeta\sim\eta_{\rm ex}}
    \left[
        \sum_{j=1}^s
        q_{z^j}\zeta(z^j)
        =
        \bm0
        \,\middle|\,
        \zeta(z^j)=v^j,\ 2\leq j\leq s
    \right]
    \leq
    \frac{1}{D-s+1}.
\end{equation}
Averaging over $v^2,\ldots,v^s$ and using $s\leq2k$, we obtain
\begin{align}
    \left\|
        M_{\nu_{\eta_{\rm ex}}^{\rm rp}}^{(k)}
        -
        P_{\rm RP}^{(k)}
    \right\|_\infty
    &=
    \max_{q\in\mathcal{Q}_k}
    \Pr_{\zeta\sim\eta_{\rm ex}}
    \left[
        \sum_{z\in\mathbb{B}_N}
        q_z\zeta(z)
        =
        \bm0
    \right]
    \nonumber\\
    &\leq
    \max_{q\in\mathcal{Q}_k}
    \frac{1}{D-|\operatorname{supp}(q)|+1}
    \nonumber\\
    &\leq
    \frac{1}{D-2k+1}
    =
    \frac{1}{2^N-2k+1}.
    \label{eq:deriv_PRP_exact_2k_wise_indep}
\end{align}

We next consider a $\delta$-approximate $2k$-wise independent ensemble $\eta$.
For a fixed $q\in\mathcal{Q}_k$ with $\operatorname{supp}(q)=\{z^1,\ldots,z^s\}$, let $\mu_{\eta,q}$ denote the distribution of $(\zeta(z^1),\ldots,\zeta(z^s))$ for $\zeta\sim\eta$ and define the event
\begin{equation}
    A_q
    \equiv
    \left\{
        (v^1,\ldots,v^s)\in\mathbb{B}_N^{\underline{s}}:
        \sum_{j=1}^s q_{z^j}v^j=\bm0
    \right\}.
\end{equation}
By $\delta$-approximate $2k$-wise independence, $d_{\rm TV}(\mu_{\eta,q},\mathsf{U}_s)\leq\delta$, and therefore
\begin{equation}
    \mu_{\eta,q}(A_q)
    \leq
    \mathsf{U}_s(A_q)+\delta.
\end{equation}
By the same argument as above, $\mathsf{U}_s(A_q)\leq1/(D-s+1)$.
We therefore obtain
\begin{equation}
    \Pr_{\zeta\sim\eta}
    \left[
        \sum_{z\in\mathbb{B}_N}
        q_z\zeta(z)
        =
        \bm0
    \right]
    \leq
    \frac{1}{D-s+1}
    +
    \delta
    \leq
    \frac{1}{D-2k+1}
    +
    \delta.
\end{equation}
Taking the maximum over $q\in\mathcal{Q}_k$ yields
\begin{equation}
    \left\|
        M_{\nu_{\eta}^{\rm rp}}^{(k)}
        -
        P_{\rm RP}^{(k)}
    \right\|_\infty
    \leq
    \frac{1}{2^N-2k+1}
    +
    \delta,
\end{equation}
which proves the second claim.
\end{proof}

Thus, generating the random phase ensemble up to the $k$-th moment reduces to generating an approximate $2k$-wise independent permutation of the computational basis.
In particular, for $2k\ll2^N$, exact $2k$-wise independence gives an error of order $O(2^{-N})$, whereas $\delta$-approximate $2k$-wise independence gives an error at most $\delta+O(2^{-N})$.

The remaining problem is how to construct such random permutations using local gates.
Efficient constructions of approximate $t$-wise independent permutation ensembles by reversible circuits have been studied in previous works \cite{brodsky2008simple,gretta2025more,gay2025pseudorandomness}.
We here state the results of these works.

We first consider geometrically local constructions.

\begin{lemma}[Construction on an $\alpha$-dimensional lattice]
\label{lem:geometric_approx_kwise_independence}
The constructions of Ref.~\cite{gay2025pseudorandomness} give $\delta$-approximate $t$-wise independent permutation ensembles using geometrically local reversible three-bit gates.
For sufficiently large $N$, the following depth and approximation-error bounds can be achieved:
\begin{align}
    d_{\rm 1D}
    &=
    O\left(Nt^2 (\log t)^5 \right),
    &
    \delta_{\rm 1D}
    &\leq
    2^{-tN},
    \\
    d_{\rm 2D}
    &=
    O\left(N^{1/2}t^3(\log t)^6\right),
    &
    \delta_{\rm 2D}
    &\leq
    2^{-tN^{1/2}},
    \\
    d_{\rm 3D}
    &=
    O\left(N^{1/3}t^3(\log t)^6\right),
    &
    \delta_{\rm 3D}
    &\leq
    2^{-N^{1/3}}.
\end{align}
The one-dimensional bound holds for $t\leq 2^N-2$, the two-dimensional bound for $t\leq 2^{cN^{1/2}}$, and the three-dimensional bound when $t\log t\leq cN^{1/3}$, for a sufficiently small constant $c>0$.
More generally, for $3\leq\alpha\leq c'\log N/\log\log N$, the same construction on an $\alpha$-dimensional lattice gives
\begin{equation}
    d_{\alpha\mathrm D}
    =
    \exp\left(O(\alpha)\right)
    N^{1/\alpha}t^3(\log t)^6, 
    \qquad
    \delta_{\alpha\mathrm D}
    \leq
    2^{-N^{1/\alpha}},
\end{equation}
provided that $t\log t\leq cN^{1/\alpha}$.
Here $c,c'>0$ are absolute constants.
Each unit of depth consists of a layer of mutually disjoint geometrically local three-bit gates.
\end{lemma}

The two-dimensional construction in Ref.~\cite{gay2025pseudorandomness} is obtained by alternating one-dimensional random circuits along different lattice directions, while the higher-dimensional construction is obtained recursively in the same manner.
Although these constructions are formulated in terms of three-bit reversible gates, this does not change the depth scaling if arbitrary two-qubit unitaries are allowed.
Indeed, every three-bit reversible gate, viewed as a three-qubit permutation unitary, can be decomposed exactly into a constant number of two-qubit unitaries acting on neighboring pairs within the three-site block \cite{barenco1995elementary}.
Thus, replacing each three-bit reversible gate by two-qubit unitaries introduces only a constant-factor depth overhead and no additional approximation error.
We emphasize that this statement uses general two-qubit unitaries rather than two-bit reversible gates.

We next consider the all-to-all construction studied in Ref.~\cite{gretta2025more}.
In this construction, each gate is a width-two reversible gate acting on three distinct wires $i,j_1,j_2$ as $x_i\mapsto x_i\oplus h(x_{j_1},x_{j_2})$, where $h:\{0,1\}^2\to\{0,1\}$ is a Boolean function.
At every step, the three wire indices and the corresponding width-two gate are sampled randomly.

\begin{lemma}[All-to-all construction]
\label{lem:all_to_all_approx_kwise_independence}
Assume $t\leq 2^{N/50}$ and $0<\delta\leq1/2$.
There exists a $\delta$-approximate $t$-wise independent permutation ensemble generated by an all-to-all reversible circuit in which a single random width-two three-bit gate is applied per layer, with depth
\begin{equation}
    d_{\rm all}^{\rm sing} =O\left( Nt\, \left[\log(Nt) \right]^3 \,[\log\log(Nt)]^2 (\log\, t)^2\log\frac{1}{\delta}\right).
\end{equation}
\end{lemma}

Lemma~\ref{lem:all_to_all_approx_kwise_independence} concerns the architecture in which only one gate is applied in each layer.
We now show that the same random circuits can be compressed by executing mutually disjoint gates in parallel.

\begin{lemma}[Parallelized all-to-all construction]
\label{lem:parallel_all_to_all_approx_kwise_independence}
Assume $2 \leq t\leq 2^{N/50}$ and $0<\delta\leq1/2$.
There exists a $\delta$-approximate $t$-wise independent permutation ensemble generated by layers of mutually disjoint all-to-all three-bit gates with depth
\begin{equation}
    d_{\rm all}^{\rm par}=O\left( t\, \left[\log(Nt) \right]^3 \,[\log\log(Nt)]^2 (\log\, t)^2\log\frac{1}{\delta}\right).
\end{equation}
\end{lemma}

\begin{proof}
We start from the sequential construction of Lemma~\ref{lem:all_to_all_approx_kwise_independence} with approximation error $\delta/2$.
Let $L$ denote the number of sequential gates in this construction, so that
\begin{equation}
\label{eq:L_upper_bound_all_rev}
    L\leq c \times \left( Nt\, \left[\log(Nt) \right]^3 \,[\log\log(Nt)]^2 (\log\, t)^2\log\frac{2}{\delta}\right) ,
\end{equation}
where $c>0$ is an absolute constant.
For a sampled circuit, we first compress all mutually compatible gates into parallel layers while preserving the implemented permutation exactly.
If the resulting parallel depth exceeds a threshold $d$, we discard that sample and replace the entire circuit by the identity.
We choose $d$ such that the probability of this exceptional event is at most $\delta/2$.
Under this choice, the replacement changes the distribution over permutations by total variation distance at most $\delta/2$.
Since the original sequential ensemble is $\delta/2$-approximate $t$-wise independent, the resulting depth-$d$ ensemble is therefore $\delta$-approximate $t$-wise independent.
It remains only to bound the probability that the sampled circuit cannot be compressed to depth $d$.

Let $T_j\subset[N]$ be the three-wire support of the $j$th gate, $1\leq j\leq L$.
In the random reversible-circuit construction of Ref.~\cite{gretta2025more}, the gate locations are sampled independently at every step, and hence $T_1,\ldots,T_L$ are independent uniformly random three-element subsets of $[N]$.

We assign the $j$th gate to the layer
\begin{equation}
    r(j)
    \equiv
    1+\max\{r(i):i<j,\ T_i\cap T_j\neq\varnothing\},
\end{equation}
where the maximum over the empty set is defined to be zero.
Two gates assigned to the same layer necessarily have disjoint supports.
Moreover, if $i<j$ and $T_i\cap T_j\neq\varnothing$, then $r(i)<r(j)$.
Therefore, executing the gates in increasing order of $r(j)$ preserves the relative order of every pair of overlapping gates.
Any pair whose relative order is changed has disjoint support and hence commutes.
Thus, the parallelized circuit implements exactly the same permutation as the original sequential circuit.

Let
\begin{equation}
    R
    \equiv
    \max_{1\leq j\leq L}r(j)
\end{equation}
denote the resulting parallel depth.
For two independently sampled supports,
\begin{equation}
    p_N
    \equiv
    \Pr[T_i\cap T_j\neq\varnothing]
    =
    1-
    \frac{\binom{N-3}{3}}{\binom{N}{3}}
    \leq
    \frac{9}{N}.
\end{equation}
If $R\geq r$, the recursive definition of the layers implies that there exist indices $1\leq i_1<\cdots<i_r\leq L$ satisfying
\begin{equation}
    T_{i_a}\cap T_{i_{a+1}}\neq \varnothing,
    \qquad
    1\leq a\leq r-1.
\end{equation}
For any fixed choice of $i_1,\ldots,i_r$, independence of the gate locations gives probability at most $p_N^{r-1}$ for this event.
A union bound therefore yields
\begin{equation}
    \Pr[R\geq r]
    \leq
    \binom{L}{r}
    \left(\frac{9}{N}\right)^{r-1}
    \leq
    \frac{N}{9}
    \left(
        \frac{9eL}{Nr}
    \right)^r.
\end{equation}

It follows that, for a sufficiently large absolute constant $C$,
\begin{equation}
    d
    =
    C\left(
        \frac{L}{N}
        +
        \log\frac{2N}{\delta}
    \right)
\end{equation}
satisfies $\Pr[R>d]\leq\delta/2$.
Indeed, if $r\geq18eL/N$, then $9eL/(Nr)\leq1/2$, and hence $\Pr[R\geq r]\leq(N/9)2^{-r}$.
Taking in addition $r\geq\log_2(2N/(9\delta))$ makes the right-hand side at most $\delta/2$.

Substituting the bound in Eq.~\eqref{eq:L_upper_bound_all_rev} into the expression for $d$, we have
\begin{align}
    d
    &\leq C \left( ct\, \left[\log(Nt) \right]^3 \,[\log\log(Nt)]^2 (\log\, t)^2\log\frac{2}{\delta} +\log\frac{2N}{\delta}\right) \nonumber \\
    &= O \left( t\, \left[\log(Nt) \right]^3 \,[\log\log(Nt)]^2 (\log\, t)^2\log\frac{1}{\delta} \right),
\end{align}
which concludes the proof.
\end{proof}

Combining these lemmas, we can now prove Lemma~\ref{lem:RP_construction_with_local_gates}.

\begin{proof}[Proof of Lemma~\ref{lem:RP_construction_with_local_gates}]
Fix any positive constant $\varepsilon$.
We choose $\delta \equiv \min\left\{\frac{\varepsilon}{2},\frac{1}{4}\right\}$. 
For sufficiently large $N$ and $k\leq O(\log N/\log\log N)$, we have $2k\leq2^N$ and $\frac{1}{2^N-2k+1} \leq \delta.$

We first consider an $\alpha$-dimensional lattice with a fixed $\alpha$.
Setting $t=2k$ in Lemma~\ref{lem:geometric_approx_kwise_independence}, its conditions are satisfied in the above range of $k$, and a $\delta$-approximate $2k$-wise independent permutation ensemble $\eta_{\alpha\mathrm D}$ can be generated with depth $O\left( N^{1/\alpha}k^3 (\log\,k)^6 \right)$. 
For $\alpha=1$, the stronger bound $O(Nk^2(\log k)^5)$ is available, but we use the uniform expression above for simplicity.

Using this permutation ensemble in the construction of Eq.~\eqref{eq:RP_implement_conjugation}, Lemma~\ref{lem:RP_ensemble_gap} gives
\begin{align}
    \left\|
        M_{\nu_{\eta_{\alpha\mathrm D}}^{\rm rp}}^{(k)}
        -
        P_{\rm RP}^{(k)}
    \right\|_\infty
    &\leq
    \frac{1}{2^N-2k+1}
    +
    \delta
    \nonumber\\
    &\leq
    2\delta
    \leq
    \varepsilon.
\end{align}
Since $P_{\zeta}$ is explicitly implemented by the reversible circuit, its inverse $P_{\zeta^{-1}}$ can be implemented by applying the inverse local gates in the reverse order, with the same circuit depth.
The intermediate random phase rotation $R(\bm\theta)$ can be implemented in a single layer.
Hence, the full unitary $U(\zeta,\bm\theta)=P_{\zeta^{-1}}R(\bm\theta)P_{\zeta}$ can be implemented with depth $O\left( N^{1/\alpha}k^3 (\log\,k)^6\right).$

We next consider the all-to-all interaction model.
Setting $t=2k$ in Lemma~\ref{lem:parallel_all_to_all_approx_kwise_independence}, its assumptions are again satisfied in the above range of $k$.
Since $\delta$ is a positive constant independent of $N$ and $k$, a $\delta$-approximate $2k$-wise independent permutation ensemble $\eta_{\rm all}$ can be generated with depth
\begin{equation*}
O \left( \, \left[\log(Nk) \right]^3 \,[\log\log(Nk)]^2 k(\log\, k)^2 \right) = O \left( \, \left(\log N \right)^3 \,(\log\log N)^2 k(\log\, k)^2 \right),
\end{equation*}
where this equality follows from the current regime $k\leq O(\log N/\log\log N)$. 
By using Lemma~\ref{lem:RP_ensemble_gap}, we can derive 
\begin{equation*}
    \left\|
        M_{\nu_{\eta_{\rm all}}^{\rm rp}}^{(k)}
        -
        P_{\rm RP}^{(k)}
    \right\|_\infty
    \leq
    \varepsilon.
\end{equation*}
As in the lattice case, implementing both $P_{\zeta}$ and $P_{\zeta^{-1}}$, together with the intermediate phase-rotation layer, changes the depth only by a constant factor.
Thus, the full RP circuit unit can be implemented with depth $O \left( \, \left(\log N \right)^3 \,(\log\log N)^2 k(\log\, k)^2 \right)$.
This completes the proof.
\end{proof}

\subsection{Constant spectral gap lower bound for doped RPmix circuit} \label{ss:gap_lower_bound_dpRPmix}

In this subsection, we give the proof of Proposition~\ref{prop:gap_LB_dp_RPmix_circuit}.
The proof of this proposition proceeds in much the same way as the proof of Proposition~\ref{prop:gap_LB_dp_2RPmix_circuit}, and many lemmas derived in Sec.~\ref{s:gap_doped_2RPmix_circuit} can be directly used here.
Since the 2RPmix group is a subgroup of the RPmix group, we have $\operatorname{Ran}P_{\mathrm{RPmix};\bm n,\overline{\bm n}}^{(k)}\subseteq\operatorname{Ran}P_{\mathrm{2RPmix};\bm n,\overline{\bm n}}^{(k)}$.
In particular, while the 2RPmix moment space allows arbitrary compatible classes $F\in\mathcal F_k(\bm n,\overline{\bm n})$, the RPmix moment space allows only permutation compatible classes $F_\sigma$ with $\sigma\in\Sigma(\bm n,\overline{\bm n})$.
For a charge sector satisfying $\Sigma(\bm n,\overline{\bm n})=\varnothing$, we have $P_{\mathrm{RPmix}}^{(k)}\Pi_{\bm n,\overline{\bm n}}=0$.

Below, we introduce the notions necessary for the analysis of the doped RPmix circuit.
First, we define the set of all type vectors whose corresponding orbit states span the RPmix subspace as
\begin{equation}
    \mathcal N_{\bm n,\overline{\bm n}}
    \equiv
    \left\{
        \bm m:
        \ket{\Omega_{\bm m}}
        \in
        \operatorname{Ran}
        P_{\mathrm{RPmix};\bm n,\overline{\bm n}}^{(k)}
    \right\}.
\end{equation}
For each $\bm m\in\mathcal N_{\bm n,\overline{\bm n}}$, define its compatible-class membership vector $\bm u(\bm m)$ by
\begin{equation}
    u_{F_\sigma}(\bm m)
    \equiv
    \begin{cases}
        1, & \operatorname{supp}(\bm m)\subseteq F_\sigma,\\
        0, & \operatorname{supp}(\bm m)\not\subseteq F_\sigma,
    \end{cases}
    \qquad
    \sigma\in\Sigma(\bm n,\overline{\bm n}),
\end{equation}
where the dimension of the vector $\bm u(\bm m)$ is $|\Sigma(\bm n,\overline{\bm n})|$.
We further denote the set of membership vectors that actually occur in this charge sector by
\begin{equation}
    \mathcal A_k(\bm n,\overline{\bm n})
    \equiv
    \left\{
        \bm u(\bm m):
        \bm m\in\mathcal N_{\bm n,\overline{\bm n}}
    \right\}.
\end{equation}
Then, for each $\bm u\in\mathcal A_k(\bm n,\overline{\bm n})$, we introduce the corresponding orthogonal projector as
\begin{equation}
    \Pi_{\bm u}
    \equiv
    \sum_{\substack{
        \bm m\in\mathcal N_{\bm n,\overline{\bm n}}\\
        \bm u(\bm m)=\bm u
    }}
    \ket{\Omega_{\bm m}}\bra{\Omega_{\bm m}}.
\end{equation}

Using these membership projectors, we obtain the disjoint decomposition of the RPmix moment space as
\begin{equation}
\label{eq:RPmix_moment_op_disjoint_decomp}
    P_{\mathrm{RPmix};\bm n,\overline{\bm n}}^{(k)}
    =
    \sum_{\bm u\in\mathcal A_k(\bm n,\overline{\bm n})}
    \Pi_{\bm u}.
\end{equation}
The projector onto a permutation compatible class $F_\sigma$, with $\sigma\in\Sigma(\bm n,\overline{\bm n})$, is also given by
\begin{equation}
    \Pi_{\bm n;\sigma}
    =
    \sum_{\substack{
        \bm u\in\mathcal A_k(\bm n,\overline{\bm n})\\
        u_{F_\sigma}=1
    }}
    \Pi_{\bm u}.
\end{equation}
Furthermore, we introduce the following maps that decompose an operator into diagonal and off-diagonal parts with respect to the membership sectors:
\begin{align}
    \mathfrak{G}_{\rm diag}(A)
    &\equiv
    \sum_{\bm u\in\mathcal A_k(\bm n,\overline{\bm n})}
    \Pi_{\bm u}A\Pi_{\bm u},
    \\
    \mathfrak{G}_{\rm off}(A)
    &\equiv
    P_{\mathrm{RPmix};\bm n,\overline{\bm n}}^{(k)}
    A
    P_{\mathrm{RPmix};\bm n,\overline{\bm n}}^{(k)}
    -
    \mathfrak{G}_{\rm diag}(A) =
    \sum_{\substack{
        \bm u,\bm u'\in\mathcal A_k(\bm n,\overline{\bm n})\\
        \bm u\neq\bm u'
    }}
    \Pi_{\bm u}A\Pi_{\bm u'}.
\end{align}
Since the projectors $\{\Pi_{\bm u}\}_{\bm u}$ are mutually orthogonal, we have
\begin{align}
    \left\|\mathfrak{G}_{\rm diag}(A)\right\|_\infty
    &=
    \max_{\bm u\in\mathcal A_k(\bm n,\overline{\bm n})}
    \left\|\Pi_{\bm u}A\Pi_{\bm u}\right\|_\infty
    \leq
    \|A\|_\infty,
    \nonumber\\
    \left\|\mathfrak{G}_{\rm off}(A)\right\|_\infty
    &\leq
    \left\|
        P_{\mathrm{RPmix};\bm n,\overline{\bm n}}^{(k)}
        A
        P_{\mathrm{RPmix};\bm n,\overline{\bm n}}^{(k)}
    \right\|_\infty
    +
    \left\|\mathfrak{G}_{\rm diag}(A)\right\|_\infty
    \leq
    2\|A\|_\infty.
    \label{eq:opineq_G_off}
\end{align}

By using these notions, we can derive Proposition~\ref{prop:gap_LB_dp_RPmix_circuit} as follows.

\begin{proof}[Proof of Proposition~\ref{prop:gap_LB_dp_RPmix_circuit}]

From the definition of the RPmix group, we only need to consider charge sectors satisfying $\Sigma(\bm n,\overline{\bm n})\neq\varnothing$.
Furthermore, after removing the trivial copies with $n_j\in\{0,N\}$ as discussed in Sec.~\ref{ss:dp2RP_setup_strategy}, we may assume $1\leq n_j\leq N-1$ for all $j\in[k]$.

In such a charge sector, we decompose the moment operators as
\begin{align}
    1-\Delta_{V\mathrm{dprpmix};\bm n,\overline{\bm n}}^{(k)}
    &\equiv
    \left\|
        P_{\mathrm{RPmix};\bm n,\overline{\bm n}}^{(k)}
        \left(
            M_{\mathrm{dope}}^{(k)}
            -
            P_{U(1)\mathrm{Haar}}^{(k)}
        \right)
        P_{\mathrm{RPmix};\bm n,\overline{\bm n}}^{(k)}
    \right\|_\infty
    \nonumber\\
    &=
    \left\|
        \sum_{\bm u,\bm u'\in\mathcal A_k(\bm n,\overline{\bm n})}
        \Pi_{\bm u}
        \left(
            M_{\mathrm{dope}}^{(k)}
            -
            P_{U(1)\mathrm{Haar}}^{(k)}
        \right)
        \Pi_{\bm u'}
    \right\|_\infty
    \nonumber\\
    &\leq
    \left\|
        \mathfrak{G}_{\rm diag}
        \left(
            M_{\mathrm{dope}}^{(k)}
            -
            P_{U(1)\mathrm{Haar}}^{(k)}
        \right)
    \right\|_\infty
    +
    \left\|
        \mathfrak{G}_{\rm off}
        \left(M_{\mathrm{dope}}^{(k)}\right)
    \right\|_\infty
    +
    \left\|
        \mathfrak{G}_{\rm off}
        \left(P_{U(1)\mathrm{Haar}}^{(k)}\right)
    \right\|_\infty.
    \label{eq:diag_offdiag_decomposition_Sigma_dprpmix}
\end{align}

First, we consider the diagonal contribution.
Since any $\bm u\in\mathcal A_k(\bm n,\overline{\bm n})$ satisfies $u_{F_\sigma}=1$ for some $\sigma\in\Sigma(\bm n,\overline{\bm n})$, we can upper bound the operator norm as
\begin{align}
    &\left\|
        \Pi_{\bm u}
        \left(
            M_{\mathrm{dope}}^{(k)}
            -
            P_{U(1)\mathrm{Haar}}^{(k)}
        \right)
        \Pi_{\bm u}
    \right\|_\infty
    \leq
    \left\|
        \Pi_{\bm n;\sigma}
        \left(
            M_{\mathrm{dope}}^{(k)}
            -
            P_{U(1)\mathrm{Haar}}^{(k)}
        \right)
        \Pi_{\bm n;\sigma}
    \right\|_\infty
    \nonumber\\
    &\leq
    \left\|
        \Pi_{\bm n;\sigma}
        \left(
            M_{\mathrm{dope}}^{(k)}
            -
            \ket{\bm n;\sigma}\bra{\bm n;\sigma}
        \right)
        \Pi_{\bm n;\sigma}
    \right\|_\infty
    +
    \left\|
        \Pi_{\bm n;\sigma}
        \left(
            P_{U(1)\mathrm{Haar}}^{(k)}
            -
            \ket{\bm n;\sigma}\bra{\bm n;\sigma}
        \right)
        \Pi_{\bm n;\sigma}
    \right\|_\infty
    \nonumber\\
    &\leq
    \frac{1}{2}
    +
    2\left[
        \exp\left(\frac{k(k-1)}{2N}\right)-1
    \right],
    \label{eq:diag_perm_bound_dprpmix}
\end{align}
where the first inequality follows from $0\leq\Pi_{\bm u}\leq\Pi_{\bm n;\sigma}$, and the final inequality follows from Lemmas~\ref{lem:approx_Haar_sector_sigma_n} and~\ref{lem:bound_each_block}$\mathrm{(i)}$.

We next upper bound the off-diagonal contribution.
For this purpose, we first derive the following element-wise inequality:
\begin{equation}
\label{eq:element_wise_inequality_dprpmix_circuit}
    \left|
        \bra{\Omega_{\bm m}}
        \mathfrak{G}_{\rm off}\left(M_{\mathrm{dope}}^{(k)}\right)
        \ket{\Omega_{\bm m'}}
    \right|
    \leq
    \left[
        \sum_{\substack{
            \{\sigma,\tau\}\subseteq\Sigma(\bm n,\overline{\bm n})\\
            \sigma\neq\tau
        }}
        T_{\rm off,abs}^{\mathrm{dp},F_\sigma F_\tau}
    \right]_{\bm m,\bm m'}.
\end{equation}
Here, each $T_{\rm off,abs}^{\mathrm{dp},F_\sigma F_\tau}$ is understood as its zero extension to $\operatorname{Ran}P_{\mathrm{RPmix};\bm n,\overline{\bm n}}^{(k)}$.
Thus, all these matrices are represented in the common orbit-state basis $\{\ket{\Omega_{\bm m}}:\bm m\in\mathcal N_{\bm n,\overline{\bm n}}\}$.

If $\bm u(\bm m)=\bm u(\bm m')$, the left-hand side vanishes, and the inequality immediately follows from the nonnegativity of the right-hand side.
If $\bm u(\bm m)\neq\bm u(\bm m')$, we have $\bra{\Omega_{\bm m}}\mathfrak{G}_{\rm off}\left(M_{\mathrm{dope}}^{(k)}\right)\ket{\Omega_{\bm m'}}=T^{\rm dp}_{\bm m,\bm m'}$.
In this case, there exists $\sigma\in\Sigma(\bm n,\overline{\bm n})$ such that either $u_{F_\sigma}(\bm m)=1$, $u_{F_\sigma}(\bm m')=0$ or $u_{F_\sigma}(\bm m)=0$, $u_{F_\sigma}(\bm m')=1$.
Without loss of generality, we assume $u_{F_\sigma}(\bm m)=1$ and $u_{F_\sigma}(\bm m')=0$.
Then, by taking any $\tau\in\Sigma(\bm n,\overline{\bm n})$ such that $u_{F_\tau}(\bm m')=1$, we have $\left[T_{\rm off,abs}^{\mathrm{dp},F_\sigma F_\tau}\right]_{\bm m,\bm m'}=\left|T^{\rm dp}_{\bm m,\bm m'}\right|$.
Therefore,
\begin{equation*}
    \left|
        \bra{\Omega_{\bm m}}
        \mathfrak{G}_{\rm off}\left(M_{\mathrm{dope}}^{(k)}\right)
        \ket{\Omega_{\bm m'}}
    \right|
    =
    \left|T^{\rm dp}_{\bm m,\bm m'}\right|
    =
    \left[T_{\rm off,abs}^{\mathrm{dp},F_\sigma F_\tau}\right]_{\bm m,\bm m'}
    \leq
    \left[
        \sum_{\substack{
            \{\sigma,\tau\}\subseteq\Sigma(\bm n,\overline{\bm n})\\
            \sigma\neq\tau
        }}
        T_{\rm off,abs}^{\mathrm{dp},F_\sigma F_\tau}
    \right]_{\bm m,\bm m'},
\end{equation*}
which shows Eq.~\eqref{eq:element_wise_inequality_dprpmix_circuit}.
By the same argument as for the doped 2RPmix circuit, this element-wise inequality gives
\begin{equation}
    \left\|
        \mathfrak{G}_{\rm off}\left(M_{\mathrm{dope}}^{(k)}\right)
    \right\|_\infty
    \leq
    \left\|
        \sum_{\substack{
            \{\sigma,\tau\}\subseteq\Sigma(\bm n,\overline{\bm n})\\
            \sigma\neq\tau
        }}
        T_{\rm off,abs}^{\mathrm{dp},F_\sigma F_\tau}
    \right\|_\infty.
    \label{eq:operator_norm_inequality_G_off_upper_bound_dprpmix}
\end{equation}
By using this inequality, we have
\begin{align}
    \left\|
        \mathfrak{G}_{\rm off}\left(M_{\mathrm{dope}}^{(k)}\right)
    \right\|_\infty
    &\leq
    \sum_{\substack{
        \{\sigma,\tau\}\subseteq\Sigma(\bm n,\overline{\bm n})\\
        \sigma\neq\tau
    }}
    \left\|
        T_{\rm off,abs}^{\mathrm{dp},F_\sigma F_\tau}
    \right\|_\infty
    \nonumber\\
    &\leq
    \frac{
        |\Sigma(\bm n,\overline{\bm n})|
        \bigl(|\Sigma(\bm n,\overline{\bm n})|-1\bigr)
    }{2}
    \cdot
    \frac{3\sqrt{2}}{\sqrt{N-1}}
    \nonumber\\
    &\leq
    \frac{3(k!)^2}{\sqrt{N-1}}
    \nonumber\\
    &\leq
    \exp\left[
        O(k\log k)-\frac{1}{2}\log(N-1)
    \right],
    \label{eq:offdiag_dope_final_bound_dprpmix}
\end{align}
where in the third inequality we used $|\Sigma(\bm n,\overline{\bm n})|\leq|S_k|=k!$.

Finally, we bound the off-diagonal contribution from the Haar moment operator.
We have
\begin{align}
    \left\|
        \mathfrak{G}_{\rm off}
        \left(P_{U(1)\mathrm{Haar}}^{(k)}\right)
    \right\|_\infty
    &\leq
    \left\|
        \mathfrak{G}_{\rm off}
        \left(
            P_{U(1)\mathrm{Haar}}^{(k)}
            -
            \sum_{\sigma\in\Sigma(\bm n,\overline{\bm n})}
            \ket{\bm n;\sigma}\bra{\bm n;\sigma}
        \right)
    \right\|_\infty+
    \sum_{\sigma\in\Sigma(\bm n,\overline{\bm n})}
    \left\|
        \mathfrak{G}_{\rm off}
        \left(
            \ket{\bm n;\sigma}\bra{\bm n;\sigma}
        \right)
    \right\|_\infty
    \nonumber\\
    &\leq
    2\left[
        \exp\left(\frac{k(k-1)}{2N}\right)-1
    \right]
    +
    \sum_{\sigma\in\Sigma(\bm n,\overline{\bm n})}
    \left\|
        \mathfrak{G}_{\rm off}
        \left(
            \ket{\bm n;\sigma}\bra{\bm n;\sigma}
        \right)
    \right\|_\infty,
    \label{eq:Haar_offdiag_reduce_rankone_dprpmix}
\end{align}
where the first inequality follows from the triangle inequality, and the second inequality follows from Lemma~\ref{lem:u1_haar_approximate_orthogonality} and Eq.~\eqref{eq:opineq_G_off}.

We now derive an operator-norm upper bound for $\mathfrak{G}_{\rm off}\left(\ket{\bm n;\sigma}\bra{\bm n;\sigma}\right)$.
For this purpose, consider the membership vector $\bm u$ satisfying $u_{F_\sigma}=1$ and $u_{F_\tau}=0$ for every $\tau\in\Sigma(\bm n,\overline{\bm n})$ with $\tau\neq\sigma$.
Such a membership vector indeed occurs in this charge sector. Since $1\leq n_j\leq N-1$ and $k<N$ in the regime considered here, we can choose mutually distinct bit strings $x^1,\ldots,x^k$ with $|x^j|=n_j$; setting $\bm y=\sigma\bm x$ then gives a type vector whose support is contained in $F_\sigma$ but in no $F_\tau$ with $\tau\neq\sigma$.
By using this $\bm u$, we decompose the state $\ket{\bm n;\sigma}$ as
\begin{equation}
    \ket{\bm n;\sigma}
    =
    \ket{\psi_\sigma}
    +
    \ket{\psi_{\overline{\sigma}}},
    \qquad
    \ket{\psi_\sigma}
    \equiv
    \Pi_{\bm u}\ket{\bm n;\sigma},
    \quad
    \ket{\psi_{\overline{\sigma}}}
    \equiv
    (\mathbb I-\Pi_{\bm u})\ket{\bm n;\sigma},
\end{equation}
and denote $q_\sigma\equiv\langle\psi_{\overline{\sigma}}|\psi_{\overline{\sigma}}\rangle$.
Then, from Lemma~\ref{lem:distinct_class_overlap_number_permutation}, we have
\begin{equation}
    q_\sigma
    \leq
    \sum_{\substack{
        \tau\in\Sigma(\bm n,\overline{\bm n})\\
        \tau\neq\sigma
    }}
    \left\|
        \Pi_{\bm n;\tau}
        \ket{\bm n;\sigma}
    \right\|^2
    \leq
    \frac{
        |\Sigma(\bm n,\overline{\bm n})|-1
    }{N}.
    \label{eq:psi_sigma_leakage_dprpmix}
\end{equation}
Furthermore, since
\begin{equation}
    \mathfrak{G}_{\rm off}
    \left(
        \ket{\bm n;\sigma}\bra{\bm n;\sigma}
    \right)
    =
    \ket{\psi_\sigma}\bra{\psi_{\overline{\sigma}}}
    +
    \ket{\psi_{\overline{\sigma}}}\bra{\psi_\sigma}
    +
    \mathfrak{G}_{\rm off}
    \left(
        \ket{\psi_{\overline{\sigma}}}
        \bra{\psi_{\overline{\sigma}}}
    \right),
\end{equation}
we can upper bound its operator norm as
\begin{align}
    \left\|
        \mathfrak{G}_{\rm off}
        \left(
            \ket{\bm n;\sigma}\bra{\bm n;\sigma}
        \right)
    \right\|_\infty
    &\leq
    \left\|
        \ket{\psi_\sigma}\bra{\psi_{\overline{\sigma}}}
    \right\|_\infty
    +
    \left\|
        \ket{\psi_{\overline{\sigma}}}\bra{\psi_\sigma}
    \right\|_\infty
    +
    \left\|
        \mathfrak{G}_{\rm off}
        \left(
            \ket{\psi_{\overline{\sigma}}}
            \bra{\psi_{\overline{\sigma}}}
        \right)
    \right\|_\infty
    \nonumber\\
    &\leq
    2\sqrt{q_\sigma}
    +
    2q_\sigma
    \nonumber\\
    &\leq
    4\sqrt{q_\sigma},
\end{align}
where we used Eq.~\eqref{eq:opineq_G_off} in the second inequality.

Therefore, using $|\Sigma(\bm n,\overline{\bm n})|\leq k!$, we obtain
\begin{align}
    \left\|
        \mathfrak{G}_{\rm off}
        \left(P_{U(1)\mathrm{Haar}}^{(k)}\right)
    \right\|_\infty
    &\leq
    2\left[
        \exp\left(\frac{k(k-1)}{2N}\right)-1
    \right]
    +
    4k!
    \sqrt{
        \frac{
            |\Sigma(\bm n,\overline{\bm n})|-1
        }{N}
    }
    \nonumber\\
    &\leq
    2\left[
        \exp\left(\frac{k(k-1)}{2N}\right)-1
    \right]
    +
    \frac{4k!\sqrt{k!}}{\sqrt N}
    \nonumber\\
    &\leq
    \exp\left[
        O(k\log k)-\frac{1}{2}\log N
    \right].
    \label{eq:offdiag_U(1)Haar_final_bound_dprpmix}
\end{align}

Combining Eqs.~\eqref{eq:diag_offdiag_decomposition_Sigma_dprpmix}, \eqref{eq:diag_perm_bound_dprpmix}, \eqref{eq:offdiag_dope_final_bound_dprpmix}, and~\eqref{eq:offdiag_U(1)Haar_final_bound_dprpmix}, we finally obtain
\begin{align}
    1-\Delta_{V\mathrm{dprpmix};\bm n,\overline{\bm n}}^{(k)}
    &\leq
    \frac{1}{2}
    +
    4\left[
        \exp\left(\frac{k(k-1)}{2N}\right)-1
    \right]
    +
    \frac{4k!\sqrt{k!}}{\sqrt N}
    +
    \frac{3(k!)^2}{\sqrt{N-1}}
    \nonumber\\
    &\leq
    \frac{1}{2}
    +
    \exp\left[
        O(k\log k)-\frac{1}{2}\log(N-1)
    \right].
    \label{eq:dprpmix_sector_final_bound}
\end{align}

Since the sectors satisfying $\Sigma(\bm n,\overline{\bm n})=\varnothing$ have vanishing RPmix moment space, and the moment operators are block diagonal with respect to the charge sectors, we obtain
\begin{equation}
    1-\Delta_{V\mathrm{dprpmix}}^{(k)}
    =
    \max_{\bm n,\overline{\bm n}}
    \left(
        1-\Delta_{V\mathrm{dprpmix};\bm n,\overline{\bm n}}^{(k)}
    \right)
    \leq
    \frac{1}{2}
    +
    \exp\left[O(k\log k)-\frac{1}{2}\log(N-1)\right].
\end{equation}
Hence, $\Delta_{V\mathrm{dprpmix}}^{(k)}$ is bounded below by a positive constant independent of $N$ and $k$, for $k\leq O(\log N/\log \log N)$.
Finally, by an argument analogous to that in Lemma~\ref{lem:gap_Vdp2RPmix_circuit_to_dp2RPmix_circuit}, we can obtain
\begin{equation}
    \Delta_{\mathrm{dprpmix}}^{(k)} \geq  \frac{1}{2}\Delta_{V\mathrm{dprpmix}}^{(k)} \geq \frac{1}{4} - \exp\left[O(k\log k)-\frac{1}{2}\log(N-1)\right],
\end{equation}
which completes the proof.

\end{proof}

\section{Time evolution of subsystem purity in the single-edge all-to-all random circuit} \label{s:purity_all}

In this section, we derive the analytical representation of the subsystem purity dynamics given in Eq.~(28) of the main text.
We consider $N\ge4$, the single-edge all-to-all ensemble $\nu\equiv\nu_{G_{\mathrm{all}}}$, and the charge sector $(\bm n,\overline{\bm n})$ with $\bm n=\overline{\bm n}=(2,2)^T$.
Since this ensemble is invariant under any permutation of sites, $[M_{\nu}^{(2)},U_{\pi}^{\otimes 2,2}]=0$ holds for every $\pi\in S_N$.
Moreover, defining $P_{\mathrm{mix};\bm n,\overline{\bm n}}^{(2)}\equiv\Pi_{\bm n,\overline{\bm n}}P_{\mathrm{mix}}^{(2)}$, we have $P_{\mathrm{mix};\bm n,\overline{\bm n}}^{(2)} \ket{\operatorname{unif}_2}=\ket{\operatorname{unif}_2}$, where $\ket{\operatorname{unif}_2} \equiv \frac{2}{N(N-1)}\sum_{1\leq i<j\leq N} \ket{i,j}^{\otimes 4}$.
Therefore, the purity dynamics can be represented as
\begin{equation}
\label{eq:purity_mixing_restriction}
    \bar p_A(t)=\bra{F_{\!A}}M_{\nu^{*t}}^{(2)}\ket{\operatorname{unif}_2}=\bra{F_{\!A}}\left(P_{\mathrm{mix};\bm n,\overline{\bm n}}^{(2)}M_{\nu}^{(2)}P_{\mathrm{mix};\bm n,\overline{\bm n}}^{(2)}\right)^t\ket{\operatorname{unif}_2}.
\end{equation}

Since $\ket{\operatorname{unif}_2}$ has equal total ket and bra occupations at each site, and this property is preserved by $M_{\nu}^{(2)}$, the evolution in Eq.~\eqref{eq:purity_mixing_restriction} is confined to the six-dimensional subspace spanned by the orbit states corresponding to the following type vectors:
\begin{align*}
    \bm m^1:&\quad m^1_{0000}=N-2,\quad m^1_{1111}=2,\\
    \bm m^2:&\quad m^2_{0000}=N-3,\quad m^2_{1111}=m^2_{1010}=m^2_{0101}=1,\\
    \bm m^3:&\quad m^3_{0000}=N-3,\quad m^3_{1111}=m^3_{1001}=m^3_{0110}=1,\\
    \bm m^4:&\quad m^4_{0000}=N-4,\quad m^4_{1010}=m^4_{0101}=2,\\
    \bm m^5:&\quad m^5_{0000}=N-4,\quad m^5_{1010}=m^5_{0101}=m^5_{1001}=m^5_{0110}=1,\\
    \bm m^6:&\quad m^6_{0000}=N-4,\quad m^6_{1001}=m^6_{0110}=2.
\end{align*}
Here, the orbit states $\ket{\Omega_{\bm m^i}}$ and their sizes $D_{\bm m^i}$ are defined as in Sec.~\ref{ss:mixing}, and $\ket{\operatorname{unif}_2}=\binom{N}{2}^{-1/2}\ket{\Omega_{\bm m^1}}$ holds.
The restriction of $P_{U(1)\mathrm{Haar};\bm n,\overline{\bm n}}^{(2)}$ to this subspace has rank two, and its range is spanned by the normalized number-permutation states
\begin{equation*}
    \ket{\bm n;\mathrm{id}}=\alpha\ket{\Omega_{\bm m^1}}+\beta\ket{\Omega_{\bm m^2}}+\gamma\ket{\Omega_{\bm m^4}},\qquad \ket{\bm n;\mathrm{sw}}=\alpha\ket{\Omega_{\bm m^1}}+\beta\ket{\Omega_{\bm m^3}}+\gamma\ket{\Omega_{\bm m^6}},
\end{equation*}
where the coefficients are given by
\begin{equation*}
    \alpha=\sqrt{2/[N(N-1)]},\quad  \beta= \sqrt{4(N-2)/[N(N-1)]}, \quad \gamma= \sqrt{(N-2)(N-3)/[N(N-1)]}.
\end{equation*}

For the analysis of this restricted moment operator, we introduce a $6\times6$ matrix $\widetilde H$ with elements
\begin{equation}
\label{eq:purity_H_definition}
\widetilde H_{ii'}\equiv N(N-1)\sqrt{\frac{D_{\bm m^i}}{D_{\bm m^{i'}}}}\times\bra{\Omega_{\bm m^i}}\left(\mathbb I-M_{\nu}^{(2)}\right)\ket{\Omega_{\bm m^{i'}}}.
\end{equation}
Counting the contributions of the two-site Haar projectors gives $\widetilde H=NH_1+H_0$, where
\begingroup
\small
\setlength{\arraycolsep}{3pt}
\begin{align}
\label{eq:purity_H_matrices}
    H_1=\frac{2}{3}
    \begin{pmatrix}
        2&0&0&0&0&0\\
        -2&1&0&0&0&0\\
        -2&0&1&0&0&0\\
        0&-1&0&0&0&0\\
        0&-1&-1&0&0&0\\
        0&0&-1&0&0&0
    \end{pmatrix}, \qquad
    H_0=\frac{2}{3}
    \begin{pmatrix}
        -4&-1&-1&0&0&0\\
        4&-2&1&-4&-1&0\\
        4&1&-2&0&-1&-4\\
        0&3&0&4&1&0\\
        0&3&3&4&14&4\\
        0&0&3&0&1&4
    \end{pmatrix}.
\end{align}
\endgroup
Although $\widetilde H$ is not Hermitian, it is similar to a Hermitian matrix and hence has real eigenvalues.
Writing $\bm e_i$ for the standard basis vectors of $\mathbb R^6$, the subsystem purity is then exactly given by
\begin{equation}
\label{eq:purity_matrix_evolution}
    \bar p_A(t)=\bm w^T\left(\mathbb I_6-\frac{\widetilde H}{N(N-1)}\right)^t\bm e_1,
\end{equation}
where $\bm w=(w_1,\ldots,w_6)^T$ is defined by $w_i\equiv\langle F_{\!A}|\Omega_{\bm m^i}\rangle/\sqrt{D_{\bm m^i}}$.

Therefore, the purity dynamics can be obtained by diagonalizing $\widetilde H$.
A direct calculation gives its six eigenvalues as
\begin{equation*}
    0,\quad 0,\quad \frac{2(N+1)}{3},\quad \frac{2z_{\mathrm{slow}}}{3},\quad \frac{2z_1}{3},\quad \frac{2z_2}{3},
\end{equation*}
where $z_{\mathrm{slow}}<z_1<z_2$ are the three roots of the cubic polynomial $f(z)\equiv z^3-(3N+13)z^2+(2N^2+38N-12)z-24N^2+24N-48$.
The two zero eigenvalues correspond to the Haar-invariant number-permutation states $\ket{\bm n;\mathrm{id}}$ and $\ket{\bm n;\mathrm{sw}}$.
A left eigenvector corresponding to the eigenvalue $2(N+1)/3$ is given by $\bm v_{\mathrm{anti}}^{\mathrm L}=(0,N-3,3-N,-4,0,4)^T$, which satisfies $(\bm v_{\mathrm{anti}}^{\mathrm L})^T\bm e_1=0$.
Therefore, this mode does not contribute to the purity dynamics, and the exact representation of $\bar p_A(t)$ is given by
\begin{equation}
\label{eq:purity_spectral_decomposition}
    \bar p_A(t)=\bar p_A^{\mathrm{Haar}}+C_{\mathrm{slow}}e^{-\lambda_{\mathrm{slow}}t}+\sum_{i=1}^{2}C_i e^{-\lambda_i t},
\end{equation}
where $\bar p_A^{\mathrm{Haar}}\equiv\bra{F_{\!A}}P_{U(1)\mathrm{Haar}}^{(2)}\ket{\operatorname{unif}_2}$.
Here, the decay rates are determined by the roots of $f(z)$ as $\lambda_j\equiv-\log[1-2z_j/(3N(N-1))]$ for $j\in\{\mathrm{slow},1,2\}$, while the coefficients $C_j$ are determined by the corresponding left and right eigenvectors of $\widetilde H$ through Eq.~\eqref{eq:purity_matrix_evolution}.

Expanding the roots of $f(z)$ and the corresponding eigenvector overlaps in powers of $N^{-1}$, we obtain the asymptotic forms of these quantities as
\begin{gather*}
    \lambda_{\mathrm{slow}}=\frac{8}{N^2}+O(N^{-3}), \qquad \lambda_1=\frac{2}{3N}+O(N^{-2}), \qquad \lambda_2=\frac{4}{3N}+O(N^{-2}), \\
    C_{\mathrm{slow}}=\frac18+O(N^{-1}), \qquad  C_1=\frac12+O(N^{-1}), \qquad  C_2=\frac14+O(N^{-1}).
\end{gather*}
The Haar-averaged purity can also be calculated exactly for even $N$, giving
\begin{equation*}
    \bar p_A^{\mathrm{Haar}}=\frac{N^3+4N^2+12N-16}{8(N-1)(N^2-N+2)}=\frac18+O(N^{-1}).
\end{equation*}

\bibliography{biblio}

\clearpage

\appendix